\documentclass[11pt]{article}
 \usepackage[margin=0.9in]{geometry}
\usepackage{jheppub}
\usepackage[utf8]{inputenc}
\usepackage{graphicx,bm,color}
\usepackage{slashed,verbatim}
\usepackage[english]{babel}
\usepackage{amsmath}
\usepackage{graphicx}
\usepackage[margin=0.9in]{geometry}
 \usepackage[colorinlistoftodos]{todonotes}
\usepackage{amssymb,graphicx,epstopdf}
 \usepackage{simplewick}
\usepackage[colorlinks=true,linktocpage=true,linkcolor=blue,citecolor=blue]{hyperref}
\usepackage{caption}
\usepackage{xcolor,mathtools}

\usepackage{bbm}
\usepackage{hyperref}

\def\sumintb{\sum\!\!\!\!\!\!\!\!\!\int\limits}
\def\nn{\nonumber\\}

\def\be {\begin{equation}}
\def\ee {\end{equation}}
\def\bea {\begin{eqnarray}}
\def\eea {\end{eqnarray}}
\def\bc {\begin{center}}
\def\ec {\end{center}}
\def\ben {\begin{enumerate}}
\def\een {\end{enumerate}}
\def\mn {\mu\nu}
\def\sp{\shortparallel}
\def\om {\omega}
\def\ti {\tilde}
\def\nn {\nonumber}

\def\eps {\epsilon}

\def\dbo{\Bigg[\!\!\Bigg[}
\def\dbc{\Bigg]\!\!\Bigg]}
\def\hmu{\hat\mu}
\def\L{\ln\frac{\hat\Lambda}{2}}
\def\Lg{\ln\frac{\hat\Lambda_g}{2}}
\def\Za{\frac{\zeta'(-1)}{\zeta(-1)}}
\def\Zc{\frac{\zeta'(-3)}{\zeta(-3)}}

\def\mn {\mu\nu}
\def\mr {\mu\rho}

\def\om {\omega}

\def\sp{\shortparallel}
\def\Tr {\mathsf{Tr}}
\def\ranglec{\rangle_{\!\!c}}
\def\rangleci{\rangle_{\!\!c_1,c_2}}
\def\lrarrow{\leftrightarrow}

\def\lb{\left(}
\def\rb{\right)}

\title{An Introduction to Thermal Field Theory and Some of its Application}
\author{Munshi G. Mustafa}
\affiliation{Theory Division, Saha Institute of Nuclear Physics, Homi Bhabha National Institute, 1/AF Bidhan Nagar,
Kolkata - 700064, India}
\emailAdd{munshigolam.mustafa@saha.ac.in}
\date{\today}

\begin{abstract}{
In this article  an introduction to the thermal field theory within imaginary time
vis-a-vis Matsubara formalism has been discussed in details. The imaginary time formalism has been introduced through both the 
operatorial and  the functional integration
method. The prescription to perform frequency sum for boson and fermion has been discussed in details. Green's
function both in Minkowski time as well as in Euclidean time has been derived.
The tadpole diagram in $\lambda \phi^4$ theory and the self-energy in $\lambda \phi^3$ theory have been
computed and their consequences have also been discussed.
The basic features  of general two point functions, such as self-energy and propagator, for both fermions and bosons in presence of a heat bath have
been discussed.  The imaginary time has also been introduced from the  relation between the functional integral and the partition function.
Then the free partition functions and thermodynamic quantities for scalar, fermion and gauge field, and interacting scalar field 
have been obtained from first principle calculation.  
 The quantum electrodynamics (QED) and gauge fixing have been discussed in details.
The one-loop self-energy for electron and photon in QED have been obtained in 
hard thermal loop (HTL) approximation. The dispersion properties and collective excitations of both electron and photon in
a material medium in presence of a heat bath have been presented. The spectral representation  of fermion and gauge boson propagators have been obtained.
In HTL approximation,  the generalisation of QED results of two point functions to quantum chromodynamics (QCD) have been outlined 
that mostly involve group theoretical factors.
Therefore, one learns about the collective excitations in a QCD plasma from the acquired knowledge of QED plasma excitations.
Then, some subtleties of finite temperature field theory have been outlined.  As an effective field theory approach  the HTL resummation  and 
the HTL perturbation theory (HTLpt) have been introduced. The leading order(LO), next-to-leading order (NLO) and 
next-to-next-leading order (NNLO) free energy and pressure for deconfined QCD medium 
created in heavy-ion collisions  have been computed within HTLpt. 
The general features of the deconfined QCD medium have also been outlined with non-perturbative effects like gluon condensate and Gribov-Zwanziger action.
The dilepton production rates from quark-gluon plasma with these non-perturbative effects have been computed and discussed in details.}
\end{abstract}

\begin{document}
\maketitle

\section{Introduction}

\begin{figure}[htb]
\begin{center}
\includegraphics[width=4cm,height=4cm]{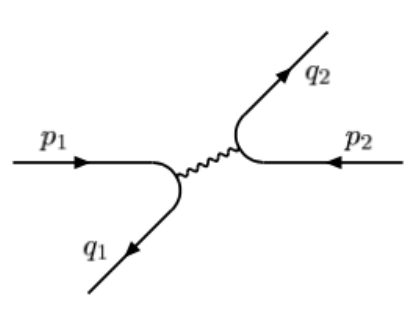}
\caption{Simple $2\rightarrow 2$ process at zero temperature.}
\label{simp_pro}
\end{center}
\end{figure}

The conventional quantum field theory is formalized at zero temperature. This is a  framework to describe a wide class of phenomena in particle physics in the energy 
range covered by all experiments, i.e., a tool to deal with  complicated many body problems or interacting system. The theoretical predictions under this framework, 
for example the cross sections of particle collisions in an accelerator, are extremely good to describe experimental data. With some modifications, it also plays a crucial 
role in atomic, nuclear and condensed matter physics. However, our real world is certainly of non-zero temperature. It is natural to wonder when and to what extent effects 
arising due to non-zero temperature are relevant, and what new phenomena could arise due to a thermal background. To understand these,  one needs a prescription of 
quantum field theory in thermal background and the general context of thermal field theory can be illustrated as below:
 
 In Fig.~\ref{simp_pro}, the simple two body process is displayed at zero temperature and it can be characterized by an observable as
 \be
 {\mathcal O} = \langle q_1q_2|p_1p_2\rangle . \label{qft1}
 \ee
 
 \begin{figure}[htb]
\begin{center}
\includegraphics[width=4cm,height=4cm]{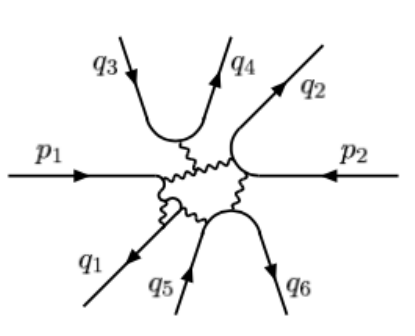}
\caption{Complicated many body process at zero temperature.}
\label{comp_pro}
\end{center}
\end{figure}

In Fig.~\ref{comp_pro}, the complicated many body process is displayed at zero temperature and it  can be characterized by
 \be
 {\mathcal O} = \langle q_1q_2\cdots q_6|p_1p_2\rangle . \label{qft2}
 \ee

 \begin{figure}[h]
\begin{center}
\includegraphics[width=4cm,height=4cm]{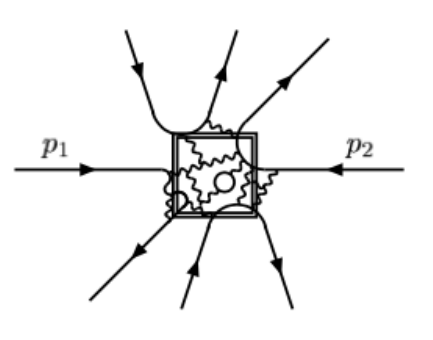}
\caption{Complicated many body process at non-zero temperature.}
\label{comp_tft}
\end{center}
\end{figure}

Now, in Fig.~\ref{comp_tft}, the complicated many body process is displayed at non-zero temperature. However, the $\Large{ \mathbf \Box}$ in Fig.~\ref{comp_tft} is simple
because the ergodic system may thermalize and their average properties can then be characterized just by thermal fluctuations in presence of temperature $T$ 
and chemical potential $\mu$ as
 \be
\langle {\mathcal O}\rangle_T ={\mathcal Z}^{-1} \textrm{Tr} \left [ e^{-\beta (H-\mu \hat N)} {\mathcal O}\right ]\, , \label{qft3}
 \ee
where $\beta=1/T$ and ${\mathcal Z}$ is the partition function,  $H$ is the Hamiltonian and $\hat N$ is the conserved number in the system. Angular braces 
$\langle \cdots \rangle_T$ indicate thermal average. The \eqref{qft3} means that though a many body scattering process is complicated but can be
addressed only through the thermal averaged properties observed over a long period of time. This indicates that the dynamics has to be ergodic that allows a 
thermodynamic treatment.
So, this brings the well-established realm of statistical mechanics and the problem becomes manageable such that each observable can be expressed in terms of $T$ and $\mu$.
Thus, the thermal field theory~ \cite{Matsubara,Schwinger, Schwinger1,Keldysh,Umezawa,Kubo,Fetter,Abrikosov,Dolan,Weinberg,Gross,Landsman,Ashok_Das,Kapusta,Le_Bellac}  
 is a combination of quantum field theory  and statistical mechanics that provides a tool to deal with complicated many body problems with interactions among its components 
 at finite $T$ and $\mu$.  Studies of physical systems at finite temperature have led, in the past, to many interesting properties such as phase transitions, blackbody radiation etc. However, the study of complicated quantum mechanical systems at finite temperature has had a systematic development only in the past few decades. There are now well developed and well understood formalisms to describe finite temperature field theories. There are now {\sf three} well defined formalisms:

\begin{enumerate}

\item Imaginary time (Matsubara) formalism~\cite{Matsubara}.

\item Real time  (Schwinger-Keldysh) formalism~\cite{Schwinger,Schwinger1,Keldysh}.

\item Thermo field dynamics  (Umezawa) formalism~\cite{Umezawa}.
\end{enumerate}
In this article, we will review the imaginary time formalism in thermal field theory and its applications.

\subsection{Need for Thermal Field Theory}

The goal of thermal field theory is to describe a large ensemble of multiple interacting particles (including gauge interactions) 
in thermal environment. It also describes creation and annihilation of new processes which were not present in vacuum field theory. 
It has been used to study questions such as phase transitions involving symmetry restoration in theories with spontaneously broken 
symmetry~\cite{Dolan,Weinberg,Kirzhnits}. One can also study the evolution of the universe at early times and cosmology~\cite{Laine, Turner,Kuzmin,Shapo,Weinberg1,Baier} 
which clearly is a system at high temperature. The finite temperature field theory has widely been applied to 
thermal neutrino production~\cite{Ghiglieri,Jackson,Ghisoiu,Salvio,Laine1,Biondini}, neutrino oscillations~\cite{GM},
leptogenesis~\cite{Biondini1,Drewes, Pradler,GM1}, ${\cal N}=4$ supersymmetric Yang-Mills theory~\cite{Solana,Hout,Hout1,Chesler,Policastro}, 
 string theory and Anti de-sitter space/Conformal Field Theory (Ads/CFT) correspondence~\cite{Maldacena,Witten,Gubser,Buchel},
 blackhole physics~\cite{Kovtun}, thermal axion production~\cite{Graf,Graf1,Salvio1}, thermal graviton production~\cite{Pradler1,Rychkov} 
 and in gravitational waves~\cite{,Ghilieri1}.  It has also been applied to  condensed matter physics~\cite{Kubo,Fetter,Abrikosov,Foster,Negele,Callen,Mahan}.

It has also been used to high energy nuclear and particle physics to describe the many body system. One of such current field of interest is
Quark-Gluon Plasma (QGP), which is a complicated many body system~\cite{Muller,Harris,Hwa,Wong} produced in high energy heavy-ion collisions at 
Relativistic Heavy-Ion Collider (RHIC) at  Brookhaven National Laboratory (BNL)  and Large Hadron Collider (LHC) at the European Organization for Nuclear Research (CERN)  
and likely to be produced at future Facility for Antiproton and Ion Research (FAIR) experiment in GSI and Nuclotron-based Ion Collider fAcility (NICA) experiment in DUBNA containing quarks (fermions) and gluons (gauge particles) that involve high temperature and/or density. 
QGP is a thermalized state of matter in which (quasi)free quarks and gluons are deconfined 
from hadrons, so that the color degrees of freedom become manifest over a large volume, than merely  in a hadronic
volume.  QGP  expands, cools,  hadronizes and hadrons reach to the detector and one needs to have unambiguous signatures to discover QGP. But, unfortunately 
 most of signals are circumstantial. Thus, in order to understand the properties of a QGP and to make unambiguous  predictions about signature of QGP formation, 
 one needs a profound description of QGP. For this purpose we have to use Quantum Chromodynamics (QCD) at finite temperature and chemical potential~\cite{Gross,Kapusta,Le_Bellac,Mike,Laine2,Blaizot,Kraemmer,Thoma,Thoma1,Schmitt,Yang,Inga}.

\subsection{Notations and Convention}

In this article,  the following notations and conventions will be followed:
\begin{enumerate}
\item[$\bullet$]  The metric in Minkowski space-time $g_{\mu\nu} ={\rm{diag}}(1,-1,-1,-1)$.
\item[$\bullet$] Will be using $\hbar=c=k_B=1$ unless  mentioned otherwise therein. So, temperature has dimension
of mass $[M]$ whereas that for 
length and time  is $[M^{-1}]$.
\item[$\bullet$] $I_m$ represents  $m \times m$ unit matrix.
\item[$\bullet$] Greek indices are used for four-vectors in space-time.
\item[$\bullet$] Einstein summation convention: repeated indices are summed over unless stated otherwise;
$A^\mu B_\mu=A_0B_0-{\mathbf A}\cdot {\mathbf B}$; $\, \, \, \, A^\mu A_\mu=A^2_0-{\mathbf A}^2$.
\item[$\bullet$] $\partial_\mu$: derivative wrt $\mu$ coordinate.
\item[$\bullet$] A four vector $P\equiv(p_0, \bm{\vec { p}}) $ and $p= |\bm{\vec { p}}|$.
\item[$\bullet$] Fermionic field ${\bar \psi}=\psi^\dagger \gamma_0$.
\end{enumerate}

This review article has been organized as follows: in section~\ref{review} we review some known facts which will be needed for our purpose. 
These are  the equilibrium statistical thermodynamics in subsec~\ref{rest}, the three pictures in quantum mechanics in subsec~\ref{brqm}, 
the functional integration in subsec~\ref{path_int} and the Grassmann variables in subsec~\ref{gras_vari}. In section~\ref{IFT} we introduce
the imaginary time formalism: we present  the connection to imaginary time and Matsubara formalism  in subsec~\ref{IFT_Matsu}, 
the operatorial method of Matsubara formalism in subsec~\ref{MF_OM}, the evolution operator and the ${\cal S}$-Matrix in subsec~\ref{smatrix}, the Green's function
at $T\ne0$ in subsec~\ref{cfgf},  the periodicity and antiperiodicity of Green's function in subsec~\ref{period}, Matsubara frequency in subsec~\ref{dfmf}, a short
summary of imaginary time formalism in tabular form in subsec~\ref{dic} and Feynman rules in subsec~\ref{fey_rule}. In section~\ref{freq_sum} we discuss discrete frequency
sum: we  present bosonic frequency sum in subsec~\ref{bos_sum}, fermionic frequency sum in subsec~\ref {fer_sum} and some examples of bosonic sum 
in subsec~\ref{example_ferq}.  We discuss scalar theory at $T\ne 0$ in section~\ref{st}: first  the tadpole diagram in $\lambda \phi^4$-theory in subsec~\ref{tad_sec}
and then one-loop self-energy in $\phi^3$-theory in section~\ref{phiq}. We discuss partition function in section~\ref{PF}: first the relation of the functional integration and the
partition function and then connection with the imaginary time in subsec~\ref{FI_Z}, the partition function for free and interacting scalar fields in subsec.~\ref{sc_part} 
and then the partition function for free fermionic 
field in subsec~\ref{FF}. In section~\ref{gse} we present the general structure of fermion two-point functions at $T\ne 0$ while in section~\ref{gsg} 
the general structure of vector boson two-point functions at $T\ne 0$  is presented. In section~\ref{qed} we discuss quantum electrodynamics (QED)
at $T\ne 0$: gauge fixing, free photon partition function, one-loop electron self-energy, effective electron propagator and its spectral representation, 
collective excitations of electrons, photon self-energy, effective photon propagator and its spectral representation and collective excitations of photons.
In section~\ref{qcd} we present quantum chromodynamics (QCD) at $T\ne 0$ by generalising the QED results of 
two point functions in HTL approximation  to QCD that mostly involve group theoretical factors and learn about the collective excitations in QCD. 
Some subtleties at finite temperature have
been discussed in section~\ref{SFTFT}.  In section~\ref{htl_app} the HTL resummation and HTL perturbation theory have been outlined along with
application to QCD thermodynamics in leading order (LO), next-to-leading order (NLO) and next-to-next-leading order (NNLO). In section~\ref{nonpert} the two-point 
functions and the collective excitations of quarks considering non-perturbative effects like gluon-condensates and Gribov-Zwanziger action have been discussed in
details. As an application of those non-perturbative effects, the dilepton production rates from QGP created in relativistic heavy-ion collisions have been calculated.
Finally, we conclude this review article in section~\ref{conc} and an appendix is presented in section~\ref{appen}.

\section{A Brief Review of Some Facts}
\label{review}
\subsection{Review of Equilibrium  Statistical Thermodynamics}
\label{rest}

In this subsection, we define some of the basic relations in equilibrium  statistical
mechanics~\cite{Reif}.
In thermal
equilibrium, the statistical behaviour of a quantum system is usually investigated through an appropriate ensemble. 
In general the density matrix for a system is defined as
\begin{equation}
\rho(\beta)= e^{-\beta {\cal H}} \, \, , \label{eq1}
\end{equation}
where $\beta=1/T$  (the
Boltzmann constant $k_B\equiv1$ is assumed) and $\cal H$ is the Hamiltonian of the system 
for a given choice of ensemble.  For canonical ensemble, ${\cal H}=H$.
For grand canonical ensemble, ${\cal H}=H-\mu N $ with $H$ is the dynamical 
Hamiltonian and $N$ is the number operator representing difference of particles and antiparticles. 
$N$ commutes with $H$
and also it is hermitian. It has simultaneous eigenstate. It is also extensive variable (scales with volume $V$)
in the thermodynamic limit.
At finite temperature the qualitative behaviours
of a system  are almost independent on the nature of the ensemble,
so the choice of the ensemble is kept arbitrary for general discussions.

With the given density matrix, the finite temperature property of any theory 
is described by the partition function
\begin{equation}
{\cal Z} ={\rm{Tr}}\rho \ = {\rm{Tr}} \left (  e^{-\beta {\cal H}} \right ) 
= \sum_n \left \langle n \left |  e^{-\beta {\cal H}} 
\right | n \right \rangle \ \ , \label{eq2} 
\end{equation}
where $|n\rangle $ is a many-particle state in the full Hilbert space. 
Trace stands for sum over a complete many-particles states in Hilbert space.

In the infinite volume limit: 
\begin{subequations}
\begin{align}
\mbox{Thermodynamic potential:} \hspace*{0.5in}& \Omega(\beta)=- T\, \ln {\cal Z}. \label{eq3} \\
\mbox{Pressure:} \hspace*{0.5in} &  P=-\frac{\partial}{\partial V} { \Omega (\beta)} \ =\ -\frac{\Omega(\beta )}{V} . \label{eq4} \\
\mbox{Number:}\hspace*{0.5in} & N_i = \frac{\partial}{\partial \mu_i} \left (T\ln {\cal Z}\right ). \label{eq5} \\
\mbox{Entropy:}\hspace*{0.5in} & S \ =\ \frac{\partial}{\partial T} \left (T\ln {\cal Z}\right ). \label{eq6}\\
\mbox{Energy:}\hspace*{0.5in} & E \ =\ - PV +TS+\mu_i N_i . \label{eq7}
\end{align}
\end{subequations}

The thermal expectation value of any physical observable can be defined as
\begin{equation}
\langle {\cal A}\rangle_\beta = {\cal Z}^{-1}(\beta) {\rm{Tr}}\left [ \rho(\beta)
{\cal A}\right ]={\cal Z}^{-1}(\beta) {\rm{Tr}}\left [
e^{-\beta {\cal H}} {\cal A} \right ] \, \, . \label{eq8}
\end{equation}

The  correlation function of any two observables is given as
\begin{equation}
\langle {\cal A}{\cal B}\rangle_\beta = {\cal Z}^{-1}(\beta) {\rm{Tr}}
\left [ \rho(\beta) {\cal A}{\cal B}\right ]={\cal Z}^{-1}(\beta)
 {\rm{Tr}}\left [
e^{-\beta {\cal H}} {\cal A} {\cal B}\right ] \, \, . \label{eq9}
\end{equation}

\subsubsection{Partition function for one bosonic degree of freedom}
 Consider a time-dependent single-particle quantum mechanical mode is occupied by bosons. Each boson in that mode has the same
 energy $\omega$. There may be $0$, $1$, $2$, or any number of bosons occupying that state without any interaction. So, it could be thought as
 a set of non-interacting quantized simple harmonic oscillators~\cite{Kapusta} and a Hamiltonian for each oscillator  is given as
 \bea
{\cal H} &=& \frac{1}{2}\omega\left(a a^\dagger +a^\dagger a \right) , \label{eq9a}
 \eea
 where $a^\dagger$ and $a$ are, respectively, the  boson creation and  annihilation operators and their action on a number eigenstate are 
 \begin{subequations}
 \begin{align}
 a^\dagger |n\rangle &= \sqrt{n+1} |n+1\rangle \, ,  \label{eq9b} \\
 a |n\rangle &= \sqrt{n} |n-1\rangle \, , \label{eq9c}
 \end{align}
 \end{subequations}
 with $a |0\rangle =0$ and  the $n$-th excited state is built as
 \be
 |n\rangle = \frac{1}{\sqrt{n!}} \left(a^\dagger \right)^n|0\rangle \, . \label{eq9d}
 \ee
  The coefficients in \eqref{eq9b} and \eqref{eq9c} follow from the requirements that $a^\dagger$ and $a$ be the hermitian conjugates
 and that $a^\dagger a$ be the number operator $\hat N$ with
 \be
 {\hat N}|n\rangle =a^\dagger a |n\rangle = n |n\rangle \, , \label{eq9e}
 \ee
 where  $a$  and $a^\dagger$ satisfy the commutation relation
 \be
 \left [ a,a^\dagger\right ] =aa^\dagger-a^\dagger a=1 . \label{eq9f}
 \ee 
Combining \eqref{eq9a} and \eqref{eq9f} the Hamiltonian becomes
\bea
{\cal H} = \omega \left ( a^\dagger a +\frac{1}{2} \right ) = \omega \left ({\hat N} +\frac{1}{2} \right ) \, . \label{eq9g}
\eea 

Now the partition function for one bosonic degree of freedom can be written from \eqref{eq2} as 
\begin{eqnarray}
 {\cal Z_B} &=& \sum^\infty_{n=0}   \langle n | e^{-\beta\omega\left ({\hat N} +\frac{1}{2} \right ) } |n\rangle 
= e^{-\frac{\beta\omega}{2}}  \sum^\infty_{n=0}   e^{-\beta\omega n} 
 = \frac{e^{-\frac{\beta\omega}{2}}}{1-e^{-\beta\omega} }\, .\label{eq9h}
\end{eqnarray}
According to \eqref{eq3}  the logarithm of the  partition function is of interest. From \eqref{eq9h} one obtains
\begin{eqnarray}
\ln {\cal Z_B} &=&- \frac{\beta\omega}{2} -\ln (1-e^{-\beta\om}), \label{eq10}
\end{eqnarray}
where the first term originates from the zero-point energy, $\frac{1}{2}\om$ in the Hamiltonian in \eqref{eq9g}.

\subsubsection{Partition function for one fermionic degree of freedom}

Like previous subsection, we consider a time-dependent single-particle quantum mechanical mode is occupied by fermions with each fermion in that mode has the same
 energy $\omega$.  We note that  the Pauli exclusion principle restricts the occupation of a single-particle mode by more than one fermion.
 Therefore, there are only two states of the system as
 \be
 |0\rangle \hspace*{0.3in} \mbox{and} \hspace*{0.3in} |1\rangle \, .  \label{eq10a}
 \ee
 
 Now the Hamiltonian for each fermion~\cite{Kapusta} is given as
 \bea
H&=& \frac{1}{2}\omega\left(\alpha^\dagger \alpha -\alpha \alpha^\dagger \right) , \label{eq10b}
 \eea
where $\alpha^\dagger$ and $\alpha$ are, respectively, the fermion creation and annihilation operators. They operate on the two states in \eqref{eq10a} as
\begin{subequations}
\begin{align}
\alpha^\dagger  |0\rangle &\, =\,   |1\rangle \, ,  \label{eq10c} \\
\alpha |1 \rangle &\, =\,   |0 \rangle \, ,  \label{eq10d} \\
\alpha^\dagger |1 \rangle &\, =\,   0  \, ,  \label{eq10e} \\
\alpha |0 \rangle &\, =\,   0  \, .  \label{eq10f} 
\end{align}
\end{subequations}
So, these operators have the properties that the operation of $\alpha^2$ and $(\alpha^\dagger)^2$ on any of the states in \eqref{eq10a} is zero.
The coefficients in \eqref{eq10c} to \eqref{eq10f} follow from the requirements that $\alpha^\dagger$ and $\alpha$ be the hermitian conjugates
 and that $\alpha^\dagger \alpha$ be the number operator $\hat N$ with
 \be
 {\hat N}|n\rangle =\alpha^\dagger \alpha |n\rangle = n |n\rangle \, , \label{eq10g}
 \ee
 where  $\alpha$  and $\alpha^\dagger$ satisfy the anticommutation relation
 \be
 \left \{ \alpha,\alpha^\dagger\right \} =\alpha \alpha^\dagger + \alpha^\dagger \alpha=1 . \label{eq10h}
 \ee 
Combining \eqref{eq10b} and \eqref{eq10h} the Hamiltonian becomes
\bea
 H= \omega \left ( \alpha^\dagger \alpha - \frac{1}{2} \right ) = \omega \left ({\hat N} -\frac{1}{2} \right ) \, . \label{eq10i}
\eea 
For grand canonical ensemble ${\cal H}=H-\mu {\hat N} $ and  the partition function for one fermionic degree of freedom can be written from \eqref{eq2} as 
\begin{eqnarray}
 {\cal Z_F} &=& \sum^{1}_{n=0}   \langle n | e^{-\beta\left ({H-\mu \hat N} \right ) } |n\rangle 
= e^{\frac{\beta\omega}{2}}  \sum^{1}_{n=0}   e^{-\beta(\omega-\mu) n}  \nonumber \\
& =& e^{\frac{\beta\omega}{2}}\left({1+e^{-\beta(\omega-\mu)} }\right)\, .\label{eq10j}
\end{eqnarray}
The logarithm of the partition function for a fermion 
\begin{eqnarray}
\ln {\cal Z_F} &=& \frac{\beta\omega}{2} +  \ln \left (1+e^{-\beta(\omega-\mu)}\right ), \label{eq11}
\end{eqnarray}
where the first term originates from the zero-point energy, $\frac{1}{2}\om$ in the Hamiltonian in \eqref{eq10i}. 

Similarly, one can obtain the partition function for an antifermion by replacing $\mu\rightarrow -\mu$ in \eqref{eq10j} as
\bea
{\cal Z}_{\bar F} & =& e^{\frac{\beta\omega}{2}}\left({1+e^{-\beta(\omega+\mu)} }\right)\, , \label{eq11a}
\eea
and logarithm reads as
\begin{eqnarray}
\ln {\cal Z}_{\bar F}&=& \frac{\beta\omega}{2} +  \ln \left (1+e^{-\beta(\omega+\mu)}\right )\, . \label{eq11b}
\end{eqnarray}

\subsubsection{Partition function for non-interacting gases of fermions and bosons}

Now we consider a gas consisting non-interacting  fermions and bosons. In principle they can interact among themselves to come to thermal
equilibrium. Once it achieves thermal equilibrium then one can slowly switch off the interactions. Such a non-interacting system may well describe~\cite{Kapusta}
the atmosphere around us, electrons in metal or white dwarf star, blackbody radiation in a heated cavity or the cosmic microwave background radiation etc.

The partition function for a non-interacting gas consisting fermions, antifermions and bosons as
\begin{eqnarray}
{\cal Z} &=& \prod_\alpha \, {\cal Z}_F^\alpha \times {\cal Z}_{\bar F}^\alpha \times {\cal Z}_B^\alpha  , \label{eq11c} 
\end{eqnarray}
where $\alpha$ represents the single particle states of each mode corresponding to fermions, antifermions and bosons. 
Now using \eqref{eq9h}, \eqref{eq10j} and \eqref{eq11a}
\begin{eqnarray}
{\cal Z} &=& \prod_\alpha \, e^{\frac{\beta\omega_\alpha}{2} } \, 
 \left( 1+ e^{-\beta(\omega_\alpha-\mu)}\right ) \ e^{\frac{\beta\omega_\alpha}{2} }  \left( 1+ e^{-\beta(\omega_\alpha+\mu)}\right ) \ 
e^{-\frac{\beta\omega_\alpha}{2} } \left( 1- e^{-\beta\omega_\alpha}\right )^{-1} \, \label{eq11d}
\end{eqnarray}
\begin{eqnarray}
\ln {\cal Z} &=& \sum_\alpha \frac{\beta\omega_\alpha}{2} + \sum_\alpha \ln \left( 1+ e^{-\beta(\omega_\alpha-\mu)}\right) \ + \sum_\alpha \frac{\beta\omega_\alpha}{2} 
 + \  \sum_\alpha \ln \left( 1+ e^{-\beta(\omega_\alpha+\mu)}\right ) \nonumber \\
&&\ -\sum_\alpha \frac{\beta\omega_\alpha}{2} \, - \ \sum_\alpha \ln \left( 1- e^{-\beta\omega_\alpha}\right ). \label{eq12}
\end{eqnarray}
The $\sum_\alpha$ is over single particles states. In infinite volume limit $\sum_\alpha \rightarrow \frac{V}{(2\pi)^3} \int d^3p$.

The logarithm of the partition function becomes
\begin{eqnarray} 
\ln {\cal Z} &=& \frac{V}{(2\pi)^3} \int d^3p \left[ \frac{\beta\omega}{2} +  \ln \left( 1+ e^{-\beta(\omega-\mu)}\right)
 +\ln \left( 1+ e^{-\beta(\omega+\mu)}\right )  \right. \nonumber \\
&& \left.  -\ln  \left( 1- e^{-\beta\omega}\right )\right ] \, , \label{eq13}
\end{eqnarray}
we note that for massless species $\omega=p$. We will also obtain this results using thermal field theory in later subsections~\ref{sc_part},
\ref{direct} and \ref{photon}.

The thermodynamic potential in \eqref{eq3} can be written as
\begin{eqnarray} 
\Omega(\beta)=- T\, \ln {\cal Z} =  &=& -\frac{VT}{(2\pi)^3} \int d^3p \left[ \frac{\beta\omega}{2} +  \ln \left( 1+ e^{-\beta(\omega-\mu)}\right)
 +\ln \left( 1+ e^{-\beta(\omega+\mu)}\right )  \right. \nonumber \\
&& \left.  -\ln  \left( 1- e^{-\beta\omega}\right )\right ] \, . \label{eq13a}
\end{eqnarray}
Therefore, various thermodynamic quantities in (\ref{eq4}) to (\ref{eq7}) can easily be computed for non-interacting gas of fermions and bosons.

Now, if there are interactions, it becomes difficult to compute  partition function for interacting system.
The partition function reads from (\ref{eq2}) as
\begin{equation}
{\cal Z} ={\rm{Tr}}\rho \ = {\rm{Tr}} \left (  e^{-\beta { \cal H}} \right ) 
= \sum_n \left \langle n \left |  e^{-\beta {\cal H}} 
\right | n \right \rangle \ \ , \label{eq14} 
\end{equation}
where $|n\rangle $ is a many-particle state in the full Hilbert space. 
Trace stands for sum over expectation values of all possible states  in Hilbert space.
There are infinite number of such states in quantum field theory (QFT).
If the particles or fields are non-interacting, then it is easier to compute ${\cal Z(\beta)}$ as we have seen above.
For an interacting system the partition function cannot be computed exactly if one expands even in perturbation series
in interaction strength in a given theory.
{\sf Matsubara (Imaginary time) formalism}\cite{Matsubara} represents  a diagrammatic way of calculating the partition function 
and other physical observables perturbatively in order by order of the coupling strength of a given theory, analogous to $T=0$ (vacuum) field theory.

\subsection{Brief Review of Quantum Mechanics}
\label{brqm}

In studying a quantum mechanical system or a system described by a quantum field theory,  one is basically interested in determining the time evolution operator. 
In the standard framework of quantum mechanics, one solves the Schr\"odinger equation to determine the energy eigenvalues and eigenstates simply 
because the time evolution operator is related to the Hamiltonian. 

Quantum systems are regarded as wave functions that satisfies the Schr\"odinger differential equation as
\be
i \frac{d}{dt} |\psi (t) \rangle = {\cal H}|\psi (t)\rangle \, , \label{qm1}
\ee
which governs the dynamics of the system in time. 

Observables are represented by hermitian operators which act on the wave function.  Thus
the Hamiltonian of the system, $ {\cal H}$, is the operator which describes the total energy of the quantum system as
\be
 {\cal H}  |\psi (t) \rangle = E |\psi (t) \rangle \, . \label{qm2}
\ee
There are three pictures in quantum mechanics due to Schr\"odinger, Heisenberg and Dirac. 
Below we briefly outline the three pictures in quantum mechanics~\cite{Sakurai}.

\subsubsection{Schr\"odinger picture (SP)}
In the Schr\"odinger picture, the operators stay fixed while the Schr\"odinger equation changes the basis with time. 

Since, all physical operators  $   {\cal O}_S$ are time independent, one can write
\be
 \dot{   {\cal O}}_S=0 \, . \label{qm3}
 \ee
The basis vector changes with time via the Schr\"odinger equation as
\be
\frac{d}{dt} |\psi (t) \rangle_S=-i   {\cal H}  |\psi (t) \rangle_S \, \label{qm4}
\ee
The differential equation leads to an expression for the wave function as
\be
|\psi(t)\rangle_{S}=e^{-i   {\cal H}t}|\psi(0)\rangle_{S}\, .\label{qm5}
\ee
indicates that  all  physical state vectors are time dependent. A quantum operator as the argument of the exponential function is defined in terms of
 its power series expansion as
 \bea
 e^{-i  {\cal H} t} & = &  \sum_{n=0}^\infty \frac{1}{n!}  \left (-i   {\cal H} t \right )^n \nonumber \\
 &=& 1 -i   {\cal H} t -\frac{1}{2} \left (   {\cal H} t \right )^2 +\cdots  \label{qm6}
 \eea
This is how the states pick up their time-dependence.

\subsubsection{Heisenberg picture (HP)}
In the Heisenberg picture, it is the operators 
which change in time while the basis of the space remains fixed.

Since the basis does not change with time which is accomplished by adding a term to the Schr\"odinger states to eliminate the time-dependence as
\begin{subequations}
\begin{align}
|\psi(0)\rangle_{H} &=e^{i   {\cal H} t} |\psi(t) \rangle_{S}=|\psi(0)\rangle_{S} \, , \label{qm7} \\
\frac{\partial}{\partial t} |\psi(0)\rangle_{H} & =0 \, . \label{qm8}
\end{align}
\end{subequations}

We may define operators in the Heisenberg picture via expectation values of a Sch\"ordinger operator as
\bea
\langle   {\cal O}_S \rangle &=& \left._S\langle\psi(t)| {\cal O}_S |\psi (t) \rangle_S \right. \nonumber \\
&=&  \left._S\langle\psi(0)| e^{i   {\cal H} t}   {\cal O}_S e^{- i   {\cal H} t} |\psi(0)\rangle_{S} \right.  \nonumber \\
&=&  \left._H\langle\psi (0) |\left( e^{i   {\cal H} t}   {\cal O}_S e^{-i   {\cal H} t} \right ) |\psi(0)\rangle_{H} \right. \, . \label{qm9a}
\eea
The operators in the Heisenberg picture, therefore, pick up time-dependence through unitary transformations as
\be
{\cal O}_H(t) = e^{i   {\cal H} t}{\cal O}_S e^{-i   {\cal H} t} = U^\dagger(t) {\cal O}_S U(t)  \, , \label{qm9}
\ee
where the Heisenberg Picture is related  to the Schr\"odinger Picture through unitary transformation $U(t)= e^{-i{\cal H}t}$, where ${\cal H}$ is the full
Hamiltonian of the system as
 \be
 {\cal H}={\cal H}_0+{\cal H}^\prime \, , \label{qm10}
 \ee
 where  ${\cal H}_0$ is the free part and ${\cal H}^\prime$ is the interacting part.
 
  We may ascertain the Heisenberg operators' time-dependence through differentiation as
\bea
\frac{d{\cal O}_H(t)} {dt} &=& i{\cal H} e^{i   {\cal H} t} {\cal O}_S e^{-i   {\cal H} t}  - i e^{i   {\cal H} t} {\cal O}_S {\cal H} e^{-i   {\cal H} t} 
+\frac{\partial {\cal O}_H}{\partial t} \nonumber \\
&=&  i \underbrace{\Big ({\cal H} {\cal O}_H - {\cal O}_H {\cal H} \Big )}_{\mbox{Commutator}}+\frac{\partial {\cal O}_H}{\partial t}  \nonumber \\
&=& i \Big [ {\cal H}, {\cal O}_H \Big ] +\frac{\partial {\cal O}_H}{\partial t} \, . \label{qm11}
\eea
The operators are thus governed by a differential equation known as Heisenberg's equation.

\subsubsection{Dirac or Interaction picture (IP)}

The Dirac (Interaction) picture is a sort of intermediary between the Schr\"odinger picture and the Heisenberg picture as both the quantum states 
and the operators carry time dependence. It is especially useful for problems including explicitly time-dependent interaction terms in the Hamiltonian.

In the interaction picture, the state vectors are again defined as transformations of the Schr\"odinger states by the free part of the Hamiltonian as
\be
|\psi(t)\rangle_{I}=e^{i{\cal H}_0t}|\psi(t)\rangle_{S} \, . \label{qm12}
\ee
The Dirac operators are transformed similarly to the Heisenberg operators as
\be
{\cal O}_I(t) = e^{i{\cal H}_0t}O_S e^{-i{\cal H}_0 t} \, . \label{qm13}
\ee
The relation between interaction picture and Schr\"odinger picture is similar to Heisenberg picture except that the unitary transformation
involves free Hamiltonian ${\cal H}_0$ instead of full one ${\cal H}$  as $U(t)=e^{-i{\cal H}_0t}$. The purpose of it  is to describe the  interaction in terms of free fields.
 
The states in the interaction picture in \eqref{qm12}  evolve in time similar to Heisenberg states as
\bea
\frac{d}{dt}|\psi(t)\rangle_{I} &=& i{\cal H}_0 |\psi(t)\rangle_{I} + e^{i{\cal H}_0t} \frac{d}{dt}|\psi(t)\rangle_{S}  \nonumber \\
&=& i{\cal H}_0 |\psi(t)\rangle_{I} + e^{i{\cal H}_0t} \left(-i {\cal H} \right) |\psi(t)\rangle_{S}  \nonumber \\
&=& i{\cal H}_0 |\psi(t)\rangle_{I} + e^{i{\cal H}_0t} \left(-i ({\cal H}_0+{\cal H}' ) \right)  e^{-i{\cal H}_0t} | \psi(t)\rangle_{I}  \nonumber \\
&=& i{\cal H}_0 |\psi(t)\rangle_{I} - i{\cal H}_0 |\psi(t)\rangle_{I} - i e^{i{\cal H}_0t} {\cal H}'  e^{-i{\cal H}_0t}  |\psi(t)\rangle_{I}  \nonumber \\
&=& - i e^{i{\cal H}_0t} {\cal H}'  e^{-i{\cal H}_0t}  |\psi(t)\rangle_{I}  \nonumber \\
&=& -i {\cal H}'(t) | \psi(t)\rangle_{I} \, , \label{qm14}
\eea
where the interacting term of the Hamiltonian is defined similarly as
 \be
{\cal H}^\prime(t) = e^{i{\cal H}_0t}{\cal H}^\prime e^{-i{\cal H}_0 t} \, . \label{qm15}
\ee
Therefore, the state vectors in the interaction picture evolve in time according to the interaction term only.

It can be easily shown through differentiation of \eqref{qm13}  that operators in the interaction picture evolve in time according only to the free Hamiltonian as
\bea
\frac{d{\cal O}_I(t)} {dt} &=& i{\cal H}_0 e^{i   {\cal H}_0 t} {\cal O}_S e^{-i   {\cal H}_0 t}  - i e^{i   {\cal H}_0 t} {\cal O}_S {\cal H}_0 e^{-i   {\cal H}_0 t} 
+\frac{\partial {\cal O}_I}{\partial t} \nonumber \\
&=&  i \underbrace{\Big ({\cal H}_0 {\cal O}_I (t)- {\cal O}_I(t) {\cal H}_0 \Big )}_{\mbox{Commutator}}+\frac{\partial {\cal O}_I}{\partial t}  \nonumber \\
&=& i \Big [ {\cal H}_0, {\cal O}_I(t) \Big ] +\frac{\partial {\cal O}_I}{\partial t} \, . \label{qm16}
\eea
The interaction picture admits  that the operators  act on the state vector
at different times and form the basis for quantum field theory and many other newer methods.

\subsection{Functional Integration}
\label{path_int}

We know that the partition function of statistical mechanics completely describes a system in equilibrium. Likewise, a system in quantum field theory is fully described by
 an integral over all space-time paths allowed, which is also known as path integral~\cite{Feynman,FH}. In this subsection we will derive the path integral formalism.

If a particle is observed in the state $|\phi_a\rangle$ at a time $t_0$, then the wave function after some time $t$ will evolve as
\be
|\phi_b\rangle =e^{-i\int\limits_{0}^{t} {\cal H} dt} |\phi_a\rangle  \, . \label{path1}
\ee
We assume that the Hamiltonian is time-independent then the transition amplitude 
for going from a state $|\phi_a\rangle$ to another state $|\phi_b\rangle$  after a time $t$ can simply be written as
\be
\langle \phi_b | e^{-i {\cal H} t} |\phi_a\rangle  \, . \label{path2}
\ee
In  statistical mechanics the most interesting cases are the ones where the system returns to its initial state after a time $t$,  then transition amplitude in \eqref{path2}  
can be written as
\be 
\langle \phi_a | e^{-i {\cal H} t} |\phi_a\rangle  \, . \label{path3}
\ee
Let $\hat\phi(\bm{\vec x},0)$ be Sch\"odinger picture field operator at time $t=0$ and let  $\hat\pi(\bm{\vec x},0)$ be its conjugate momentum operator.
The eigenstates of the field operator are denoted by $|\phi\rangle$ and those for momentum operator are $|\pi\rangle$. They satisfy the eigenvalue equations as
\begin{subequations}
\begin{align}
\hat\phi(\bm{\vec x},0) |\phi\rangle & = \phi(\bm{\vec x}) |\phi\rangle \, , \label{path3a} \\
\hat\pi(\bm{\vec x},0) |\pi\rangle & = \pi(\bm{\vec x}) |\pi\rangle \, , \label{path3b}
\end{align}
\end{subequations}
where $\phi(\bm{\vec x})$ and $\pi(\bm{\vec x})$ are corresponding eigenvalues.
Now we define the completeness and orthogonality relations for field $\phi$ 
\bea
\int d\phi(x) |\phi\rangle \langle \phi | &=& 1  \nonumber \\
\langle \phi_a |\phi_b \rangle &=& \prod_{x}  \delta \left ( \phi_a(\bm{\vec x})-  \phi_b(\bm{\vec x}) \right ),  \label{path4} 
\eea
and for momentum density $\pi$  as
\bea
\int \frac{1}{2\pi} \ d\pi(x) |\pi\rangle \langle \pi | &= &1  \nonumber \\
\langle \pi_a |\pi_b \rangle &=& \prod_{x}  \delta \left ( \pi_a(\bm{\vec x})-  \pi_b(\bm{\vec x}) \right ) , \label{path5} 
\eea
and their overlap is defined as
\be
\langle \phi |\pi\rangle = \exp \left (i\int d^3x \ \pi(\bm{\vec x}) \ \phi(\bm{\vec x}) \right ) .\label{path6}
\ee
The probability amplitude is described by the time evolution from the initial state to final state through all intermediate states at the time
intervals $\Delta t$.  Now, splitting the time interval ($0$, $t$) into $N$ equal steps of size $\Delta t = t/N$. With this the probability amplitude in \eqref{path3} can
be written as
\be
\langle \phi_a | e^{-i {\cal H} t} |\phi_a\rangle = \langle \phi_a | \underbrace{e^{-i{\cal H} \Delta t} \times \cdots \times  e^{-i{\cal H} \Delta t}}_{N \, \mbox{ times}}  |\phi_a\rangle \, . \label{path7}
\ee
Now a complete set of states is inserted after each time interval in \eqref{path7}, alternating between the one in \eqref{path4} and \eqref{path5} as
\bea
\langle \phi_a | e^{-i {\cal H} t} |\phi_a\rangle &=& \lim_{N\rightarrow \infty} \int \left (\prod_{i=1}^N \frac{1}{2\pi} d\pi_i \ d\phi_i \right ) 
\langle \phi_a|\pi_N\rangle \langle \pi_N |e^{-i{\cal H} \Delta t} |\phi_N\rangle \nonumber \\
&&\times \langle \phi_N |\pi_{N-1}\rangle \langle \pi_{N-i} | e^{-i{\cal H} \Delta t} |\phi_{N-i} \rangle \cdots \cdots \nonumber \\ 
&& \times \langle \phi_2 |\pi_1\rangle \langle \pi_1 | e^{-i{\cal H} \Delta t} |\phi_1 \rangle \langle\phi_1 |\phi_a\rangle \, , \label{path8}
\eea
where every second term can be rewritten using the  overlap in \eqref{path6} which  always appear on the form
\be
\langle \phi_{i+1}|\pi_i\rangle = \exp \left( i\int d^3x \ \pi_i(\bm{\vec x}) \phi_{i+1} (\bm{\vec x}) \right )\, , \label{path9}
\ee
and just following \eqref{path3}  the last term becomes
\be
\langle \phi_1| \phi_a \rangle = \delta(\phi_1-\phi_a) \, . \label{path10}
\ee
Now the exponential in \eqref{path8} can be expanded for small time interval $\Delta t$ as
\be
 \langle \pi_i |e^{-i{\cal H}_i \Delta t} |\phi_i\rangle \approx  \langle \pi_i |\left(1-i{\cal H}_i \Delta t + \cdots \right ) |\phi_i\rangle \, . \label{path11}
\ee
In \eqref{path8}, the Hamiltonian ${\cal H}_i$ always appears between two states with same index $i$ which indicates that the Hamiltonian is always evaluated
at the same point in time. Now one can write \eqref{path11} as
\bea
 \langle \pi_i |\left(1-i{\cal H}_i \Delta t \right ) |\phi_i\rangle &=&   \langle \pi_i |\phi_i\rangle \left(1-i{\cal H}_i \Delta t \right )\nonumber \\
 &=&  \left(1-i{\cal H}_i \Delta t \right )  \exp \left(- i\int d^3x \ \pi_i(\bm{\vec x}) \phi_i (\bm{\vec x}) \right ) \, . \label{path12}
\eea
Now, the expansion in first order in \eqref{path12} can be changed back to exponential as
\be
 \langle \pi_i | e^{-i{\cal H}_i \Delta t } |\phi_i\rangle \approx e^{-i{\cal H}_i \Delta t}  \exp \left(- i\int d^3x \ \pi_i(\bm{\vec x}) \phi_i (\bm{\vec x}) \right ) \, . \label{path13}
\ee
The Hamiltonian ${\cal H}$ is the integral of Hamiltonian density ${\cal H}_d$ as
\be
{\cal H}_i = \int d^3x \ {\cal H}_d \left (  \pi_i(\bm{\vec x}) \phi_i (\bm{\vec x}) \right ) \, . \label{path14}
\ee
Now, the transition amplitude in \eqref{path8} can be written as
\bea
\langle \phi_a | e^{-i {\cal H} t} |\phi_a\rangle &=&\frac{1}{2\pi}  \lim_{N\rightarrow \infty} \int  \prod_{i=1}^N  \ d\pi_i \ d\phi_i \  \delta(\phi_1-\phi_a) \nonumber \\
&&\times \exp\left ( -i\Delta t  \sum_{j=1}^N \int d^3x\ \left [ {\cal H}_d \left (  \pi_j, \phi_i  \right ) - \frac{ \pi_j\left (\phi_{j+1}-\phi_j\right )}{\Delta t} \right ]\right ) \, . \label{path15}
\eea
where $\phi_{N+1}=\phi_a=\phi_1$. In the continuum limit in time, one can write
\bea
&& \lim_{N\rightarrow \infty} i\Delta t  \sum_{j=1}^N \left [ \pi_j \frac{ \left (\phi_{j+1}-\phi_j\right )}{\Delta t} - {\cal H}_d \left (  \pi_j, \phi_i  \right ) \right ]  \nonumber \\
&& \rightarrow i \int_0^{t_f} dt  \left ( \pi(\bm{\vec x},t) \frac{\partial \phi(\bm{\vec x},t)}{\partial t} - {\cal H}_d \left (  \pi(\bm{\vec x},t), \phi(\bm{\vec x},t)  \right )\right ) \, . \label{path16a}
\eea
The transition amplitude can now be written as
\bea
\langle \phi_a | e^{-i {\cal H} t} |\phi_a\rangle &=& \int {\cal D}\pi \int\limits_{\phi(\bm{\vec x},0) =\phi_a(\bm{\vec x})}^{\phi(\bm{\vec x},t)=\phi_a(\bm{\vec x})} {\cal D}\phi  \nonumber \\
&&\times \exp \left [  i \int_0^{t_f} dt \int d^3x  \left ( \pi(\bm{\vec x},t) \frac{\partial \phi(\bm{\vec x},t)}{\partial t} - {\cal H}_d \left (  \pi(\bm{\vec x},t), \phi(\bm{\vec x},t)  \right )\right )
 \right ] , \label{path16}
\eea
where the functional integration is denoted by $\cal D$. The integration runs over all possible momenta $\pi(\bm{\vec x},t)$ whereas $\phi(\bm{\vec x},t)$ is restricted  by
the boundary conditions, starts at $\phi_a(\bm{\vec x})$  at  initial time $t=0$ and ends at $\phi_a(\bm{\vec x})$ final time $t=t_f$. 

The Hamiltonian density of a system is given as
\begin{equation}
 {\cal H}_d = \pi(\bm{\vec x}, t) \frac{\partial \phi(\bm{\vec x}, t))}{\partial t} -{\cal L} (\phi(\bm{\vec x}, t) {\dot \phi(\bm{\vec x}, t))}), \label{path17}
\end{equation}
where ${\cal L}(\phi(\bm{\vec x}, t), {\dot \phi(\bm{\vec x}, t)})$ is the Lagrangian density of a system. 
Now combining \eqref{path16} and \eqref{path17}, the transition amplitude can be written as
\be
\langle \phi_a | e^{-i {\cal H} t} |\phi_a\rangle = \int\limits_{\phi(\bm{\vec x},0) =\phi_a(\bm{\vec x})}^{\phi(\bm{\vec x},t)=\phi_a(\bm{\vec x})}  {\cal D}\phi 
\ e^{i\int\limits_{0}^{t_f} dt \int  d^3x \ {\cal L} (\phi(\bm{\vec x}, t), {\dot \phi(\bm{\vec x}, t)})}
=\int\limits_{\phi(\bm{\vec x},0) =\phi_a(\bm{\vec x})}^{\phi(\bm{\vec x},t)=\phi_a(\bm{\vec x})}  {\cal D}\phi  \ e^{i S [\phi]}\, , \label{path18}
\ee
where $S[\phi]$ is the action of a system. This is the so-called path integral, and here  the transition amplitude for a system is
simply the sum over all possible paths it may take in going from its initial to its final state.

\subsection{Grassmann Variables}
\label{gras_vari}

The basic feature of Grassmann variables~\cite{Chandlin,Berezin} is that they anticommute, so integrals over Grassmann variables are very convenient for dealing with fermionic fields, being described by anticommutation relations. In this subsec, the Grassmann algebra is defined and some integrals which will be needed later on are calculated.

A single Grassmann variable $\xi$ is defined by anticommutation relation as
\be
{\Big \{}\xi,\xi {\Big \}} =0 \, . \label{GV1}
\ee
It can be generalized to a set of $N$ variables $\xi_i$ and a paired set $\xi_i^\dagger$.  The algebra is defined by
\be
{\Big \{}\xi_i,\xi_j {\Big \}} = \left \{\xi_i,\xi^\dagger_j \right \} =\left \{\xi^\dagger_i,\xi^\dagger_j \right \} = 0 \, . \label{GV2}
\ee
In particular, from \eqref{GV1} the square of any Grassmann number is zero: 
\be 
\xi^2=0 .\label{GV2a} 
\ee
Because of this the most general function of $\xi$ is defined by using a Taylor series
expansion as
\be
\xi=a+b\xi \, , \label{GV3}
\ee
where $a$ and $b$ are $c$-numbers. Using anticommutation rules one can obtain the ordering as
\be
\xi_1\eta_1 \cdots \xi_N\eta_N = (-1)^{\frac{1}{2} N(N-1)} \xi_1\cdots \xi_N\eta_1\cdots \eta_N \, , \label{GV4}
\ee
for two Grassmann variables $\xi$ and $\eta$. The integration  is  defined as
\begin{subequations}
\begin{align}
\int d\xi &=0 \, , \label{GV5} \\
\int d\xi \ \xi &=1 \, , \label{GV6} 
\end{align}
\end{subequations}
and when performing an integral over multiple Grassmann variables, the following sign convention will be used
\be
\int d\xi_1 \int d\xi_2\ \xi_2 \ \xi_1 = +1, \label{GV7}
\ee
that is, doing the inner integral first. The Gaussian integral over a complex Grassmann variable is defined as
\be
\int d\xi^\dagger d\xi \ e^{-\xi^\dagger b \ \xi} = \int d\xi^\dagger d\xi \left (1 - \xi^\dagger b \xi  \right ), \label{GV8}
\ee
through Taylor expansion and all higher orders vanish. Using anticommutation of $\xi$ and $\xi^\dagger$ one gets
\be
\int d\xi^\dagger d\xi \ e^{-\xi^\dagger b \ \xi} = \int d\xi^\dagger d\xi \left (1 +  \xi \xi^\dagger b \right ) = b, \label{GV9}
\ee
It can be generalized to $N$ Grassmann variables that results in a Gaussian integral involving  $N\times N$ matrix $D$ as
\be
\int d\xi_1^\dagger d\xi_1 \cdots d\xi_N^\dagger d\xi_N  \ e^{-\xi^\dagger  D \xi} . \label{GV10}
\ee
This can be calculated by considering $N$ Grassmann variables and components $D_{ij}$ of the matrix 
which make the exponent hermitian:$ (\xi^*D_{ij}\xi)^* = \xi D_{ij}\xi^*$, and expanding the integral, keeping in mind that due to \eqref{GV2a} 
only one term will be non-zero
\be
\int \prod d\xi^*_i \ d\xi_i \ e^{-\xi_i^*D_{ij}\xi_j} =\int  d\xi^*_1 \ d\xi_1\cdots  d\xi^*_N \ d\xi_N \frac{1}{N!} \left (-\xi_{i_1}^*D_{i_1j_1}\xi_{j_1} \right)
\cdots  \left (-\xi_{i_N}^*D_{i_N j_N}\xi_{j_N} \right) \, . \label{GV11}
\ee
Now ordering $\xi$ and $d\xi$ following \eqref{GV4} one can write
\bea
\int \prod d\xi^*_i \ d\xi_i \ e^{-\xi_i^*D_{ij}\xi_j} &=&\frac{1}{N!} \int d\xi^*_1 \cdots d\xi^*_N \xi^*_{i_1} \cdots \xi^*_{i_N} 
\int  d\xi_1 \cdots d\xi_N \xi_{j_1} \cdots \xi_{j_N} D_{i_1 j_1}\cdots D_{i_N j_N} \nonumber \\
&=& \frac{1}{N!} \varepsilon_{i_1\cdots i_N}  \varepsilon_{j_1\cdots j_N}  D_{i_1 j_1}\cdots D_{i_N j_N} \nonumber \\
&=&{\mbox{det}}D \, , \label{GV12}
\eea
where $\varepsilon$ appears from permuting $\xi^*_{i_1} \cdots \xi^*_{i_N} $ to $\xi^*_{1} \cdots \xi^*_{N} $, then $\xi_{j_1} \cdots \xi_{j_N}$.  Now using the ordering 
in \eqref{GV4} twice for the integrals, one obtains
\be
\int d\xi^\dagger_1 d\xi_1 \cdots  d\xi^\dagger_N d\xi_N \ e^{-\xi^\dagger  D \ \xi} = {\mbox{det}}D \, . \label{GV13}
\ee
which we will be needed for computing the partition function in functional integration approach for fermion and ghost fields later.

\section{Imaginary Time Formalism}
\label{IFT}
\subsection{Connection to Imaginary Time and Matsubara Formalism}
\label{IFT_Matsu}

For a given Schr\"odinger operator, ${\cal A}_S$, the Heisenberg operator,
${\cal A}_H(t)$ can be written  from \eqref{qm9} as
\begin{equation}
{\cal A}_H(t) = e^{i{\cal H}t} \, {\cal A}_S \,  e^{-i{\cal H}t} \, \, .
\label{eq15}
\end{equation}

The thermal correlation function of two operators 
can also be written from \eqref{eq9} as
\begin{eqnarray}
\langle {\cal A}_H(t) {\cal B}_H(t') \rangle_\beta  
&=& {\cal Z}^{-1}(\beta) {\rm{Tr}}\left [ e^{-\beta {\cal H}} {\cal A}_H(t) 
{\cal B}_H(t') \right ] \nonumber \\
&=& {\cal Z}^{-1}(\beta) {\rm{Tr}}\left [ e^{-\beta {\cal H}} e^{i{\cal H}t}{\cal A}_S  e^{-i{\cal H}t}
 e^{i{\cal H}t'}{\cal B}_S e^{i{\cal H}t'} \right ] \nonumber \\ 
&=& {\cal Z}^{-1}(\beta) {\rm{Tr}}\left [ e^{i{\cal H}(t+i\beta)}{\cal A}_S  e^{-i{\cal H}t} e^{\beta {\cal H}} e^{-\beta {\cal H}}
 e^{i{\cal H}t'}{\cal B}_S e^{i{\cal H}t'} \right ] \nonumber \\ 
&=& {\cal Z}^{-1}(\beta) {\rm{Tr}}\left [ e^{i{\cal H}(t+i\beta)}{\cal A}_S  e^{-i{\cal H}(t+i\beta)} \  e^{-\beta {\cal H}}
 e^{i{\cal H}t'}{\cal B}_S e^{i{\cal H}t'} \right ] \nonumber \\ 
&=& {\cal Z}^{-1}(\beta) {\rm{Tr}}\left [ e^{-\beta {\cal H}} e^{i{\cal H}t'}{\cal B}_S e^{i{\cal H}t'} \ 
e^{i{\cal H}(t+i\beta)}{\cal A}_S  e^{-i{\cal H}(t+i\beta)} \ \right ] \nonumber \\ 
&=& {\cal Z}^{-1}(\beta) {\rm{Tr}}\left [ e^{-\beta {\cal H}} {\cal B}_H(t') 
{\cal A}_H(t+i\beta) \right ] \nonumber \\
&=& \langle {\cal B}_H(t') {\cal A}_H(t+i\beta) \rangle_\beta
\, \, , \label{eq16}
\end{eqnarray}
This is called Kubo-Martin-Schwinger (KMS) relation. This relation holds irrespective of Grassmann parities of the operators, {\it viz.}, for 
bosonic as well as fermionic operator.  In the following we have number of points to emphasize:

\begin{enumerate}
\item This KMS relation will lead to periodicity for boson and anti-periodicity for fermions because of commutation and 
anti-commutation relations, respectively.

\item The imaginary temperature $i\beta =i/T$ is connected to the time $t$, which means that the temperature is related to the  
 imaginary time as $\beta=it$  as shown in Fig.~\ref{wick_rot}. This is called  {\it Wick rotation}.

\begin{figure}[h]
\begin{center}
\includegraphics[width=6cm,height=5cm]{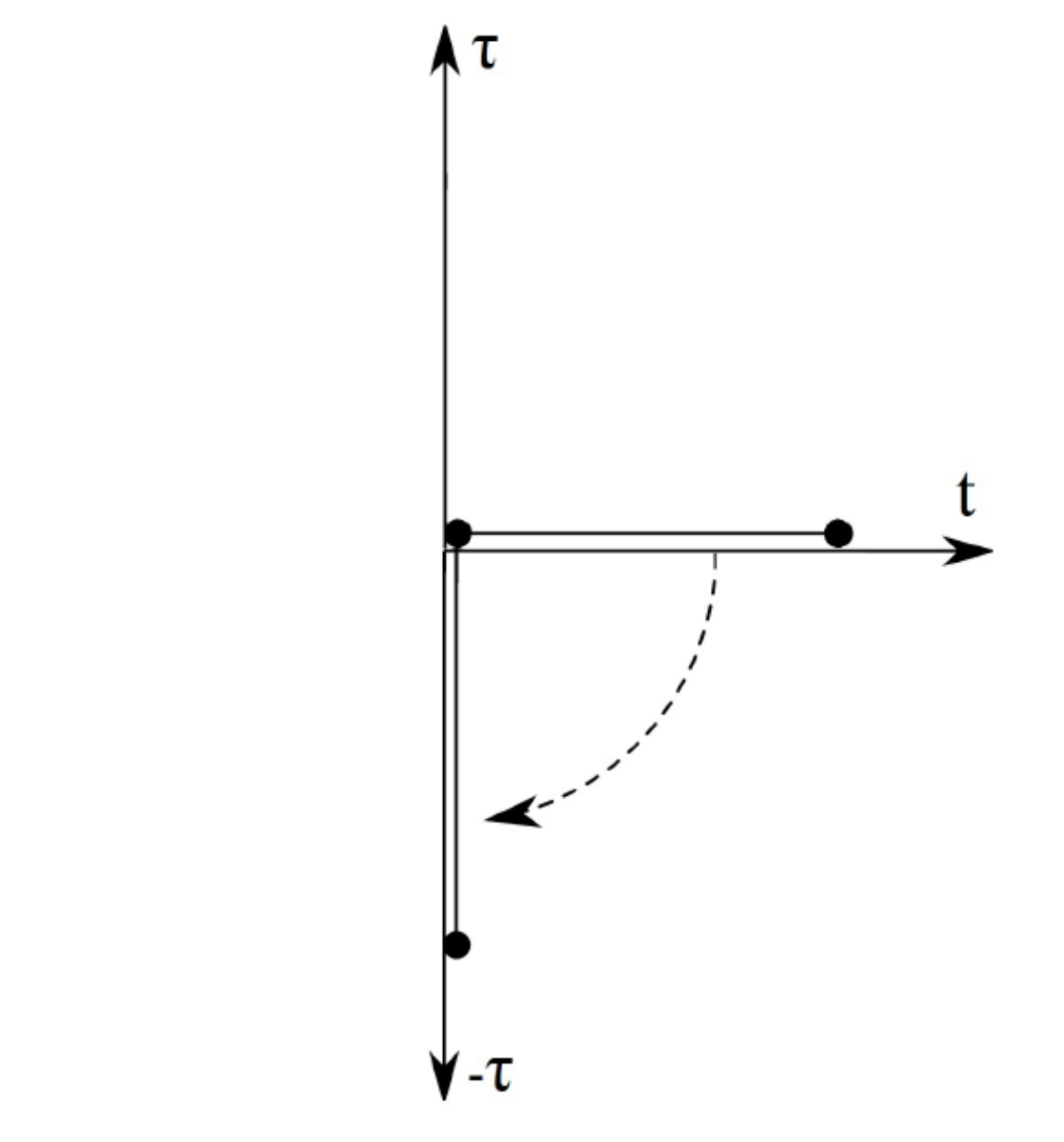}
\caption{The Wick rotation in the  imaginary time axis: $t=-i\tau$}
\label{wick_rot}
\end{center}
\end{figure}

\item
It is important to note that the Boltzmann factor $e^{-\beta {\cal H}}$ acquires the form of a time 
evaluation operator ($e^{-i{\cal H}t}$) for imaginary time ($\beta=it$) through analytic continuation. 
It may be a mere coincidence but there may be some deeper connection which is not known yet! 

\item $\beta=1/T=it \Rightarrow \beta$ becomes finite.
\item
Using the time evolution operator, one can  obtain ${\cal S}-$Matrix, and thus Feynman rules and diagrams.

\item
The Matsubara (imaginary time) formalism~\cite{Matsubara} yields a way of evaluating partition function and other
quantities through a diagrammatic method which is similar to that in zero temperature field theory.
\end{enumerate}
We note that there are two prescriptions for
Matsubara formalism :  i) Operatorial Method  \ \ ii) Path Integral Method.
In the next subsec~\ref{MF_OM}, we will discuss operatorial formalism while the path integral or functional integral formalism will be illustrated
when we discuss partition function of a system later in subsec~\ref{FI_Z} .

\subsection{Matsubara Formalism (Operatorial Method)}
\label{MF_OM}
The Hamiltonian of a system can be decomposed as
\begin{equation}
 {\cal H} = { \cal H}_0 + {\cal H}' \, \, , \label{eq17}
\end{equation}
 where ${\cal H}_0$ and ${\cal H}'$ are the free and interaction parts, 
respectively. 
The purpose of doing so is to describe the interaction in terms of free field (free theory). However,
${\cal H}_0=H_0-\mu N$ for grand canonical ensemble but we do it in general.

Now one can write the density matrix from \eqref{eq1} as
\begin{eqnarray}
\rho (\beta) 
&=& e^{-\beta {\cal H}} = e^{-\beta {\cal H}_0} e^{-\beta {\cal H}'}
= \rho_0 (\beta) {\cal S}(\beta)  \nonumber \\
{\rm{with}} \, \, \, && \rho_0 (\beta) \equiv e^{-\beta {\cal H}_0} ,
 \, \, \, \,  {\rm {and}} \ \ \ \ \ \
{\cal S}(\beta) = e^{\beta {\cal H}_0} \ e^{-\beta {\cal H}}=\rho_0^{-1}(\beta)\rho(\beta). \label{eq18}
\end{eqnarray}

The density matrix can have evolution equation~\cite{Ashok_Das} with $0\leq\tau\leq\beta$:
\begin{eqnarray}
 \frac{\partial \rho_0(\tau)}{\partial \tau} &=&  \frac{\partial}{\partial \tau }(e^{-\tau {\cal H}_0}) \ 
= -{\cal H}_0 \ \rho_0(\tau) \, \nonumber \\
\frac{\partial \rho(\tau)}{\partial \tau} &=&\frac{\partial}{\partial \tau }(e^{-\tau {\cal H}}) \  = -{\cal H} \ \rho (\tau)
= - \left ({ \cal H}_0 + {\cal H}'\right ) \rho(\tau) \, . \label{eq19}
\end{eqnarray}

Now ${\cal S}(\tau)$ satisfies the evolution equation with $0\leq\tau\leq\beta$, following 
(\ref{eq18}) and (\ref{eq19}),  as
\begin{eqnarray}
 \frac{\partial {\cal S} (\tau)}{\partial \tau} &=& \frac{\partial}{\partial \tau } \left[ \rho_0^{-1}(\tau)\rho(\tau)\right]
\ = \ \frac{\partial \rho_0^{-1}(\tau)}{\partial \tau} \rho(\tau) 
+ \rho_0^{-1}\frac{\partial \rho(\tau)}{\partial \tau} \nonumber \\
&=& \frac{\partial}{\partial \tau } \left( e^{\tau {\cal H}_0}\right) \rho(\tau) +\rho_0^{-1}(\tau) \left[-({\cal H}_0+{\cal H}')\right] \rho(\tau)
\nonumber \\
&=&\rho_0^{-1}(\tau) {\cal H}_0 \rho(\tau)  - \rho_0^{-1}(\tau) {\cal H}_0 \rho(\tau) - \rho_0^{-1}(\tau) {\cal H}' \rho(\tau) \nonumber \\
&=& -\rho_0^{-1}(\tau) {\cal H}' \rho(\tau) = -\rho_0^{-1}(\tau) {\cal H}'\rho_0(\tau) {\cal S}(\tau) \nonumber \\
&=& - e^{\tau{\cal H}_0}{\cal H}' e^{-\tau{\cal H}_0} {\cal S}(\tau) \nn\\
&=& -{\cal H}'_{\rm {I}}(\tau) {\cal S}(\tau)\, \, , \label{eq20}
\end{eqnarray}
with modified interaction Hamiltonian  is related to Schr\"odinger picture as 
${\cal H}'_{\rm {I}}(\tau) = e^{\tau {\cal H}_0}{\cal H}' e^{-\tau {\cal H}_0}$ in quantum field theory. We note following points in general:

\begin{enumerate}
\item[] 
Operator in interaction picture in quantum field theory:  ${\cal A}(t)=e^{i{\cal H}_0t}{\cal A}e^{-i{\cal H}_0t}$; 
Adjoint operator: ${\cal A}^\dagger(t)=e^{i{\cal H}_0t}{\cal A}^\dagger e^{-i{\cal H}_0t}$; Transformed adjoint operator:
${\cal A}^T(t)=e^{i{\cal H}_0t}{\cal A}^\dagger e^{-i{\cal H}_0t}$. This implies that ${\cal A}^\dagger(t) = {\cal A}^T(t) $.

\item[] 
On the other hand, the operator in  $\tau$ space:  ${\cal A}(\tau)=e^{\tau{\cal H}_0}{\cal A}e^{-\tau{\cal H}_0}$; 
Adjoint operator: ${\cal A}^\dagger(\tau)=e^{-\tau{\cal H}_0}{\cal A}^\dagger e^{\tau{\cal H}_0}$; 
Transformed adjoint operator: ${\cal A}^T(\tau)=e^{\tau{\cal H}_0}{\cal A}^\dagger e^{-\tau{\cal H}_0}$.  This indicates that 
$ {\cal A}^\dagger (\tau) \ne {\cal A}^T(\tau)$. Thus,
the transformation in $\tau$ is not unitary,  ${\cal A}^\dagger (\tau)\ne {\cal A}(\tau)$, for real $\tau$ because the adjoint of an operator does 
not coincide with the transformed adjoint operator.
This can be avoided if $\tau$ is imaginary, $\tau=it$. Under such rotation a hermitian field remains hermitian
with the {\it appropriate definition of 
hermiticity for complex coordinates}, $\phi^\dagger(\tau=it)=\phi(\tau^*)$,
because the argument becomes complex.
This makes Matsubara formalism an {\it imaginary 
time} formalism as shown pictorially in Fig.~\ref{wick_rot}.
The Matsubara formalism becomes  almost equivalent to zero temperature  field theory with exception that
$\tau$ becomes finite $0\le \tau\le \beta$.
\end{enumerate}

\subsection{Evolution Operator and ${\cal S}$-Matrix}
\label{smatrix}
The evaluation operator in (\ref{eq20}) in a finite interval $0\le \tau\le \beta$  reads (dropped  superscript 'I') as~\cite{Ashok_Das} 
\begin{eqnarray}
 \frac{\partial {\cal S} (\tau)}{\partial \tau} &=& -{\cal H}'(\tau) {\cal S}(\tau)\, \, , \label{eq21}
\end{eqnarray}
With boundary condition ${\cal S}(0) = 1$, ${\cal H}' \rightarrow 0$. At $\tau=0$ all fields are free.
This implies that interaction builds on with the evolution and  leads to ${\cal S}(\tau)$.
Other way, when the interaction is turned on, the free particles move in, interact and then move away 
in the finite interval $0\le \tau\le \beta$ .
 
Integrating (\ref{eq21}) in the interval $0\le \tau\le \beta$, one can get
\begin{equation}
 {\cal S}(\beta)-{\cal S}(0) = - \int_{0}^{\beta} d\tau \, {\cal H}' \, {\cal S}(\tau) \label{eq22}
\end{equation}
This is an exact equation obeyed by ${\cal S}(\tau)$ but cannot be solved.
 Using the same iterative method as zero temperature case~\cite{Ashok_Das}:
\begin{equation}
 {\cal S}(\beta)= {\cal T}\left [ \exp \left ({-\int_0^\beta} {\cal H}' \ d\tau \right) \right ] , \label{eq23}
\end{equation}
where ${\cal T}$ is the time ordered product in imaginary time $\tau=it$.
This (\ref{eq23}) is same as zero temperature field theory with the exception that the imaginary time
integration is in finite interval $0\le \tau \le \beta$.
For a given interaction ${\cal H}'$, one can expand the exponential  and each term in the expansion will lead 
to Feynman diagram of various orders in interaction strength (coupling of the theory). We can now make following comments on Feynman rules:
\begin{enumerate}
\item The interaction vertex is  same as those of zero temperature.
\item The symmetry  factor for a given loop diagram is same as those of zero temperature.
\item What should be the structure of finite temperature propagator in imaginary time is not clear yet!.
 \item What should be the form of finite temperature loop integral in imaginary time is not clear yet.
\end{enumerate}

\subsection{Two-Point Correlation Function: Green's Function}          
\label{cfgf}
\subsubsection{Green's function at $T=0$ in Minkowski time (real time)}
\label{gft0}
The Greens function for $T=0$ in Minkowski time is defined~\cite{pbpal}  as
\begin{eqnarray}
G(X,X') &=& 
\left \langle 0| {\cal T}_t \left [  \Phi(X) \Phi(X') \right ] |0 \right \rangle,
 \label{eqgt0}
\end{eqnarray}
where $X\equiv (x_0,\bm{\vec x})$ and  the time ordered (${\cal T}_t$) product of two fields in real time:
\begin{eqnarray}
  {\cal T}_t \left [  \Phi(X) \Phi(X') \right ]= \Theta(t-t') \Phi(X)\Phi (X') + \Theta(t'-t) \Phi(X')\Phi (X), \label{eqgt01}
\end{eqnarray}
where the scalar field can be expressed as
\begin{eqnarray}
 \Phi(X)&=& \int \frac{d^3k}{(2\pi)^{3/2}} \frac{1}{{(2\omega_k})^{1/2}} \left[a({k})e^{-iK \cdot X} + a^\dagger({k})e^{iK \cdot X}\right],
 \label{eqgt02}
\end{eqnarray}
with $K\equiv(k_0,\bm{\vec k})$, $\omega_k=\sqrt{{k}^2+m^2}$, $a({k})$ is the annihilation operator and $a^\dagger({k})$ is 
the creation operator.
 
 The vacuum is defined as $a({k})|0\rangle=0$.
  
  The multiparticle states can be written from \eqref{eq9d} as
  \begin{subequations}
  \begin{align}
   |n\rangle = |n_1({k}_1), n_2({k}_2), n_3({k}_3) \cdots\cdots\rangle &= \prod_i \,
   \frac{\left [ a^\dagger({k}_i)\right ]^{n_i({k}_i)}}{\sqrt{n_i({k}_i) !}}|0\rangle, \label{eqgt03}\\
   a({k_i}) |n_i(k_i)\rangle &= \sqrt{n(k_i)} |n_i(k_i)-1\rangle, \label{eqgt03a} \\
   a^\dagger({k_i}) |n_i(k_i)\rangle &= \sqrt{n(k_i)+1} |n_i(k_i)+1\rangle .
   \label{eqgt03b}
   \end{align}
  \end{subequations}
Using (\ref{eqgt0}), \eqref{eqgt02} , \eqref{eqgt03},   \eqref{eqgt03a} and  \eqref{eqgt03b}, 
the Green's function\footnote{One can also get it as a solution of Klein Gordon equation with a unit source term.} 
can be written as
\begin{eqnarray}
G(X-X') &=& \int \frac{d^4K}{(2\pi)^4} \frac{e^{-iK\cdot (X-X')}}{K^2-m^2} = \int \frac{d^4K}{(2\pi)^4} \, e^{-iK\cdot (X-X')} \, G(K) , \label{eqgt04}
\end{eqnarray}
where the momentum space Green's function is given as
\begin{equation}
 G(K)=\frac{1}{K^2-m^2} = \frac{1}{k_0^2-\omega_k^2}\, .  \label{eqgt05}
\end{equation}
We note that though Green's function is a two-point function, it depends on the differences of the two 
endpoints because of translational invariance.
Now, this $G(X-X')$ describes the free propagation of scalar particle from $X'$ to 
$X$ for $x_0=t > x'_0=t'$ implying creation at $X'$ and destruction at $X$. 

Now $G(K)$ has poles at $k_0=\pm\omega_k$ on the real axis. 
With Feynman prescription  one can write $G(K)$ in (\ref{eqgt05}) by shifting its poles in complex plane  as
\bea
\Delta_F(K) &=& \frac{1}{K^2-m^2+i\epsilon'} = \frac{1}{k_0^2- (\omega_k-i\epsilon)^2} \nn \\
&=& \frac{1}{2\omega_k}\left[ \frac{1}{k_0-\omega_k+i\epsilon} - \frac{1}{k_0+\omega_k-i\epsilon} \right],
\eea
where $\Delta_F(K)$ is complex and  $\epsilon$ and $\epsilon'$ are small number and related by $\epsilon'=2\epsilon \omega_k$.
However,  we do not distinguish them  as $\epsilon\rightarrow 0$ at the end of the calculation.
Now (\ref{eqgt04}) can be written with Feynman prescription as
\begin{eqnarray}
\Delta_F(X-X') &=& \int \frac{d^4K}{(2\pi)^4} \frac{e^{-iK\cdot (X-X')}}{K^2-m^2+i\epsilon'} = \int \frac{d^4K}{(2\pi)^4} \, e^{-iK\cdot (X-X')} \, \Delta_F(K) , \label{eqgt05a}
\end{eqnarray}
Integrating over $k_0$ in complex $k_0$ plane~\cite{pbpal}, one finds
\begin{eqnarray}
 \Delta_F(X-X')&=&  -i \int \frac{d^3k}{(2\pi)^3} \frac{1}{2\omega_k} \left. \left[ e^{-iK\cdot (X-X')} \Theta(t-t') +   e^{iK\cdot (X-X')} 
 \Theta(t'-t) \right ]  \right |_{k_0=\omega_k} \nn \\
  i\Delta_F(X-X')&=&  \int \frac{d^3k}{(2\pi)^3} \frac{1}{2\omega_k} \left. \left[ e^{-iK\cdot (X-X')} \Theta(t-t') +   e^{iK\cdot (X-X')} 
 \Theta(t'-t) \right ]  \right |_{k_0=\omega_k}  \nn \\
&=& \left \langle 0| {\cal T}_t \left [  \Phi(X) \Phi(X') \right ] |0 \right \rangle 
 , \label{eqgt05b}
\end{eqnarray}
which is the Feynman propagator.
Now comparing (\ref{eqgt05b}) and (\ref{eqgt0}), the Green's function becomes
\begin{eqnarray}
  G(X-X')&=&  \int \frac{d^3k}{(2\pi)^3} \frac{1}{2\omega_k} \left. \left[ e^{-iK\cdot (X-X')} \Theta(t-t') +   e^{iK\cdot (X-X')} 
 \Theta(t'-t) \right ]  \right |_{k_0=\omega_k}   
 , \label{eqgt06}
\end{eqnarray}
where the first part in right side  is for retarded time ($t> t'$) 
whereas the second part is for advanced time ($t'>t)$.

\subsubsection{Green's function at $T\ne0$ in Minkowski time (real time)}
\label{gftne0}
Greens function in thermal environment in Minkowski time can be written as
\begin{eqnarray}
G_\beta(X,X') =  i\Delta_F(X-X') &=&
\left \langle | {\cal T}_t \left [  \Phi(X) \Phi(X') \right ] | \right \rangle_\beta \nn \\
&=& \frac{1}{{\cal Z}(\beta)} {\rm{Tr}}\left ( e^{-\beta {\cal H}} {\cal T}_t \left [  \Phi(X) \Phi(X') \right ] \right ) \nn \\
&=& \frac{1}{{\cal Z}(\beta)} \sum_n 
\left \langle n | {\cal T}_t \left [  \Phi(X) \Phi(X') \right ] | n \right  \rangle e^{-\beta E_n},
\label{eqgt1}
\end{eqnarray}
where $\langle \rangle_\beta$ is the thermal expectation value and ${\cal Z}(\beta)$ is the partition function. 
Trace stands for sum over a complete many-particles states in Hilbert space, which is replaced by a sum that 
runs over a complete many-particle states $|n\rangle$  weighted by the Boltzmann factor $e^{-\beta E_n}$ in thermal 
environment\footnote{We note that the Lorentz invariance is broken at $T\ne 0$ which we will discuss later in details. However,
$E$ can be written  in the rest frame of the medium (heat bath) as $E=u_\mu K^\mu = u\cdot K$, where $u$ is four velocity of the  
medium (heat bath) in its rest frame with $u_\mu =(1,0,0,0)$.}, a little different than $T=0$ case.

We would now like to compute the Green's function~\cite{Thoma,Thoma1} for $t>t'$ using (\ref{eqgt01}) and (\ref{eqgt02}) as
\begin{eqnarray}
 G_\beta^>(X,X') &=& \frac{1}{{\cal Z}(\beta)} \int \frac{d^3k}{(2\pi)^{3/2}} \ \frac{d^3k'}{(2\pi)^{3/2}} \ 
 \frac{1}{{(2\omega_k})^{1/2}} \ \frac{1}{{(2\omega_{k'}})^{1/2}} \sum_n e^{-\beta E_n} \nn \\
 &&\times  \left \langle n \left | \Big\{ a({k})e^{-iK\cdot X} + a^\dagger({k})e^{iK\cdot X} \Big \}
\left \{ a({k})e^{-iK'\cdot X'} + a^\dagger({k})e^{iK'\cdot X'} \right \} \right | n \right  \rangle \nn \\
&=& \frac{1}{{\cal Z}(\beta)} \int \frac{d^3k}{(2\pi)^{3/2}} \ \frac{d^3k'}{(2\pi)^{3/2}} \ 
 \frac{1}{{(2\omega_k})^{1/2}} \ \frac{1}{{(2\omega_{k'}})^{1/2}} \sum_n e^{-\beta E_n}  \nn \\
  &&   \left \langle n \left |  a({k})a({k'})e^{-iK\cdot X} e^{-iK'\cdot X'} 
 + a^\dagger({k})a({k'})e^{iK\cdot X} e^{-iK'\cdot X'} \right.\right. \nn \\
&& \left.\left. + a({k})a^\dagger ({k'})e^{-iK\cdot X} e^{iK'\cdot X'} 
+a^\dagger({k})a^\dagger({k'})e^{iK\cdot X} e^{iK'\cdot X'}\right | n \right  \rangle . \label{eqgt2}
\end{eqnarray}

On using (\ref{eqgt03a}), (\ref{eqgt03b}) and the orthonormal conditions, the second and third terms in (\ref{eqgt2}) would survive as 
\begin{eqnarray}
 G_\beta^>(X-X')\!\!
 &=&\!\!\frac{1}{{\cal Z}(\beta)}\int \frac{d^3k}{(2\pi)^{3/2}}\!\frac{d^3k'}{(2\pi)^{3/2}} \ 
 \frac{1}{{(2\omega_k})^{1/2}} \ \frac{1}{{(2\omega_{k'}})^{1/2}} \sum_n e^{-\beta E_n} \nn \\
 &&\times \Big [ \sqrt{n(k) n(k')} \ \delta^3 (\bm{\vec k}-\bm{\vec k'}) e^{i (K \cdot X-K'\cdot X')}  \nn \\
 && +
  \sqrt{[n(k)+1][ n(k')+1]} \ \delta^3 (\bm{\vec k}-\bm{\vec k'}) e^{-i(K \cdot X+K'\cdot X')}
 \Big  ] \nn \\
&=& \frac{1}{{\cal Z}(\beta)} \int \frac{d^3k}{(2\pi)^{3}} \ \frac{1}{2\omega_k} \ \sum_n e^{-\beta E_n}
\Big [(n(k)+1) \ e^{-iK\cdot(X-X')} \nn \\
 &&+ n(k)\ e^{iK\cdot(X-X')} \Big] . \label{eqgt3}
\end{eqnarray}
Now we use
\begin{eqnarray}
 \frac{1}{{\cal Z}(\beta)} \ \sum_n n(k) e^{-\beta E_n}  &=& \frac{1}{{\cal Z}(\beta)} \ \sum_n n(k) e^{-\beta \omega_k n(k)} 
 = \frac{1}{{\cal Z}(\beta)} \ \sum_n n \left [e^{-\beta \omega_k}\right]^n  \nn \\
 &=& \frac{1}{e^{\beta\omega_k}-1} \equiv n_B(\omega(k)), \label{eqgt4} \\
 \frac{1}{{\cal Z}(\beta)} \ \sum_n e^{-\beta E_n} &=& 1 , \label{eqgt5}
\end{eqnarray}
where $n_B(\omega_k)$ is the Bose-Einstein distribution.

Combining (\ref{eqgt4}) and (\ref{eqgt5}) with (\ref{eqgt3}), we can have 
\begin{eqnarray}
  G_\beta^>(X-X') &=& \int \frac{d^3k}{(2\pi)^{3}} \ \frac{1}{2\omega_k} \
  \Big [(1+n_B) \ e^{-iK\cdot(X-X')} + n_B\ e^{iK\cdot(X-X')} \Big], \label{eqgt6}
\end{eqnarray}
which is the Green's function for $t>t'$.  At $T=0$, $n_B(\omega_k)=0$ and (\ref{eqgt6}) reduces to
\begin{eqnarray}
  G_\beta^>(X-X') &=& \int \frac{d^3k}{(2\pi)^{3}} \ \frac{1}{2\omega_k} \  e^{-iK\cdot(X-X')}, \label{eqgt7}
\end{eqnarray}
which agrees with the first term in (\ref{eqgt06}).

Now the physical interpretations of Eq.(\ref{eqgt6}) are given below:
\begin{enumerate}
 \item[$\bullet$] At finite $T$, like zero temperature ($T=0$),  a scalar field which is created at $X'$, i.e., $t'$ propagates
 to $X$  at $t$ and then annihilated.
 \item[$\bullet$] Besides spontaneous creation at $X'$, there will also be induced creation (first term in (\ref{eqgt6}) involving $n_B$) 
 at $X'$ and absorption (second term involving $n_B$) at $X$ due to the presence of heat bath.
\end{enumerate}

Similarly, the Green's function for $t'>t$ can be obtained from the symmetry property ($ K \rightarrow -K$) as
\begin{eqnarray}
  G_\beta^<(X-X') &=& \int \frac{d^3k}{(2\pi)^{3}} \ \frac{1}{2\omega_k} \
  \Big [n_B \ e^{-iK\cdot(X-X')} + (1+ n_B) \ e^{iK\cdot(X-X')} \Big], \label{eqgt8}
\end{eqnarray}
which reduces to the second term in (\ref{eqgt06}) at $T=0$ as
\begin{eqnarray}
  G_\beta^<(X-X') &=& \int \frac{d^3k}{(2\pi)^{3}} \ \frac{1}{2\omega_k} \  e^{iK\cdot(X-X')}. \label{eqgt9}
\end{eqnarray}

\subsubsection{Green's function at $T\ne0$ in Euclidean time (imaginary time)}
\label{gfet}

Following (\ref{eq16}), the thermal Green's function can be written as
\begin{eqnarray}
 G_\beta(\tau,\tau') &=& \left \langle {\cal T} \left [  \Phi_H(\tau) \Phi_H(\tau') \right ] \right \rangle_\beta \nonumber \\
&=& {\cal Z}^{-1}(\beta) \ {\rm{Tr}} \left ( e^{-\beta {\cal H}} {\cal T}\left [ \Phi_H(\tau) \Phi_H(\tau') \right]\right ) \label{eq24}
\end{eqnarray}
with $\Phi_H(\tau) = e^{\tau {\cal H}} \, \Phi \, e^{-\tau{\cal H}}$. The time variables $\tau$, $\tau'$ lie between 
$0\le \tau\le \beta$ and $0\le \tau'\le \beta$. ${\cal T}$ is imaginary time ordering. 
We have suppressed spatial dependence as it could be included any time.
The field $\Phi$ can represent bosonic/fermionic filed. The spinor indices are also suppressed here but can be taken care wherever
needed.
Imaginary time ordering ${\cal T}$ is same as zero temperature field theory. Then
\begin{eqnarray}
 {\cal T}\left [ \Phi_H(\tau) \ \Phi_H(\tau') \right ] = \Theta(\tau-\tau') \Phi_H(\tau) \ \Phi_H(\tau') \pm 
\Theta(\tau'-\tau) \Phi_H(\tau') \ \Phi_H(\tau), \label{eq25}
\end{eqnarray}
where $\pm$ refer to boson/fermion, respectively.
 
Though Green's function is a two-point function, it depends on the differences of the two endpoints because of 
translational invariance as  $  G_\beta(\bm{\vec x}-\bm{\vec x}';\tau -\tau')$.
 The time variables $\tau$, $\tau'$ lie between 
$0\le \tau\le \beta$ and $0\le \tau'\le \beta$. This implies that two-point function has $-\beta\le\tau-\tau'\le \beta$.

\subsection{Periodicity (Anti-periodicity) of the Green's Function}
\label{period}        
The thermal Green's function~\cite{Ashok_Das}  from (\ref{eq24}) for $\tau > \tau'$:
\begin{eqnarray}
G_\beta(\bm{\vec x},\bm{\vec x}';\tau,\tau')&=& {\cal Z}^{-1}(\beta) \ {\rm{Tr}} 
\left ( e^{-\beta {\cal H}} {\cal T}\left [ \Phi_H(\bm{\vec x},\tau) \Phi_H(\bm{\vec x}',\tau') \right]\right ) \nonumber \\
&{=}\atop {\tau >\tau'} &{\cal Z}^{-1}(\beta) \ {\rm{Tr}}\left [ e^{-\beta {\cal H}} \Phi_H(\bm{\vec x},\tau) \Phi_H(\bm{\vec x'},\tau') 
\Theta(\tau-\tau') \right] \nonumber \\
&{=}\atop {\tau >\tau'} &{\cal Z}^{-1}(\beta) \ {\rm{Tr}}\left [ \Theta(\tau-\tau')  \Phi_H(\bm{\vec x'},\tau') e^{-\beta {\cal H}} \Phi_H(\bm{\vec x},\tau)
 \right] \nonumber \\
&{=}\atop {\tau >\tau'} &{\cal Z}^{-1}(\beta) \ {\rm{Tr}}\left [ \Theta(\tau-\tau') e^{-\beta {\cal H}} e^{\beta {\cal H}}  
\Phi_H(\bm{\vec x'},\tau') e^{-\beta {\cal H}} \Phi_H(\bm{\vec x},\tau)
 \right] \nonumber \\
&{=}\atop {\tau >\tau'} &{\cal Z}^{-1}(\beta) \ {\rm{Tr}}\left [ \Theta(\tau-\tau')  e^{-\beta {\cal H}}
\Phi_H(\bm{\vec x'},\tau'+\beta)  \Phi_H(\bm{\vec x},\tau)
 \right] \nonumber \\
&{=}\atop {\tau >\tau'} & \pm \ {\cal Z}^{-1}(\beta) \ {\rm{Tr}}\left [e^{-\beta {\cal H}}  \Theta(\tau-\tau')   \Phi_H(\bm{\vec x},\tau) \Phi_H(\bm{\vec x'},\tau'+\beta) 
 \right] \nonumber \\
&=& \pm \ G_\beta(\bm{\vec x},\bm{\vec x'};\tau ,\tau'+\beta).
\label{eq26}
\end{eqnarray}
We have used the cyclic properties of the trace, inserted the unit operator $1 =  e^{-\beta {\cal H}} e^{\beta {\cal H}}$, and used the time
evolution of the state: $\Phi_H(\bm{\vec x'},\tau'+\beta)= e^{\beta {\cal H}} \Phi_H(\bm{\vec x'},\tau') e^{-\beta {\cal H}} $.

Since the Green's function alters sign for Dirac field after one period of $\beta$, this means that 
the Dirac fields must be antiperiodic in imaginary time as $\Phi(\bm{\vec x}, \tau)= - \Phi(\bm{\vec x}, \tau+\beta)$, whereas bosonic fields are periodic 
as they do not change sign as $\Phi(\bm{\vec x}, \tau)= \Phi(\bm{\vec x}, \tau+\beta)$.    
Since we didn't touch upon space direction, it remains unaffected as: $-\infty \le {x}\le \infty \Rightarrow$ open.

In $T=0$ (Minkowski space-time) both space and time remain open: $-\infty \le {x}\le \infty$ and $-\infty\le t \le \infty$.
Topology: Structure of space-time at $R^4=R^3\times R^1$  where both space and time are open and  in equal footing.
In $T\ne 0 $ (Euclidean space; imaginary time):  space remains open $-\infty \le {x}\le \infty \Rightarrow R^3$.  Time remains closed:  
$0\le \tau \le \beta$ $\Rightarrow R^1 \rightarrow S^1$ (circle).
Topology: Structure of space-time at $T\ne0$ is transformed as $R^4=R^3\times R^1 \Rightarrow R^3\times S^1$. This changes the temporal components leaving
the spatial components unaffected. {\it It amounts to decoupling of  space and time and the theory is no longer Lorentz invariant.}

Further, the chemical potential can also be inducted by transforming the temporal component of 
the gauge field, through a substitution $\partial_0-i\mu$ in the Lagrangian. Such substitution  changes  the temporal 
component while leaving the spatial components of the gauge field unaltered. It also decouples space and time 
and the theory is no more Lorentz invariant. {\it In addition to explicitly breaking the Lorentz invariance, the 
presence of a chemical potential may additionally break other internal symmetries.}

At $T \, {\mbox{and}} \, \mu \ne 0$  field theory is equivalent to quantising a quantum system in finite box, i.e., one dimensional 
box in $\tau$ direction ($0\le \tau \le \beta$) but space remains open, i.e., $ R^3\times S^1$ in which Lorentz invariance is broken.

\subsection{Discrete Frequency (Matsubara Frequency)}       
\label{dfmf}
The Fourier decomposition of Green's function~\cite{Ashok_Das}  in real time (Minkowski space-time)

\begin{eqnarray}
 G(\bm{\vec x}, \bm{\vec x'};t, t' )& = & \int \frac{d^4K}{(2\pi)^4} \, e^{-iK.(X-X')} G(K) \nn \\
  &=& \int \frac{d^3k}{(2\pi)^3} \, 
 e^{i \bm{\vec k}\cdot (\bm{\vec x} -\bm{\vec x}')}
\int \frac{dk_0}{2\pi} \, e^{-i k_0 (x_0 - x'_0)}  G(k_0,\bm{\vec k}), \label{eq27}
\end{eqnarray}
where a four vector is defined as $ Q\equiv (q_0, {q})$ and $G(K)$ is momentum space Green's function. 

For convenience we assume $X'=0 \Rightarrow x'=0,\, x_0'=t'=0$:
\begin{eqnarray}
 G(\bm{x},0;t,0)&=&\int \frac{d^3k}{(2\pi)^3} \, 
 e^{i \bm{\vec k} \cdot \bm{\vec x} }
\int \frac{dk_0}{2\pi} \, e^{-i k_0 t}  G(k_0,\bm{\vec k}). \label{eq28}
\end{eqnarray}
Now switching over from Minkowski space (real time) to Euclidean space (imaginary time):
 $t\rightarrow -i\tau$; \, \, $k_0 \rightarrow ik_4$, \, \, $\bm{\vec x} \rightarrow \bm{\vec x}$ and 
 $\bm{\vec k} \rightarrow \bm{\vec k}$ .
With this (\ref{eq28}) reads as
\begin{eqnarray}
 G_\beta(\bm{\vec x}, \tau)& = &  \int \frac{d^3k}{(2\pi)^3} \, e^{i \bm{\vec k}\cdot \bm{\vec x} }
\int \frac{d (ik_4)}{2\pi} \, e^{-i(i k_4) (-i\tau)}  G(k_0=ik_4,\bm{\vec k}). \label{eq29}
\end{eqnarray}

 Since $\tau$ is finite ($0\le \tau\le \beta$), the corresponding Fourier transform reads as
 \be
 \int \frac{d (ik_4)}{2\pi} f(ik_4,\bm{\vec k}) \rightarrow \frac{1}{\beta} \sum_{n=-\infty}^{n=+\infty} f(k_0=ik_4=i\omega_n, \bm{\vec k}). \label{eq29a}
 \ee
The finiteness of Euclidean time $\tau$ will result in discrete frequency with a $\sum$ over it. 

Now (\ref{eq29}) reads as
\begin{eqnarray}
 G_\beta(\bm{\vec x}, \tau)& = &\frac{1}{\beta} \ \sum_{n=-\infty}^{n=+\infty} \int \frac{d^3k}{(2\pi)^3} \, e^{i \bm{\vec k}\cdot \bm{\vec x} }
\ e^{- i\omega_n \tau}  G_\beta(k_0=i\omega_n,\bm{\vec k}). \label{eq30}
\end{eqnarray}
This indicates that going from Minkowski to Euclidean time, one needs to replace
 \be
 \int \frac{d^4K}{(2\pi)^4}  \rightarrow \frac{1}{\beta} \sum_{n=-\infty}^{n=+\infty}\int \frac{d^3k}{(2\pi)^3} . \label{eq29b}
 \ee
We can now drop the space dependent part in (\ref{eq30}) as it is irrelevant for the time being but can be put back when required:
\begin{eqnarray}
 G_\beta(\tau)& = & \frac{1}{\beta} \ \sum_{m=-\infty}^{m=+\infty} 
\ e^{- i \omega_m \tau}  G_\beta(i\omega_m), \label{eq31}
\end{eqnarray}
where the inverse transformation is given as
\be
G_\beta(i\omega_m)= \frac{1}{2} \int\limits_{-\beta}^{+\beta}\ d\tau \ e^{i\omega_m \tau} \ G_\beta (\tau), \label{eq32}
\ee
with $\omega_m=m\pi/\beta ; \, \, m=0, \pm 1, \, \pm 2, \, \pm 3, \cdots $ as all integer modes are allowed in Fourier transformation.
However, the value of $m$ will be restricted according to (\ref{eq26}) because the two points function for bosonic fields satisfies periodicity 
condition as , $G_\beta(\tau)= G_\beta(\tau+\beta)$, whereas  the two points function for fermionic fields satisfies anti-periodicity condition as 
 $G_\beta(\tau)= -G_\beta(\tau+\beta)$.

We now split (\ref{eq32}) as 
\begin{eqnarray}
G_\beta(i\omega_m)&=& \frac{1}{2} \int\limits_{-\beta}^{0} d\tau \ e^{i\omega_m \tau} \ G_\beta (\tau)
+  \frac{1}{2} \int\limits_{0}^{\beta} d\tau \ e^{i\omega_m \tau} \ G_\beta (\tau) \nn \\
&=& \pm \frac{1}{2} \int\limits_{-\beta}^{0} d\tau \  e^{i\omega_m \tau} \ G_\beta (\tau+\beta)
+  \frac{1}{2} \int\limits_{0}^{\beta} d\tau \ e^{i\omega_m \tau} \ G_\beta (\tau) \label{eq33}
\end{eqnarray}
where in the first term $`+$' refers to boson whereas $`-$' refers to fermion. Now we make changes $\tau \rightarrow -\tau$ in 
the first term as
\begin{eqnarray}
G_\beta(i\omega_m)&=& \pm \frac{1}{2} \int\limits_{0}^{\beta} d\tau \ e^{-i\omega_m \tau} \ G_\beta (-\tau+\beta)
+  \frac{1}{2} \int\limits_{0}^{\beta} d\tau \ e^{i\omega_m \tau} \ G_\beta (\tau). \label{eq34}
\end{eqnarray}
Again in the first term of (\ref{eq34}) we change a variable as $-\tau+\beta=\tau \Rightarrow $ upper limit: $ 0$ and lower limit: $\beta$,
then 
\begin{eqnarray}
G_\beta(i\omega_m)&=& \pm \frac{1}{2} \int\limits_{0}^{\beta} d\tau \ e^{i\omega_m (\tau-\beta)} \ G_\beta (\tau)
+  \frac{1}{2} \int\limits_{0}^{\beta} d\tau \ e^{i\omega_m \tau} \ G_\beta (\tau) \nn\\
&=& \frac{1}{2} \left(1\pm e^{-i\omega_m \beta}\right )\ \int\limits_{0}^{\beta} d\tau \ e^{i\omega_m \tau} \ G_\beta (\tau) 
= \frac{1}{2} \left(1\pm e^{-i \frac{m\pi}{\beta} \beta}\right )\ \int\limits_{0}^{\beta} d\tau \ e^{i\omega_m \tau} \ G_\beta (\tau) \nn \\
&=& \frac{1}{2} \Big (1 \pm (-1)^m \Big )\ \int\limits_{0}^{\beta} d\tau \ e^{i\omega_m \tau} \ G_\beta (\tau) . \label{eq35}
\end{eqnarray}
We note that $`+$' is for boson. This implies that  $m$ has to be  even $\Rightarrow \omega_n=\frac{2\pi n}{\beta}$ with $n\in Z$.
Now, $`-$' is for fermion which implies that  $m$ has to be odd $\Rightarrow \omega_n=\frac{(2 n+1)\pi}{\beta}$ with $n\in Z$.
Thus,
\begin{eqnarray}
 G_\beta(\tau)& = &  \frac{1}{\beta} \ \sum_{n=-\infty}^{n=+\infty} 
\ e^{- i \omega_n \tau}  G_\beta(i\omega_n) \nn \\
G_\beta(i\omega_n) &= & \int\limits_{0}^{\beta}\ d\tau \ e^{i\omega_n \tau} \ G_\beta (\tau), \label{eq36}
\end{eqnarray}
with the discrete Matsubara frequency $\omega_n$ as the characteristics of the imaginary time and represented by 
\[
k_0=ik_4=i\omega_n= \left \{ \begin{array}{ll}
                  \frac{2n \pi i }{\beta} & \mbox{for boson ,} \nonumber \\
                  \frac{(2n+1)\pi i}{\beta} & \mbox{for fermion.}\nonumber
		   \end{array} 
		\right.  \label{eq37} 
\]
Now, the complete Green's function:
\begin{eqnarray}
 G_\beta(\bm{\vec x}, \tau) &=& \frac{1}{\beta} \sum_{n=-\infty}^{+\infty} \int\frac{d^3k}{(2\pi)^3} 
 e^{-i(\omega_n\tau-\bm{\vec k} \cdot \bm{\vec x})} G_\beta(i\omega_n, \bm{\vec k}),  \label{eq38a} \\
 G_\beta(i\omega_n,\bm{\vec k}) &=& \int\limits_0^\beta d\tau \int d^3 {x} \,
 e^{i(\omega_n\tau-\bm{\vec k} \cdot \bm{\vec x})} G_\beta(\bm{\vec x}, \tau). \label{eq38b}
\end{eqnarray}

\newpage
\subsection{Dictionary: $T=0$ to $T\ne 0$ Field Theory (Imaginary Time)}
\label{dic}
\def\multic{ \multicolumn{1}{|c|} }
\def\multboth{ \multicolumn{1}{|c|} }

\def\beudp{\noindent
            \begin{center}
            \begin{tabular}{|c|c|}
            \hline
             \multic{  ${\bm{ T=0; \,\, -\infty\le t\le \infty} }$}
             &\multic{  ${\bm{T\ne 0; \, \, \, \, 0\le \tau (=it) \le \beta}}$ }   
                 \\
            \hline}

\def\eeud{\hline
            \end{tabular}
            \end{center} }
\beudp
Topology: $R^4=R^3\times R^1$
& Topology: $R^4=R^3\times S $\\
{ $-\infty\le \bm{\vec x},\, t \le \infty$ (open)} & { $-\infty\le \bm{\vec x} \le \infty$ (open);  $0\le \tau  \le \beta$ (closed)}\\
\hline
Operator in Interaction Picture & Operator in Interaction Picture \\
${\cal A}_I(t)=e^{i{\cal H}_0t}{\cal A}_Se^{-i{\cal H}_0t}$ & ${\cal A}_I(\tau)=e^{\tau{\cal H}_0}{\cal A}_Se^{-\tau{\cal H}_0}$ \\
${\cal A}_I^T(t)=e^{i{\cal H}_0t}{\cal A}_S^\dagger e^{-i{\cal H}_0t} \Rightarrow {\cal A}_I^\dagger(t) = {\cal A}_I^T(t) $
& ${\cal A}_I^T(\tau)=e^{\tau{\cal H}_0}{\cal A}_S^\dagger e^{-\tau{\cal H}_0} \Rightarrow {\cal A}_I^\dagger (\tau) \ne {\cal A}_I^T(\tau)$\\
Transformation is Unitary & It's not Unitary \\ 
\hline
Interaction Hamiltonian in IP& Interaction Hamiltonian in IP\\
 ${\cal H}_I'(t)=e^{i{\cal H}_0t}{\cal H}_S'e^{-i{\cal H}_0t}$ & ${\cal H}'_I(\tau)=e^{\tau {\cal H}_0}{\cal H}_S'e^{-\tau {\cal H}_0}$\\
\hline
Time evolution & Time evolution \\
  $i \frac{\partial \Phi(t)}{\partial t}={\cal H}'_I(t) \Phi(t)$ & $\frac{\partial {\cal S}(\tau)}{\partial \tau}=-{\cal H}_I'(\tau) {\cal S}(\tau)$ \\
{ $\Phi(t)$ is built up from $\Phi(0)$}& { $ S(\tau)$ is built up from ${\cal S}(0)$}\\
\hline
{ ${\cal S}$-Matrix}& { ${\cal S}$-Matrix}\\
 ${\cal S} = {\cal T}\left[\exp\left(-i\int_{-\infty}^{\infty} {\cal H}'_I(t) dt \right)\right]$ 
&$ {\cal S}(\beta) = {\cal T}\left[\exp\left(-\int_{0}^{\beta} {\cal H}'_I(\tau) d\tau \right)\right]$ \\
${\cal T}$: Time ordering &${\cal T}$: Imaginary time ordering \\
\hline
Wicks Theorem &  Same \\
\hline
Vertex &  Same \\
\hline
Symmetry factor &  Same \\
\hline
{ Boundary Conditions}& { Boundary Conditions}\\
 $G(X,X')$ &$G_\beta(\bm{\vec x},\bm{\vec x'};\tau,\tau') 
= \pm G_\beta(\bm{\vec x},\bm{\vec x}';\tau,\tau'+\beta) $ \\
$-\infty<X,X'<\infty$ & {(+) for  boson; (-) for fermion }\\
\hline
{ Green's Function}& {Green's Function}\\
 & $G_\beta(\bm{\vec x},\bm{\vec x'};\tau,\tau') = \langle |{\cal T}\left[\Phi_H(\bm{\vec x},\tau)\Phi_H^\dagger(\bm{\vec x'},\tau')\right]|\rangle_{\beta}$ \\
 $G(X,X') = \langle 0|{\cal T}\left[\Phi(X)\Phi(X')\right]|0\rangle$  
 & $G_\beta(\bm{\vec x},\bm{\vec x'};\tau,\tau') = \frac{\langle |{\cal T}\left[\Phi_I(\bm{\vec x},\tau)\Phi_I^\dagger(\bm{\vec x'},\tau'){\cal S}(\beta)\right]|\rangle_{\beta ,0}}{\langle {\cal S}(\beta)\rangle_{\beta,0}}$ \\
& $\beta, 0$ indicates the thermal expectation with free theory~\cite{Ashok_Das}\\
\hline
{ Propagator (Momentum Space GF)}& { Propagator (Momentum Space GF)}\\
 $i\Delta_F(K)=G(K), \, \, \,  \,   K \equiv (k_0,\bm{\vec k})$, \, $k = |\bm{\vec{ k}}|$  & $i\Delta_F(k_0,\bm{\vec k})=G(k_0,\bm{\vec k})$, \,\, \, $k = |\bm{\vec{ k}}|$ \\
{ $k_0$ is continuous} &  
$
k_0=ik_4=i\omega_n= \left \{ \begin{array}{ll}
                  \frac{2n \pi i }{\beta} & \mbox{for boson ,} \nonumber \\
                  \frac{(2n+1)\pi i}{\beta} & \mbox{for fermion.}\nonumber
		   \end{array} 
		\right.  
$
\\
{ $k$ is continuous}& { $k$ is continuous}
\\
\hline
{ Loop integral}& { Loop integral}\\
  $\int \frac{d^4K}{(2\pi)^4}; $ 
& $\frac{1}{\beta}\sum\!\!\!\!\!\!\!\int\limits_{k_0}  \frac{d^3{k}}{(2\pi)^3}$; \\
{ $k_0$ is continuous} &  
$
k_0=ik_4=i\omega_n= \left \{ \begin{array}{ll}
                  \frac{2n \pi i }{\beta} & \mbox{for boson ,} \nonumber \\
                  \frac{(2n+1)\pi i}{\beta} & \mbox{for fermion.}\nonumber
		   \end{array} 
		\right.  
$
\\
{ $k$ is continuous}& { $k$ is continuous}
\\
\hline
\eeud
\subsection{Feynman Rules}
\label{fey_rule}
Now following the table we can write the Feynman rules for $T\ne 0$ as
\begin{enumerate}
 \item[$\bullet$] The propagator is same as $T=0$ but with the fourth component of Minkowski momentum is now discrete 
\[
k_0=ik_4=i\omega_n= \left \{ \begin{array}{ll}
                  \frac{2n \pi i }{\beta} & \mbox{for boson ,} \nonumber \\
                  \frac{(2n+1)\pi i}{\beta} & \mbox{for fermion.}\nonumber
		   \end{array} 
		\right.  
\]

 \item[$\bullet$] The loop integral at $T=0$ should be replaced as
 \begin{equation}
 \int \frac{d^4K}{(2\pi)^4} \rightarrow  \frac{1}{\beta}\sumintb_{k_0}  \frac{d^3{k}}{(2\pi)^3},
 \end{equation}
 where the fourth component of Minkowski momentum is replace by discrete frequency
 \[
k_0=ik_4=i\omega_n= \left \{ \begin{array}{ll}
                  \frac{2n \pi i }{\beta} & \mbox{for boson ,} \nonumber \\
                  \frac{(2n+1)\pi i}{\beta} & \mbox{for fermion.}\nonumber
		   \end{array} 
		\right.  
\]

\item[$\bullet$] Vertex is same as the $T=0$ field theory.
 
 \item[$\bullet$] Symmetry factor for a given diagram is same as the $T=0$ field theory.
\end{enumerate}

Once the Feynman amplitudes are written using these Feynman rules, one needs now to compute the discrete frequency sum.
Below we discuss the techniques to evaluate the frequency sum at finite $T$.

\section{Frequency Sum}
\label{freq_sum}

The Euclidean time Green's function in coordinate space:
\begin{eqnarray}
 G_\beta(\bm{\vec x}, \tau)& = \atop{\tau >0}& -\frac{1}{\beta} \sum_{n=-\infty}^{n=+\infty} \int \frac{{d^3}k}{(2\pi)^3}
\, e^{i\bm{\vec k}\cdot \bm{\vec x}} \, \frac{e^{-i\omega_n\tau}}{\omega_n^2+\omega_k^2} 
\end{eqnarray}
with $\omega_n=2\pi nT$ and $\omega_k=\sqrt{k^2+m^2}$. 

 We need to perform the discrete frequency sum:
 \be
 T \sum_{n=-\infty}^{n=+\infty} \frac{e^{-i\omega_n\tau}}{\omega_n^2+\omega_k^2}   \label{sum_int0}
 \ee
Also in order to calculate matrix element corresponding to a given Feynman diagram in theory, 
we need to perform frequency sums. There are two types of frequency sums: bosonic and fermionic.

\subsection{Bosonic Frequency Sum}
\label{bos_sum}

In general the form of the bosonic frequency sum can be written as
\be
\frac{1}{\beta} \sum_{n=-\infty}^{n=+\infty} f(k_0=i\omega_n=2\pi i n T) \label{sum_intb0}
\ee
where $k_0$ is the fourth (temporal) component of momentum in Minkowski space-time and $f$ is a meromorphic function
\footnote{A meromorphic function is a ratio of two well-behaved (holomorphic) functions in complex plane as $f(z)=g(z)/h(z)$ 
with $h(z)\ne 0$. However such a function will still be well-behaved  if it has finite order, 
isolated poles and zeros and no essential singularities or branch cuts in its domain.}.

\begin{figure}[h]
\hspace{.5in}
\includegraphics[width=5cm,height=6cm]{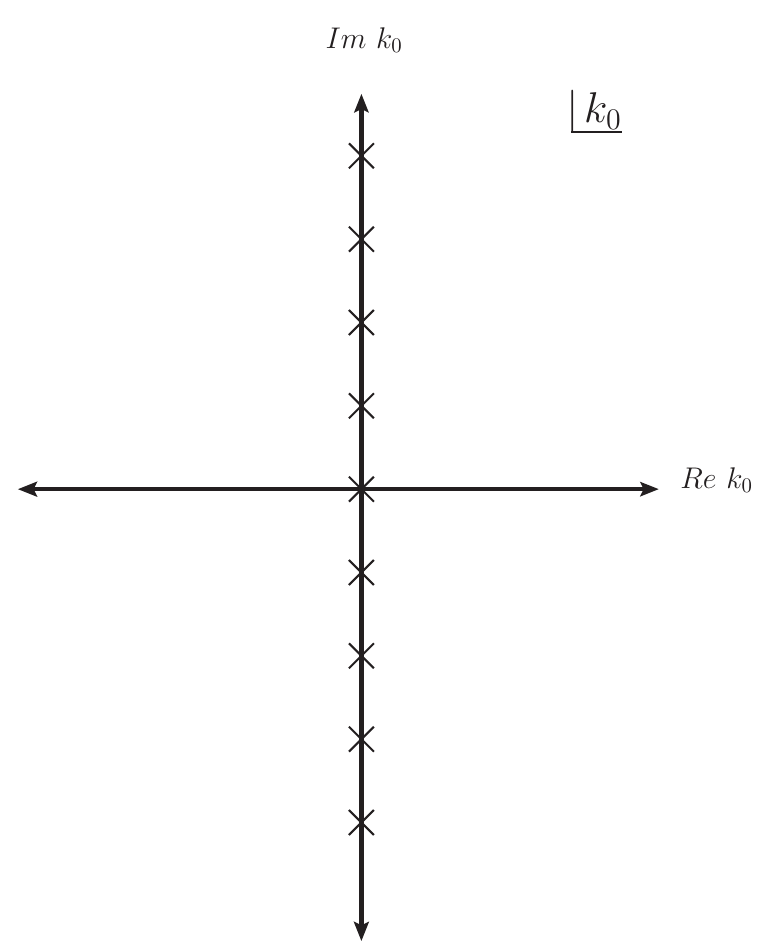}
\hspace*{1.5in} \includegraphics[width=5cm,height=6cm]{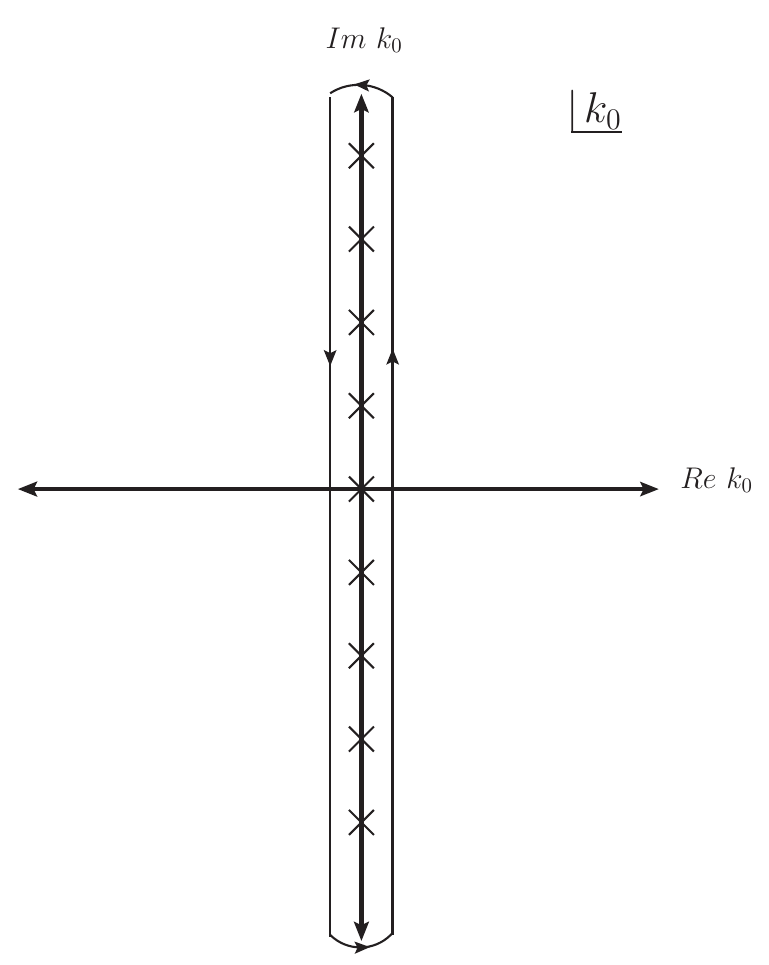}
\caption{Poles of $\coth(\beta k_0/2)$  at $k_0=2\pi i n T; n=0, 1, 2 \cdots $, in complex $k_0$ plane}
\label{coth_poles_fig}
\end{figure}

We know hyperbolic cotangent has poles at $\coth(n\pi i)$ with residue unity.   
Therefore,  one can insert  hyperbolic cotangent with suitable argument as~\cite{Kapusta,Le_Bellac}
\be
\frac{1}{\beta} \sum_{n=-\infty}^{n=+\infty} f(k_0=i\omega_n=2\pi i n T) \,{\rm{Res}}\left[ \frac{\beta}{2} \coth\left ( \frac{\beta k_0}{2}\right )\right], \label{sum_intb1}
\ee
where hyperbolic cotangent corresponds to poles (see Fig.~\ref{coth_poles_fig}) at 
\be
\coth\left ( \frac{\beta k_0}{2}\right ) = \coth(n\pi i) \, \, \Rightarrow \, \, k_0= \frac{2\pi i n }{\beta} = i\omega_n
\ee
with residues $2/\beta$, and ${\rm{Res}}\left[\frac{\beta}{2} \coth\left ( \frac{\beta k_0}{2}\right )\right]$ will lead to unity. 

Then one can write (\ref{sum_intb1}) without any loss of generality as 
\begin{eqnarray}
&&\frac{1}{\beta} \sum_{n=-\infty}^{n=+\infty} f(k_0) \,\textrm{Res}\left[  \frac{\beta}{2} \coth\left ( \frac{\beta k_0}{2}\right )\right] \nn \\
&&=\frac{1}{\beta} \sum_{n=-\infty}^{n=+\infty} \frac{\beta}{2} \ {\rm{Res}}\left[ f(k_0) \,  \coth\left ( \frac{\beta k_0}{2}\right );
 \, \, \Rightarrow \, {\rm {poles:}} \, \, k_0= i\omega_n  = \frac{2\pi i n }{\beta} \right ].
\end{eqnarray}

Employing the residue theorem in reverse the sum over 
residues can now be expressed as an integral over a contour 
$C$  in $\angle k_0 $ enclosing the poles of the meromorphic function $f(k_0)$
but excluding the poles of 
the hyperbolic cotangent ($k_0=i\omega_n=2\pi i T$)  
as
\begin{eqnarray}
 \frac{1}{\beta} \sum_{n=-\infty}^{n=+\infty} {\rm{Res}} \left[f(k_0) \, \frac{\beta}{2} \coth\left ( \frac{\beta k_0}{2}\right )\right] &=&
 \frac{T}{2\pi i} \oint\limits_{C_1 \cup C_2} dk_0 \, f(k_0) \frac{\beta}{2} \, \coth\left ( \frac{\beta k_0}{2}\right )\nn\\
 &=& \frac{1}{2\pi i} \oint\limits_{C_1 \cup C_2} dk_0 \, f(k_0) \frac{1}{2} \, \coth\left ( \frac{\beta k_0}{2}\right ),
 \label{sum_intb2}
\end{eqnarray}
where $i$ in the numerator of RHS of (\ref{sum_intb2}) is absorbed as $i\, dk_4=d(ik_4) =dk_0$.

Now, some important points to note on Eq.(\ref{sum_intb2}):
\begin{enumerate}
\item [$\bullet$] $\left[\exp(\beta k_0) - 1\right ]^{-1}$ vis-a-vis $\coth(\beta k_0/2)$ has series of poles 
at $k_0=i\omega_n=2\pi i n T$ and is bounded and  analytic everywhere except at poles.
 
\item [$\bullet$] $f(k_0=i\omega_n)$ is a meromorphic function which has simple poles but no essential singularities or branch cuts.

\item [$\bullet$] The simple poles of $f(k_0=i\omega_n)$ should not coincide the series of poles of $\left[\exp(\beta k_0) - 1\right ]^{-1}$
$\Rightarrow \, f(k_0=i\omega_n)$ should not have singularity along the imaginary $k_0$ axis.

\item[$\bullet$] The contour $C$ can be divided into two half circles in complex $k_0$ plane  $C_1$ and $C_2$ ~\cite{Kapusta} 
without enclosing the poles of the  $\left[\exp(\beta k_0) - 1\right ]^{-1}$ vis-a-vis $\coth(\beta k_0/2)$ but
the contours $C_1$ and $C_2$ should enclose poles of $f(k_0)$ as shown in Fig.~\ref{coth_poles_fig}, provided 
the meromorphic function $f(k_0)$ should decrease fast to achieve convergence.

\item [$\bullet$] If all these properties are satisfied then $ T \sum_{n=-\infty}^{n=+\infty} f(k_0=i\omega_n)$ can be replaced by
contour integration and this is equivalent to  switching (analytically continuing) from Euclidean time (discrete frequency in Euclidean space) to real time
 (continuous frequency in Minkowski space-time).
\end{enumerate}

\begin{figure}[h]
\begin{center}
\includegraphics[width=8cm,height=6cm]{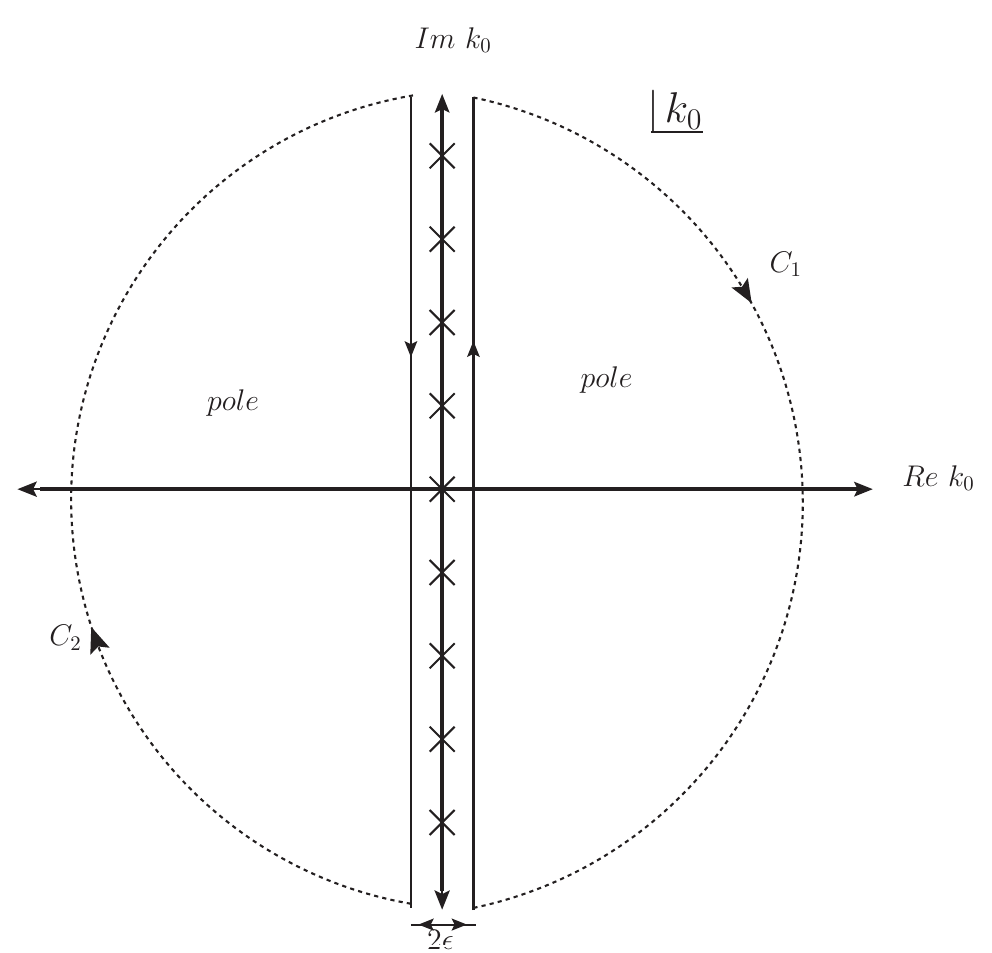}
\vspace*{.5in}
\caption{Contours  $C_1$ and $C_2$ that include the poles of the meromorphic function  $f(k_0)$ 
in complex $k_0$ plane. The contours are also shifted
by an amount  $\pm \epsilon$ from the  ${\rm{Im}} k_0$ line to exclude the poles of $\coth(\beta k_0/2)$
at $k_0=2\pi i n T$. }
\label{coth_poles_fig1}
\end{center}
\end{figure}

\subsubsection{Separation of vacuum and matter part}
We know~\cite{Kapusta}
\bea
\frac{1}{2} \coth\left ( \frac{\beta k_0}{2}\right ) 
&=& \frac{1}{2} \frac{\exp(\beta k_0/2) + \exp(-\beta k_0/2)}{\exp(\beta k_0/2) - \exp(-\beta k_0/2)}
= \frac{1}{2} \frac{\exp(\beta k_0) + 1}{\exp(\beta k_0) - 1} \nn\\
&=&\frac{1}{2}\left[ 1 +\frac{2}{{\exp(\beta k_0) - 1}} \right ]. \label{sum_intb3}
\eea

Using (\ref{sum_intb3}) and (\ref{sum_intb2}) in (\ref{sum_intb0}) the bosonic sum integral can be written as

\be
\frac{1}{\beta} \sum_{n=-\infty}^{n=+\infty} f(k_0=i\omega_n) = \frac{1}{2\pi i} \oint\limits_{C_1 \cup C_2}\,  dk_0 \, f(k_0=i\omega_n) \, 
\left[ \frac{1}{2} +\frac{1}{{\exp(\beta k_0) - 1}} \right ]
\label{sum_intb4}
\ee
which separates $T=0$ (vacuum) and $T\ne0$ (medium) part. 

\subsubsection{Choice of contour:}

As discussed above, the contour $C$ can be divided into two half circles $C_1$ and $C_2$ in complex $k_0$ plane
that excludes  the poles of the  $\left[\exp(\beta k_0) - 1\right ]^{-1}$ vis-a-vis $\coth(\beta k_0/2)$ 
but includes the poles of the meromorphic function $f(k_0)$ as shown in Fig.~\ref{coth_poles_fig1},  the integrand converges.
Lets choose the contour $C_1$ which goes from $\epsilon-i\infty$ to $\epsilon+i\infty$ whereas
the contour $C_2$ goes from $-\epsilon+i\infty$ to $-\epsilon-i\infty$.
 
Now (\ref{sum_intb4}) can be decomposed as
\begin{eqnarray}
\frac{1}{\beta} \sum_{n=-\infty}^{n=+\infty} f(k_0=i\omega_n) &=& \frac{1}{2\pi i} \int\limits_{\epsilon-i\infty}^{\epsilon+i\infty} \,  dk_0 \, f(k_0=i\omega_n) \, 
\left[ \frac{1}{2} +\frac{1}{{e^{\beta k_0} - 1}} \right ] { \Rightarrow \, {\rm {along}} \,\,  C_1 } \nn \\
&+& \frac{1}{2\pi i} \int\limits_{-\epsilon+i\infty}^{-\epsilon-i\infty} \!\!\!  dk_0 \, f(k_0=i\omega_n) \, 
\left[ -\frac{1}{2} -\frac{1}{{e^{-\beta k_0} - 1}} \right ] { \Rightarrow \, {\rm {along}} \, \,  C_2 }, \label{sum_intb5}
\end{eqnarray}
where we note the following:
\begin{enumerate}
 \item[$\bullet$] The first term contains pole for $k_0>0$ in the contour $C_1$.
 \item[$\bullet$] The overall negative sign in the second term is due to pole $k_0<0$ in the contour $C_2$.
 \item[$\bullet$] Also in the second term the argument of the exponential is negative because it has to converge
 as $k_0\rightarrow \,  -\infty$, since the contour $C_2$ is in the left half plane.
 \end{enumerate}

Now we make a substitution $k_0 \rightarrow -k_0$ in the second term in (\ref{sum_intb5})  and can be written as 
\begin{eqnarray}
 \frac{1}{2\pi i} \int\limits_{\epsilon-i\infty}^{\epsilon+i\infty} \,  d(-k_0) \, f(-k_0) \, 
\left[ -\frac{1}{2} -\frac{1}{{\exp(\beta k_0) - 1}} \right ]
&=& \frac{1}{2\pi i} \int\limits_{-i\infty}^{+i\infty} \,  dk_0 \, f(-k_0) \, \frac{1}{2} \nn \\
&& +  \frac{1}{2\pi i} \int\limits_{\epsilon-i\infty}^{\epsilon+i\infty} \,  dk_0 \, f(-k_0) \, 
\frac{1}{{\exp(\beta k_0) - 1}} . \label{sum_intb6}
\end{eqnarray}

Using (\ref{sum_intb6}) in (\ref{sum_intb5}), one gets
\begin{eqnarray}
\frac{1}{\beta}\sum_{n=-\infty}^{n=+\infty} f(k_0=i\omega_n) &=& 
\frac{1}{2\pi i} \int\limits_{-i\infty}^{+i\infty} \,  dk_0 \, \frac{1}{2}\Big [f(k_0)+f(-k_0)\Big] \nn \\
&& +  \frac{1}{2\pi i} \int\limits_{\epsilon-i\infty}^{\epsilon+i\infty} \,  dk_0 \Big [f(k_0)+f(-k_0)\Big] 
\frac{1}{{\exp(\beta k_0) - 1}} . \label{sum_intb7}
\end{eqnarray}
We note that the Contour is in right half plane. The first term in RHS is vacuum contribution whereas the second term is matter contribution.
 One can use either (\ref{sum_intb2}) or (\ref{sum_intb4}) or (\ref{sum_intb7}) conveniently.
  
 \subsection{Fermionic Frequency Sum for Zero Chemical Potential ($\mu=0$)}.   
\label{fer_sum}
In general the form of the fermionic frequency sum can be written as
\be
\frac{1}{\beta} \sum_{n=-\infty}^{n=+\infty} f(k_0=i\omega_n=(2n+1) i \pi  T), \label{sum_intf0}
\ee
where $k_0$ is the fourth (temporal) component of momentum in Minkowski space-time and the $f(k_0)$  is a meromorphic function.
We know hyperbolic tangent has poles at $\tanh(\frac{\pi i}{2}+ n\pi i)$ with residue unity.   
Therefore,  one can insert  hyperbolic tangent with suitable argument as~\cite{Kapusta,Le_Bellac}
\be
\frac{1}{\beta} \sum_{n=-\infty}^{n=+\infty} f(k_0=i\omega_n= (2n+1) i \pi  T) \, \textrm{Res}\left[ \frac{\beta}{2} \tanh\left ( \frac{\beta k_0}{2}\right )\right], \label{sum_intf1}
\ee
where hyperbolic tangent corresponds to poles at 
\be
\tanh\left ( \frac{\beta k_0}{2}\right ) = \tanh(\frac{\pi i}{2}+ n\pi i) \, \, \Rightarrow \, \, k_0= \frac{(2n+1) i\pi n }{\beta} = i\omega_n
\ee
with residues $2/\beta$. 
Then one can write (\ref{sum_intf1}) without any loss of generality as 
\begin{eqnarray}
&&\frac{1}{\beta} \sum_{n=-\infty}^{n=+\infty} f(k_0) \,\textrm{Res}\left[  \frac{\beta}{2} \tanh\left ( \frac{\beta k_0}{2}\right ) \right] \nn \\
&&=
\frac{1}{\beta} \sum_{n=-\infty}^{n=+\infty} \frac{\beta}{2} \ {\rm{Res}}\left[ f(k_0) \,  \tanh\left ( \frac{\beta k_0}{2}\right );
  \Rightarrow {\rm {poles:}}   k_0= i\omega_n  = \frac{(2n+1)i \pi }{\beta} \right ]. \label{sum_intf2}
\end{eqnarray}

\begin{figure}[h]
\begin{center}
\includegraphics[width=9cm,height=7cm]{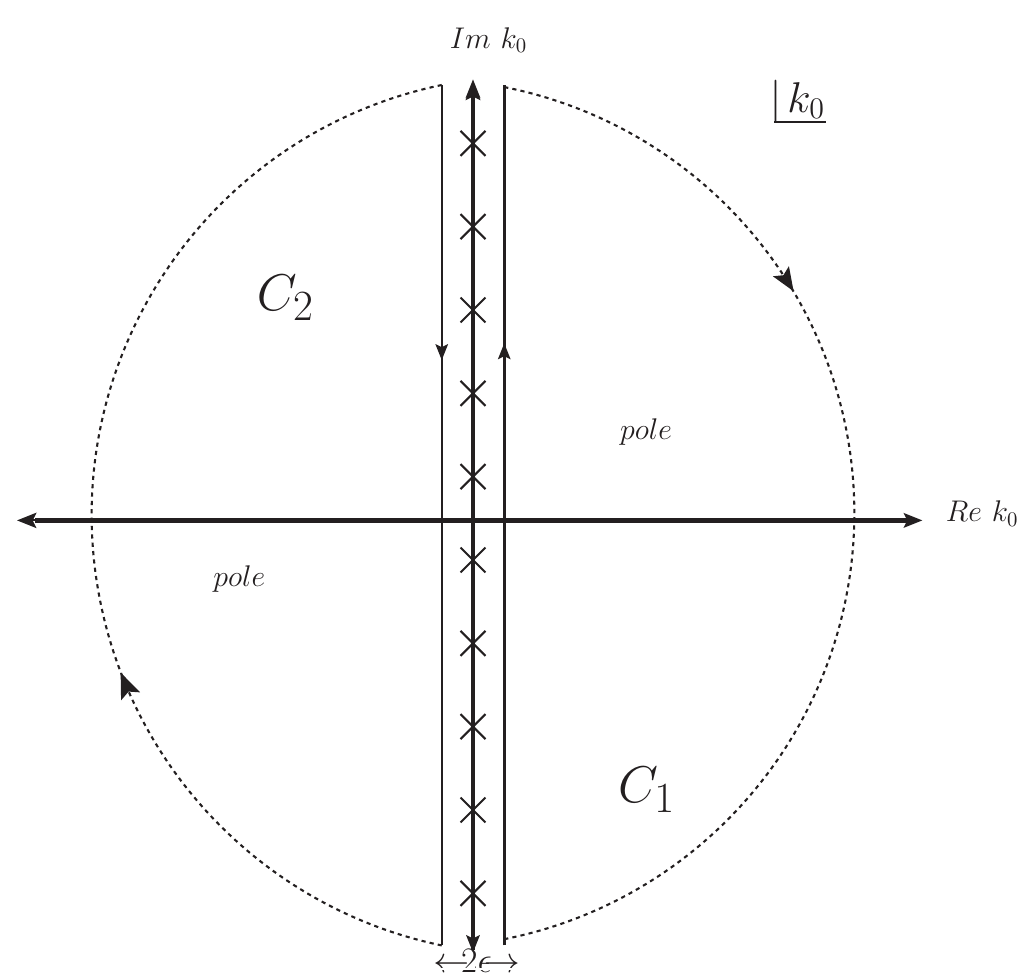}
\vspace*{.5in}
\caption{Contours  $C_1$ and $C_2$ that include the poles of the meromorphic function  $f(k_0)$ 
in complex $k_0$ plane for fermion with zero chemical potential ($\mu=0$). The contours are also shifted
by an amount  $\pm \epsilon$ from the  ${\rm{Im}} k_0$ line to exclude the poles of $\tanh(\beta k_0/2)$ at
$k_0=(2n+1)i\pi T; \, n=0, 1,2,\cdots$. }
\label{tanh_poles}
\end{center}
\end{figure}

Employing the residue theorem, as before,  in reverse the sum over 
residues can now be expressed as an integral over contours $C_1$ and $C_2$ 
in $\angle k_0 $ enclosing the poles of the meromorphic function $f(k_0)$
but excluding the poles of  hyperbolic tangent ($k_0=i\omega_n=(2n+1) \pi i T$)  
as
\begin{eqnarray}
\frac{1}{\beta} \sum_{n=-\infty}^{n=+\infty} f(k_0) \, \frac{\beta}{2} \tanh\left ( \frac{\beta k_0}{2}\right ) &=&
 \frac{T}{2\pi i} \oint\limits_{C_1 \cup C_2} dk_0 \, f(k_0) \frac{\beta}{2} \, \tanh\left ( \frac{\beta k_0}{2}\right )\nn\\
 &=& \frac{1}{2\pi i} \oint\limits_{C_1 \cup C_2} dk_0 \, f(k_0) \frac{1}{2} \, \tanh\left ( \frac{\beta k_0}{2}\right ),
 \label{sum_intf3}
\end{eqnarray}
where contours $C_1$ and $C_2$ are represented in Fig.~\ref{tanh_poles}

\subsubsection{Separation of vacuum and matter part}
We know
\bea
\frac{1}{2} \tanh\left ( \frac{\beta k_0}{2}\right ) 
&=& \frac{1}{2} \frac{\exp(\beta k_0/2) - \exp(-\beta k_0/2)}{\exp(\beta k_0/2) + \exp(-\beta k_0/2)}
= \frac{1}{2} \frac{\exp(\beta k_0) - 1}{\exp(\beta k_0) + 1} \nn\\
&=& \frac{1}{2}\left[ 1 - \frac{2}{{\exp(\beta k_0) + 1}} \right ]. \label{sum_intf4}
\eea 
Using (\ref{sum_intf4}) and (\ref{sum_intf3}) in (\ref{sum_intf0}) the fermionic sum integral can be written as
\be
\frac{1}{\beta} \sum_{n=-\infty}^{n=+\infty} f(k_0=i\omega_n) = \frac{1}{2\pi i} \oint\limits_{C_1 \cup C_2} \,  dk_0 \, f(k_0=i\omega_n) \, 
\left[ \frac{1}{2} - \frac{1}{{\exp(\beta k_0) + 1}} \right ]
\label{sum_intf5}
\ee
which separates $T=0$ (vacuum) and $T\ne0$ (medium) part. 

\subsubsection{Choice of contour}
Proceeding same way as the bosonic case before, one can write
\begin{eqnarray}
T \sum_{n=-\infty}^{n=+\infty} f(k_0=i\omega_n) &=& 
\frac{1}{2\pi i} \int\limits_{-i\infty}^{+i\infty} \,  d(k_0) \, \frac{1}{2}\Big [f(k_0)+f(-k_0)\Big] \nn\\
&& +  \frac{1}{2\pi i} \int\limits_{\epsilon-i\infty}^{\epsilon+i\infty} \,  d(k_0) \Big [f(k_0)+f(-k_0)\Big] 
\frac{1}{{\exp(\beta k_0) + 1}} . \label{sum_intf6}
\end{eqnarray}
We note that the Contour is in right half plane. The first term in RHS is vacuum contribution whereas the second term is matter contribution.
 One can use either (\ref{sum_intf3}) or (\ref{sum_intf5}) or (\ref{sum_intf6}) conveniently.

\subsection{Examples of Frequency Sum for Bosonic Case}
\label{example_ferq}

 The Euclidean time bosonic Green's function in coordinate space can be written as
 \begin{eqnarray}
  G_\beta(\bm{\vec x}, \tau)& = \atop{\tau >0}&\, \,  -\frac{1}{\beta} \sum_{n=-\infty}^{n=+\infty} \int \frac{{d^3}k}{(2\pi)^3}
 \, e^{i\bm{\vec k} \cdot \bm{\vec x}} \, \frac{e^{-i\omega_n\tau}}{\omega_n^2+\omega_k^2} , \label{ex0}
 \end{eqnarray}
with $\omega_n=2\pi nT$ and $\omega_k=\sqrt{k^2+m^2}$. Now we  put back $i\omega_n=k_0$, the fourth component of Minkowski
momentum to write the standard form as given in (\ref{sum_intb0}) without any loss generality:
 \begin{eqnarray}
  G_\beta(\bm{\vec x}, \tau)& = \atop{\tau >0}& \, \,   \int \frac{{d^3}k}{(2\pi)^3} \, e^{i\bm{\vec k}\cdot \bm{\vec x}} \, 
  \ \frac{1}{\beta} \sum_{n=-\infty}^{n=+\infty}\, \frac{e^{-k_0\tau}}{k_0^2-\omega_k^2} \, \, \nn  \\
   &= \atop{\tau >0}& \, \,  \int \frac{{d^3}k}{(2\pi)^3} \, e^{i\bm{\vec k}\cdot \bm{\vec x}} \, 
   \ \frac{1}{\beta} \sum_{n=-\infty}^{n=+\infty}\, f(k_0=i\omega_n)  . \label{ex1}
 \end{eqnarray}
Now we perform the frequency sum.
As discussed the sum integration can be performed using either (\ref{sum_intb2}) or (\ref{sum_intb4}) or (\ref{sum_intb7}) conveniently. 
However, we  would use (\ref{sum_intb2}) for the purpose 
\begin{eqnarray}  
 \frac{1}{\beta} \sum_{n=-\infty}^{n=+\infty}\, f(k_0=i\omega_n)  & = \atop{\tau >0}&  
  \  \frac{1}{\beta} \sum_{n=-\infty}^{n= +\infty}\, \frac{e^{-k_0\tau}}{k_0^2-\omega_k^2} \nn  \\
 &=\atop {\tau >0}& \, \frac{1}{2\pi i} \oint\limits_{C} dk_0 \, \frac{e^{-k_0\tau}}{k_0^2-\omega_k^2} \, 
 \frac{1}{2} \, \coth\left ( \frac{\beta k_0}{2}\right ). \label{ex2}
 \end{eqnarray}
We now rewrite (\ref{ex2}) here
\begin{eqnarray}  
 \frac{1}{\beta} \sum_{n=-\infty}^{n=+\infty}\, f(k_0=i\omega_n)  & = \atop{\tau >0}&  
  \  \frac{1}{\beta} \sum_{n=-\infty}^{n= +\infty}\, \frac{e^{-k_0\tau}}{k_0^2-\omega_k^2} \nn  \\
 &=\atop {\tau >0}& \, \frac{1}{2\pi i} \oint\limits_{C} dk_0 \, \frac{e^{-k_0\tau}}{k_0^2-\omega_k^2} \, 
 \frac{1}{2} \, \coth\left ( \frac{\beta k_0}{2}\right ). \label{aex1}
 \end{eqnarray}
Using (\ref{sum_intb7}), one gets (we drop $\tau>0$ from all equations below)
\begin{eqnarray}
&&\frac{1}{\beta}\sum_{n=-\infty}^{n=+\infty} f(k_0=i\omega_n) = 
\frac{1}{2\pi i} \int\limits_{-i\infty}^{+i\infty} \,  dk_0 \, \frac{1}{2}\Big [f(k_0)+f(-k_0)\Big]
+\frac{1}{2\pi i} \int\limits_{\epsilon-i\infty}^{\epsilon+i\infty} \,  dk_0 \Big [f(k_0)+f(-k_0)\Big] 
\frac{1}{{e^{\beta k_0} - 1}} \nn \\
&=& 
\frac{1}{2\pi i} \int\limits_{-i\infty}^{+i\infty} \,  dk_0 \, \frac{1}{2}\left [\frac{e^{-k_0\tau}}{k_0^2-\omega_k^2}
+\frac{e^{k_0\tau}}{k_0^2-\omega_k^2}\right]
+\frac{1}{2\pi i} \int\limits_{\epsilon-i\infty}^{\epsilon+i\infty} \,  dk_0
\left [\frac{e^{-k_0\tau}}{k_0^2-\omega_k^2} +\frac{e^{k_0\tau}}{k_0^2-\omega_k^2}\right]
\frac{1}{{e^{\beta k_0} - 1}}\nn \\
&=& I_1^0+I_2^0+I_3^\beta+I_4^\beta ,\label{aex2}
\end{eqnarray}
where $I_1^0$ and $I_2^0$ are for $T=0$ parts whereas $I_3^\beta$ and $I_4^\beta $ are for $T\ne0$ parts. We now evaluate them below:

\begin{figure}
\begin{center}
 \vspace*{-0.2in}
\includegraphics[width=7cm,height=6cm]{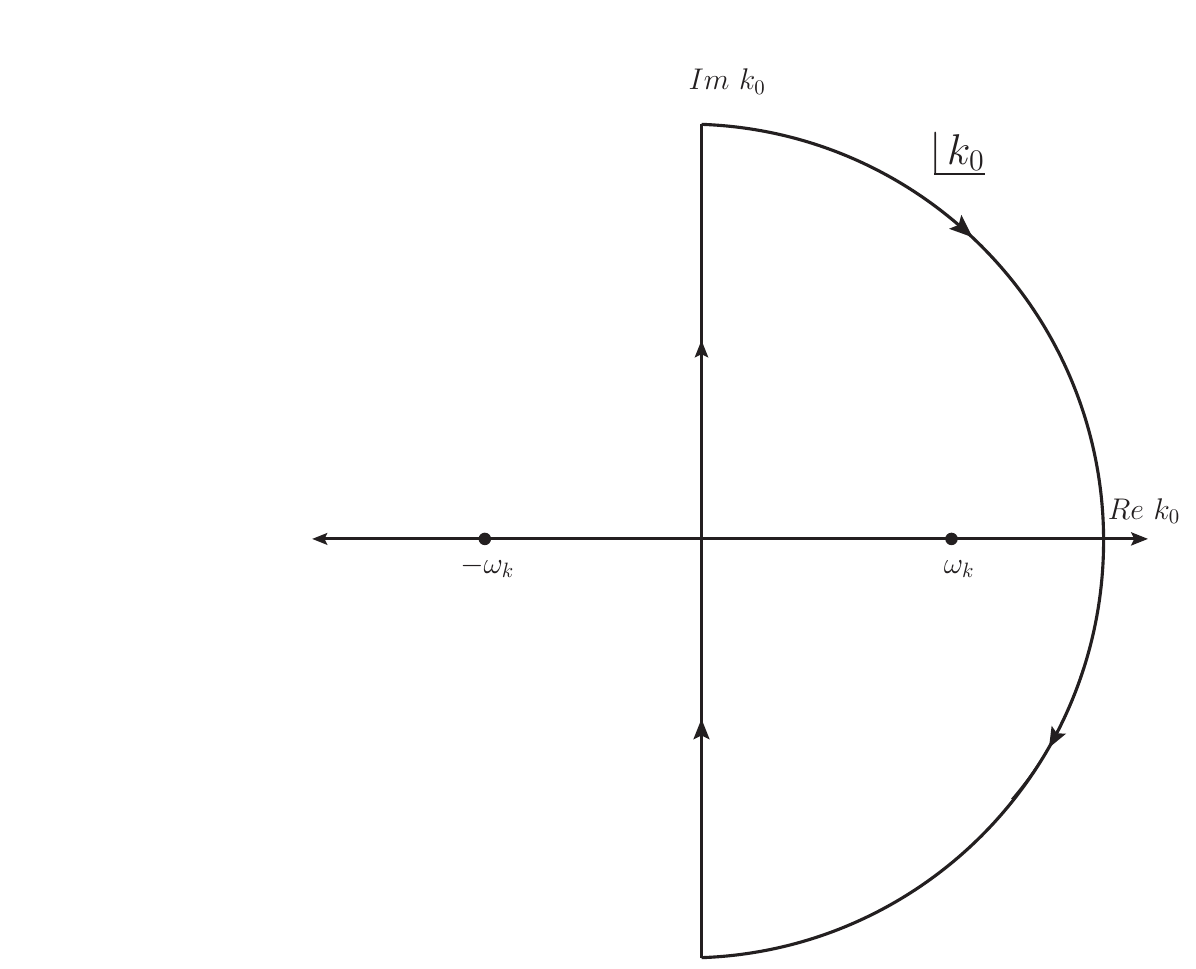}
\caption{Contour corresponding to the Integral $I_1^0$ in the complex $k_0$ plane.}
\label{contour_i10}
\end{center}
\end{figure}
 \begin{eqnarray}
  I_1^0 &=& \frac{1}{2\pi i} \int\limits_{-i\infty}^{+i\infty} \,  dk_0 \, \frac{1}{2} \, \, \frac{e^{-k_0\tau}}{k_0^2-\omega_k^2}
 \end{eqnarray}
\begin{enumerate}
 \item [$\bullet$] Poles: $k_0=\pm \omega_k$
 \item [$\bullet$] Convergence: for $\tau>0$,  $e^{-k_0\tau}$ converges in the domain $0\le\tau\le \beta$ only for $k_0=\omega_k$. 
 The relevant contour is given in Fig.~\ref{contour_i10}.
 
 \begin{equation}
 I_1^0 = \frac{1}{2} \, \frac{1}{2\pi i} \times (2\pi i) \, \lim_{k_0\rightarrow \omega_k} \, \,  \frac{(k_0-\omega_k)}{(k_0-\omega_k)(k_0+\omega_k)} e^{-k_0\tau}
 =\frac{1}{2} \frac{e^{-\omega_k\tau}}{2\omega_k}. \label{aex3} 
 \end{equation}

\end{enumerate}

 \begin{eqnarray}
  I_2^0 &=& \frac{1}{2\pi i} \int\limits_{-i\infty}^{+i\infty} \,  dk_0 \, \frac{1}{2} \, \, \frac{e^{k_0\tau}}{k_0^2-\omega_k^2}
 \end{eqnarray}
\begin{figure}[h]
\begin{center}
\includegraphics[width=6cm,height=6cm]{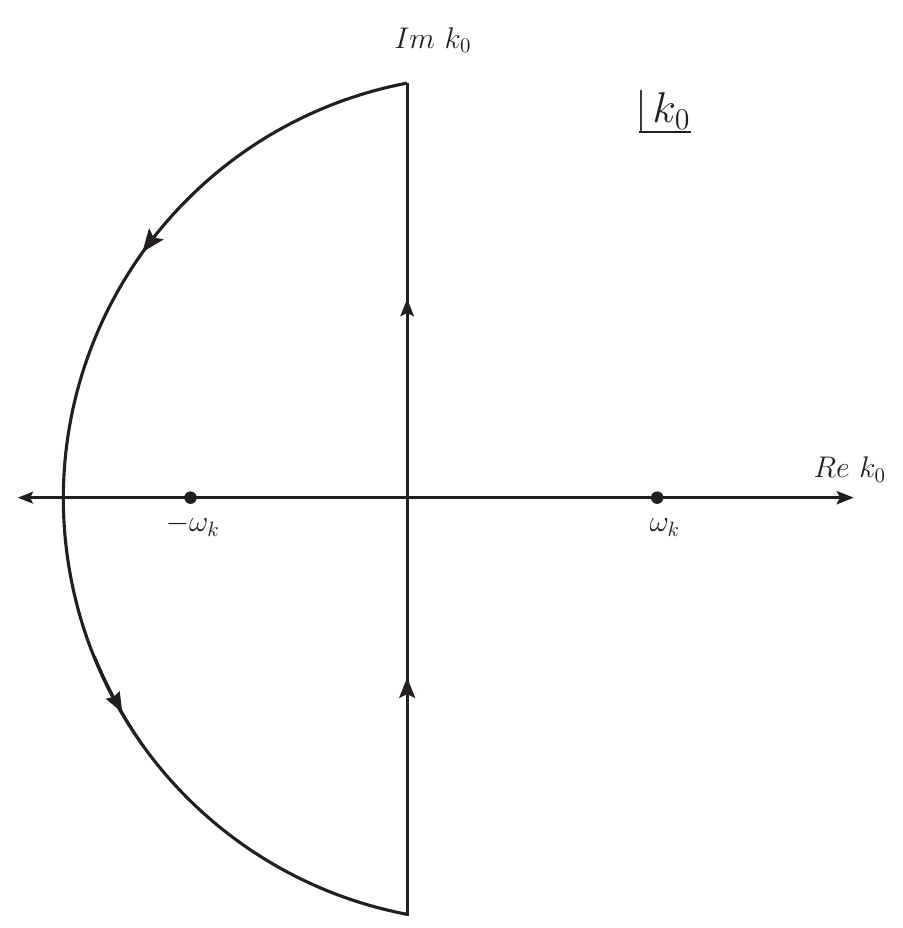}
\caption{Contour corresponding to the Integral $I_2^0$ in complex $k_0$ plane.}
\label{contour_i20}
\end{center}
\end{figure} 

\begin{enumerate}
 \item [$\bullet$] Poles: $k_0=\pm \omega_k$
 \item [$\bullet$] Convergence: for $\tau>0$,  $e^{k_0\tau}$ converges in the domain $0\le\tau\le \beta$ only for $k_0=-\omega_k$. 
 The relevant contour is given in Fig.~\ref{contour_i20}. Note that the contour is anticlockwise so it will induct a negative sign.
 
 \begin{equation}
 I_2^0 = \frac{1}{2} \, \frac{1}{2\pi i} \times (- 2\pi i) \, \lim_{k_0\rightarrow - \omega_k} \, \, 
 \frac{(k_0+\omega_k)}{(k_0-\omega_k)(k_0+\omega_k)} e^{k_0\tau}
 =\frac{1}{2} \frac{e^{-\omega_k\tau}}{2\omega_k}=I_1^0. \label{aex4} 
 \end{equation}
\end{enumerate}

 \begin{eqnarray}
 I_3^\beta &=& \frac{1}{2\pi i} \int\limits_{\epsilon-i\infty}^{\epsilon+i\infty} \,  dk_0 
\frac{e^{-k_0\tau}}{k_0^2-\omega_k^2} 
\frac{1}{{e^{\beta k_0} - 1}}.
\end{eqnarray}

\begin{enumerate}

\item [$\bullet$] Poles: $k_0=\pm \omega_k$
\item [$\bullet$] Convergence: for $\tau > 0$,  $e^{-k_0\tau} / (e^{\beta k_0} - 1) \sim \ e^{-k_0\tau} \, e^{-\beta k_0}$ 
 converges in the domain $0\le\tau\le \beta$ only for $k_0=\omega_k$. 
 The relevant contour is given in Fig.~\ref{contour_i3T}. 

\begin{figure}[h]
\begin{center}
\includegraphics[width=6cm,height=6cm]{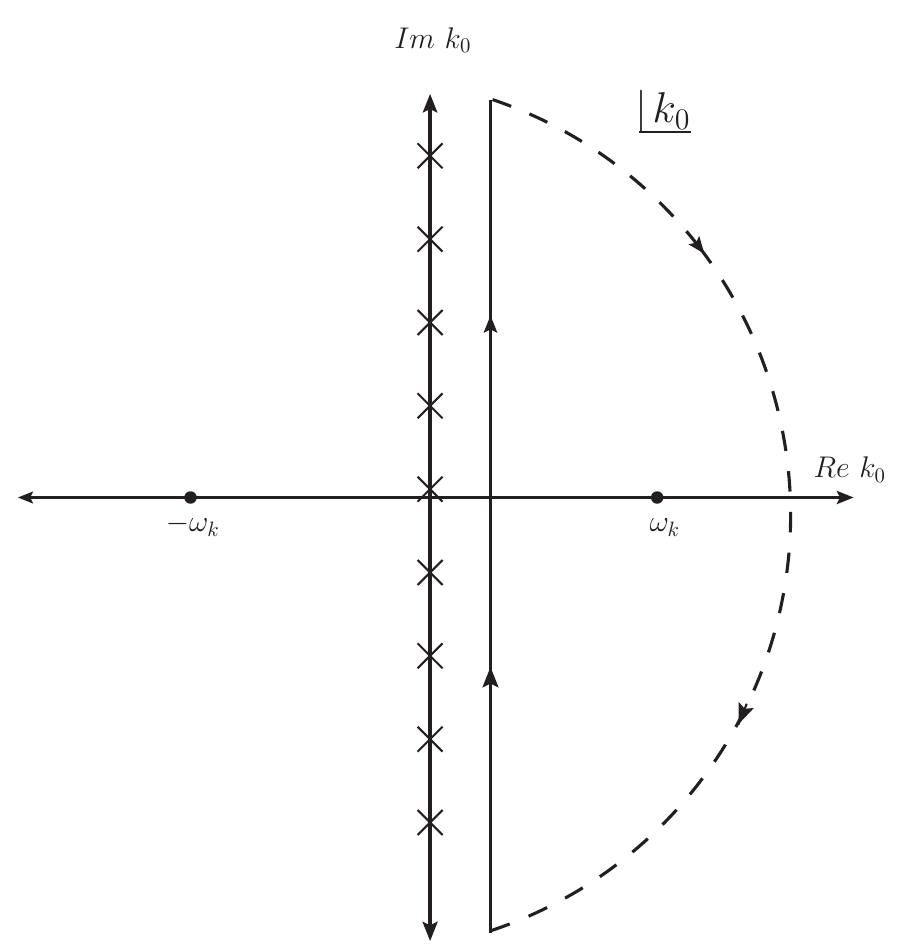}
\caption{Contour corresponding to the Integral $I_3^\beta$ in complex $k_0$ plane.}
\label{contour_i3T}
\end{center}
\end{figure} 
 
 \begin{equation}
 I_3^0 =  \frac{1}{2\pi i} \times ( 2\pi i) \, \lim_{k_0\rightarrow  \omega_k} \, \, 
 \frac{(k_0+\omega_k)}{(k_0-\omega_k)(k_0-\omega_k)} e^{k_0\tau}
 =\frac{1}{2\omega_k} \, \frac{e^{-\omega_k\tau}}{{e^{\beta k_0} - 1}} 
 = \frac{e^{-\omega_k\tau}}{2\omega_k}\, \, n_B(\omega_k). \label{aex5} 
 \end{equation}
 
\end{enumerate}

\begin{eqnarray}
 I_4^\beta &=& \frac{1}{2\pi i} \int\limits_{\epsilon-i\infty}^{\epsilon+i\infty} \,  dk_0 
\frac{e^{k_0\tau}}{k_0^2-\omega_k^2} 
\frac{1}{{e^{\beta k_0} - 1}}.
\end{eqnarray}

\begin{figure}[h]
\begin{center}
\includegraphics[width=6cm,height=6cm]{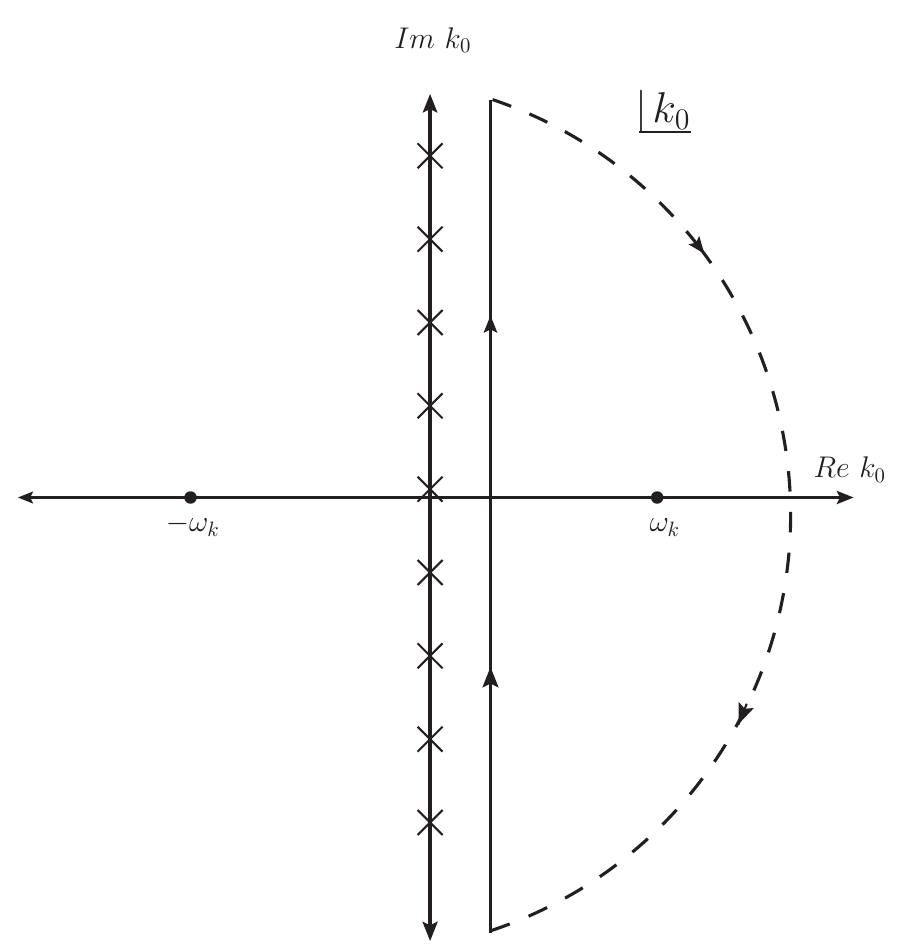}
\caption{Contour corresponding to the Integral $I_4^\beta$ in complex $k_0$ plane.}
\label{contour_i4T}
\end{center}
\end{figure} 

\begin{enumerate}
\item [$\bullet$] Poles: $k_0=\pm \omega_k$
\item [$\bullet$] Convergence: for $\tau > 0$,  $e^{k_0\tau} / (e^{\beta k_0} - 1) \sim \ e^{k_0\tau} \, e^{-\beta k_0}$ 
 converges in the domain $0\le\tau\le \beta$ only for $k_0=\omega_k$ and $\beta > \tau $. 
 The relevant contour is given in Fig.~\ref{contour_i4T} 

 \begin{equation}
 I_4^0 = \frac{1}{2\pi i} \times ( 2\pi i) \, \lim_{k_0\rightarrow  \omega_k} \, \, 
 \frac{(k_0+\omega_k)}{(k_0-\omega_k)(k_0-\omega_k)} e^{k_0\tau}
 =\frac{1}{2\omega_k} \, \frac{e^{\omega_k\tau}}{{e^{\beta k_0} - 1}} 
 = \frac{e^{\omega_k\tau}}{2\omega_k}\, \, n_B(\omega_k). \label{aex6} 
 \end{equation}

At this point it is worth noting that the contours for $I_1^0, \, \, I_3^\beta$ and $I_4^\beta$ are in right half plane whereas
that for $I_2^0$ is in the left half plane.
\end{enumerate}

Now combining (\ref{aex3}), (\ref{aex4}), (\ref{aex5}) and (\ref{aex6}) with (\ref{aex2}), one gets for $\tau>0$ as 
\begin{eqnarray}  
 \frac{1}{\beta} \sum_{n=-\infty}^{n=+\infty}\, f(k_0=i\omega_n) &= \atop{\tau >0}&\, \,  I_1^0+I_2^0+I_3^\beta+I_4^\beta \nn \\
&= \atop{\tau >0}&\, \frac{1}{2\pi i} \oint\limits_{C} dk_0 \, \frac{e^{-k_0\tau}}{k_0^2-\omega_k^2} \, 
\frac{1}{2} \, \coth\left ( \frac{\beta k_0}{2}\right ) \nn\\
& = \atop{\tau >0}& \frac{1}{2\pi i} \times (2\pi i) \,  \left [\left. R_1\right |_{k_0=\omega_k}+\left. R_2\right |_{k_0=-\omega_k}\right ] \nn\\
& = \atop{\tau >0}&  \, \frac{e^{-\omega_k\tau}}{2\omega_k} + \frac{e^{-\omega_k\tau}}{2\omega_k} n_B(\omega_k ) 
+ \frac{e^{\omega_k\tau}}{2\omega_k} n_B(\omega_k ) \nn \\
& = \atop{\tau >0}&  \, \frac{e^{-\omega_k\tau}}{2\omega_k} (1 +  n_B(\omega_k )) 
+ \frac{e^{\omega_k\tau}}{2\omega_k} n_B(\omega_k )
. \label{aex7}
 \end{eqnarray}
 
Again putting (\ref{aex7}) in (\ref{ex0}) on gets Euclidean time Green's function for $\tau >0$ as
\begin{eqnarray}
G_\beta(\bm{\vec x}, \tau) &= \atop{\tau >0}& \, \,  \int \frac{{d^3}k}{(2\pi)^3} \, \frac{1}{2\omega_k} \, 
  \Big [ (1 +  n_B(\omega_k )) \,   \,  e^{-\omega_k\tau} \, e^{i\bm{\vec k}\cdot \bm{\vec x}} \,
+ n_B(\omega_k ) \, \, e^{\omega_k\tau} \, \, e^{i\bm{\vec k}\cdot \bm{\vec x}} \, \Big ]  . \label{aex8}
 \end{eqnarray}
 Now changing  $\bm{\vec k} \rightarrow -\bm{\vec k}$ in the second term inside the square braces, leads to the Euclidean time green's 
 function as
  \begin{eqnarray}
G_\beta(\bm{\vec x}, \tau) &= \atop{\tau >0}& \, \,  \int \frac{{d^3}k}{(2\pi)^3} \, \frac{1}{2\omega_k} \, 
  \Big [ (1 +  n_B(\omega_k )) \, \,  e^{-\omega_k\tau}  \, \, e^{i\bm{\vec k}\cdot \bm{\vec x}} \,
+ n_B(\omega_k ) \, \, e^{\omega_k\tau} \, \, e^{-i\bm{\vec k}\cdot \bm{\vec x}} \, \Big ]  . \label{aex9}
 \end{eqnarray}
 
 If one makes a Wick rotation $\tau \rightarrow it$ to Minkowski time, it becomes
   \begin{eqnarray}
G_\beta(\bm{\vec x}, t) 
&= \atop{t >0}& \, \,  \int \frac{{d^3}k}{(2\pi)^3} \, \frac{1}{2\omega_k} \, 
  \Big [ (1 +  n_B(\omega_k )) \, \,  e^{-i\omega_k t}  \, \, e^{i\bm{\vec k}\cdot \bm{\vec x}} \,
+ n_B(\omega_k ) \, \, e^{i \omega_k t} \, \, e^{-i\bm{\vec k}\cdot \bm{\vec x}} \, \Big ] \nn \\
&= \atop{t >0}& \, \,  \int \frac{{d^3}k}{(2\pi)^3} \, \frac{1}{2\omega_k} \, 
  \Big [ (1 +  n_B(\omega_k )) \, \,  e^{-i K\cdot X}  \, \,
+ n_B(\omega_k ) \, \, e^{i K\cdot X} \, \Big ], \label{aex10}
 \end{eqnarray}
which agrees with the real time Green's function in \eqref{eqgt6}.

\section{Scalar Theory}
\label{st}

\subsection{Tadpole Diagram in $\lambda \phi^4$ Theory}
\label{tad_sec}

\begin{figure}[h]
 \begin{center}
 \includegraphics[width=5cm,height=4cm]{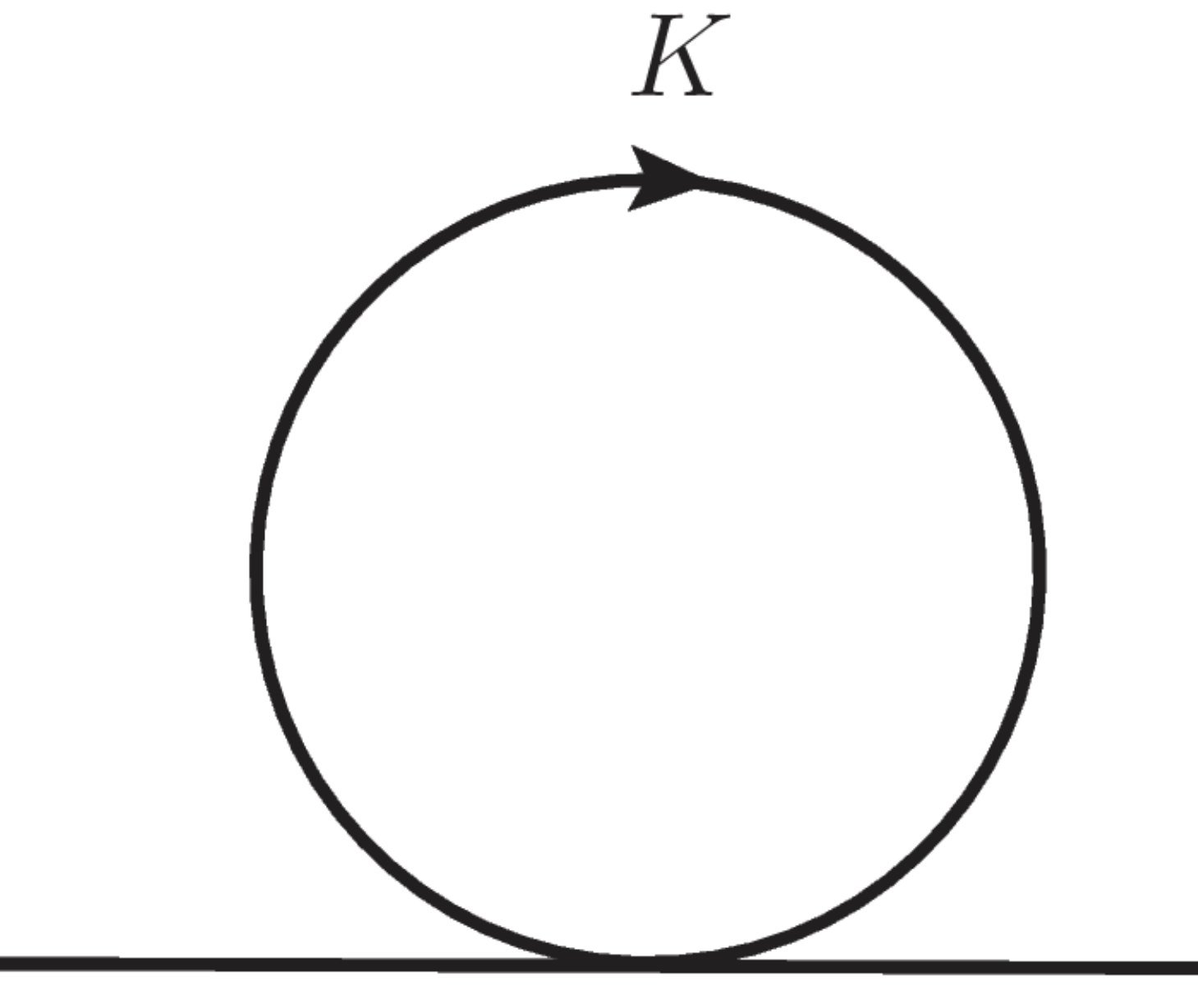}
 \caption{Tadpole diagram in $\lambda \phi^4$ theory.}
 \label{tadpole}
 \end{center}
 \end{figure}

The ${\cal S}$-matrix in Euclidean time as given in (\ref{eq23})
\begin{equation}
 {\cal S}(\beta)= {\cal T}\left [ \exp \left ({-\int_0^\beta} {\cal H}' \ d\tau \right) \right ] = \sum_N \frac{[\cdots]^N}{N!} , \label{lphi1}
\end{equation}
where ${\cal T}$ is the time ordered product in imaginary time $\tau=it$. Following the same procedure as $T=0$ field theory
the scalar Lagrangian is given as
\begin{eqnarray}
 {\cal L} &=& \frac{1}{2} \left[ \partial_\mu\phi \partial^\mu\phi-\frac{1}{2}m^2 \phi^2\right ] -\frac{\lambda}{4!} \phi^4
 ={\cal L}_0 \, + \, {\cal L}_I, \label{spf0}
\end{eqnarray}
where where $\phi$ is a scalar field, $\lambda$ is the coupling in the theory and ${\cal L}_0$ is the free scalar field Lagrangian. 
The interaction Lagrangian is 
\begin{eqnarray}
 {\cal L}_I= -\,\frac{1}{4!} \lambda \phi^4. \label{lphi2}
\end{eqnarray}
$4!$ comes from how many ways the $\phi$ fields can be arranged.
Vertex is $\,\,-i\lambda$ (same as vacuum).  The interaction Lagrangian in \eqref{lphi2} will lead to the tadpole diagram as shown in Fig~\ref{tadpole}.
 
 Symmetry factor: After contraction how the remaining legs (fields) are connected to the interaction
 vertex. The topology of the tadpole diagram in $\phi^4$ theory is given in Fig.~\ref{tadpole}, which results from one contraction
 involving two $\phi$ fields  and remaining two $\phi$ fields can then be connected to the vertex in two ways.  So, the symmetry factor is $1/2$
 (same as vacuum).
 
The amplitude corresponding to the tadpole diagram in Fig~\ref{tadpole} can be written following the Feynman rules defined
earlier as
\begin{eqnarray}
 \Pi &=& \frac{1}{2} \, (-i\lambda) \,  \int \frac{d^4K}{(2\pi)^4} \, \frac{i}{K^2-m^2} 
 = \frac{1}{2} \, (-i\lambda) \, \frac{1}{\beta} \sum_{k_0} \int \frac{d^3k}{(2\pi)^3}\, \frac{i}{k_0^2-\omega_k^2} \nn\\
 &=& \frac{1}{2} \, \lambda \, \int \frac{d^3k}{(2\pi)^3} \, \frac{1}{\beta} \sum_{k_0} \, \frac{1}{k_0^2-\omega_k^2}, \label{lf1}
\end{eqnarray}
 where $\omega_k=\sqrt{k^2+m^2}$ and $\Pi$ is independent of external momentum.
 
  Now the function under the frequency sum is same as those in (\ref{ex1})
 for $\tau=0$. So one can write
\begin{eqnarray}
\frac{1}{\beta} \sum_{k_0} \, \frac{1}{k_0^2-\omega_k^2}&=& \left. \left [I_1^0+I_2^0+I_3^\beta+I_4^\beta \right ]\right |_{\tau=0} \nn\\
&=& \left. \left [  \frac{e^{-\omega_k\tau}}{2\omega_k} + \frac{e^{-\omega_k\tau}}{2\omega_k} n_B(\omega_k ) 
+ \frac{e^{\omega_k\tau}}{2\omega_k} n_B(\omega_k ) \right ] \right |_{\tau=0} \nn\\
& = &   \frac{1}{2\omega_k} \Big [1 +  2 n_B(\omega_k )\Big ] . \label{lf2}
\end{eqnarray}
Using (\ref{lf2}) in (\ref{lf1}), one can write
\begin{eqnarray}
\Pi &=& \frac{1}{4} \, \lambda \, \int \frac{d^3k}{(2\pi)^3} \,  \frac{1}{\omega_k} \Big [1 +  2 n_B(\omega_k )\Big ]. \label{lf3}
\end{eqnarray}
Now, we note that
\begin{enumerate}
 \item[$\bullet$]  the first term is the vacuum ($T=0$) contribution and it is ultraviolate divergent. This could be regulated using
dimensional regularisation at $T=0$ and it vanishes.
\item[$\bullet$]the second term is finite as it involves the Bose-Einstein distribution, which falls of exponentially for 
large $\omega_k$ or momentum. The finite temperature does not cause any ultraviolate divergence but induct infrared 
divergence\footnote{At $m=0$ and $|k| \rightarrow 0 \, \Rightarrow n_B(\omega_k=0) =1/(1-1)\rightarrow \infty$ there is an 
infrared divergence due to zero bosonic mode caused by $\omega_n=2\pi  n T$ for $n=0$. We will come back later how  can this
infrared divergence be regulated.}.  
\end{enumerate}
The second term can be written for ($m=0$) as
\begin{eqnarray}
\Pi &=& \frac{1}{4} \, \lambda \, \frac {4\pi}{(2\pi)^3} \, \int\limits_{0}^{\infty}\,  dk \, \, k^2 \,  \frac{1}{k}   \frac{2}{e^{\beta k}-1}
= \frac{1}{4\pi^2} \, \lambda \, \int\limits_{0}^{\infty} \, dk \, k \, \frac{1}{e^{\beta k}-1}\nn \\
&=& \frac{\lambda T^2}{4 \pi^2} \, \int\limits_{0}^{\infty} \, \frac{dx\,\,x}{e^{x}-1} \, \, \, \, \, \, \, \,  [{\rm{Assumed}} \, \, \, x=\beta k] \nn \\
&=& \frac{\lambda T^2}{4 \pi^2} \, \zeta(2) 
= \frac{\lambda T^2}{24}. \label{lf4}
\end{eqnarray}
where we have used 
\begin{equation}
\int\limits_{0}^{\infty} \, \frac{dx\,\,x^{n-1}}{e^{x}-1} = (n-1)! \zeta(n). \label{zeta_fun}
\end{equation}
We note that $\Pi= \lambda T^2/24$ will act as a thermal mass of the scalar field at finite temperature. We will discuss this in details
when the dispersion property of a particle at finite $T$ will be discussed later.

\subsection{One-Loop Self-Energy in $\phi^3$-Theory}.  
\label{phiq}
 We consider three scalar fields $\phi$, $\phi_1$ and $\phi_2$ which differ by masses. The interaction Lagrangian density  is given by
\be
{\cal L}_I = -\lambda \phi(x)\phi_1(x)\phi_2(x) \, ,\label{phiq1} 
\ee
where $\lambda$ is the interaction strength. 
 
 \begin{figure}[hbt]
 \begin{center}
 \includegraphics[width=12cm,height=6cm]{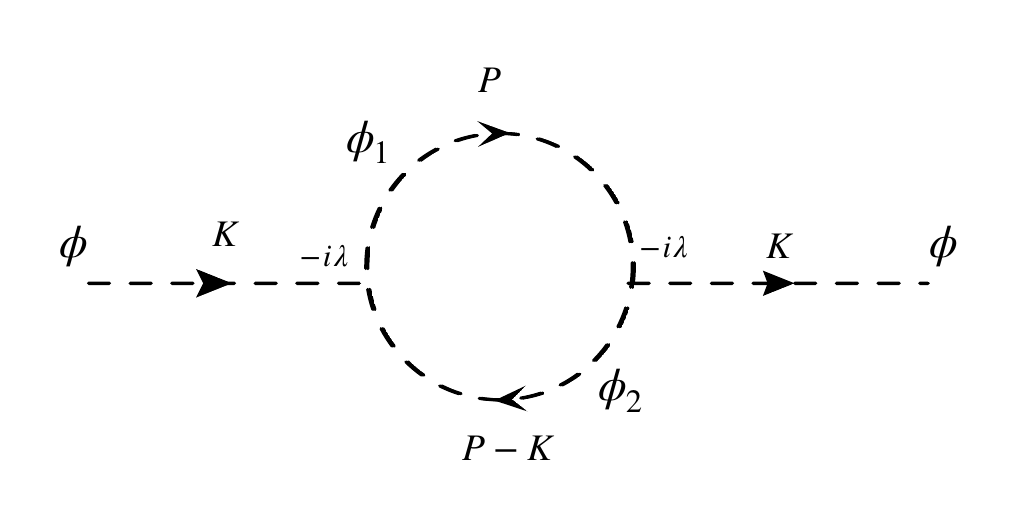}
 \caption{Scalar self-energy diagram in $\phi^3$-theory at $T\ne 0$.}
 \label{scalar_se}
 \end{center}
 \end{figure} 
 
Our aim is to compute the diagram in Fig.~\ref{scalar_se}, which occurs typically in one-loop approximation 
of the self-energy of the field $\phi$. The self-energy can be written from Fig.~\ref{scalar_se}  as
\be
\Pi_2(K) = \int \frac{d^4P}{(2\pi)^4} (-i\lambda) \frac{i}{P^2-m_1^2}(-i\lambda) \frac{i}{(P-K)^2-m_2^2}\, , \label{phiq2}
\ee
where $P$ is the momentum of the $\phi_1$ field with mass $m_1$, $P-K$ is the momentum of the $\phi_2$ field with mass $m_2$. The self-energy can be
written as
\bea
\Pi_2(k_0,k) &=& \lambda^2 \int \frac{d^3p}{(2\pi)^3} T\sum_{p_0}  \frac{1}{(p_0^2-E_p^2)[(p_0-k_0)^2-E_{p-k}^2]}  \nn \\
&=& \lambda^2 \int \frac{d^3p}{(2\pi)^3} T\sum_{p_0} f(p_0=i\om_n, k_0=i\om_m) \, , \label{phiq3}
\eea
where $E_p^2=p^2+m_1^2$ and $E^2_{p-k}=(p-k)^2+m_2^2$.
Now the frequency sum over $p_0$ should be replaced by the contour integral given in \eqref{sum_intb7} as
\bea
T\sum_{p_0} f(p_0=i\om_n, k_0=i\om_m) &=& \frac{1}{2\pi i} \int_{-i\infty}^{i\infty} dp_0 \frac{1}{2} \left[ f(p_0)+f(-p_0) \right ] \nn \\
&&+ \frac{1}{2\pi i} \int_{\eps-i\infty}^{\eps+i\infty} dp_0 \frac{f(p_0)+f(-p_0)}{e^{\beta p_0}-1}  \nn \\
&=& I^0_1+I^0_2+I^\beta_3+I^\beta_4 \, . \label{phiq4}
\eea
Calculation of $I^0_1$:
 \begin{figure}[h]
 \begin{center}
 \includegraphics[width=7cm,height=6cm]{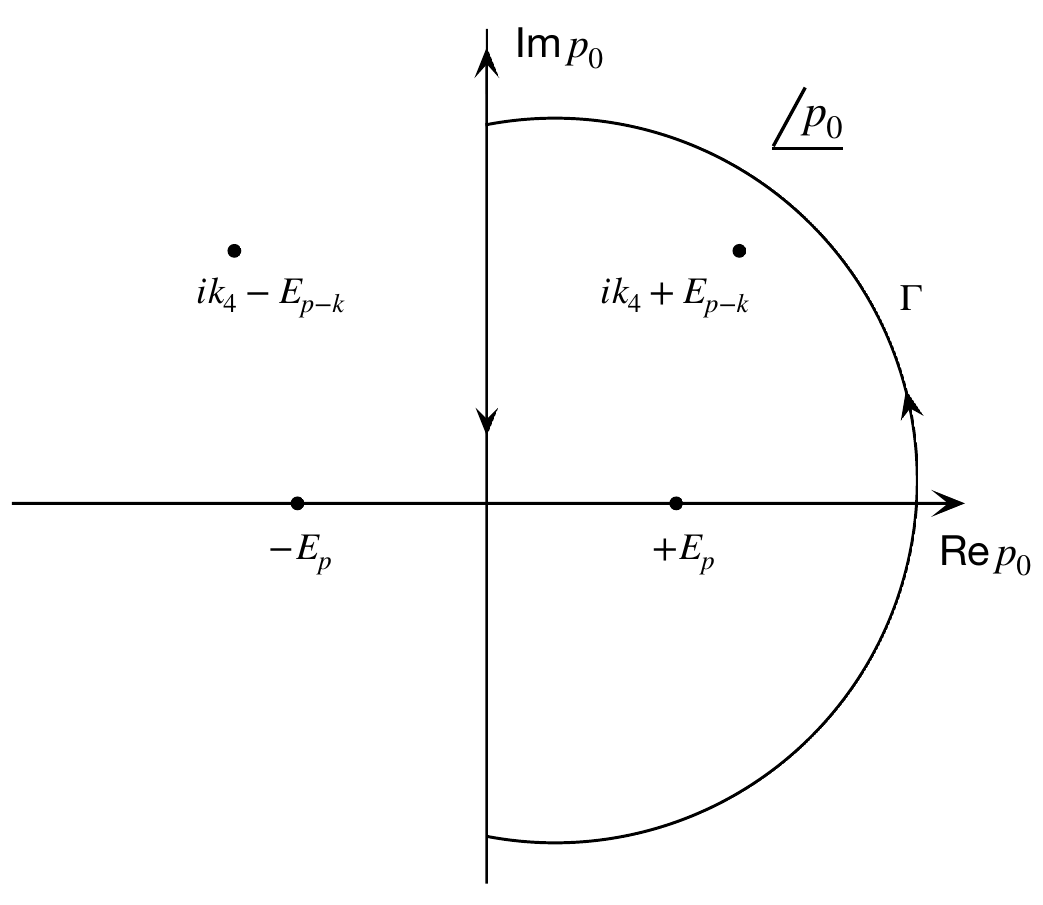}
 \caption{Contour corresponding to the integral $I_1^0$ in complex $p_0$ plane.}
 \label{contour1}
 \end{center}
 \end{figure} 

\bea
I^0_1&=&  \frac{1}{2\pi i} \frac{1}{2}  \int_{-i\infty}^{i\infty} dp_0  f(p_0) = \frac{1}{2\pi i} \frac{1}{2}  \int_{-i\infty}^{i\infty} dp_0 
\frac{1}{(p_0^2-E_p^2)[(p_0-k_0)^2-E_{p-k}^2]}\, , \label{phiq5}
\eea
which has four poles at $p_0=\pm E_p$ and  $k_0 \pm E_{p-k}$ with $k_0=ik_4=i\om_m$. The contour is in right half plane as shown in Fig.~\ref{contour1} 
from the definition of the conversion of frequency sum to contour integral. Therefore,
\bea
I^0_1&=&  \frac{1}{2\pi i} \frac{1}{2}  \int_{-i\infty}^{i\infty} dp_0  f(p_0) 
=  \frac{1}{2\pi i} \times \frac{1}{2} \times 2\pi i \times {\textrm{sum of residues}}  \nn \\ 
&=& \frac{1}{2} \left [ \left. R_1^0\right |_{p_0=E_p} +\left. R^0_2\right |_{p_0=k_0+E_{p-k}} \right] \nn \\
&=&  -\frac{1}{2} \left[ \frac{1}{2E_p}\, \frac{1}{(E_p-k_0)^2-E^2_{p-k}}+\frac{1}{2E_{p-k}}\, \frac{1}{(E_{p-k}+k_0)^2-E^2_{p}}\right] \, .
\label{phiq6}
\eea
Calculation of $I^0_2$:
 \begin{figure}[h]
 \begin{center}
 \includegraphics[width=7cm,height=6cm]{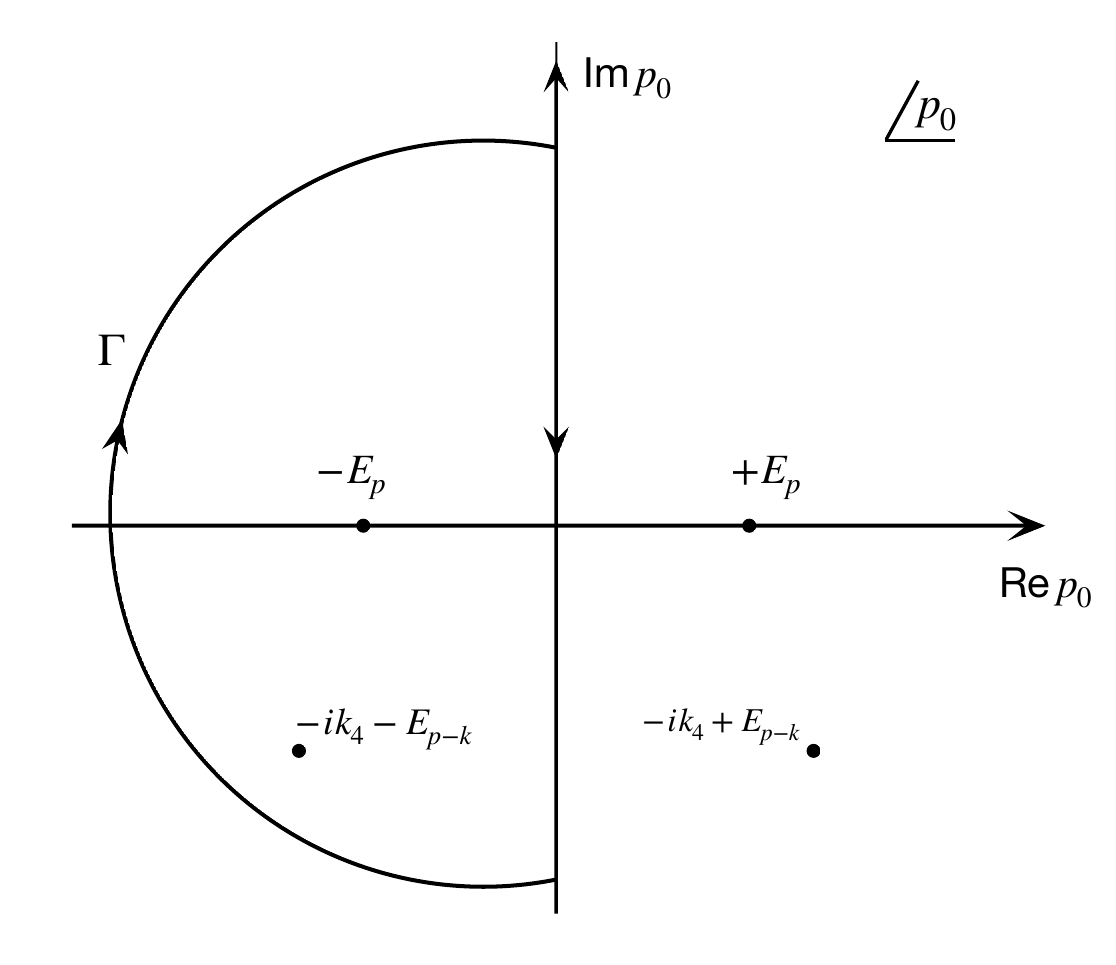}
 \caption{Contour corresponding to the integral $I_2^0$ in complex $p_0$ plane.}
 \label{contour2}
 \end{center}
 \end{figure} 
\bea
I^0_2&=&  \frac{1}{2\pi i} \frac{1}{2}  \int_{-i\infty}^{i\infty} dp_0  f(-p_0) = \frac{1}{2\pi i} \frac{1}{2}  \int_{-i\infty}^{i\infty} dp_0 
\frac{1}{(p_0^2-E_p^2)[(p_0+k_0)^2-E_{p-k}^2]}\, , \label{phiq7}
\eea
which has four poles at $p_0=\pm E_p$ and  $-k_0 \pm E_{p-k}$. The contour is in left half plane as shown in Fig.~\ref{contour2}.
 Therefore,
\bea
I^0_2&=&  \frac{1}{2\pi i} \frac{1}{2}  \int_{-i\infty}^{i\infty} dp_0 \frac{1}{2} f(-p_0) 
=  \frac{1}{2\pi i} \times \frac{1}{2} \times (-2\pi i) \times {\textrm{sum of residues}} \, , \label{phiq8} 
\eea
where negative sign in the right hand side is due to the contour in clockwise direction. We can now write
\bea
I^0_2&=&- \frac{1}{2} \left [ \left. R_1^{0'}\right |_{p_0=-E_p} +\left. R^{0'}_2\right |_{p_0=-k_0-E_{p-k}} \right] \nn \\
&=&  -\frac{1}{2} \left[ \frac{1}{2E_p}\, \frac{1}{(E_p-k_0)^2-E^2_{p-k}}+\frac{1}{2E_{p-k}}\, \frac{1}{(E_{p-k}+k_0)^2-E^2_{p}}\right] \, .
\label{phiq9}
\eea
Calculation of $I^\beta_3$:
 \begin{figure}[h]
 \begin{center}
 \includegraphics[width=7cm,height=6cm]{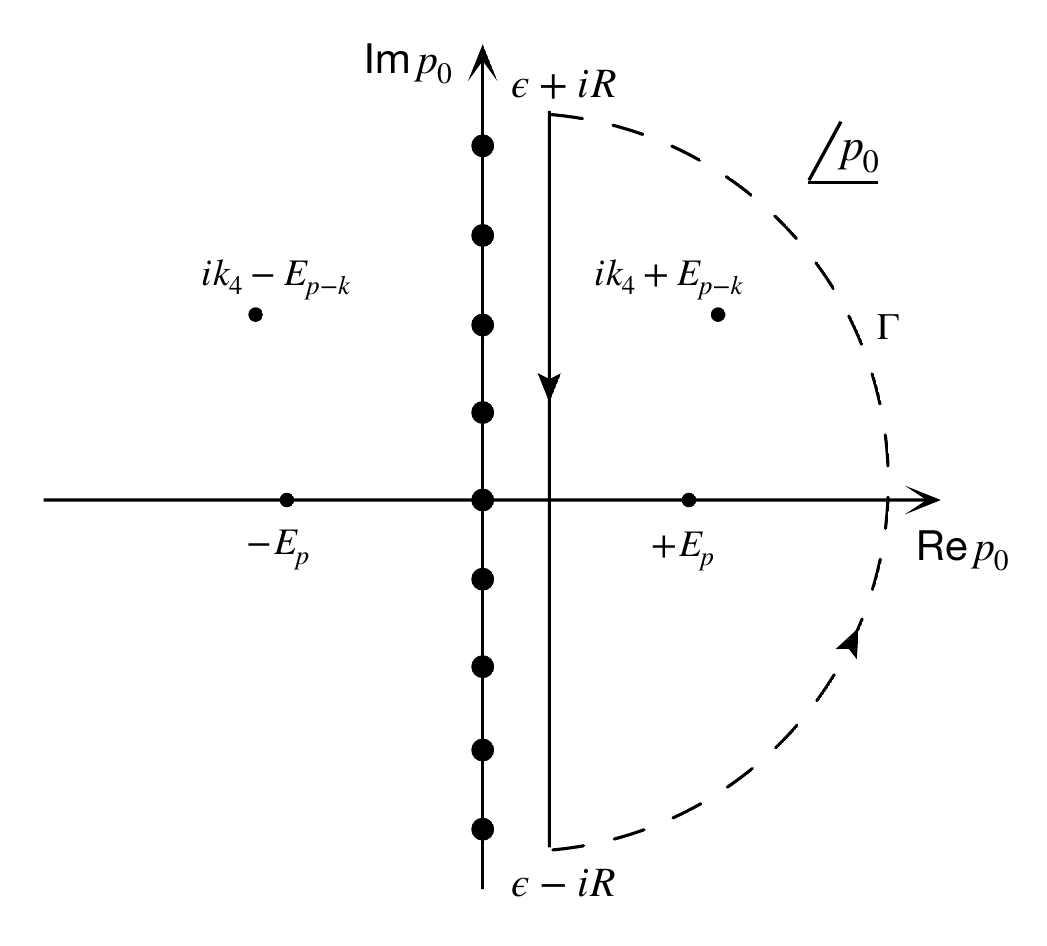}
 \caption{Contour corresponding to the integral $I_3^\beta$ in complex $p_0$ plane.}
 \label{contour3}
 \end{center}
 \end{figure} 
\bea
I^\beta_3&=&  \frac{1}{2\pi i}   \int_{\eps-i\infty}^{\eps+i\infty} dp_0  \frac{ f(p_0)}{e^{\beta p_0}-1}
= \frac{1}{2\pi i}   \int_{\eps-i\infty}^{\eps+i\infty} dp_0 
\frac{1}{(p_0^2-E_p^2)[(p_0-k_0)^2-E_{p-k}^2]} \frac{1}{e^{\beta p_0}-1}\, , \label{phiq10}
\eea
which has four poles at $p_0=\pm E_p$ and  $k_0 \pm E_{p-k}$. The contour is in right half plane as shown Fig.~\ref{contour3}. Therefore,
\bea
I^\beta_3&=&  \frac{1}{2\pi i}   \int_{\eps-i\infty}^{\eps+i\infty} dp_0  \frac{f(p_0)} {e^{\beta p_0}-1}
=  \frac{1}{2\pi i} \times  2\pi i \times {\textrm{sum of residues}}  \nn \\ 
&=&  \left [ \left. R^\beta_3\right |_{p_0=E_p} +\left. R^\beta_4\right |_{p_0=k_0+E_{p-k}} \right] \nn \\
&=&  - \left[ \frac{1}{2E_p}\, \frac{1}{(E_p-k_0)^2-E^2_{p-k}}\, \frac{1} {e^{\beta E_p}-1}
+\frac{1}{2E_{p-k}}\, \frac{1}{(E_{p-k}+k_0)^2-E^2_{p}} \, \frac{1} {e^{\beta(k_0+E_{p-k}) }-1}\right] \, .
\label{phiq11}
\eea
Using $k_0=ik_4=i\om_m=2\pi i m T$  we get 
\be
e^{\beta(k_0+E_{p-k}) }-1= e^{\beta E_{p-k} }-1 , \label{phiq12} 
\ee
as $e^{k_0\beta}=e^{2\pi i m }=1$. Now we can write
\bea
I^\beta_3&=&  - \left[ \frac{1}{2E_p}\, \frac{1}{(E_p-k_0)^2-E^2_{p-k}}\, \frac{1} {e^{\beta E_p}-1}
+\frac{1}{2E_{p-k}}\, \frac{1}{(E_{p-k}+k_0)^2-E^2_{p}} \, \frac{1} {e^{\beta E_{p-k} }-1}\right] \, .
\label{phiq13}
\eea

Calculation of $I^\beta_4$:
 \begin{figure}[h]
 \begin{center}
 \includegraphics[width=7cm,height=6cm]{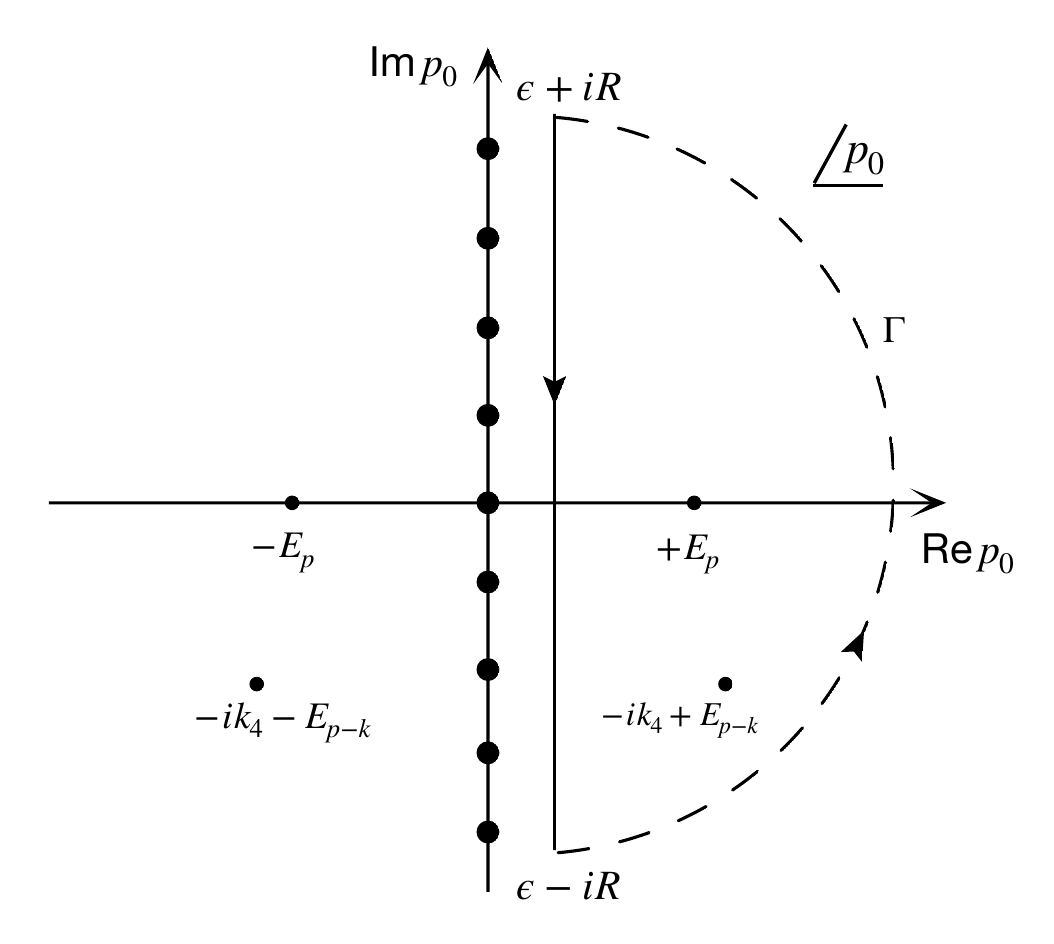}
 \caption{Contour corresponding to the integral $I_4^\beta$ in complex $p_0$ plane.}
 \label{contour4}
 \end{center}
 \end{figure} 
\bea
I^\beta_4&=&  \frac{1}{2\pi i}   \int_{\eps-i\infty}^{\eps+i\infty} dp_0   \frac{ f(-p_0)}{e^{\beta p_0}-1}= \frac{1}{2\pi i}   \int_{\eps-i\infty}^{\eps+i\infty} dp_0 
\frac{1}{(p_0^2-E_p^2)[(p_0+k_0)^2-E_{p-k}^2]} \frac{1}{e^{\beta p_0}-1}\, , \label{phiq14}
\eea
which has four poles at $p_0=\pm E_p$ and  $-k_0 \pm E_{p-k}$. The contour is in right half plane as shown in Fig.~\ref{contour4}. Therefore,
\bea
I^\beta_4&=&  \frac{1}{2\pi i}   \int_{\eps-i\infty}^{\eps+i\infty} dp_0  \frac{f(-p_0)} {e^{\beta p_0}-1}
=  \frac{1}{2\pi i}  \times 2\pi i \times {\textrm{sum of residues}}  \nn \\ 
&=&  \left [ \left. R_3^{\beta '} \right |_{p_0=E_p} +\left. R^{\beta '}_4\right |_{p_0=-k_0+E_{p-k}} \right] \nn \\
&=&  - \left[ \frac{1}{2E_p}\, \frac{1}{(E_p+k_0)^2-E^2_{p-k}}\, \frac{1} {e^{\beta E_p}-1}
+\frac{1}{2E_{p-k}}\, \frac{1}{(E_{p-k}-k_0)^2-E^2_{p}} \, \frac{1} {e^{\beta(E_{p-k}-k_0) }-1}\right] \nn \\
&=&  - \left[ \frac{1}{2E_p}\, \frac{1}{(E_p+k_0)^2-E^2_{p-k}}\, \frac{1} {e^{\beta E_p}-1}
+\frac{1}{2E_{p-k}}\, \frac{1}{(E_{p-k}-k_0)^2-E^2_{p}} \, \frac{1} {e^{\beta E_{p-k} }-1}\right] \, . 
\label{phiq15}
\eea
Using \eqref{phiq4}, \eqref{phiq6}, \eqref{phiq9}, \eqref{phiq13} and \eqref{phiq15} in \eqref{phiq3}, one can have the self-energy expression as
\bea
\Pi_2(k_0=i\om_m,k)&=&-\lambda^2 \int \frac{d^3p}{(2\pi)^3} \frac{1}{4E_pE_{p-k}} \nn \\
&& \left[ \left(1+n_B(E_p)+n_B(E_{p-k})\right) \left(\frac{1}{k_0-E_p-E_{p-k}}
-\frac{1}{k_0+E_p+E_{p-k}}\right)\right. \nn \\
&& \left. -(n_B(E_p)-n_B(E_{p-k}))\left(\frac{1}{k_0-E_p+E_{p-k}}
-\frac{1}{k_0+E_p-E_{p-k}}\right) \right ] \, , \label{phiq16}
\eea
where $n_B(E_i)=1/(e^{\beta E_i}-1)$. The terms $1/(k_0\pm E_p \pm E_{p-k})$ and $1/(k_0\mp E_p \pm E_{p-k})$ are the Landau damping factors.
We note the following points on \eqref{phiq16}:
\begin{enumerate}
\item[$\bullet$] It is to be noted that the Bose-Einstein distribution function $n_B(E_i)$ appearing in $\Pi_2(k_0,k)$ involves on-shell energies $E_P$ and $E_{p-k}$
of the internal lines of the self-energy but the energies of the internal lines should be off-shell. This implies that there should be cut or discontinuity in $\Pi_2(k_0,k)$.
\item[$\bullet$]  $\Pi_2(k_0=i\om_m,k)$ is defined for discrete imaginary values of $k_0=i\om_m=2\pi i mT$.
 One could make analytic continuation in whole complex plane by putting $k_0=i\om_m\rightarrow \om+i\eta$ if $\Pi^*_2(k_0=i\om_m,k)=\Pi_2(k^*_0,k)$.
\item[$\bullet$] It is easy to see that the analytic extension has cuts along the real axis and the discontinuity along the cuts is pure imaginary:
\be
\textrm{Disc} \, \Pi_2(k_0=i\om_m,k)=\Pi_2(k_0=\om+i\eta,k)-\Pi_2(k_0=\om-i\eta,k)=2i\,\textrm{Im} \, \Pi_2(k_0=\om+i\eta,k) \, . \label{phiq17}
\ee
\item[$\bullet$]  The discontinuity can easily be obtained by finding out $\textrm{Im} \, \Pi_2(k_0=\om+i\eta,k) $.  Using the relation
\be
\textrm{Im}\, f(q_0=\om+i\eta,q) =\textrm{Im}\, \left(\frac{1}{q_0+q\mp\eta}\right) = \pm\pi\delta(q_0+q) = - \textrm{Sgn}(\eta)\pi \delta(q_0+q) \, , \label{phiq18}
\ee
one can find the $\textrm{Im} \, \Pi_2(k_0=\om+i\eta,k) $ as
\bea
&&\textrm{Im} \, \Pi_2(\om,k) =\pi \lambda^2 \int \frac{d^3p}{(2\pi)^3} \frac{1}{4E_pE_{p-k}} \nn \\
&\times& \Big[ \Big\{(1+n_B(E_p))(1+n_B(E_{p-k}))-n_B(E_p)n_B(E_{p-k})\Big\} \nn \\
&\times& \Big\{\delta(\om-E_p-E_{p-k})-\delta(\om+E_p+E_{p-k})\Big\}\nn\\
&-& \Big\{n_B(E_p)(1+n_B(E_{p-k})) -n_B(E_{p-k})(1+n_B(E_p))\Big\}\nn\\
&\times& \Big\{\delta(\om-E_p+E_{p-k})-\delta(\om+E_p-E_{p-k})\Big\}\Big] . \nn\\
\label{phiq19}
\eea
\begin{figure}[h]
 \begin{center}
 \includegraphics[width=6cm,height=4cm]{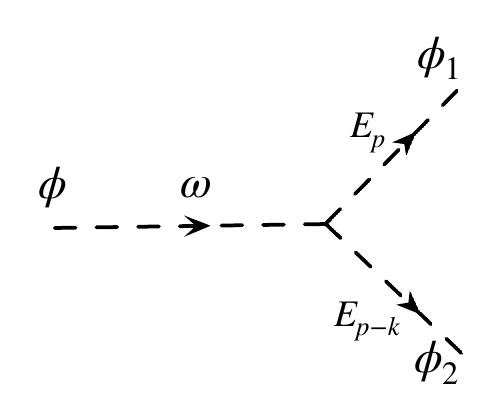}
 \caption{Feynman diagram for decay process in $\phi^3$-theory at $T= 0$.}
 \label{decay0}
 \end{center}
 \end{figure} 
\item[$\bullet$] At $T=0$ 
\be
\textrm{Im} \, \Pi_2(\om,k) =\pi \lambda^2 \int \frac{d^3p}{(2\pi)^3} \frac{1}{4E_pE_{p-k}} \Big\{\delta(\om-E_p-E_{p-k})-\delta(\om+E_p+E_{p-k})\Big\} .\nn
\ee
The first term with energy conserving $\delta(\om-E_p-E_{p-k})$ indicates a decay process $\phi\rightarrow \phi_1+\phi_2$ as shown in Fig.~\ref{decay0}.
The energy conserving $\delta(\om+E_p+E_{p-k})$ in the second term will never be satisfied and hence does not correspond to any physical process.

\item[$\bullet$] At $T\ne 0$ the available phase space is weighted by the distribution function. 
There will also be additional processes compared to $T=0$ case above. The processes are related by principle of detailed balance. 
  \begin{figure}[h]
 \begin{center}
 \includegraphics[width=10cm,height=5cm]{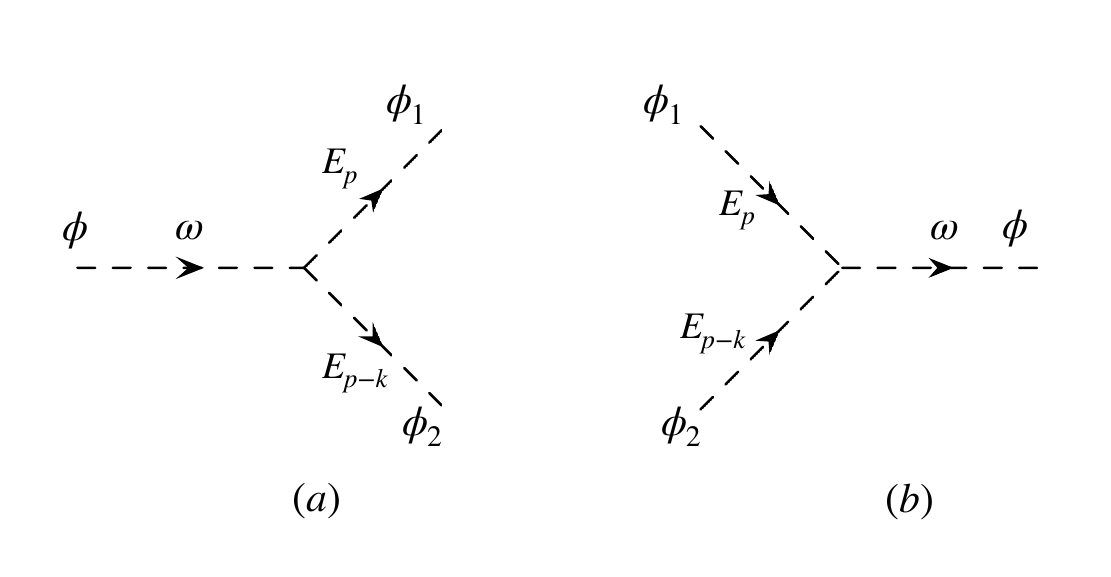}
 \caption{Feynman diagram for various processes in $\phi^3$-theory at $T\ne 0$.}
 \label{decay1}
 \end{center}
 \end{figure} 
 
 (a) Consider the  first term in \eqref{phiq19}:
 \be
 \Big[(1+n_B(E_p))(1+n_B(E_{p-k}))-n_B(E_p)n_B(E_{p-k})\Big ]\delta(\om-E_p-E_{p-k}) \, .\nn
\ee
The  term $(1+n_B(E_p))(1+n_B(E_{p-k}))\delta(\om-E_p-E_{p-k})$ indicates a decay process $\phi\rightarrow \phi_1+\phi_2$ in Fig.~\ref{decay1}(a) similar to $T=0$ case but modified by thermal weight factor. The term $n_B(E_p)n_B(E_{p-k})\delta(\om-E_p-E_{p-k})$ indicates a reverse process $\phi_1+\phi_2\rightarrow \phi$ 
in Fig.~\ref{decay1}(b) which was not there in $T=0$ case.

  \begin{figure}[h]
 \begin{center}
 \includegraphics[width=10cm,height=5cm]{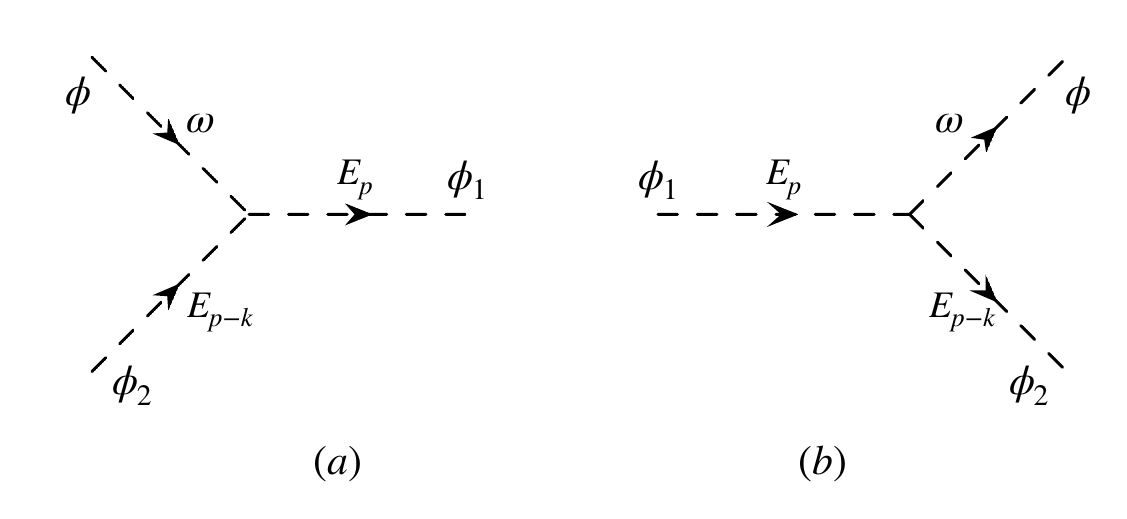}
 \caption{Feynman diagram for various processes in $\phi^3$-theory at $T\ne 0$.}
 \label{decay2}
 \end{center}
 \end{figure} 
(b) Consider the third term  in \eqref{phiq19}:
\be
\Big[n_B(E_p)(1+n_B(E_{p-k})) -n_B(E_{p-k})(1+n_B(E_p))\Big]\delta(\om-E_p+E_{p-k}) \, . \nn
\ee
The  term $n_B(E_p)(1+n_B(E_{p-k})) \delta(\om-E_p+E_{p-k})$ indicates a absorption process $\phi+\phi_2\rightarrow \phi_1$ in Fig.~\ref{decay2}(a) . The  term 
$n_B(E_{p-k})(1+n_B(E_p))\delta(\om-E_p+E_{p-k})$ indicates an emission process $\phi_1\rightarrow \phi+\phi_2$ in Fig.~\ref{decay2}(b) . These process were not there in $T=0$ case.

(c) Consider the fourth term in \eqref{phiq19}:
\be
\Big[n_B(E_p)(1+n_B(E_{p-k})) -n_B(E_{p-k})(1+n_B(E_p))\Big]\delta(\om+E_p-E_{p-k}) \, . \nn
\ee
The term $n_B(E_p)(1+n_B(E_{p-k})) \delta(\om+E_p-E_{p-k}) $ indicates a absorption process $\phi+\phi_1\rightarrow \phi_2$ as shown in Fig.~\ref{decay3}(a) whereas
the term $n_B(E_{p-k})(1+n_B(E_p))\delta(\om+E_p-E_{p-k}) $ implies an emission process $\phi_2\rightarrow \phi+\phi_1$ in Fig.~\ref{decay3}(b) . These process were not there in $T=0$ case.
  \begin{figure}[h]
 \begin{center}
 \includegraphics[width=10cm,height=5cm]{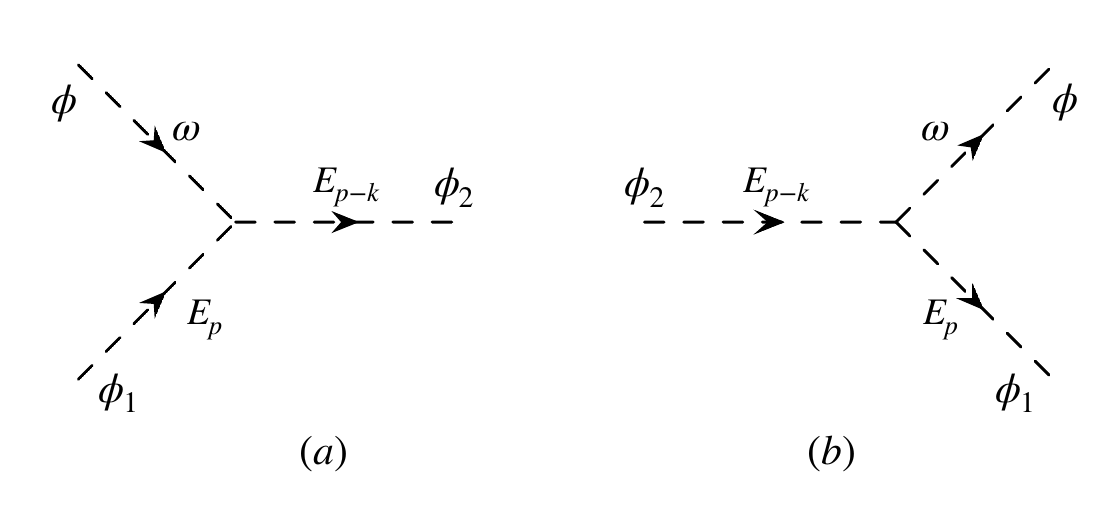}
 \caption{Feynman diagram for various processes in $\phi^3$-theory at $T\ne 0$.}
 \label{decay3}
 \end{center}
 \end{figure} 
\end{enumerate}

\section{Partition Function}
\label{PF}
Using (\ref{eq23}) the density matrix of the system in (\ref{eq18}) becomes
\begin{eqnarray}
\rho (\beta) &\equiv& e^{-\beta {\cal H}} = \rho_0(\beta) S(\beta) = e^{-\beta {\cal H}_0 } \, {\cal T} \left ( e^{{-\int_0^\beta} {\cal H}' \ d\tau } \right )\nn\\
&=&  e^{-\beta {\cal H}_0 } \, \sum_{l=0}^\infty \frac{1}{l!}\,  {\cal T} \left (-\int\limits_0^\beta {\cal H}' \ d\tau \right)^l .\label{eqpa1} 
\end{eqnarray}
Using this the partition function in (\ref{eq14}) becomes
\begin{eqnarray}
{\cal Z}(\beta) = {\rm{Tr}}\rho (\beta) \ &=& {\rm{Tr}} \left[ e^{-\beta {\cal H}_0 } \, \sum_{l=0}^\infty \frac{1}{l!}\,  
{\cal T} \left (-\int\limits_0^\beta {\cal H}' \ d\tau \right)^l \, \, \right ] \nn\\
&=& {\rm{Tr}} \left[ e^{-\beta {\cal H}_0 } \, + \, e^{-\beta {\cal H}_0 } \ \sum_{l=1}^\infty \frac{1}{l!}\,  
{\cal T} \left (-\int\limits_0^\beta {\cal H}' \ d\tau \right)^l \, \, \right ] \nn \\
&=& {\rm{Tr}} \left[ e^{-\beta {\cal H}_0 } \right ]  \, + {\rm{Tr}} \left[ \, e^{-\beta {\cal H}_0 } \ \sum_{l=1}^\infty \frac{1}{l!}\,  
{\cal T} \left (-\int\limits_0^\beta {\cal H}' \ d\tau \right)^l \, \, \right ] \nn\\
&=& {\cal Z}_0 + {\cal Z}_I^l, \label{eqp2}
\end{eqnarray}
where the free (${\cal Z}_0$) and the interaction (${\cal Z}_I^l $) pieces are separated out by expanding around the free piece.

The logarithm of the partition function is of interest as far as thermodynamic quantities are concerned. We can write as
\begin{eqnarray}
\ln {\cal Z}(\beta)&=& \ln \left[{\cal Z}_0 + {\cal Z}_I^l \right ]
= \ln \left \{{\cal Z}_0 \left[1 + \frac{{\cal Z}_I^l}{{\cal Z}_0} \right ] \right \} 
= \ln {\cal Z}_0 + \ln \left[1 + \frac{{\cal Z}_I^l}{{\cal Z}_0} \right ],\nn\\
&=& \ln {\cal Z}_0 +  \left[ \frac{{\cal Z}_I^l}{{\cal Z}_0} \, \, -\frac{1}{2} \left(\frac{{\cal Z}_I^l}{{\cal Z}_0}\right )^2\, + \cdots  \right ].
\label{eqp3}
\end{eqnarray}
where
\bea
\ln {\cal Z}_I &=& \ln \left[1 + \frac{{\cal Z}_I^l}{{\cal Z}_0} \right ] = \ln \left(1+ 
\frac{{\rm{Tr}} \left[ \, e^{-\beta {\cal H}_0 } \ \sum_{l=1}^\infty \frac{1}{l!}\, {\cal T} \left (-\int\limits_0^\beta {\cal H}' \ d\tau \right)^l \, \, \right ]}
{{\rm{Tr}} \left[ e^{-\beta {\cal H}_0 } \right ]}\right) \nn\\
&=&\ln \left (1+ \sum_{l} \frac{1}{l!}\, \left \langle {\cal T} \left (-\int\limits_0^\beta {\cal H}' \ d\tau \right)^l \right \rangle_0 \right )
= \ln \left[1 +\sum_l \frac{\left\langle {\cal  F}_I^l \right \rangle_0 }{l!}\right ] , \label{eqp3a}
\eea
where $\langle \cdots\rangle_0$ is the expectation value with respect to non-interacting ensemble. We know that ${\cal H}'$ contains one power of coupling in the theory.
Now an expansion of  $\left\langle {\cal  F}_I^l \right \rangle_0$ up to 3rd order in $l$ ($=1,2, 3$) means an expansion of  $\ln {\cal Z}_I$ up to third order in coupling, 
which can then be written as
\bea
\ln {\cal Z}_I &\approx& \ln \left [1+ \left\langle {\cal  F}_I \right \rangle_0 +  \frac{1}{2} \left\langle {\cal  F}_I^2 \right \rangle_0 
+ \frac{1}{6}\left\langle {\cal  F}_I^3 \right \rangle_0\right ]=\ln(1+x). \label{eqp3b}
\eea
Again expanding $\ln(1+x)= \sum_{j=1}^\infty \ (-1)^{j+1} x^j/j$, one can write
\bea
\ln {\cal Z}_I &\approx& x-\frac{1}{2}x^2 +\frac{1}{3}x^3 +\cdots  \nn\\
&\approx& \left\langle {\cal  F}_I \right \rangle_0 +  \frac{1}{2} \left ( \left\langle {\cal  F}_I^2 \right \rangle_0 - 
\left\langle {\cal  F}_I \right \rangle_0^2\right ) 
+ \frac{1}{6} \left ( \left\langle {\cal  F}_I^3 \right \rangle_0
-3 \left\langle {\cal  F}_I \right \rangle_0 \, \left\langle {\cal  F}_I^2 \right \rangle_0 
+ 2 \left\langle {\cal  F}_I \right \rangle_0^3 
\right ) \nn\\
&\approx& \left (\ln {\cal Z}_I \right )_1 + \left (\ln {\cal Z}_I \right )_2 + \left (\ln {\cal Z}_I \right )_3 ,
\label{eqp3c}
\eea
where we have assembled the terms according to the power of coupling. Using (\ref{eqp3c}) in (\ref{eqp3}), the perturbative expansion
of the partition function becomes
\bea
\ln {\cal Z} &\approx& \ln {\cal Z}_0  + \left (\ln {\cal Z}_I \right )_1 + \left (\ln {\cal Z}_I \right )_2 + \left (\ln {\cal Z}_I \right )_3 ,
\label{eqp3d} 
\eea

where 
\bea
\left ( \ln {\cal Z}_I \right )_1 &=& \left\langle{\cal  F}_I \right \rangle_0 \, , \label{eqp3d1}  \\
\left (\ln {\cal Z}_I \right )_2 &=& \frac{1}{2} \left ( \left\langle {\cal  F}_I^2 \right \rangle_0 - 
\left\langle {\cal  F}_I \right \rangle_0^2\right )\, , \label{eqp3d2} \\ 
\left (\ln {\cal Z}_I \right )_3 &=& \frac{1}{6} \left ( \left\langle {\cal  F}_I^3 \right \rangle_0
-3 \left\langle {\cal  F}_I \right \rangle_0 \, \left\langle {\cal  F}_I^2 \right \rangle_0 
+ 2 \left\langle {\cal  F}_I \right \rangle_0^3 
\right ). \label{eqp3d3}  
\eea

This is a perturbative expansion of the partition function around the free theory and one needs to compute it in order by order
of the coupling strength of a given theory. As discussed earlier that the Trace in (\ref{eqp2}) stands for sum over expectation values of all 
possible states in Hilbert space and there are  infinite number of such states in quantum field theory. So, for an interacting system the partition 
function will extremely be tedious to compute  even if one expands in perturbation series in interaction strength in a given theory. 
It would be convenient to compute 
the partition function in functional or path  integral approach.

\subsection{Relation of Functional Integration and the Partition Function}
\label{FI_Z}
  
Since we will be dealing  with statistical thermodynamics problem when the system returns to its initial state after a time evolution from $t=0$ to $t$, the corresponding
transition can be written in a functional form as $\langle \phi_a|e^{-i{\cal H}t}|\phi_a\rangle$, assuming the Hamiltonian is time independent that simplifies
the transition amplitude from one state to other.

The transition amplitude in Minkowski space-time is obtained in functional integration approach in \eqref{path18} as  
\bea
\langle \phi_a | e^{-i {\cal H} t} |\phi_a\rangle &=& \int\limits_{\phi(\bm{\vec x},0) = \phi_a(\bm{\vec x})}^{\phi(\bm{\vec x},t)=\phi_a(\bm{\vec x})}  {\cal D}\phi 
\ e^{i\int\limits_{0}^{t_f} dt \int  d^3x \ {\cal L} (\phi(\bm{\vec x}, t), {\dot \phi(\bm{\vec x}, t)})} 
=\int\limits_{\phi(\bm{\vec x},0) =\phi_a(\bm{\vec x})}^{\phi(\bm{\vec x},t)=\phi_a(\bm{\vec x})}  {\cal D}\phi  \ e^{i S [\phi]}\, , \label{fi1}
\eea
where $S[\phi]$ is the action of a system. This is the so-called path integral, and 
where $\cal D$ is the functional or path integral runs over all possible paths of the field $\phi(x)$. These fields are  restricted by
boundary conditions while going from initial time $t=0$ to final time $t_f=t$ as discussed in subsec~\ref{path_int}.
The action in Minkowski space-time is written as
\begin{equation}
 {\cal S}[\phi] = \int  d^4 {X} \,  {\cal L} = \int\limits_0^t\, dt \int \, d^3 {x}\, {\cal L} . \label{fi4} 
\end{equation}

Now, the partition function in \eqref{eq2} reads as
\begin{equation}
{\cal Z} ={\rm{Tr}}\rho \ = {\rm{Tr}} \left (  e^{-\beta {\cal H}} \right ) 
= \sum_n \ \  \left \langle  n  \left |  e^{-\beta {\cal H}} 
\right | n  \right \rangle \ \ , \label{fi3a} 
\end{equation}
where the summation over $n$ includes all the possible energy eigenstates of the system in Hilbert space. 
In the continuum case the summation becomes an integral, and the eigenstates $|\phi\rangle$
form a complete set, each with energy $E_\phi$ . Thus, the partition function becomes
\begin{equation}
{\cal Z} = \int \ d\phi \  \left \langle \phi  \left |  e^{-\beta {\cal H}} 
\right | \phi  \right \rangle  \left (= \int d\phi e^{-\beta E_\phi} \right ) \ . \label{fi4a} 
\end{equation}
If one compares (\ref{fi1}) and (\ref{fi4a}), there is a striking similarity between the path integral formulation of the transition amplitude
in quantum field theory and the partition function in statistical mechanics provided  that
\begin{enumerate}
\item the Boltzmann factor $e^{-\beta {\cal H}}$ acquires the form of a time 
evaluation operator ($e^{-i{\cal H}t}$) for imaginary time ($\beta=it$) through analytic continuation. 

\item  the time interval $[0,t]$ in the transition amplitude in (\ref{fi1}) takes the role of $\beta$ in the
 partition function with interval $[0,\beta$] along with $\tau=it$ . This is known as {\it Wick rotation}  
 that rotates the integration by $90^\circ$ in complex plane  as displayed in Fig.~\ref{wick_rot}.
 
\item  the field $\phi$ obey  periodic or anti-periodic boundary condition, $\phi(\bm{\vec x},0)=\pm\phi(\bm{\vec x}, \beta)$, 
as discussed earlier in subsec.~\ref{period}.
\end{enumerate}

With this the transition amplitude can be regarded as the partition function in path integral approach as
\begin{eqnarray}
 {\cal Z} ={\rm{Tr}}\rho \ &=& {\rm{Tr}} \left (  e^{-\beta {\cal H}} \right )  
= \int \ d\phi \  \left \langle \phi  \left |  e^{-\beta {\cal H}}  \right | \phi  \right \rangle 
=  \int {\cal D}\phi\ 
   e^{i\int\limits_0^t \, dt \int \, d^3 {x}\, {\cal L}}  \label{fi5} \nn\\
&{=\atop t\rightarrow -i\tau}& \, \int\limits_{\phi(\bm{\vec x},0)=\pm\phi(\bm{\vec x}, \beta)} {\cal D}\phi\ 
  e^{\int\limits_0^\beta\, d(it) \int \, d^3 {x}\, {\cal L}(t \rightarrow -i\tau)}  \nn \\
& =& \int\limits_{\phi(\bm{\vec x},0)=\pm\phi(\bm{\vec x}, \beta)} {\cal D}\phi\ 
   e^{\int\limits_0^\beta\, d\tau \int \, d^3 {x}\, {\cal L}(t \rightarrow -i\tau)}
  . \label{fi6}
\end{eqnarray}
 One can  compute the partition function in Euclidean time $\tau$ and discrete frequency $i\omega_n$ directly using (\ref{fi6}). 

\subsection{Scalar Field Partition Function}   
\label{sc_part}

\subsubsection{Partition function for free real scalar field}

A real  non-interacting scalar field Lagrangian is given in \eqref{spf0}  by
\begin{eqnarray}
 {\cal L}_0 &=& \frac{1}{2} \partial_\mu\phi \partial^\mu\phi-\frac{1}{2}m^2 \phi^2 
 = \frac{1}{2} \left[\left (\frac{\partial \phi}{\partial t}\right)^2
 -({\mathbf \nabla} \phi )^2-m^2 \phi^2 \right], \label{spf1}
\end{eqnarray}
in Minkowski space-time. 
 This can be written only in Euclidean time ~\cite{Kapusta,Schmitt,Yang,Inga} as
 \begin{eqnarray}
  {\cal L}_0(t\rightarrow -i\tau) &{=\atop{t\rightarrow -i\tau}} & \frac{1}{2} \left[\left ({\partial_t \phi}\right)^2
  -({\mathbf \nabla} \phi )^2 -m^2 \phi^2 \right] 
   =- \frac{1}{2} \left[\left ({\partial_\tau \phi}\right)^2
  + ({\mathbf \nabla} \phi )^2+m^2 \phi^2 \right]\, . \label{spf2}
 \end{eqnarray}

The Fourier transform of the field $\phi(X)$ can be written as
\begin{eqnarray}
\phi(X)= \phi(x,\tau) &{=\atop {t\rightarrow -i\tau}} \atop {k_0\rightarrow i\omega_n}& \frac{1}{\sqrt{V\beta}} \sum_K \, 
 e^{-i K\cdot  X }\, \phi (K)
=   \frac{1}{\sqrt{V\beta}} \sum_K \, 
e^{i \bm{\vec k \cdot \vec x } }\,
\, e^{- i (i\omega_n) (-i\tau)  } \, \phi (\omega_n,\bm{\vec k})\nn\\
&=&\frac{1}{\sqrt{V\beta}} \sum_K \, 
e^{i \bm{\vec k \cdot \vec x } }
\, e^{- i \omega_n \tau  } \, \phi (\omega_n,\bm{\vec k}),
\label{spf3}
\end{eqnarray}
where $\sum_K=\sum_{n, \bm{\vec k}}$ and $V$ is the three volume. 

Using \eqref{spf2} in  (\ref{fi6}), one can write the partition function for free scalar field as 
\begin{eqnarray}
 {\cal Z}_0 
  &= &\, \int\limits_{\phi (\bm{\vec x}, 0)=\phi(\bm{\vec x},\beta)} {\cal D}\phi\ 
  e^{-\frac{1}{2}  
  \int\limits_0^\beta \, d\tau \int \, d^3 {x}\,
 \left[\left ({\partial_\tau \phi} \right)^2
 + ({\mathbf \nabla} \phi )^2 + m^2 \phi^2 \right]}, \label{spf4} 
\end{eqnarray}
where scalar field obeys periodicity condition.

Now we calculate explicitly the terms in the exponential of \eqref{spf4}:
\begin{enumerate}
 \item[] First term:
 \begin{eqnarray}
&&\int\limits_0^\beta d\tau \int  d^3 {x}\, \left (\partial_\tau \phi \right)^2 =  \int d^4X \,
\frac{1}{V\beta} \sum_K \,  \sum_{K'} \,
 \partial_\tau \left[ e^{-i\omega_n \tau}\, e^{i \bm{\vec k \cdot \vec x } } \phi(\omega_n,\bm{\vec k}) \right ]\nn\\
&&\times  \partial_\tau \left[ e^{-i \omega_m \tau }\, e^{i \bm{\vec k' \cdot \vec x } }\phi(\omega_m,\bm{\vec k'}) \right ] \nn\\
  &=&  \int  d^4X \, \frac{1}{V\beta} \sum_K \,  \sum_{K'} \,
    e^{i (\bm{\vec k} + \bm{\vec k'}) \cdot \bm {\bm x} } e^{-i(\omega_n+\omega_m)\tau}
 \phi(\omega_n,\bm{\vec k}) (- i\omega_n)(-i\omega_m) \phi(\omega_m,\bm{\vec k'}) \nn\\
 &=&  \frac{1}{V\beta} \sum_K   \sum_{K'}  \left [ V \delta^3(\bm{\vec k} +\bm{\vec k'})\right ] \left[ \beta  \delta (-\omega_n-\omega_m) \right]  
 \phi(\omega_n,\bm{\vec k})  (- i\omega_n)(-i\omega_m) \phi(\omega_m,\bm{\vec k'}) \nn\\
& = &  \sum_K \ \,\phi(\omega_n,\bm{\vec k}) \,  \omega_n^2  \phi(-\omega_n,-\bm{\vec k}) 
 =   \sum_K  \, \omega_n^2  \, \phi(K) \,   \phi(-K) \nn \\
 &=&  \sum_K  \, \omega_n^2  \, \phi^* (K) \, \phi(K),
  \label{spf6}
 \end{eqnarray}
where we have  used  $\phi(K)=\phi^*(-K)$ since $\phi(X)$ is real in (\ref{spf3}). We have also used of the following relations:
\begin{eqnarray}
\int\limits_0^\beta d \tau \ e^{i(-\omega_n-\omega_m)\tau} &=& \beta \delta(-\omega_n-\omega_m) \, ,  \label{spf7a} \\
\int d^3x \ e^{i(\bm{\vec k} + \bm{\vec k'})\cdot x} &=&  V \delta^3(\bm{\vec k} + \bm{\vec k'}) \, ,   \label{spf7b} \\
 \omega_{-n} &=& -\frac{2\pi n}{\beta} = -\omega_n \, .  \label{spf7c} 
\end{eqnarray}

 \item[] Second  term:
 \begin{eqnarray}
 \int\limits_0^\beta d\tau \int  d^3 {x}\, \left (\mathbf \nabla \phi \right)^2 &=& 
  \,  \sum_K {k}^2 \phi^* (K) \, \phi(K). \label{spf8}
 \end{eqnarray}
 
  \item[] Third  term:
 \begin{eqnarray}
 \int\limits_0^\beta d\tau \int  d^3 {x}\,\,  m^2\phi^2 &=& 
  \,  \sum_K \, m^2 \, 
 \phi^*(K) \,  \phi(K) \, . \label{spf9}
 \end{eqnarray}
 \end{enumerate}
 Using (\ref{spf6}),  (\ref{spf8}) and (\ref{spf9}) one can write
 \begin{eqnarray}
{-\frac{1}{2}  
  \int\limits_0^\beta \, d\tau \int \, d^3 {x}\,
 \left[\left ({\partial_\tau \phi} \right)^2
 + ({\mathbf \nabla} \phi )^2 + m^2 \phi^2 \right]} &=& -\, \frac{1}{2} 
 \,   \sum_K \, \,\phi^*(K) \Big [k_0^2+{k}^2 +m^2 \Big ] \phi(K) \nn\\
 &=&  -\, \frac{1}{2} 
 \,   \sum_K \, \,\phi^*(K) \Big [k_0^2+\omega_k^2 \Big ] \phi(K) \nn\\
 &=& -\frac{1}{2} 
 \, \sum_K \,\phi^*(K)\, {\cal G}^{-1}_0(K) \phi(K) \, ,
\label{spf10}
\end{eqnarray}
where ${\cal G}^{-1}_0=(\omega_n^2+\omega_k^2)$ is the free inverse propagator in Euclidean time with energy $\omega_k=\sqrt{k^2+m^2}$. 

Using (\ref{spf10}) in (\ref{spf4}) one can write the free scalar field partition function as
\begin{eqnarray}
{\cal Z}_0\,\,  &=& \, \int\limits_{\phi(\bm{\vec x},0)=\phi(\bm{\vec x}, \beta)} {\cal D}\phi\ 
\exp \left [ - \frac{1}{2} 
 \, \sum_K\,\phi^*(K)\,[{\cal G}^{-1}_0(K) ] \phi(K) \right ] \nn\\
 &=& \prod_K \int\limits_{\phi(\bm{\vec x},0)=\phi(\bm{\vec x}, \beta)} {\cal D}\phi\ 
\exp \left [ -\frac{1}{2} 
 \, \phi^*(K)\, \,  [{\cal G}^{-1}_0(K) ] \,\, \phi(K) \right ] 
. \label{spf11}
\end{eqnarray} 
We also note that inverse propagator is in general a diagonal matrix. 

The integral can be performed using the standard
identity\footnote{This identity  is a generalisation  of the one dimensional Gaussian integral 
$\int\limits_{-\infty}^\infty dx \ \ \exp(-\frac{1}{2} a y^2) =\sqrt{2\pi /a}$ and can be shown by expressing 
the bilinear $ {\mathbf y}\cdot {\widehat A} {\mathbf y}$ in terms of eigenvalues of $\widehat A$.}
\begin{eqnarray}
 \int d^Dy\, \,  e^{-\frac{1}{2} \, {\mathbf y} \cdot {\widehat A} {\mathbf y}} = (2\pi )^{D/2} ({\mbox{det}} {\widehat A})^{-1/2}, \label{spf12}
\end{eqnarray}
for hermitian positive definite matrix $\widehat A$ and $D$ is the dimension of the system. The partition function can now be written as
\begin{eqnarray}
 {\cal Z}_0 &=& N \left [ {\mbox{det}}{\cal G}^{-1}_0 \right ]^{-\frac{1}{2}} , \label{spf13}
\end{eqnarray}
where the determinant is taken over momentum space as ${\cal G}^{-1}_0(K)$ is diagonal.
We also note that the constant factor is absorbed in $N$, which is irrelevant as it is temperature independent. Now the 
logarithm of a partition function up to a constant, becomes
\begin{eqnarray}
\ln {\cal Z}_0\,  &=& \ln   \left [ {\mbox{det}}[{\cal G}^{-1}_0(K) ] \right ]^{-\frac{1}{2}} \nn\\
&=& -\frac{1}{2} \ln \prod_{K} [{\cal G}^{-1}_0(K) ] = -\frac{1}{2} \sum_K \ln [{\cal G}^{-1}_0(K) ] \nn\\
&=& -\frac{1}{2} \sum_{n, {k}}  \ln (\omega_n^2+\omega_k^2)
=-\frac{1}{2}\, V \int \frac{d^3{k}}{(2\pi)^3} \,  \sum_{n}  \ln (\omega_n^2+\omega_k^2). \label{spf14}
\end{eqnarray}

Below we perform the sum integral as 
\bea
T\sum_n\frac{1}{T} \ln\left(\omega_n^2+\omega_k^2 \right)
= T\sum_{k-0} \frac{1}{T} \ln\left(-k_0^2+\omega_k^2\right)
= T\sum_{k_0}\frac{1}{T} \left[\ln\left(\omega_k-k_0\right)+\ln\left(\omega_k+k_0 \right )\right], \label{spf15}
\eea
where $\omega=i\omega_n=k_0$. 

Now concentrating on the first term:  using \eqref{sum_intb2} we can write
\bea
T\sum_{k_0} \frac{1}{T} \ln\left(\omega_k-k_0\right) &=& \frac{1}{2\pi i T}\oint dk_0  \ln\left(\omega_k-k_0 \right)
\frac{1}{2} \ \coth\left(\frac{\beta k_0}{2}\right) \nn\\
&=& -\frac{1}{4\pi i T} \oint dk_0 \left(\frac{-1}{\omega_k-k_0}\right)\int dk_0 \coth\left(\frac{\beta k_0}{2}\right)\nn\\
&=& \frac{1}{4\pi i T} \oint dk_0 \left(\frac{1}{\omega_k-k_0}\right)\int dk_0 \coth\left(\frac{\beta k_0}{2}\right)\nn\\
&=& \frac{1}{4\pi i T} \oint dk_0 \left(\frac{1}{\omega_k-k_0}\right)\frac{2}{\beta}\ln\left[\sinh\left(\frac{\beta k_0}{2}\right)\right]\nn\\
&=& \frac{1}{2\pi i} \oint dk_0 \left(\frac{1}{\omega_k-k_0}\right)\ln\left[\sinh\left(\frac{\beta k_0}{2}\right)\right] \nn\\
&=& \ln\left[\sinh\left(\frac{\beta \omega_k}{2}\right)\right].  \label{spf16}
\eea
Similarly for the second term:  choosing the opposite contour as the first one  we get,
\bea
T\sum_{k_0}\ln\left(\omega_k+k_0 \right) &=& \ln\left[\sinh\left(\frac{\beta \omega_k}{2}\right)\right], \label{spf17}
\eea
we note here that there will be a term of $i\pi$ in (\ref{spf17}), which we neglect as it would make partition function imaginary.
So, the results of frequency sum in (\ref{spf15})  can be written as,
\bea
T\sum_n\frac{1}{T} \ln\left(\omega_n^2+\omega_k^2 \right) &=& 2\ln\left[\sinh\left(\frac{\beta \omega_k}{2}\right)\right]\nn\\
&=& \beta \omega_k-2\ln 2 + 2\ln\left(1-e^{-\beta \omega_k}\right). \label{spf19}
\eea
We used 
\bea
\sinh\left(\frac{\beta \omega_k}{2}\right)&=&\frac{1}{2}\left(e^{\frac{\beta \omega_k}{2}}-e^{\frac{-\beta \omega_k}{2}}\right)
= \frac{1}{2}e^{\frac{\beta \omega_k}{2}}\left(1-e^{-\beta \omega_k}\right)\nn\\
\therefore \ln\left[\sinh\left(\frac{\beta \omega_k}{2}\right)\right] &=& \frac{\beta \omega_k}{2}-\ln 2 +\ln\left(1-e^{-\beta \omega_k}\right).
\label{spf18}
\eea

Now using (\ref{spf19}) in (\ref{spf14}) one gets logarithm of the partition function 
\bea
\ln {\cal Z}_0\, &=& - V \int \frac{d^3{k}}{(2\pi)^3} \,
\left [ \frac{\beta \omega_k}{2} - \ln 2 + \ln\left(1-e^{-\beta \omega_k}\right) \right ] , \label{spf20}
\eea
where again $\ln 2$ can be neglected as it is independent of temperature. This agrees with the result obtained in (\ref{eq10}).

The free energy density for free scalar field can be obtained from \eqref{eq3} as
\bea
F_0 &=&  -T \ln {\cal Z}_0\, =  V T \, \int \frac{d^3{k}}{(2\pi)^3} \,
\left [ \frac{\beta \omega_k}{2} + \ln\left(1-e^{-\beta \omega_k}\right) \right ] , \label{spf21}
\eea
In the infinite volume limit the pressure for free scalar field  can be obtained from \eqref{eq4} as
\bea
{\cal P}_0 &=& -\frac{{\Omega}_0}{V} =    T \, \int \frac{d^3{k}}{(2\pi)^3} \,
\left [ -\frac{\beta \omega_k}{2}  - \ln\left(1-e^{-\beta \omega_k}\right) \right ] , \label{spf22}
\eea
where the first term is the zero temperature part. 

\subsubsection{Partition function for interacting scalar field}

The interaction Lagrangian density as given in (\ref{spf0}) as 
\begin{eqnarray}
 {\cal L}_I &=&  \frac{\lambda}{4!} \phi^4, \label{spfi1}
\end{eqnarray}

The partition function in first order $\lambda$ as given in (\ref{eqp3d1})
\begin{eqnarray}
\left ( \ln {\cal Z}_I \right )_1 &=& \left\langle{\cal  F}_I \right \rangle_0 =\frac{1}{{\cal Z}_0} \, {\rm{Tr}} \left[ \, e^{-\beta {\cal H}_0 } \   
{\cal T} \left (-\int\limits_0^\beta {\cal H}' \ d\tau \right)  \right ]. \label{spfi2}
\end{eqnarray}
Using path integral it becomes~\cite{Kapusta}
\begin{eqnarray}
\left ( \ln {\cal Z}_I \right )_1 &=&
 \frac{\int {\cal D}\phi\ {\cal S}_I \, 
  e^{i {\cal S}_0[\phi]}} {\int {\cal D}\phi\  e^{i {\cal S}_0[\phi]}} \nn\\
   &=&  \, \frac{\int_{\tiny{\mbox{\sf periodic}}} {\cal D}\phi\ 
   \left(-\int\limits_0^\beta\, d \tau \int \, d^3 {x}\,{\cal L}_I ({t\rightarrow -i\tau})\right )  \,  
   e^{-\int\limits_0^\beta\, d \tau \int \, d^3 {x}\, {\cal L}_0 ({t\rightarrow -i\tau})}}
  { \int_{\tiny{\mbox{\sf periodic}}} {\cal D}\phi\ 
  e^{-\int\limits_0^\beta\, d \tau \int \, d^3 {x}\, {\cal L}_0 ({t\rightarrow -i\tau})}} 
  , \label{spfi3}
\end{eqnarray}
where free partition function is already calculated in (\ref{spf11}) as 
\begin{eqnarray}
{\cal Z}_0\,  &=& \, \int_{\tiny{\mbox{\sf periodic}}} {\cal D}\phi_{n,\bm{\vec k}}\ 
\exp \left [ - \frac{1}{2} 
 \, \sum_{n,k} \,[\omega^2_n+\omega_k^2]  \left |\phi_{n,\bm{\vec k}}\right |^2 \right ]. \label{spfi4}
\end{eqnarray}
Now we need to compute the numerator of (\ref{spfi3}). Using the same method as the free case one can proceed 
as ~\cite{Kapusta,Yang} 
\bea
 \left ( \ln {\cal Z}_I \right )_1 &=& \frac{1}{{\cal Z}_0} \, \int_{\tiny{\mbox{\sf periodic}}} {\cal D}\phi_{n,\bm{\vec k}} \ 
 \left(-\lambda \int\limits_0^\beta\, d \tau \int \, d^3 {x}\, \phi^4  \right )  \,
 \exp \left [ - \frac{1}{2} \, \sum_{n,k}\,[\omega^2_n+\omega_k^2] \left |\phi_{n,\bm{\vec k}}\right |^2 \right ] .
 \label{spfi5}
\eea
Using the Fourier decomposition of fields in (\ref{spf3}), one can get
\bea
\lambda \int\limits_0^\beta\, d \tau \int \, d^3 {x}\, \phi^4 &=& \frac{\lambda}{(V\beta)^2} \, \sum_{m_1,\bm{\vec q_1}}
\sum_{m_2,\bm{\vec q_2}} \sum_{m_3,\bm{\vec q_3}} \sum_{m_4,\bm{\vec q_4}} \, \phi_{m_1,\bm{\vec q_1}} \, \phi_{m_2,\bm{\vec q_2}}
\, \phi_{m_3,\bm{\vec q_3}} \, \phi_{m_4,\bm{\vec q_4}} \nn \\
&& \times \int\limits_0^\beta\, d \tau \int \, d^3 {x}\, \,
e^{i(\bm{\vec q_1} + \bm{\vec q_2} + \bm{\vec q_3} +\bm{\vec q_4})\cdot {x}} \, \, 
e^{-i(\omega_{m_1} + \omega_{m_2} + \omega_{m_3} +\omega_{m_4})\tau} \nn\\
&=& \frac{\lambda}{(V\beta)^2} \,  \sum_{m_1,\bm{\vec q_1}}
\sum_{m_2,\bm{\vec q_2}} \sum_{m_3,\bm{\vec q_3}} \sum_{m_4,\bm{\vec q_4}} \, \phi_{m_1,\bm{\vec q_1}} \, \phi_{m_2,\bm{\vec q_2}}
\, \phi_{m_3,\bm{\vec q_3}} \, \phi_{m_4,\bm{\vec q_4}}  \nn \\
&& \times \beta \delta (-\omega_{m_1} - \omega_{m_2} - \omega_{m_3} - \omega_{m_4}) \ V \ \delta^3 (\bm{\vec q_1} + \bm{\vec q_2} + \bm{\vec q_3} +\bm{\vec q_4}). \label{spfi6}
\eea
We have used two delta functions following (\ref{spf7a}) and (\ref{spf7b}) from  $\tau$ and $x$ integrations, respectively and
they guarantee the energy momentum conservation in the interaction vertex. Now, the non-zero contribution comes 
when $\omega_{m_1}=-\omega_{m_2}$, $\omega_{m_3}=-\omega_{m_4}$ and  $\bm{\vec q_1}=-\bm{\vec q_2}$, $\bm{\vec q_3}=-\bm{\vec q_4}$.
Obviously, there are another two combination that would give non-zero contributions.
This allows 3 non-zero permutations among $\omega_{m_i}$ and ${q}_i$. Then one can write
\bea
\lambda \, \int\limits_0^\beta\, d \tau \int \, d^3 {x}\, \phi^4 &=& 3\lambda \, (\beta V) \, \left (\frac{1}{V\beta}\right )^2 \,
\sum_{m,\bm{\vec q}} \sum_{l,\bm{\vec p}} \, \left | \phi_{m,\bm{\vec q}}\right |^2 \, \left | \phi_{l,\bm{\vec p}}\right |^2 .  \label{spfi7}
\eea

Using (\ref{spfi7}) in (\ref{spfi5}), one gets
\bea
\left ( \ln {\cal Z}_I \right )_1 &=& -3 \lambda  \beta  V   \left (\frac{1}{V\beta}\right )^2  \sum_{m,\bm{\vec q}} \sum_{l,\bm{\vec p}} 
 \frac{ {\prod\atop{n,\bm{\vec k}}} \int_{\tiny{\mbox{\sf periodic}}} {\cal D}\phi_{n,\bm{\vec k}} \ 
 \left | \phi_{m,\bm{\vec q}}\right |^2 \, \left | \phi_{l,\bm{\vec p}}\right |^2 \,
 \exp \left [ - \frac{1}{2} \,(\omega^2_n+\omega_k^2) \left |\phi_{n,\bm{\vec k}}\right |^2 \right ]}
 { {\prod\atop{n,\bm{\vec k}}} \int_{\tiny{\mbox{\sf periodic}}} {\cal D}\phi_{n,\bm{\vec k}}\ 
\exp \left [ -\frac{1}{2} \,(\omega^2_n+\omega_k^2)  \left |\phi_{n,\bm{\vec k}}\right |^2 \right ]}\nn\\
&=& -3 \lambda  \beta  V   \left (\frac{1}{V\beta}\right )^2
\left [ \sum_{n\bm{\vec ,k}}
 \frac{ \int_{\tiny{\mbox{\sf periodic}}} {\cal D}\phi_{n,\bm{\vec k}} 
 \left | \phi_{n,\bm{\vec k}}\right |^2  \,
 \exp \left [ - \frac{1}{2} (\omega^2_n+\omega_k^2) \left |\phi_{n,\bm{\vec k}}\right |^2 \right ]}
 {  \int_{\tiny{\mbox{\sf periodic}}} {\cal D}\phi_{n,\bm{\vec k}} 
\exp \left [ -\frac{1}{2} \,(\omega^2_n+\omega_k^2)  \left |\phi_{n,\bm{\vec k}}\right |^2 \right ]}
\right ]^2, \label{spfi18a}
\eea
if $m=l=n$ and  $\bm{\vec q} =\bm{\vec p}=\bm{\vec k}$, the integral over $\phi_{n,\bm{\vec k}}$ gets factorised as above. 
When $m\ne l \ne n$ and  $\bm{\vec q}\ne \bm{\vec p}\ne \bm{\vec k}$, the $\phi_{n,\bm{\vec k}}$
integral in the numerator and denominator are identical and they cancel out, thus the integral disappears.

Using Gaussian integral $\int dy \ y^{2n} \, e^{-\alpha y^2}= \Gamma(n+1/2)/\alpha^{n+1/2}$, one gets
\bea
\left ( \ln {\cal Z}_I \right )_1 &=& -3\, \lambda \, \beta V  \ \left (\frac{1}{V\beta}\right )^2
\left [\frac{\Gamma(1+\frac{1}{2})}{\alpha^{3/2}} \times \frac{\alpha^{1/2}}{\Gamma(\frac{1}{2})}\right ]^2 
= -3\, \lambda \, \beta V  \  \left (\frac{1}{V\beta}\right )^2
\left [\frac{1}{2\alpha} \right ]^2 \nn \\
&=&- 3\, \lambda \, \beta V  \ \left (\frac{1}{V\beta}\right )^2 \left [ \sum_{n,k} \frac{1}{(\omega_n^2+\omega_k^2)} \right ]^2
 = -3\, \lambda \, \beta V  \left [\frac{1}{V\beta} \sum_{n,k} \frac{1}{\omega_n^2+\omega_k^2} \right ]^2.
 \label{spfi8b}
\eea
Taking continuum limit $\sum_k\rightarrow V\, \int d^3{k}/(2\pi)^3$, one gets
\bea
\left ( \ln {\cal Z}_I \right )_1 &=& - 3\, \lambda \, \beta V  \left [\frac{1}{\beta} \sum_n \int 
\frac{d^3k}{(2\pi)^3} \,  \frac{1}{\omega_n^2+\omega_k^2} \right ]^2. \label{spfi10}
\eea
 where the $\sum_n$ is  Euclidean. Casting this into Minkowski $\omega_n=k_4=-ik_0$, one can write  
 \bea
\left ( \ln {\cal Z}_I \right )_1 &=& -\ 3\, \lambda \, \beta V  \left [\frac{1}{\beta} \sum_{k_0} \int 
 \frac{d^3k}{(2\pi)^3} \,  \frac{1}{k_0^2-\omega_k^2} \right ]^2 , \label{spfi11}
 \eea
 where  one can use contour integration in (\ref{sum_intb1}) vis-a-vis (\ref{sum_intb2}) and (\ref{sum_intb4}).
 We also note the following points:
 \begin{enumerate}
 \item[$\bullet$] The term inside the square braces are the sum-integral that comes from the loop integral $\int d^4K/(2\pi)^4$ and
 the scalar propagator ($1/(k_0^2-{k}^2-m^2)=1/(k_0^2-\omega_k^2)$ with energy $\omega_k=\sqrt{{k}^2+m^2}$.
 \item[$\bullet$] $\lambda$ is the interaction vertex.
 \item[$\bullet$] $\beta V$ is the left out factor that comes from the energy-momentum conservation (e.g., (\ref{spfi6})) in the vertex.
 \item[$\bullet$] The square, $[\cdots]^2$, indicates that two loops connected to one interaction point $\lambda$.
 \item[$\Rightarrow$] All these together correspond to a topologically distinct diagram \Big[{\includegraphics[scale=0.3]{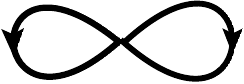}}\Big].
 \item[$\bullet$] $3$ is the symmetry factor that comes from 3 different permutation of contraction allowed by the energy-momentum conservation.
 \end{enumerate}
 Now the first order correction to the scalar partition function in (\ref{spfi11}) can be represented in Feynman diagram  as
 \bea
\left ( \ln {\cal Z}_I \right )_1 &=& -\ 3\, \beta V  \times {\includegraphics[height=0.5cm,width=1cm]{vaccum_bubble.pdf}}. \label{spfi12}
 \eea 
The frequency sum in (\ref{spfi11}) is exactly similar to that of tadpole diagram as done in Subsec.~\ref{tad_sec} but with different symmetry 
factor ($1/2$). Excluding this $1/2$ factor  and $\lambda$ and the zero temperature part, the result of the sum integral can be obtained 
from (\ref{lf4}) as $T^2/12$ . Thus, the first 
order correction to the scalar partition function in (\ref{spfi11}) becomes
 \bea
\left ( \ln {\cal Z}_I \right )_1 &=& -\ 3\, \lambda \, \beta V  \left [\frac{1}{\beta} \sum_n \int 
 \frac{d^3k}{(2\pi)^3} \,  \frac{1}{k_0^2-\omega_k^2} \right ]^2 
 = -\ 3\, \lambda \, \beta V \ \frac{T^4}{144}= -\  \lambda \, \frac{VT^3}{48}. \label{spfi13}
 \eea
 
\subsubsection{Pressure}
The logarithm of the scalar partition function up to first order in coupling can now be written 
using (\ref{eqp3}), (\ref{spf20}) and (\ref{spfi13}) as
\bea
\ln {\cal Z}(\beta)&=& \ln {\cal Z}_0 +  \left ( \ln {\cal Z}_I \right )_1 \nn\\
&=& - V \int \frac{d^3{k}}{(2\pi)^3} \, \ln\left(1-e^{-\beta \omega_k}\right)  
-\  \lambda \, \frac{VT^3}{48}  +{\cal O}(\lambda^2) , \label{spfi14}
\eea
where we have also dropped the $T=0$ contribution in $\ln{\cal Z}_0$. In the infinite volume limit the pressure up to 
first order can be obtained as
\bea
{\cal P} = -\frac{\Omega}{V} &=&   - T \, \int \frac{d^3{k}}{(2\pi)^3} \,
 \ln\left(1-e^{-\beta \omega_k}\right) 
-\  \lambda \, \frac{T^4}{48}  +{\cal O}(\lambda^2) \nn\\
&=& \frac{\pi^2T^4}{90} -\  \lambda \, \frac{T^4}{48} +{\cal O}(\lambda^2), \label{spfi15}
\eea
where $\Omega$ is the thermodynamic potential. The other thermodynamic quantities can be obtained from pressure.
One can also compute the higher order corrections to the partition function following (\ref{eqp3d2}) and (\ref{eqp3d3}), 
and thus higher order thermodynamic quantities.

\subsection{Fermion Field}
\label{FF}
Until now we have  discussed partition function for real  scalar field and it's pressure up to first order 
in coupling. In this section we will compute the partition function for free fermionic fields. The computation of 
interacting fermionic partition function is postponed until we introduce gauge theory, quantum electrodynamics (QED).
Since fermions are anticommuting fields, they are Grassmann variables. We start this section by reviewing some of 
the properties of Fermionic fields.

\subsubsection{Fermionic Lagrangian and conserved charge}
\label{global}

The Lagrangian density that describes the non-interacting fermion is given in Minkowski space-time~\cite{pbpal,Peskin} as
\bea
{\cal L} &=& {\bar \psi} \left (i\partial \!\!\! \slash -m \right ) \psi , \label{fpf1}
\eea
where $m$ is the mass of the fermion and the fields $\psi$ and $\bar \psi$ are to be treated independently. 
The $\gamma$-matrices in Dirac-Pauli representation are given as

\bea
\gamma^0= \left ( \begin{array}{ll}
                  I & 0  \\
                  0 & -I 
		   \end{array} 
		\right ) \hspace*{0.3in}  {\mbox{and}} 
		\hspace*{0.3in}  
		\gamma^i= \left ( \begin{array}{ll}
                  0 & \sigma^i  \\
                  -\sigma^i & 0 
		   \end{array} 
		\right ) \, , \label{fpf1a}
\eea
where $I$ is a $2\times 2$ unit matrix and  the Pauli matrices $\sigma^i$'s are
\bea
\sigma^1= \left ( \begin{array}{ll}
                  0 & 1  \\
                  1 & 0 
		   \end{array} 
		\right ) \hspace*{0.2in} 
\sigma^2= \left ( \begin{array}{ll}
                  0 & -i  \\
                  i & 0 
		   \end{array} 
		\right ) \hspace*{0.2in}
		{\mbox{and}} 
		\hspace*{0.2in}  
\sigma^3= \left ( \begin{array}{ll}
                  1 & 0  \\
                  0 & -1  
		   \end{array} 
		\right ) \, .  \label{fpf1b}
\eea

Using the Euler-Lagrange equation for $\bar \psi$ field
\bea
\frac{\partial {\cal L}}{\partial \bar \psi} &=& {\partial_\mu} \left ( \frac{\partial {\cal L}}
{\partial (\partial^\mu \bar \psi )} \right ) \Rightarrow
\left (i\gamma^\mu\partial_\mu -m \right ) \psi =\partial_\mu[0]=0 , 
\label{fpf2}
\eea
one gets the Dirac equation for $\psi$ field. The Hamiltonian density is given as
\bea
{\cal H}_d &=& \pi\partial_0\psi +\partial_0\bar\psi\bar\pi -{\cal L}. \label{fpf3}
\eea
The conjugate momenta can be obtained as
\bea
\pi&=& \frac{\partial {\cal L}}{\partial (\partial_0\psi)} = i\psi^\dagger , \label{fpf4a} \\
{\bar \pi}&=& \frac{\partial {\cal L}}{\partial (\partial_0{\bar \psi})} = 0 . \label{fpf4b} 
\eea
Using (\ref{fpf4a}) and (\ref{fpf4b}) in (\ref{fpf3}), the Hamiltonian density becomes
\bea
{\cal H}_d &=& i\psi^\dagger\partial_0\psi  - {\bar \psi} \left (i\partial \!\!\! \slash -m \right ) \psi 
= i{\bar \psi} \gamma^0 \partial_0\psi - i{\bar \psi} \gamma^0 \partial_0\psi + i{\bar \psi} \gamma^i \partial_i\psi + {\bar \psi}m\psi \nn\\
&=& i\bar \psi \left (\gamma^i\partial_i +m  \right ) \psi . \label{fpf5}
\eea
This is the Hamiltonian density for canonical ensemble. 
However, allowing a local transformation, {\it i.e.}, $\alpha$ 
depends on $X$, one gets 
\bea
{\cal L} \rightarrow {\cal L}'&=& \bar\psi e^{i\alpha(X)} \left (i\partial \!\!\! \slash -m \right ) \psi e^{-i\alpha(X)} 
= \bar\psi \left (i\partial \!\!\! \slash -m \right ) \psi \, + \, \bar\psi \partial \!\!\! \slash \alpha(X) \psi \nn\\
&=& {\cal L} \,  + \, \bar\psi \partial \!\!\! \slash \alpha(X) \psi . \label{fpf5a}
\eea
As seen if $\alpha(X)=\alpha $,  is a constant, then ${\cal L}'={\cal L}$, the Lagrangian density is invariant under global 
symmetry. This symmetry will lead to a conserved current according to Noether's theorem. By solving the equation of 
motion for $\alpha$, one gets
\bea
 {\partial_\mu} \left ( \frac{\partial {\cal L}'}{\partial (\partial^\mu \alpha )} \right ) &=&\frac{\partial {\cal L}'}{\partial \alpha}  \nn\\
 \partial_\mu \left [\bar \psi \gamma^\mu \psi \right ]&=& \partial_\mu j^\mu = 0 , 
\label{fpf6}
\eea
where the conserved current is found as 
\bea
j^\mu&=& \bar \psi \gamma^\mu \psi \, . \label{fpf6a}
\eea
Now the temporal component $j^0 =\bar \psi \gamma^0\psi = \bar \psi^\dagger \psi= \rho$ is associated with conserved number density. The conserved 
number can be obtained as
\bea
N &=& \int d^3x \, j^0 = \int d^3x \, \, \bar \psi^\dagger \psi = \int d^3x \, \, \rho \, . \label{fpf6b}
\eea
Now, the new Hamiltonian density in presence of chemical potential associated with a conserved number becomes
\bea
{\cal H}_d -\mu \rho &=& i\bar \psi \left (\gamma^i\partial_i +m  \right ) \psi -\mu \rho 
=\bar \psi \left (i\gamma^i\partial_i +m  -\mu\gamma^0 \right ) \psi . \label{fpf7}
\eea
The corresponding new Lagrangian density becomes
\bea
{\cal L} &=& \bar \psi \left (i\gamma^\mu\partial_\mu - m  + \mu\gamma^0 \right ) \psi , \label{fpf8}
\eea
which indicates that the presence of the chemical potential is like  changing the zeroth component
of the gauge field (external field), through the substitution $\partial_0 -i\mu$ in the Lagrangian.

\subsubsection{Partition function and pressure for free fermions}
\label{direct}
The partition function for free fermionic field reads from (\ref{fi6}) as
\begin{eqnarray}
 {\cal Z}_0 &=& \, \int\limits_{\psi (\bm{\vec x}, 0)=-\psi(\bm{\vec x},\beta)} {\cal D}[{\bar \psi}] \ {\cal D}[{\psi}] \
  e^{\int\limits_0^\beta\, d \tau \int \, d^3 {x}\, {\cal L} ({t\rightarrow -i\tau})}.  \label{fpf9} 
\end{eqnarray}
As before, the Fourier transform of the fermionic fields  $\psi(X)$ and $\bar \psi(X)$ can be written as
\begin{eqnarray}
\psi(X)= \psi(\bm{\vec x},\tau) &{=\atop {t\rightarrow -i\tau}} \atop {k_0\rightarrow i\omega_n}& \frac{1}{\sqrt{V\beta}} \sum_K \, 
 e^{-i K\cdot  X }\, \psi (K)  =   \frac{1}{\sqrt{V\beta}} \sum_{n,\bm{\vec k}} \, 
e^{i \bm{\vec k \cdot \vec x} } \, \, e^{- i \omega_n \tau  } \, \psi (\omega_n,\bm{\vec k})\, , \nn\\
\bar\psi(X)= \bar \psi(\bm{\vec x},\tau) &{=\atop {t\rightarrow -i\tau}} \atop {k_0\rightarrow i\omega_n}& \frac{1}{\sqrt{V\beta}} \sum_K \, 
 e^{i K\cdot  X }\, \bar \psi (K)  =   \frac{1}{\sqrt{V\beta}} \sum_{n,\bm{\vec k}} \, 
e^{-i \bm{\vec k \cdot \vec x} } \, \, e^{i \omega_n \tau  } \, \bar \psi (\omega_n,\bm{\vec k}) \, .
\label{fpf10}
\end{eqnarray}
where  $V$ is the three volume.  The Lagrangian density in (\ref{fpf8}) can now be written in Euclidean time as
\bea
{\cal L} ({t\rightarrow -i\tau})&{=\atop{t\rightarrow -i\tau}}& \bar \psi \left (i\gamma^\mu\partial_\mu - m  + \mu\gamma^0 \right ) \psi 
{=\atop{t\rightarrow -i\tau}} \bar \psi \left (i\gamma^0\partial_0 -i\gamma^i\partial_i - m  + \mu\gamma^0 \right ) \psi \nn \\
&=& - \bar \psi \left (\gamma^0\partial_\tau +i\gamma^i\partial_i + m  - \mu\gamma^0 \right ) \psi \, . \label{fpf11} 
\eea
Using the Fourier transformed of fermionic fields in (\ref{fpf10})  we compute
\bea
\int\limits_0^\beta\, d \tau \int \, d^3 {x}\, {\cal L} ({t\rightarrow -i\tau}) &=&- \frac{1}{V\beta}
\int\limits_0^\beta\, d \tau \int \, d^3 {x}\,\sum_{n,\bm{\vec k}}\sum_{m,\bm{\vec k'}}
e^{-i \bm{\vec k \cdot \vec x} } \, \, e^{i \omega_n \tau  } \, \bar \psi (\omega_n,\bm{\vec k}) \nn\\
&& \hspace*{0.2in} \left [\gamma^0\partial_\tau +i\gamma^i\partial_i + m  - \mu\gamma^0 \right ]
e^{i \bm{\vec k' \cdot \vec x} } \, \, e^{- i \omega_m \tau  } \, \psi (\omega_m,\bm{\vec k'}) \nn\\
&=& - \frac{1}{V\beta}
\int\limits_0^\beta\, d \tau \int \, d^3 {x}\,\sum_{n,\bm{\vec k}}\sum_{m,\bm{\vec k'}}
e^{-i \bm{\vec k \cdot \vec x} } \, \, e^{i \omega_n \tau  } \, \bar \psi (\omega_n,\bm {\vec k}) \nn\\
&& \hspace*{0.2in} \left [\gamma^0(-i\omega_m) +i\gamma^i (i k^\prime_i) + m  - \mu\gamma^0 \right ]
e^{i \bm{\vec k' \cdot \vec x} }\, \, e^{- i \omega_m \tau  } \, \psi (\omega_m,\bm{\vec k'}) \nn\\
&=& - \frac{1}{V\beta}\sum_{n,\bm{\vec k}}\sum_{m,\bm{\vec k'}}
\bar \psi (\omega_n,\bm{\vec k}) \left [\gamma^0(-i\omega_m) +i\gamma^i (i k^\prime_i) + m  - \mu\gamma^0 \right ]
 \psi (\omega_m,\bm{\vec k'}) \nn\\
&& \times \beta \delta(\omega_n-\omega_m) \, V \delta^3(\bm{\vec k'}-\bm{\vec k}) \nn\\
 &=& - \sum_{n,\bm{\vec k}}
\bar \psi (\omega_n,\bm{\vec k}) \left [-\gamma^0(i\omega_n +\mu) - \gamma^i k_i + m  \right ]
 \psi (\omega_n,\bm{\vec k}) \nn\\
 &=& - \sum_{n,\bm{\vec k}}
\bar \psi (\omega_n,\bm{\vec k}) \left [S_0^{-1}((i\omega_n+\mu),\bm{\vec k})   \right ] \psi (\omega_n,\bm{\vec k}), \label{fpf12}
\eea
where $S_0^{-1}(i\omega_n+\mu,\bm{\vec k})$ is the inverse of free fermionic propagator in presence of chemical potential $\mu$.
The partition function in (\ref{fpf9}) becomes
\bea
{\cal Z}_0 &=& \prod_{n,\bm{\vec k}} \int\limits_{\tiny\mbox{\sf antiperiodic}} {\cal D}[{\bar \psi}] \ {\cal D}[{\psi}] \
e^{-\bar \psi (\omega_n,\bm{\vec k}) \left [S_0^{-1}((i\omega_n+\mu),\bm{\vec k})   \right ] \psi (\omega_n,\bm{\vec k})}\nn\\
&=& \prod_{n,{k}} \mbox{det} \left [S_0^{-1}((i\omega_n+\mu),\bm{\vec k})   \right ] 
, \label{fpf12a}
\eea
where we have used the result of the functional integral in \eqref{GV13} involving Grassman variables.
The logarithm of ${\cal Z}_0$ becomes
\bea
\ln {\cal Z}_0 &=& \sum_{n,k} \ln \, \mbox{det} \left [S_0^{-1}((i\omega_n+\mu),\bm{\vec k})   \right ] \nn \\
&=& \sum_{n,k} \ln \mbox{det} \left [-\gamma^0(i\omega_n +\mu) - \gamma^i k_i + m   \right ]  = \sum_{n,k} \ln \mbox{det} [Y]. \label{fpf13}
\eea
Now one can write
\bea
\mbox{det} [Y] &=& \mbox{det}  \left\{  
                 \left ( \begin{array}{ll}
                  -(i\omega_n+\mu)& \hspace*{0.4in} 0  \\
                  0 & (i\omega_n+\mu) 
		   \end{array} 
		\right ) 
+            \left ( \begin{array}{ll}
                  0 & -\bm{\vec \sigma \cdot \vec k}  \\
                  \bm{\vec \sigma \cdot \vec k}& \hspace*{0.2in} 0 
		   \end{array} 
		\right )   
+  \left ( \begin{array}{ll}
                  m & 0  \\
                  0 & m  \nonumber
		   \end{array} 
		\right ) \right\}  
\eea
\bea
\hspace*{-1.3in} &=& \mbox{det} \left \{ 
                 \left ( \begin{array}{ll}
                  -(i\omega_n+\mu)+m& \hspace*{0.3in} -\bm{\vec \sigma \cdot \vec k}\\
                  \bm{\vec \sigma \cdot \vec k}& (i\omega_n+\mu)+m 
		   \end{array} 
		\right ) \right \} \, ,
		\label{fpf14}
\eea
where each element is a $2\times 2$ matrix. Using the identity $(\bm{\vec \sigma \cdot \vec k})^2=k^2$, one can compute the
determinant as
\bea
\mbox{det} [Y] &=&  \left [ k^2+m^2 -(i\omega_n+\mu)^2 \right ]^2 . \label{fpf15}
\eea
Combining \eqref{fpf15} and \eqref{fpf13}, one can write
\bea
\ln {\cal Z}_0 &=& 2\, \sum_{n,{k}} \ln \left [ k^2+m^2 -(i\omega_n+\mu)^2 \right ] \nn\\
&=& 2V\sum_n \int\frac{d^3k}{(2\pi)^3}\left [\ln ((\omega_k-\mu) -i\omega_n) +
\ln  ((\omega_k+\mu) + i\omega_n)\right ]\ , \label{fpf16}
\eea
where $\omega_k=\sqrt{k^2+m^2}$ and $\sum_k$ is replaced by $V \int\frac{d^3k}{(2\pi)^3}$.  

Now we will perform the frequency sum in (\ref{fpf16}):
\begin{enumerate}
 \item [] First term: using  \eqref{sum_intf3}, one can write
 \bea
 \sum_n \ln [(\omega_k-\mu) -i\omega_n] &=& T\sum_n \frac{1}{T} \ln  [(\omega_k-\mu) -k_0] \nn \\
&=& \frac{1}{2\pi i T} \oint  dk_0 \ln [(\omega_k-\mu) -k_0]\, \frac{1}{2} \, \tanh \left (\frac{\beta k_0}{2}\right ) \nn\\
&=& -\frac{1}{4\pi i T} \oint dk_0 \left(\frac{-1}{(\omega_k-\mu)-k_0}\right) \int dk_0 \tanh \left(\frac{\beta k_0}{2}\right)\nn\\
&=& \frac{1}{4\pi i T} \oint dk_0 \left(\frac{1}{(\omega_k-\mu)-k_0}\right)\int dk_0 \tanh\left(\frac{\beta k_0}{2}\right)\nn
\eea
\bea
&=& \frac{1}{4\pi i T} \oint dk_0 \left(\frac{1}{(\omega_k-\mu)-k_0}\right)\frac{2}{\beta}\ln\left[\cosh\left(\frac{\beta k_0}{2}\right)\right]\nn \\
&=& \frac{1}{2\pi i} \oint dk_0 \left(\frac{1}{(\omega_k-\mu)-k_0}\right)\ln\left[\cosh\left(\frac{\beta k_0}{2}\right)\right]\nn \\
&=& \ln\left[\cosh\left(\frac{\beta (\omega_k-\mu)}{2}\right)\right] \nn\\
&=&\ln \left[\frac{1}{2}\left(e^{{\beta (\omega_k-\mu)}/{2}}+e^{{-\beta (\omega_k-\mu)}/{2}}\right)\right] \nn\\
&=& \ln \left[\frac{1}{2}e^{{\beta (\omega_k-\mu)}/{2}}\left(1-e^{-\beta (\omega_k-\mu)}\right)\right]\nn\\
&=&  \left[ \frac{\beta (\omega_k-\mu)}{2}-\ln 2 +\ln\left(1-e^{-\beta (\omega_k-\mu)}\right)\right] \nn\\
&=& \left[ \frac{\beta (\omega_k-\mu)}{2} +\ln\left(1-e^{-\beta (\omega_k-\mu)}\right)\right],
\label{fpf17a}
\eea
where again $\ln 2$ is neglected as it is independent of temperature.

\item [] Second term:
\bea
\sum_n \ln [(\omega_k+\mu) +i\omega_n] &=& T\sum_n \frac{1}{T} \ln [ (\omega_k-\mu) -k_0] 
= \ln\left[\cosh\left(\frac{\beta (\omega_k+\mu)}{2}\right)\right]\nn\\
&=& \left[ \frac{\beta (\omega_k+\mu)}{2}-\ln 2 +\ln\left(1-e^{-\beta (\omega_k+\mu)}\right)\right] \nn\\
&=& \left[ \frac{\beta (\omega_k+\mu)}{2} +\ln\left(1-e^{-\beta (\omega_k+\mu)}\right)\right].\label{fpf17b}
\eea
\end{enumerate}

 Now using (\ref{fpf17a}) and (\ref{fpf17b}) in (\ref{fpf16}), one gets logarithm of the partition function 
 \bea
 \ln {\cal Z}_0\, &=&  2V \int \frac{d^3{k}}{(2\pi)^3} \,
 \left [ {\beta \omega_k} + \ln\left(1+e^{-\beta (\omega_k-\mu)}\right) + \ln\left(1+e^{+\beta (\omega_k+\mu)}\right) \right ] , 
 \label{fpf18}
 \eea
 which agrees with that obtained in quantum statistical mechanics
 in (\ref{eq11}) and \eqref{eq11b}.
 
 The free energy density for free fermionic field  can be obtained as
 \bea
 F_0 &=&  -\frac{T \ln {\cal Z}_0}{V}\, =  -2T \, \int \frac{d^3{k}}{(2\pi)^3} \,
  \left [ {\beta \omega_k} + \ln\left(1+e^{-\beta (\omega_k-\mu)}\right) + \ln\left(1+e^{+\beta (\omega_k+\mu)}\right) \right ]. \label{fpf19}
 \eea
 In the infinite volume limit the pressure for  fermionic field  can be obtained as
 \bea
 {\cal P}_0 &=& -F_0 =   2 T \, \int \frac{d^3{k}}{(2\pi)^3} \,
  \left [ {\beta \omega_k} + \ln\left(1+e^{-\beta (\omega_k-\mu)}\right) + \ln\left(1+e^{+\beta (\omega_k+\mu)}\right) \right ] , \label{fpf20}
 \eea
 where the first term is the zero temperature part which should be dropped as it only shifts the vacuum energy.
  After performing the integration of temperature dependent part, one obtains pressure for free massless fermion
 \bea
 {\cal P}_0 &=& \frac{7\pi^2 T^4}{180} + \frac{\mu^2T^2}{6}+\frac{\mu^4}{12\pi^2} \, . \label{fpf21}
 \eea

\subsubsection{A reverse way: first number density and then pressure and entropy density for fermions}
\label{reverse}

The partition function in Minkowski space time can be written 
\begin{eqnarray}
{\cal Z}(\beta,[\mu])=\int {\cal D}[\bar \psi] 
{\cal D}[{\psi}] 
e^{i\int d^4X\mathcal{L} (\psi,{\bar \psi};[\mu])}.
\label{i1}
\end{eqnarray}
The pressure can be written as
\begin{equation}
{\cal P}(\beta; [\mu])=\frac{1}{\cal V} \ln{\cal Z}(T;[\mu])= \frac{T}{ V} \ln{\cal Z}(T;[\mu])
\ , \label{i2}
\end{equation}
where the four-volume, ${\cal V}=\beta V $ with $V$ is the three-volume.
 
 \begin{figure}[htb]
 \begin{center}
\includegraphics[width=12cm,height=4cm]{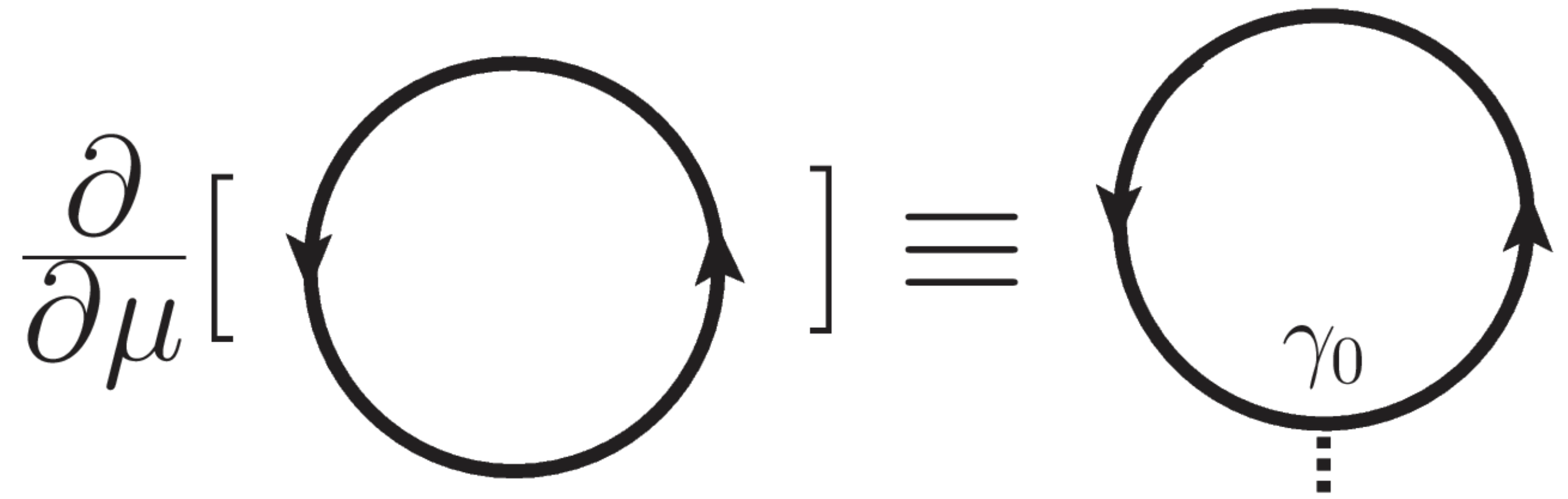}
 \caption{The Feynman diagram for number density originating from $\mu$ derivative of the pressure diagram that brings 
 a $\gamma_0$.}
 \label{diagr_nd}
 \end{center}
 \end{figure}

Now the number density $\rho$ can be obtained~\cite{HM,HMT}  as
\begin{eqnarray}
\!\!\!\!\!
\rho &\equiv&
\frac{\partial{\cal P}(\beta; [\mu])}{\partial \mu}
=\frac{i}{{\cal V}{\cal Z}[\beta;j]}\ {\int {\cal D}[\bar \psi] 
{\cal D}[\psi] 
\int d^4x \ {\bar \psi(x)} \gamma_0[\mu] \psi (x)} \,
e^{ \left ({i\int d^4x {\cal L}
(\psi,{\bar{\psi}}; [\mu])}\right)} \ .
\label{i3} 
\end{eqnarray}
The full fermionic propagator in presence  of uniform $\mu$ can be written as
\begin{eqnarray}
i{\cal S}_{\alpha\sigma}[\mu](x,x')&=&\frac{
\int {\cal D}[\bar \psi] {\cal D}[\psi] 
\psi_{\alpha}(x){\bar \psi_\sigma(x')} \exp \left ({i\int d^4x {\cal L}
(\psi,{\bar{\psi}}; [\mu])}\right)}
{ \int {\cal D}[\bar \psi] {\cal D}[\psi] 
 \exp \left ({i\int d^4x {\cal L} 
(\psi,{\bar{\psi}};[\mu])}\right)} \ .
\label{i4}
\end{eqnarray}
Now using (\ref{i4}) and performing the traces over 
Dirac and coordinate indices in (\ref{i3}) one can write
\begin{eqnarray}
\rho &=&  \, \int\! \frac{d^4K}{(2\pi)^4} 
\mbox{tr}\left [iS[\mu](K)\  (-i) \gamma_0[\mu](K,-K;0) \right ] \nn\\
&=& \, \int\! \frac{d^4K}{(2\pi)^4} 
\mbox{tr}\left [S[\mu](K)\  \gamma_0[\mu](K,-K;0) \right ] \, ,
\label{i5}
\end{eqnarray}
where '{\rm{tr}}' indicates the trace over the  Dirac indices.

Now if we consider free fermions, then the full propagator  $S[\mu]$ can be replaced by the free fermion
propagator $S_0[\mu]$ and the number density in (\ref{i5}) can be written as
\begin{eqnarray}
\rho_0 &=& \ \, \int\! \frac{d^4K}{(2\pi)^4} 
\mbox{tr}\left [S_0[\mu](K)\ \gamma_0[\mu](K,-K;0) \right ] , 
\label{i6}
\end{eqnarray}
The expression for number density in (\ref{i6}) corresponds to
Feynman diagram in Fig.~\ref{diagr_nd}. 
The free fermionic propagator  for momentum $K$ in helicity representation is given in \eqref{gse25}  as
\begin{equation}
S_0(K)=\frac{\gamma_0-\vec\gamma\cdot\hat k} {2d_+(k_0,k)}\ + \frac{\gamma_0+
\vec\gamma\cdot\hat k}{2d_-(k_0,k)}, \label{f2}
\end{equation}
\begin{equation}
\mbox{with}\,\,\,d_\pm=k_0\mp k \ \ . \label{f3}
\end{equation}
Using (\ref{f2}) in (\ref{i6}) and performing the trace over Dirac matrices, 
we get
\begin{eqnarray}
\rho_0(T,\mu)=2\int\frac{d^3k}{(2\pi)^3}\, \frac{1}{\beta} \sum_{k_0=(2n+1)\pi i T+\mu}
\left[\frac{1}{k_0-k}+ \frac{1}{k_0+k}\right] .\label{f4}
\end{eqnarray}
We note here that the chemical potential is not considered in the propagator but
considered in the discrete frequency as $k_0=i\omega_n+\mu$. This shifts the pole of
$\tanh$ by an amount $\mu$ in (\ref{sum_intf3}).  This  will lead to
same result as will see below. 

For computing the frequency sum in (\ref{f4}), we use the standard technique of contour 
integration as given in (\ref{sum_intf5}) in presence of $\mu$ as
\begin{equation}
\frac{1}{2\pi i}\oint\limits_C\left[\frac{1}{k_0-k}+
\frac{1}{k_0+k}\right]\frac{\beta}{2}\mbox{tanh}\left(\frac{\beta (k_0-\mu)}{2}
\right)dk_0 =\frac{\beta}{2}\ \frac{1}{2\pi i}\times 
(-2\pi i)\sum {\mbox{Residues}}\ . 
\label{f5}
\end{equation}
It is noted that the first term of (\ref{f5}) has a simple pole at 
$k_0=k$. On the other hand  the second term also has a simple pole at $k_0=-k$. After calculating 
the residues, one obtains the number density~\cite{HM,HMT} 
\begin{eqnarray}
\rho_0(T,\mu)&=&-\int \frac{d^3k}{(2\pi)^3}
\left[\tanh\frac{\beta(k-\mu)}{2}
-\tanh\frac{\beta(k+\mu)}{2}\right]
\nonumber\\
&=&2\int\frac{d^3k}{(2\pi)^3}\left[n(k-\mu)-n(k+\mu)\right], 
\label{f8}
\end{eqnarray}
where $n(x)=1/(e^{\beta x}+1)$, is the Fermi-Dirac distribution function.

Now, by integrating the first line of (\ref{f8}) w.r.t. $\mu$,  one obtains the pressure for non-interacting fermion gas as 
\begin{eqnarray}
{\mathcal P}_0(T,\mu) =2 T \int \frac{d^3k}{(2\pi)^3} \left [ 
\beta k + \ln\left(1+e^{-\beta(k-\mu)}\right) 
+ \ln\left(1+e^{-\beta(k+\mu)}\right) \right ], \label{fp}
\end{eqnarray}
where the first term is the
zero-point energy that produces  usual vacuum divergence. It also agrees with that obtained in (\ref{fpf20}).
The entropy density for non-interacting can be obtained from pressure as
\begin{eqnarray}
{\cal S}_0(T,\mu)&=&\frac{\partial {\cal P}_0}{\partial T}=2 \int\frac{d^3k}{(2\pi)^3}
\Big [\ln\left(1+e^{-\beta(k-\mu)}\right)
+ \ln\left(1+e^{-\beta(k+\mu)}\right)
\nonumber\\
&& \left. 
+\frac{\beta (k-\mu)}{e^{\beta(k-\mu)}+1}+\frac{\beta (k+\mu)}
{e^{\beta(k+\mu)}+1}\right].
\label{fs}
\end{eqnarray} 

\section{General Structure of Fermionic Two-point Functions at $T\ne 0$}
\label{gse}
\subsection{Fermion Self-Energy}
\label{fgse}
A theory possessing only fermions and gauge bosons with no bare masses for fermions  is chirally invariant for all orders. It is to 
be noted that the theory is also parity invariant. At $T=0$, chiral invariance has two implications: i) there are no ${\bar \psi}\psi$ coupling in 
any finite order of perturbation theory,  ii) the general form of the fermion self-energy can be written as~\cite{HAW}
\be
\Sigma (P)=-{\cal A} P\!\!\!\!\slash, \label{gse0}
\ee
for particle momentum $P\equiv(p_0=\omega, \bm{\vec p})$, $p=|\bm{\vec p}|$ and ${\cal A}$ is Lorentz invariant structure function which is function of $P^2$.

The effective fermion propagator can be written from \eqref{gse14}  as
\be
S^\star(P) = \frac{1}{P\!\!\!\! \slash -\Sigma(P)} = \frac{P \!\!\!\! \slash}{(1+{\cal A})P^2}. \label{gse1}
\ee
The poles, $P^2=0$ are on the light cone, $\omega=p$ and $(1+{\cal A})$ modifies the residues.

At $T\ne0$, the above item i) still holds whereas item ii) does not. At $T\ne 0$, the system will not be in a vacuum because at such high temperature there will
be antiparticles present in equal numbers as the particles. This constitutes a heat bath which introduces a special Lorentz frame. So, the heat bath has 
four velocity $u^\mu =(1,0,0,0)$ with $u^\mu u_\mu=1$. The presence of four velocity  means that the most general ansatz for fermion self-energy~\cite{HAW} will be of the form
\bea
\Sigma(P) &=& -{\cal A} P\!\!\!\! \slash - {\cal B}u\!\!\!\slash \, ,\label{gse2} 
\eea
where ${\cal B}$ is another Lorentz invariant structure function in addition to ${\cal A}$.
Since $P^2=\omega^2-p^2$, one can interpret $\omega=p_0=P^\mu u_\mu$ and $p=\left (P^\mu u_\mu -P^2\right)^{1/2}$ as Lorentz invariant energy and
momentum, respectively. The Lorentz invariant structure functions are obtained as follows:

We now write from \eqref{gse2} as
\begin{subequations}
 \begin{align}
\Sigma P\!\!\!\! \slash &= -{\cal A}P^2 - {\cal B} P\!\!\!\! \slash u\!\!\! \slash , \label{gse3}\\
\Sigma u\!\!\! \slash &= - {\cal A} P\!\!\!\! \slash u\!\!\! \slash -{\cal B} . \label{gse4}
\end{align}
\end{subequations}
Taking trace of \eqref{gse3} and \eqref{gse4}, we get
\begin{subequations}
 \begin{align}
{\rm{Tr}} \left [ \Sigma P\!\!\!\! \slash \, \right ] &= -4{\cal A}P^2 - 4{\cal B} \left( P\cdot u\right ), \label{gse5}\\
\left( P\cdot u\right ) {\rm{Tr}} \left [\Sigma u\!\!\! \slash\right ]  &= -4 {\cal A} \left( P\cdot u\right )^2 -4 {\cal B} \left( P\cdot u\right ). \label{gse6}
\end{align}
\end{subequations}
Solving \eqref{gse5} and \eqref{gse6}, one obtains
\be
{\cal A}(\omega,p) = \frac{1}{4} \frac {{\rm{Tr}} \left [ \Sigma P\!\!\!\! \slash \, \right ] - \left( P\cdot u\right ) {\rm{Tr}} \left [\Sigma u\!\!\! \slash\right ]  }
{\left (P\cdot u\right )^2 - P^2} .\label{gse7}
\ee
Further one can write
\begin{subequations}
 \begin{align}
P^2 {\rm{Tr}} \left [\Sigma u\!\!\! \slash\right ] &= -4{\cal A}P^2 \left( P\cdot u\right ) -4{\cal B} P^2 , \label{gse8} \\
\left( P\cdot u\right ) {\rm{Tr}} \left [ \Sigma P\!\!\!\! \slash \, \right ] &= -4{\cal A}P^2 \left( P\cdot u\right ) -4{\cal B} \left( P\cdot u\right )^2 .\label{gse9}
\end{align}
\end{subequations}
Solving \eqref{gse8} and \eqref{gse9}, one obtains
\be
{\cal B}(\omega,p) = \frac{1}{4} \frac  {P^2 {\rm{Tr}} \left [\Sigma u\!\!\! \slash\right ] - \left( P\cdot u\right ) {\rm{Tr}} \left [ \Sigma P\!\!\!\! \slash\,\right ]  }
{\left (P\cdot u\right )^2 - P^2} .\label{gse10}
\ee
Now in the rest frame of the heat bath, $u^\mu=(1,0,0,0)$, the most general ansatz for fermionic self-energy reads~\cite{HAW} as 
\bea
\Sigma(P) &=& -{\cal A}(\omega,p) P\!\!\!\! \slash - {\cal B}(\omega,p)\gamma_0 , \label{gse11} 
\eea
with structure functions
\begin{subequations}
 \begin{align}
{\cal A}(\omega,p) &= \frac{1}{4p^2} \left ({\rm{Tr}} \left [ \Sigma P\!\!\!\! \slash \, \right ] - \omega {\rm{Tr}} \left [\Sigma \gamma_0\right ] \right ), \label{gse12}\\
{\cal B}(\omega,p) &= \frac{1}{4p^2} \left ( P^2 {\rm{Tr}} \left [\Sigma\gamma_0 \right ] - \omega {\rm{Tr}} \left [ \Sigma P\!\!\!\! \slash \, \right ]  \right ) .\label{gse13}
\end{align}
\end{subequations}

\subsection{Fermion Propagator}
\label{fgp}
\begin{figure}[h]
 \vspace*{-0.15in}
\includegraphics[width=16cm,height=2cm]{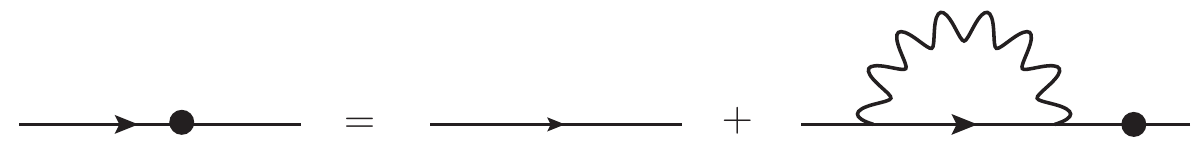}
 \caption{Pictorial representation of Dyson-Schwinger equation for effective fermion propagator.}
 \label{dyson}
 \end{figure}
 
In  Fig.\ref{dyson}  we represent the full propagator by $S^*(P)$ and the bare propagator by $S_0(P)$ and self-energy as $\Sigma(P)$, then the full propagator is written as,
\bea
S^*(P) &=& S_0(P) +S_0(P) \Sigma (P) S^*(P)  \nn \\
S^*(P) S^{*-1}(P) &=& S_0(P) S^{*-1}(P) +S_0(P) \Sigma (P) S^*(P) S^{*-1}(P)  \nn \\
1 &=& S_0(P) S^{*-1}(P) +S_0(P) \Sigma (P) \nn \\
S_0^{-1}(P) &=& S_0^{-1}(P) S_0(P) S^{*-1}(P) +S_0^{-1}(P)S_0(P) \Sigma (P) \nn \\
S^{*-1}(P) &=& S_0^{-1}(P) -\Sigma (P) \nn \\
S^{*-1}(P) &=& \slashed{P} -\Sigma (P) \, , \label{gse15}
\eea
which is known as fermionic Dyson-Schwinger equation.
 The effective fermion propagator  can be obtained as
\be
S^\star(P)=\frac{1}{P\!\!\!\! \slash -\Sigma(P)} .\label{gse14}
\ee
Using \eqref{gse2} one can write the effective propagator as
\be
S^\star(P)=\frac{1}{(1+{\cal A})P\!\!\!\! \slash+{\cal B}u\!\!\! \slash} =\frac{(1+{\cal A})P\!\!\!\! \slash+{\cal B}u\!\!\! \slash}{[(1+{\cal A})P\!\!\!\! \slash+{\cal B}u\!\!\! \slash]^2}
= \frac{P\!\!\!\! \slash -\Sigma(P)} {\cal D} =\frac{S^{\star-1}(P)}{\cal D} ,\label{gse16}
\ee
where the Lorentz invariant quantity ${\cal D}$ is given as
\bea
{\cal D}{(p,u)} &=& \left [(1+{\cal A})P\!\!\!\! \slash+{\cal B}u\!\!\! \slash\right ]^2 
= (1+{\cal A})^2 {P\!\!\!\! \slash}^2 +2(1+{\cal A}){\cal B} P\cdot u + {\cal B}^2 .\label{gse17}
\eea
In the rest frame of heat bath, Eq.\eqref{gse17} reads as
\bea
{\cal D}{(p,\omega)} &=&  (1+{\cal A})^2 (\omega^2-p^2) +2(1+{\cal A}){\cal B} \omega + {\cal B}^2 
= \left [ (1+{\cal A})\omega +{\cal B}\right ]^2 - (1+{\cal A})^2p^2 \nn \\
&=& \left [(1+{\cal A})(\omega-p)+{\cal B} \right ]  \left [(1+{\cal A})(\omega+p)+{\cal B} \right ] 
= {\cal D}_+{\cal D}_- , \label{gse18}
\eea
where
\be
{\cal D}_\pm (p,\omega)= (1+{\cal A})(\omega\mp p)+{\cal B} . \label{gse19}
\ee
In free case ${\cal A}={\cal B}=0$ and Eq.\eqref{gse19} becomes
\be
d_\pm (p,\omega)= \omega\mp p. \label{gse19a}
\ee
Combining \eqref{gse18} and \eqref{gse16}, one can write the effective propagator as
\be
S^\star(P)=\frac{S^{\star-1}(P)}{{\cal D}_+{\cal D}_-} .\label{gse20}
\ee
We can write the self energy in \eqref{gse11} as
\bea
\Sigma(P) &=& -{\cal A}(\omega,p) P\!\!\!\! \slash - {\cal B}(\omega,p)\gamma_0 \nn \\
&=& -({\cal A}\omega+{\cal B})\gamma_0 + {\cal A} p \  {\vec  \gamma} \cdot\bm{\hat { p}} \nn \\
&=& \frac{1}{2} \left[  -({\cal A}\omega+{\cal B})\gamma_0 - ({\cal A}\omega+{\cal B})\gamma_0 -{\cal A}p\gamma_0 +{\cal A}p\gamma_0 
+{\cal A} p \  {\vec \gamma}\cdot \bm{\hat { p}}+{\cal A} p \  {\vec \gamma}\cdot \bm{\hat { p}} \right. \nn \\
&&\ \ \ \  \left. -({\cal A}\omega+{\cal B}) {\vec \gamma}\cdot \bm{\hat { p}} +({\cal A}\omega+{\cal B}) {\vec \gamma}\cdot \bm{\hat { p}} \right]\nn\\
&=& \frac{1}{2}\left [-({\cal A}\omega+{\cal B})(\gamma_0 -  {\vec \gamma}\cdot \bm{\hat {p}}) 
-({\cal A}\omega+{\cal B})(\gamma_0 + {\vec \gamma}\cdot \bm{\hat { p}}) -{\cal A}p (\gamma_0 - {\vec \gamma}\cdot \bm{\hat { p}}) 
+{\cal A}p (\gamma_0 + {\vec \gamma}\cdot \bm{\hat { p}}) \right ]\nn\\
&=&-\frac{1}{2} \left[\left({\cal A}(\omega+p)+{\cal B} \right )(\gamma_0 - {\vec \gamma}\cdot \bm{\hat { p}}) 
+\left({\cal A}(\omega-p)+{\cal B} \right )(\gamma_0 + {\vec \gamma}\cdot \bm{\hat { p}}) \right ].\label{gse21} 
\eea
Now we can write
\bea
P\!\!\!\! \slash &=& \gamma_0\omega -p {\vec \gamma}\cdot \bm{\hat { p}}  
=\frac{1}{2} \left [ \gamma_0 \omega +\gamma_0\omega -\gamma_0 p +\gamma_0 p -p {\vec \gamma}\cdot \bm{\hat { p}} -p {\vec \gamma}\cdot \bm{\hat { p}} 
+\omega {\vec \gamma}\cdot \bm{\hat { p}}-\omega {\vec \gamma}\cdot \bm{\hat { p}} \right ]\nn\\
&=&\frac{1}{2}\left [(\omega-p)(\gamma_0 + {\vec \gamma}\cdot \bm{\hat { p}}) +(\omega+p)(\gamma_0 - {\vec \gamma}\cdot \bm{\hat { p}})  \right ]. \label{gse22}
\eea
In the rest frame of heat bath the inverse of the effective propagator in \eqref{gse15} can now be written as
\bea
S^{\star-1}(P)&=&{P\!\!\!\! \slash -\Sigma(P)}\nn\\
&=& \frac{1}{2}\left [\left \{(\omega-p)+{\cal A}(\omega-p)+{\cal B}\right \} (\gamma_0 + {\vec \gamma}\cdot \bm{\hat {  p}}) 
+ \left \{(\omega+p)+{\cal A}(\omega+p)+{\cal B}\right \} (\gamma_0 - {\vec \gamma}\cdot \bm{\hat { p}})  \right ]\nn\\
&=& \frac{1}{2}\left [(1+{\cal A})(\omega-p)+{\cal B}\right] (\gamma_0 + {\vec \gamma}\cdot \bm{\hat { p}})
+ \frac{1}{2}\left [(1+{\cal A})(\omega+p)+{\cal B}\right] (\gamma_0 - {\vec \gamma}\cdot \bm{\hat { p}})\nn \\
&=& \frac{1}{2} (\gamma_0 + {\vec \gamma}\cdot \bm{\hat  p}) {\cal D}_+ + \frac{1}{2} (\gamma_0 - {\vec \gamma}\cdot \bm{\hat { p}}) {\cal D}_-
 \label{gse23}
\eea
Using \eqref{gse23} in \eqref{gse20}, one finally obtains the effective fermion propagator as
\be
S^\star(P) =  \frac{1}{2} \frac{(\gamma_0 - {\vec \gamma}\cdot \bm {\hat { p}})} {{\cal D}_+(\omega,p)}+ \frac{1}{2} \frac{(\gamma_0 + {\vec \gamma}\cdot \bm {\hat { p}})} {{\cal D}_- (\omega,p)} ,
 \label{gse24}
 \ee
which is decomposed in helicity eigenstates. 

In free fermion case, the  propagator becomes
\be
S(P) =  \frac{1}{2} \frac{(\gamma_0 - {\vec \gamma}\cdot \bm{\hat { p}})} {d_+(\omega,p)}+ \frac{1}{2} \frac{(\gamma_0 + {\vec \gamma}\cdot \bm{\hat { p}})} {d_-(\omega,p) } ,
 \label{gse25}
 \ee
where $d_\pm(\omega,p)$ are given in \eqref{gse19a} and has already been used in Sec.\ref{reverse} in \eqref{f2}.

\section{General structure of Gauge Boson Two-point Functions at $T\ne 0$}
 \label{gsg}
\subsection{Covariant Description}.      
\label{cd}
The general structure of the gauge boson self-energy in vacuum~\cite{Peskin} is given as
\bea
\Pi^{\mn}(P^2) = V^{\mn}\Pi(P^2),
\eea
where the form factor  $\Pi(P^2)$ is  Lorentz invariant and 
depends only on the four scalar $P^2$. The vacuum projection operator is given by
\bea
V^{\mu\nu} = \eta^{\mu\nu}-\frac{P^\mu P^\nu}{P^2},
\eea
which  satisfies the gauge invariance through the transversality condition 
\bea
P_\mu \Pi^{\mu\nu} = 0 ,
\label{trans_cond}
\eea
with $\eta^{\mu\nu}\equiv(1,-1,-1,-1)$ and $P\equiv(\omega, \bm{\vec p})$.  It is also symmetric under the exchange of $\mu\leftrightarrow \nu$ as
\be
\Pi_{\mn}(P^2)=\Pi_{\nu\mu}(P^2) .
\ee
The presence of the heat bath  or the finite temperature ($\beta=1/T$)  breaks the Lorentz invariance
of the system.  One collects all the  four vectors 
and tensors in order to construct a general  covariant structure of the gauge boson 
self-energy at finite temperature.  These are 
$P^\mu$ and  $ \eta^{\mu\nu}$ from vacuum, and the four-velocity $u^\mu$ of the heat bath. With these one can form four types of tensors, 
namely $P^\mu P^\nu, P^\mu u^\nu + u^\mu P^\nu, u^\mu u^\nu$ and $\eta^{\mu\nu}$~\cite{Ashok_Das,Weldon:1982aq}. These 
four tensors can form two independent tensors by virtue of two constraints provided by 
the transversality condition in (\ref{trans_cond}).  One can form two
mutually orthogonal projection tensors from these two independent tensors in order to construct  
 Lorentz-invariant structure of the gauge boson two point functions
at finite temperature.

Now, we define the Lorentz scalars, vectors and tensors that characterise 
the heat bath:
\bea
u^\mu &=&(1,0,0,0), \nn\\
P^\mu u_\mu&=&P\cdot u=\omega ,\label{scal1}
\eea
\subsection{Tensor Decomposition}
\label{tds}
Similar to vaccum, we can define $\tilde{\eta}^{\mn}$ transverse to $u^{\mu}$ as
 \begin{subequations}
 \begin{align}
\ti{\eta}^{\mn} &= \eta^{\mn} - u^\mu u^\nu \, \label{td1} \\
 u_{\mu}\ti{\eta}^{\mn} &= u_{\mu}\eta^{\mn} - u_{\mu}u^\mu u^\nu  = u^\nu - u^\nu = 0 \, . \label{td2}
\end{align}
\end{subequations}
 So $u^\mu$ and $\ti{\eta}^{\mn}$ are transverse.

 Any four vector can be decomposed parallel and orthogonal component with respect to $u^\mu$:
  \begin{subequations}
 \begin{align}
 P^\mu_\sp &=(P\cdot u)u^\mu = \omega u^\mu, \label{td2a} \\
 P^\mu_\perp &=\ti{P}^\mu = P^\mu -  P^\mu_\sp = P^\mu-\omega u^\mu \, . \label{td2b}
\end{align}
\end{subequations} 
Now,
\bea
\ti{P}^2 &=& \left(P^\mu - \om u^\mu\right)\left(P_{\mu} - \om u_{\mu}\right)
 = P^2 - \om^2 - \om^2 + \om^2
 = P^2 - \om^2
 = -p^2 \, . \label{td2c}
\eea 
 We can also define any four vector parallel and perpendicular to $P^\mu$
 \begin{subequations}
 \begin{align}
u^\mu_\sp & = \frac{(P\cdot u)P^\mu}{P^2} = \frac{\om P^\mu}{P^2} \, , \label{td3} \\
\bar{u}^\mu & \equiv u^\mu_\perp = u^\mu - u^\mu_\sp = u^\mu - \frac{\om P^\mu}{P^2} \, . \label{td4}
\end{align}
\end{subequations}
So, $P^\mu \bar{u}_{\mu} = 0$.

Again,
 \begin{subequations}
 \begin{align}
V^{\mn} &= \eta^{\mn} - \frac{P^\mu P^\nu}{P^2}\, ,\label{td5} \\
P_\mu V^{\mn} &= 0 \, . \label{td6}
\end{align}
\end{subequations}
Given these, it is possible to construct only two independent second rank symmetric tensors at finite 
temperature from $\eta^{\mn},P^\mu P^\nu,u^\mu u^\nu,P^\mu u^\nu + u^\mu P^\nu$ which are orthogonal to $P^\mu$.
These two tensors are~\cite{Ashok_Das}
\be
A^{\mn} = \ti{\eta}^{\mn} - \frac{\ti{P}^\mu \ti{P}^\nu}{\ti{P}^2} \, , \label{td6a}
\ee
and
\be
B^{\mn} = \frac{P^2}{\ti{P}^2} \bar{u}^\mu \bar{u}^\nu  = \frac{\bar{u}^\mu\bar{u}^\nu}{\bar{u}^2} \, , \label{td7}
\ee
where
\be
A^{\mn} + B^{\mn} = V^{\mn} = \eta^{\mn} - \frac{P^\mu P^\nu}{P^2} \, . \label{td8}
\ee
We show that $A^{\mn}$ and $B^{\mn}$ are orthogonal to $P^\mu$:
\bea
P_{\mu}A^{\mn} &=& P_{\mu}\ti{\eta}^{\mn} - P_{\mu}\frac{\ti{P}^\mu \ti{P}^\nu}{\ti{P}^2} 
 \nn\\
 &=& P_\mu\left(\eta^{\mn} - u^\mu u^\nu\right) - \frac{P_\mu\left(P^\mu - \om u^\mu\right)\ti{P}^\nu}{\ti{P}^2}\nn\\
 &=& \left(P^\nu - \om u^\nu\right) - \left(P^2 - \om^2\right)\frac{\ti{P}^\nu}{\ti{P}^2}\nn\\
 &=& \ti{P}^\nu - \frac{\ti{P}^2\ti{P}^\nu}{\ti{P}^2}\nn\\
 &=& 0 \, . \label{td9}
\eea

\bea
P_{\mu}B^{\mn} &=& \frac{P^2}{\ti{P}^2}P_\mu\left(u^\mu - \frac{\om P^\mu}{P^2}\right)\left(u^\nu - \frac{\om P^\nu}{P^2}\right)
 \nn\\
 &=& \frac{P^2}{\ti{P}^2}\left[\left(\om - \frac{\om P^2}{P^2}\right)\left(u^\nu - \frac{\om P^\nu}{P^2}\right)\right]\nn\\
 &=&  0 \, . \label{td10}
\eea
$A^{\mn}$ and $B^{\mn}$ also satisfy following relations:
 \begin{subequations}
 \begin{align}
A^{\mn} B_{\mn} &= \left(\ti{\eta}_{\mn} - \frac{\ti{P}_{\mu}\ti{P}_{\nu}}{\ti{P}^2}\right)\frac{P^2}{\ti{P}^2}\bar{u}^\mu\bar{u}^\nu = 0 \, \label{td11} \\
A_{\mu\nu}A^{\nu\rho} &=\left(\tilde{\eta}_{\mu\rho}-\frac{\tilde{P}_\mu \tilde{P}_\rho}{\tilde{P}^2}\right)
\left(\tilde{\eta}^{\rho\nu}-\frac{\tilde{P}^\rho \tilde{P}^\nu}{\tilde{P}^2}\right)= A_{\mu}^{\rho} \, ,\label{td12} \\
B_{\mu\nu}B^{\nu\rho} & =  \left(V_{\mu\rho}-A_{\mu\rho}\right)\left(V^{\rho\nu}-A^{\rho\nu}\right) \nn \\
&= V_{\mu\rho}V^{\rho\nu} - 2A_{\mu\rho}V^{\rho\nu} + A_{\mu\rho}A^{\rho\nu} \nn \\
& =V_\mu^\nu -2A_\mu^\nu + A_\mu^\nu= B_{\mu}^{\rho} \, , \label{td13}\\
A^{\mn}A_{\mn} &= 2\, , \label{td14} \\
B^{\mn}B_{\mn} &= 1\, , \label{td15} \\
A_{\mn}(P) &= A_{\nu\mu}(P) = A_{\mn}(-P), \label{td16}\\
B_{\mn}(P) &= B_{\nu\mu}(P) = B_{\mn}(-P) \, . \label{td17}
\end{align}
\end{subequations}
Finally, one can obtain
 \begin{subequations}
 \begin{align}
A^{\mn} &= \eta^{\mn} - u^\mu u^\nu - \frac{\left(P^\mu - \om u^\mu\right)\left(P^\nu - \om u^\nu\right)}{P^2 - \om^2}\nn\\
&= \frac{1}{P^2 - \om^2}\left[(P^2 - \om^2)(\eta^{\mn} - u^\mu u^\nu) - P^\mu P^\nu - \om^2 u^\mu u^\nu + \om(P^\mu u^\nu + u^\mu P^\nu)\right] \, , \label{A_exp}\\
B^{\mn} &= \frac{P^2}{\ti{P}^2}\bar{u}^\mu\bar{u}^\nu
 = \frac{P^2}{P^2 - \om^2}\left[(u^\mu - \frac{\om P^\mu}{P^2})(u^\nu - \frac{\om P^\nu}{P^2})\right]\nn\\
 &= \frac{P^2}{P^2 - \om^2}\left[u^\mu u^\nu - \frac{\om P^\mu u^\nu}{P^2} - \frac{\om u^\mu P^\nu}{P^2} + \frac{\om^2 P^\mu P^\nu}{P^4}\right]\nn\\
&= \frac{1}{P^2\left(P^2 - \om^2\right)}\left[P^4 u^\mu u^\nu + \om^2 P^\mu P^\nu - \om P^2(P^\mu u^\nu + u^\mu P^\nu)\right]\, . \label{B_exp}
\end{align}
\end{subequations}

\subsection{General Structure of Self-energy of a Vector Particle in a Thermal Medium}
\label{gbs}
The self-energy of a vector particle in a medium (finite temperature/density) can be written as
\be
\Pi_{\mn}= \Pi_T(\om,p)A_{\mn} + \Pi_L(\om,p)B_{\mn} , \label{gen_exp}
\ee
It obeys the current conservation or transversality condition as
\bea
P^\mu\Pi_{\mn} &=& \Pi_T P^\mu A_{\mn} + \Pi_L P^\mu B_{\mn}\nn\\
&=& 0 \, . \label{gbs1}
\eea

Also note that at zero temperature
\be
\Pi^{(0)}_{\mn}(\om,p) = \Pi^{(0)}(P^2)\left(\eta_{\mn} - \frac{P_\mu P_\nu}{P^2}\right) \, , \label{gbs2}
\ee
where 
\be
\Pi^{(0)}_L(\om,p) = \Pi^{(0)}_T(\om,p) = \Pi^{(0)}(P^2) \, . \label{gbs3}
\ee

Using \eqref{gen_exp} we can write
\bea
\Pi_{00}(\om,p) &=& \Pi_T(\om,p)A_{00} + \Pi_L(\om,p)B_{00} \, . \label{gbs5}
\eea
We obtain from \eqref{A_exp} and \eqref{B_exp}, respectively, as
 \begin{subequations}
 \begin{align}
A_{00} &= 0\, , \label{gbs6} \\
B_{00} &= -\frac{p^2}{P^2} \, . \label{gbs7}
\end{align}
\end{subequations}
Using \eqref{gbs6} and \eqref{gbs7} in \eqref{gbs5}, one obtains~\cite{Ashok_Das}
\bea
\Pi_L(\om,p) &=& \left(-\frac{P^2}{p^2}\right)\Pi_{00}(\om,p) \label{pi_L} \, .
\eea

Again, we can write 
\be
\eta^{\mn} \Pi_{\mn} = \Pi_T\eta^{\mn}A_{\mn} + \Pi_L\eta^{\mn}B_{\mn} \, . \label{gbs8}
\ee
One obtains
 \bea
 \eta^{\mn}A_{\mn} &=& \eta^{\mn}A_{\mn} - u^{\mu}u_{\mu} - \frac{\ti{P}^{\mu}\ti{P}_{\nu}}{\ti{P}^2}\nn\\
 = D - 1 - 1
 = D - 2, \label{gbs9}
 \eea
 where $D=4$ is the dimension of the system. Also, one can obtain
 \bea
 \eta^{\mn}B_{\mn} &=& \eta^{\mn}\frac{P^2}{\ti{P}^2} \bar{u}_\mu \bar{u}_\nu 
 = \frac{P^2}{\ti{P}^2} \bar{u}_\mu \bar{u}^\mu 
 = \frac{P^2}{\ti{P}^2} \left(u_{\mu} - \frac{\om P_{\mu}}{P^2}\right)\left(u^{\mu} - \frac{\om P^{\mu}}{P^2}\right)\nn\\
 &=& \frac{P^2}{\ti{P}^2} \left[u_{\mu} u^{\mu} - \frac{\om}{P^2}\left(P_{\mu}u^{\mu} + P^{\mu}u_{\mu}\right) + \frac{\om^2}{P^4}P_{\mu}P^{\mu}\right]\nn\\
 &=& \frac{P^2}{\ti{P}^2} \left(1 - \frac{2\om^2}{P^2} + \frac{\om^2}{P^2}\right)
 = \frac{P^2}{P^2 - \om^2}\left(\frac{P^2 - \om^2}{P^2}\right)
 = 1 \, . \label{gbs10}
 \eea
Equation \eqref{gbs9} denotes the presence of two transverse modes whereas \eqref{gbs10} corresponds to  a longitudinal mode of the gauge boson.

 Using \eqref{gbs9} and \eqref{gbs10} in \eqref{gbs8} we obtain~\cite{Ashok_Das} 
\be
\Pi_T(\om,p) = \frac{1}{D-2}\left[{\Pi}^{\mu}_{\mu}(\om,p) - \Pi_L(\om,p)\right] =\frac{1}{2}\left[{\Pi}^{\mu}_{\mu}(\om,p) - \Pi_L(\om,p)\right].\label{pi_T}
\ee
\subsection{Massless Vector Gauge Boson Propagator in Covariant Gauge}
\label{gbp}

\begin{center}
 \begin{figure}[htb]
 \begin{center}
  \includegraphics[scale=1.0]{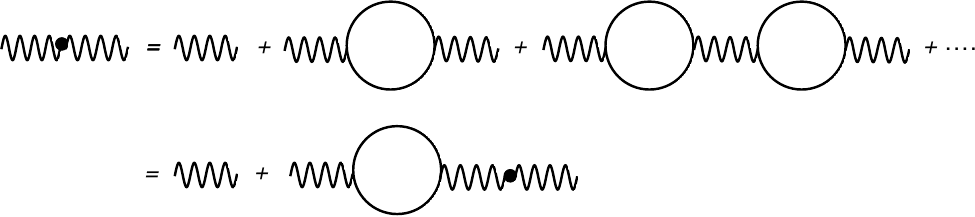}
  \caption{Effective gauge boson propagator.}
  \label{dyson_eqn_pic}
  \end{center}
 \end{figure}
\end{center} 

 We represent the full propagator by $D_{mn}$ and the bare propagator by $D^0_{\mn}$ and each blob that appears in the summation as $\Pi_{\mn}$.  
 One can work out this summation in Fig.~ \ref{dyson_eqn_pic} in tensorial form for the effective gauge boson propagator in Fig.~\ref{dyson_eqn_pic}  as
\bea
D^{\mr} &=& {D^{0}}^{\mr} + {D^{0}}^{\mu\alpha} \Pi_{\alpha\beta}{D}^{\beta\rho} \nn\\
D^{\mu\rho}D^{-1}_{\rho \nu} &=& {D^{0}}^{\mu\rho}D^{-1}_{\rho \nu} + {D^{0}}^{\mu\alpha}\Pi_{\alpha\beta}D^{\beta\rho}D^{-1}_{\rho\nu}\nn\\
\delta^{\mu}_\nu &=& {D^{0}}^{\mu\rho}D^{-1}_{\rho\nu} + {D^{0}}^{\mu\alpha}\Pi_{\alpha\beta}\delta^{\beta}_{\nu}\nn\\
\delta^{\mu}_{\nu} &=& {D^{0}}^{\mu\rho}D^{-1}_{\rho\nu} + {D^{0}}^{\mu\alpha}\Pi_{\alpha\nu}\nn\\
\delta^{\mu}_\nu (D^{0}_{\mu \gamma})^{-1} &=& {D^{0}}^{\mr}(D^{0}_{\mu\gamma})^{-1}D^{-1}_{\rho\nu} + {D^{0}}^{\mu\alpha}(D^{0}_{\mu\gamma})^{-1}\Pi_{\alpha\nu}\nn\\
(D^{0}_{\nu\gamma})^{-1} &=& \delta^{\rho}_{\gamma}D^{-1}_{\rho\nu} + \delta^{\alpha}_{\gamma}\Pi_{\alpha\nu}\nn\\
D^{-1}_{\nu\gamma} &=& (D^{0}_{\nu\gamma})^{-1} - \Pi_{\nu\gamma} . \label{dys}
\eea
Effective photon propagator is given by
\be
{D^{-1}_{\mn} = (D^0_{\mn})^{-1} -\Pi_{\mn}} \, , \label{dyson_eqn}
\ee
which is known as Dyson-Schwinger equation.

For massless gauge boson propagator is given in \eqref{qed21} as
\be
D^0_{\mn} = -\frac{\eta_{\mn}}{P^2} + {(1-\xi)} \frac{P_\mu P_\nu}{P^4} \, \label{gsp1}
\ee
where $\xi$ is the gauge fixing parameter with $\xi=1$ in Feynman gauge and $\xi=0$ in Landau gauge. We will discuss the gauge fixing and gauge boson propagator
in subsec~\ref{gfix}.

Dyson-Schwinger equation is given in \eqref{dyson_eqn} as
\be
D^{-1}_{\mn} = (D^0_{\mn})^{-1} - \Pi_{\mn} \, . \label{gsp2}
\ee
So we have to calculate the inverse of $D^0_{\mn}$.

Lets define
\be
(D^0_{\mu\rho})^{-1} = a\eta_{\mu\rho} + bP_\mu P_\rho \, . \label{gsp3}
\ee
We know 
\bea
D^0_{\mu\rho}({D^0}^{\rho\nu})^{-1} &=& \delta_\mu^\nu  \nn \\
\frac{1}{P^2}\left(-\eta_{\mu\rho} +(1-\xi) \frac{ P_\mu P_\rho}{P^2}\right)\left(a\eta^{\rho\nu} + bP^\rho P^\nu\right) &=& \delta_\mu^{\nu}\nn\\
 \frac{1}{P^2}\left(-a\delta_\mu^\nu + \frac{a(1-\xi)}{P^2}P_\mu P^\nu - bP_\mu P^\nu + b(1-\xi){P_\mu P^\nu}\right) &=& \delta_\mu^\nu \, . \label{gsp4}
\eea
Equating coefficients of both side, we get
\be
{a=-P^2}\,\,,\,\,{b=-\frac{1-\xi}{\xi}} \, . \label{gsp5}
\ee
Using \eqref{gsp5} in \eqref{gsp3}, one obtains
\be
 (D^0_{\mn})^{-1} =- P^2\eta_{\mn} - \frac{1-\xi}{\xi}P_\mu P_\nu .\label{gsp6}
\ee
We know from \eqref{td8}
\bea
P_\mu P_\nu &=& P^2(\eta_{\mn} - A_{\mn} - B_{\mn}) \, . \label{gsp7}
\eea
Substituting this in $(D^0_{\mn})^{-1}$, one can have
\bea
(D^0_{\mn})^{-1} &=& -P^2\eta_{\mn} - \frac{1-\xi}{\xi}P^2(\eta_{\mn} - A_{\mn} - B_{\mn})\nn\\
&=& -\frac{P^2}{\xi}\eta_{\mn} + \frac{1-\xi}{\xi} P^2(A_{\mn} + B_{\mn}) \, , \label{gsp8}
\eea
which is the inverse of $D^0_{\mn}$.

Using  \eqref{gsp8} and \eqref{gen_exp}  in \eqref{dyson_eqn} one can find
\bea
D_{\mn}^{-1} &=&-\frac{P^2}{\xi}\eta_{\mn} + \frac{1-\xi}{\xi} P^2(A_{\mn} + B_{\mn})  - \Pi_T A_{\mn} - \Pi_L B_{\mn} \nn \\
 &=& -\frac{P^2}{\xi}\eta_{\mn} + (P_m^2 - \Pi_T)A_{\mn} + (P_m^2 - \Pi_L)B_{\mn}\label{gsp8a} \, ,
\eea
where $P_m^2=\frac{(1-\xi)}{\xi}P^2$. Eq.({\ref{gsp8a}}) is inverse of $D_{\mn}$.
Now we have to find out $D_{\mn}$ from $D_{\mn}^{-1}$.

Lets define
\be
D_{\mu\rho} = cP_\mu P_\rho + d A_{\mu\rho} + e B_{\mu\rho} \, . \label{gsp9}
\ee
We know
\be
D_{\mu\rho}(D^{\rho\nu})^{-1} = \delta_\mu^\nu \,  \nn
\ee
\be
\frac{cP^2}{\xi}P_\mu P^\nu+d\left(P^2+\Pi_T\right)A_\mu^\nu+e\left(P^2+\Pi_L\right)B_\mu^\nu = -\delta_\mu^\nu \, .  \label{gsp10}
\ee
Substituting $A_\mu^\nu$ and $B_\mu^\nu$ from \eqref{A_exp} and \eqref{B_exp}, the above equation becomes
\bea
&&\frac{cP^2}{\xi}P_\mu P^\nu +d\left(P^2+\Pi_T\ \right ) \left [ \delta_\mu^\nu -\frac{P^2}{P^2-\om^2} u_\mu u^\nu  -\frac{P_\mu P^\nu}{P^2-\om^2} 
+\frac{\om}{P^2-\om^2} \left(u_\mu P^\nu +P_\mu u^\nu \right)\right] \nn \\
&+& e\left(P^2+\Pi_L\right)\frac{P^2}{P^2-\om^2} \left [ u_\mu u^\nu -\frac{\om}{P^2}  \left(u_\mu P^\nu +P_\mu u^\nu \right) +\frac{\om^2}{P^4}P_\mu P^\nu\right ] 
=-\delta_\mu^\nu \, . \label{gsp10a} 
\eea 
Now  equating coefficients on both side,  we get
 \begin{subequations}
 \begin{align}
{\mbox{Coefficients of}}\,\, \delta_\mu^\nu:\hspace{2cm} d&=-\frac{1}{P^2+\Pi_T}\, , \label{gsp11} \\
{\mbox{Coefficients of}}\,\, u_\mu P^\nu+P_\mu u^\nu : \hspace{2cm}e&=-\frac{1}{P^2+\Pi_L}\, , \label{gsp12} \\
{\mbox{Coefficients of}}\,\, P_\mu P^\nu: \hspace{2cm}c&=-\frac{\xi}{P^4} \, . \label{gsp13}
\end{align}
\end{subequations}
Using \eqref{gsp11}, \eqref{gsp12} and \eqref{gsp13} in \eqref{gsp9}, one obtains the effective propagator of interacting photon~\cite{Ashok_Das} in presence of thermal medium as
\be
{D_{\mn} = -\frac{\xi}{P^4}P_\mu P_\nu - \frac{1}{P^2+\Pi_T}A_{\mn} - \frac{1}{P^2+\Pi_L}B_{\mn}} \, . \label{gsp14}
\ee

 \subsection{Massive Vector Boson Propagator}
 \label{vec_boson_mass}
 The free propagator for massive vector boson is given as
 \bea
 D^0_{\mn} &=& \frac{-\eta_{\mn} + \frac{P_\mu P_\nu}{m_v^2}}{P^2-m_v^2}\nn\\
&\equiv&\frac{-\eta_{\mn} + \frac{P_\mu P_\nu}{m_v^2}}{X} \, , \label{mvb1}
 \eea
 where $X=P^2-m_v^2$ and $m_v$ is the mass of the vector boson.
 
 The inverse of $ D^0_{\mn} $ can be written as
 \be
(D^0_{\rho\nu})^{-1} = a\eta_{\rho\nu} + bP_{\rho}P_{\nu} \, . \label{mvb2}
 \ee
 We have 
 \bea
D^0_{\mu\rho}{(D^{0\rho\nu}})^{-1} &=& \delta_\mu^\nu \nn \\
\left(\frac{-\eta_{\mu\rho} + \frac{P_\mu P_\rho}{m_v^2}}{X}\right)\left(a\eta^{\rho\nu} + bP^{\rho}P^{\nu}\right) &=& \delta_\mu^\nu\nn\\
 -a\delta_\mu^\nu - bP_\mu P^\nu + a\frac{P_\mu P^\nu}{m_v^2} + b\frac{P^2 P_\mu P^\nu}{m_v^2} &=& X\delta_\mu^\nu\, . \label{mvb3}
\eea
 Equating coefficients of $\delta_\mu^\nu,$ and $P_\mu P^\nu$ yields,\\
 \be
 a=-X=-(P^2-m_v^2) \, , \label{mvb4}
\ee
and
 \bea
 \frac{a}{m_v^2}-b+\frac{bP^2}{m_v^2} &=& 0 \nn \\
b&=&1 \, . \label{mvb5}
\eea
Now the inverse of the free propagator in \eqref{mvb2} becomes
\be
 (D^0_{\mu\nu})^{-1}=-(P^2-m_v^2)\eta_{\mn}+P_\mu P_\nu\, . \label{mvb6}
  \ee
Using \eqref{td8}  we get
\be
P_\mu P_\nu = P^2\left(\eta_{\mn} - A_{\mn} - B_{\mn}\right) \, , \label{mvb7}
 \ee
 one gets
 \bea
 (D^0_{\mu\nu})^{-1} &=& -(P^2-m_v^2)\eta_{\mn}+P^2\left(\eta_{\mn}-A_{\mn}-B_{\mn}\right)\nn\\
 &=& m_v^2\eta_{\mn}-P^2\left(A_{\mn}+B_{\mn}\right) \, . \label{mvb8}
 \eea
Putting back this value in Dyson equation in \eqref{dyson_eqn}
 \be
D_{\mn}^{-1} = m_v^2\eta_{\mn}-P^2\left(A_{\mn}+B_{\mn}\right) - \Pi_{\mn} \, . \label{mvb9}
\ee
Using Eq.({\ref{gen_exp}}) we get
\bea
D_{\mn}^{-1} &=& m_v^2\eta_{\mn}-P^2\left(A_{\mn}+B_{\mn}\right) - \Pi_T A_{\mn} - \Pi_L B_{\mn}\nn\\
&=& m_v^2\eta_{\mn} - \left(P^2+\Pi_T\right)A_{\mn} - \left(P^2+\Pi_L\right)B_{\mn}  \label{mvb10}
\eea
 Now we have to find $D_{\mn}$. Lets define
 \be
D^{\rho\nu} = \alpha P^\rho P^\nu + \beta A^{\rho\nu} + \gamma B^{\rho\nu} \, . \label{mvb11}
 \ee
 We can write
 \bea
D_{\mu\rho}(D^{\rho\nu})^{-1} &=& \delta_\mu^\nu \nn \\
 \left(\alpha P_\mu P_\rho + \beta A_{\mu\rho} + \gamma B_{\mu\rho}\right)\left(m_v^2\eta^{\rho\nu}-P^2\left(A^{\rho\nu}+B^{\rho\nu}\right) 
 - \Pi_T A^{\rho\nu} - \Pi_L B^{\rho\nu}\right) &=& \delta_\mu^\nu \nn \\
\alpha m_v^2 P_\mu P^\nu + \beta m_v^2 A_\mu^\nu - \beta\left(P^2+\Pi_T\right)A_\mu^\nu + \gamma m_v^2B_\mu^\nu 
- \gamma\left(P^2+\Pi_L\right)B_\mu^\nu &=& \delta_\mu^\nu\nn\\
\alpha m_v^2 P_\mu P^\nu + \beta\left(m_v^2-P^2-\Pi_T\right)A_\mu^\nu + \gamma\left(m_v^2-P^2-\Pi_L\right) &=& \delta_\mu^\nu \, . \label{mvb12}
 \eea
 Substituting $A_{\mu}^\nu,B_{\mu}^\nu$ from Eq.({\ref{A_exp}}),Eq.({\ref{B_exp}}) and equating coefficients we get,
 \begin{subequations}
 \begin{align}
 {\mbox{Coefficients of}\,\,\, \delta_\mu^\nu}:\hspace{2cm}\beta &= \frac{1}{m_v^2-P^2-\Pi_T}\, , \label{mvb13}\\
 {\mbox{Coefficients of}\,\,\, u_\mu u^\nu}:\hspace{2cm} \gamma &= \frac{1}{m_v^2-P^2-\Pi_L}\, , \label{mvb14}\\
{\mbox{Coefficients of}\,\,\, P_\mu P^\nu}:\hspace{2cm}\alpha &= \frac{1}{P^2m_v^2} \, . \label{mvb15}
 \end{align}
 \end{subequations}
 Using \eqref{mvb13} to \eqref{mvb15} in \eqref{mvb11} one gets the propagator for massive vector boson in a thermal medium
\be
D_{\mn} = \frac{P_\mu P_\nu}{P^2 m_v^2} - \frac{A_{\mn}}{{P^2-m_v^2+\Pi_T}} - \frac{B_{\mn}}{P^2-m_v^2+\Pi_L} \, . \label{mvb16}
 \ee

\section{Quantum Electrodynamics (QED)}
\label{qed}
Quantum electrodynamics (QED) is an abelian gauge theory. The symmetry group is  $U(1)$ 
abelian group which are also commutative group. In QED, the interaction between two spin $1/2$
fermionic fields is mediated by electromagnetic field photon, $A_\mu$, 
which is a gauge field. Before doing anything else it is essential to introduce gauge and gauge fixing first.

\subsection{Dirac Field}
The Dirac Lagrangian density in (\ref{fpf1}) describes the free fermion and given as
\bea
{\cal L}_D &=& {\bar \psi} \left (i\partial \!\!\! \slash -m \right ) \psi . \label{qed1}
\eea
As we have seen in subsec.~\ref{global} that it is invariant under a global phase transformation,  $e^{-ie\alpha}\psi(X) $, with a fixed 
phase parameter which does not depend upon space and time.  This is a global transformation.
If the phase factor $\alpha$ is any differentiable function of space-time, $\alpha(X)$, i.e.,  
at each space-time point  it is  different, then the transformation, 
\be
\psi(X) \rightarrow e^{-ie\alpha(X)}\psi(X), \label{qed1a}
\ee
is called local transformation.
The Lagrangian density in (\ref{qed1}) is no longer invariant under such local
transformation as seen in (\ref{fpf5a}):
\bea
{\cal L}_D \rightarrow {\cal L}^\prime_D&=& \bar\psi e^{ie\alpha(X)} \left (i\partial \!\!\! \slash -m \right ) \psi e^{-ie\alpha(X)} \nn \\
&=& \bar\psi \left (i\partial \!\!\! \slash -m \right ) \psi \, + \, e\bar\psi \partial \!\!\! \slash \alpha(X) \psi \nn\\
&=& {\cal L}_D \,  + \, e\bar\psi \partial \!\!\! \slash \alpha(X) \psi . \label{qed2}
\eea
One needs to include a gauge potential (field)  $A_\mu$ in the theory. As we will see below that 
this gauge field $A_\mu$ together with the original fermionic fields  make the Lagrangian invariant under such
local phase transformation. This is also called local gauge transformation.

Under the local gauge transformation, the modified Dirac invariant Lagrangian in (\ref{qed1}) now reads as
\bea
{\cal L}_D &=& {\bar \psi} \left (i\partial \!\!\! \slash -m \right ) \psi -e {\bar \psi}\gamma^\mu A_\mu \psi 
= {\bar \psi} \left (i D \! \!\!\! \slash -m \right ) \psi,  \label{qed3}
\eea
where the original partial differential operator is replaced by the covariant differential
operator
\bea
\partial_\mu \rightarrow D_\mu&=& \partial_\mu+ieA_\mu , \label{qed3a}
\eea
along with the transformation of the gauge field as
\bea
A_\mu \rightarrow A_\mu + \partial_\mu \alpha(X). \label{qed3b}
\eea
{\Huge [}Now lets check the invariance of (\ref{qed3}) under local gauge transformation: $\psi(X)\rightarrow \psi'= e^{-ie\alpha(X)}\psi$
\bea
\bar \psi' D_\mu \psi' &=& \bar \psi' [\partial_\mu+ieA_\mu]  e^{-ie\alpha(X)}\psi \nn\\
&=&  \bar \psi' e^{-ie\alpha(X)} \partial_\mu \psi -ie \bar\psi' (\partial_\mu \alpha(X)) e^{-ie\alpha(X)}  \psi +ie \bar \psi' A_\mu \psi e^{-ie\alpha(X)}.
\label{qed4}
\eea
If (\ref{qed4}) vis-a-vis (\ref{qed3}) is to be invariant under local transformation of the fermionic field 
$\psi(X)\rightarrow \psi'=e^{-ie\alpha(X)}\psi $, the gauge field 
has also to be transformed as $A_\mu \rightarrow A_\mu + \partial_\mu \alpha(X)$ as given in (\ref{qed3b}). So
\bea
\bar \psi' D_\mu \psi'
&=&  \bar \psi' e^{-ie\alpha(X)} \partial_\mu \psi -ie \bar\psi' (\partial_\mu \alpha(X)) e^{-ie\alpha(X)}  \psi +ie \bar \psi' (A_\mu+\partial_\mu\alpha(X)) 
 e^{-ie\alpha(X)} \psi \nn\\
&=& \bar \psi  (\partial_\mu + ie A_\mu) = \bar \psi D_\mu \psi.
\label{qed5}
\eea
This suggests that (\ref{qed3}) is invariant under local gauge transformation.{\Huge ]}

\subsection {Pure Gauge Field}
 
Following Maxwell's equations  both  electric field ${\mathbf E}$ and the magnetic 
field ${\mathbf B}$ can be written from pure gauge field $A_\mu$ in a manifestly covariant form as
\bea
F_{\mu\nu}= \partial_\mu A_\nu -\partial_\nu A_\mu, \label{qed6}
\eea
which is known as electromagnetic field tensor. This is  a gauge invariant quantity
and also 
antisymmetric, $F_{\mu\nu}=-F_{\nu\mu}$, under the exchange of the Lorentz 
indices $\mu\leftrightarrow \nu$.
One can now construct a Lorentz scalar out of $F_{\mu\nu}$, which can be included in the Lagrangian density for 
pure gauge field as
\bea
{\cal L}_{\gamma} &=& -\frac{1}{4} F_{\mu\nu}F^{\mu\nu}, \label{qed7}
\eea
where the normalisation factor $1/4$ is chosen in such a way that it gives the correct equation of motion for electromagnetic field.

\subsection{Electromagnetic Lagrangian}
Now one can write the total electromagnetic Lagrangian density~\cite{pbpal,Peskin} describing fermions, electromagnetic field and
interaction between them as
\bea
{\cal L}_{em}&=& {\cal L}_\gamma + {\cal L}_D  \nn\\
&=& -\frac{1}{4} F_{\mu\nu}F^{\mu\nu} + {\bar \psi} \left (i D \! \!\!\! \slash -m \right ) \psi  \nn\\
&=& -\frac{1}{4} F_{\mu\nu}F^{\mu\nu} + {\bar \psi} \left (i \gamma^\mu \partial_\mu -m \right ) \psi  -e\bar \psi \gamma^\mu\psi A_\mu.
\label{qed8}
\eea

\begin{figure}
\begin{center}
\includegraphics[scale=0.4]{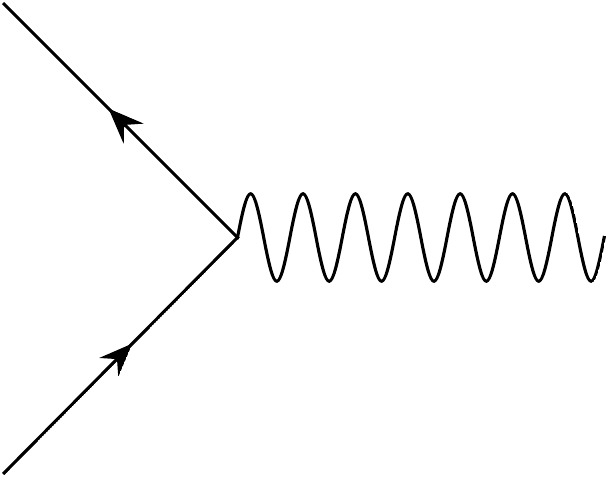}
\end{center}
\caption{Electron-photon interaction vertex.}
\label{electron_photon_vertex}
\end{figure}
We note the following:
\begin{enumerate}
\item[$\bullet$] The first term corresponds to the free Lagrangian density of a gauge field (photon)
\item[$\bullet$] The second term is the Lagrangian density that describes the free fermions with mass $m$. Now if one wants to include
chemical potential $\mu$, one should follow as done in (\ref{fpf8}) in Subsec.~\ref{global} . The presence of the chemical potential is 
like shifting  the temporal component of the gauge field by $\partial_0-i\mu$.
\item[$\bullet$] The third term originates from the local $U(1)$ gauge symmetry and corresponds to the interaction Lagrangian density of fermionic 
and gauge field in which gauge field interact with fermionic field through the dimensionless coupling parameter $e$. This interaction is represented
in Fig.~\ref{electron_photon_vertex}. Using this interaction 
in perturbative techniques, one  can compute Feynman diagrams for the theory.
\item[$\bullet$] The second and third term together will lead to the equation of motion for gauge field $\partial_\mu F^{\mu\nu}= -j^\nu$ .
\item[$\bullet$] If one checks the gauge invariance of second and third term together, then it will lead to a conserved current density and 
the Lagrangian density will then differ by a total derivative,  but leaves the equation of motion unchanged .
\end{enumerate}

\subsection{Gauge Fixing}
\label{gfix}

One of the problem of (\ref{qed8}) is that it can not be quantised.  One way to see this is that the propagator for photon does not exist.
As we have already experienced from scalar and fermionic fields that the propagator is obtained from free theory, 
we start with the free Lagrangian density for gauge field (photon)
\bea
{\cal L}_{\gamma} &=& -\frac{1}{4} F_{\mu\nu}F^{\mu\nu}  
= -\frac{1}{4} \left[ \partial_\mu A_\nu -\partial_\nu A_\mu \right ] \left[ \partial^\mu A^\nu -\partial^\nu A^\mu \right ] \nn\\
&=& -\frac{1}{2} \partial_\mu A_\nu \left[ \partial^\mu A^\nu -\partial^\nu A^\mu \right ] +\partial_\mu(\cdots) \nn\\
&=& \frac{1}{2} A_\nu \left[\eta^{\mu\nu} \Box  -\partial^\mu \partial^\nu  \right ] A_\mu,
\label{qed9}
\eea
where we have interchanged $\mu\rightarrow \nu$ in the second term inside the first square braces in the second line. We have done the 
integration by parts to arrive at third line. Now using the Fourier decomposition of the gauge field
one can get
\bea
{\cal L}_{\gamma} &=& -\frac{1}{2} A_\nu \left[-\eta^{\mu\nu} P^2  + P^\mu P^\nu  \right ] A^\mu ,
\label{qed10}
\eea
where the quantity inside the square braces should in principle be the inverse of photon propagator say $(D_0^{\mu\nu})^{-1}$ as
\bea
(D_0^{\mu\nu})^{-1}&=&-\eta^{\mu\nu} P^2  +P^\mu P^\nu . \label{qed11}
\eea
The inverse of $(D_0^{\mu\nu})^{-1}$ should give the photon propagator as
\bea
D_0^{\mu\nu}= a \eta^{\mu\nu} +b P^\mu P^\nu . \label{qed12}
\eea
Now
\bea
D_0^{\mu\lambda} (D_0)^{-1} _{\lambda\nu}&=&\delta^\mu_\nu \nn\\
\left ( a \eta^{\mu\lambda} +b P^\mu P^\lambda\right ) \left ( - \eta_{\lambda\nu}P^2 + P_\lambda P_\nu \right )  &=& \delta^\mu_\nu \nn\\
 - a P^2 \delta^\mu_\nu + a  P^\mu P_\nu - b P^2  P^\mu P_\nu  + bP^2 P^\mu P^\nu  &=& \delta^\mu_\nu \nn\\
-a P^2 \delta^\mu_\nu + a  P^\mu P_\nu   &=& \delta^\mu_\nu . \label{qed13}
\eea
Comparing the coefficients in both sides, one gets $a=-1/P^2$ and $a=0$ but $b$ remains undetermined. The matrix $(D^{\mu\nu}_0)^{-1}$ 
is not invertible, so the propagator $D^{\mu\nu}_0$ does not exist. 

\noindent {\Huge[}Another way: In general any second rank tensor can be written as
\be
{\cal M}^{\mu\nu} = c{\cal P}^{\mu\nu}_L +d {\cal P}^{\mu\nu}_T , \label{qed13a}
\ee
where the longitudinal and transverse projection operator in orthogonal subspace are
\bea
{\cal P}_L^{\mu\nu} &=& P^\mu P^\nu/P^2 , \nn\\
{\cal P}_T^{\mu\nu}&=&  \left (\eta^{\mu\nu}-{P^\mu P^\nu}/{P^2} \right ), \label{qed13b}
\eea
which satisfy the properties of the projection operator as 
\be
{\cal P}_L^{\mu\nu}+ {\cal P}_T^{\mu\nu}=\eta^{\mu\nu}, \, \, \,
{\cal P}_L^2={\cal P}_L,\, \, \, {\cal P}_T^2={\cal P}_T, \, \, \,  {\cal P}_L{\cal P}_T=0  
\ee
The inverse of (\ref{qed13a}) can be written as
\be
({\cal M}^{\mu\nu})^{-1} = c^{-1}{\cal P}^{\mu\nu}_L +d^{-1}{\cal P}^{\mu\nu}_T . \label{qed13c}
\ee
Now (\ref{qed11}) can be written in terms of projection operator
as 
\bea
(D^{\mu\nu})^{-1}&=&-\eta^{\mu\nu} P^2  +P^\mu P^\nu = (0){\cal P}_L^{\mu\nu}+(-P^2) {\cal P}_T^{\mu\nu}. \label{qed13d}
\eea
Now comparing (\ref{qed13d}) with (\ref{qed13a}) one gets  $c=0$ and $d=-P^2$. Therefore,
(\ref{qed13c}) suggests that $d^{-1}=-1/P^2$ and $c^{-1}$ does not exist as $c^{-1} =1/0$.
Thus the inverse of (\ref{qed13d}) also does not exist.
{\Huge]}

Some of the reasons are noted below:
\begin{enumerate}
\item [$\bullet$] The momenta conjugate to the $A^\mu$ are given by
\bea
\pi^\mu &=& \frac{\delta {\cal L}_{em}}{\delta {\dot A}_\mu} =   \frac{\delta  {\cal L}_{em}}{\delta (\partial_0 A_\mu)} =F^{\mu 0} \,
\label{qed15}
\eea
which give $\pi^0=0$, as the diagonal components of $F^{\mu\nu}$ are zero. This clearly indicates that one of the canonical
momenta does not exist. The equations defining canonical momenta cannot therefore be inverted to express quantities $\dot A_\mu$ 
in terms of the momenta. So, the Hamiltonian formalism also does not exist.

\item [$\bullet$] In Maxwell's equations there are six field  variables from both $\mathbf E$ and $\mathbf B$ but the two
homogeneous equations imply four constraints on the electromagnetic field components so that there only are two independent
component of electromagnetic fields. Therefore, all components of the gauge field $A^\mu$ are not independent but they are 
connected by gauge transformation, even though only two components are independent.

 \item [$\bullet$] The matrix in the transverse projection operator in (\ref{qed11}) has zero eigenvalue as can be seen
 \bea
 \left ( -\eta^{\mu\nu} P^2  +P^\mu P^\nu \right ) P_\nu &=&0  \nn\\
 \left (\eta^{\mu\nu}   -\frac{P^\mu P^\nu}{P^2} \right ) P_\nu &=& 0 \nn\\
{\cal P}_T^{\mu\nu}  P_\nu &=&0 . \label{qed14}
 \eea

The (\ref{qed14}) is the transversality condition that projects on to orthogonal subspace. This also indicates that
the presence of zero eigenvalues of the ${\cal P}_T^{\mu\nu}$ is a direct consequence of the gauge invariance of the theory.
Gauge invariance implies that the theory contains fewer degrees of freedom (only transverse degrees of freedom),
which reflects itself the presence of zero eigenvalues  in the quadratic part of the Lagrangian 
density in (\ref{qed7}) vis-a-vis (\ref{qed9}). However, gauge field $A^\mu$ is represented by four components $A^\mu$,  
all of which are not independent but connected by gauge transformation. There are only two independent components and one 
needs to eliminate or restrict the additional degrees of gauge freedom.  It is also to be noted that the degrees of gauge 
freedom absent in (\ref{qed7}) should not reappear through the interaction. That is ensured through the interaction of the
gauge field with a conserved fermionic current as noted in the last item after (\ref{qed8}) in previous page. 
\end{enumerate}
Before quantisation this redundancy is dealt by fixing the gauge that restrict the gauge degrees 
of freedom in the theory. The Lagrangian density for the gauge field (photon) in $U(1)$ gauge theory can be 
rewritten along with an addition term (referred as gauge fixing term with a gauge parameter $\xi$) as 
\bea
{\cal L}_{\gamma} &=& -\frac{1}{4} F_{\mu\nu}F^{\mu\nu} -\frac{1}{2\xi} (\partial_\nu A^\nu) (\partial_\mu A^\mu)  \nn\\
&=& -\frac{1}{4} \left[ \partial_\mu A_\nu -\partial_\nu A_\mu \right ] \left[ \partial^\mu A^\nu -\partial^\nu A^\mu \right ] 
-\frac{1}{2\xi} (\partial_\nu A^\nu) (\partial_\mu A^\mu) \nn\\
&=& -\frac{1}{2} \partial_\mu A_\nu \left[ \partial^\mu A^\nu -\partial^\nu A^\mu \right ] 
-\frac{1}{2\xi} (\partial_\nu A^\nu)  (\partial_\mu A^\mu)  +\partial_\mu(\cdots)  \nn\\
&=& \frac{1}{2} A_\nu \left[\eta^{\mu\nu} \Box  -\left(1-\frac{1}{\xi}\right )\partial^\mu \partial^\nu  \right ] A_\mu,
\label{qed16}
\eea
where the inverse of the propagator can be written as
\bea
(D_0^{\mu\nu})^{-1} = \left[-\eta^{\mu\nu} P^2  +\left(1-\frac{1}{\xi}\right )P^\mu P^\nu  \right ], \label{qed17}
\eea
and the inverse of which should give us the correct form the propagator as
\bea
D_0^{\mu\nu} = \left[ a \eta^{\mu\nu}  + b \left(1-\frac{1}{\xi}\right )P^\mu P^\nu  \right ] \, . \label{qed18}
\eea
As before solving for coefficients $a$ and $b$:
\bea
D_0^{\mu\lambda} (D_0)^{-1} _{\lambda\nu}&=&\delta^\mu_\nu \nn\\
\left [ a \eta^{\mu\lambda}  +b \left(1-\frac{1}{\xi}\right ) P^\mu P^\lambda\right ] \left [ - \eta_{\lambda\nu} P^2 + \left(1-\frac{1}{\xi}\right )
P_\lambda P_\nu \right ]  &=& \delta^\mu_\nu \nn\\
-a P^2 \delta^\mu_\nu +  \left(1-\frac{1}{\xi}\right ) \left [a  - b  P^2  
+ b \left(1-\frac{1}{\xi}\right ) P^2 \right ] P^\mu P_\nu  &=& \delta^\mu_\nu 
, \label{qed20}
\eea
where comparing the coefficients, one gets
\bea
a &=& -\frac{1}{P^2}, \nn\\
\left [ a  - b  P^2  
+ b \left(1-\frac{1}{\xi}\right ) P^2 \right ]&=& 0 \nn\\
\Rightarrow \, \, \, b &=&  -\frac{\xi}{P^4} .
\eea
With these the photon propagator from (\ref{qed18}) reads as
\bea
D_0^{\mu\nu} =- \frac{1}{P^2}\left[  \eta^{\mu\nu}  - \left(1-\xi\right )\frac{P^\mu P^\nu}{P^2} \right ] . \label{qed21}
\eea
The Feynman propagator will read as 
\bea
{\cal D}_0^{\mu\nu}=i D_0^{\mu\nu} = -\frac{i}{P^2}\left[  \eta^{\mu\nu}  - \left(1-\xi\right )\frac{P^\mu P^\nu}{P^2} \right ] . \label{qed22}
\eea
It is worth noting at this point if a gauge field has mass then the gauge fixing is not required, as we will see later
when the massive vector boson discussed in subsec~\ref{vec_boson_mass}. In different gauges $\xi$ takes different values as  $\xi=1$ (Feynman gauge) and
$\xi=0$ (Landau gauge). However, the final result is independent of gauge choice, so one can choose it conveniently for the purpose.

After gauge fixing the QED Lagrangian reads as
\bea
{\cal L}_{em}
&=& -\frac{1}{4} F_{\mu\nu}F^{\mu\nu} -\frac{1}{2\xi} (\partial_\mu A^\mu)^2 + {\bar \psi} 
\left (i \gamma^\mu \partial_\mu -m \right ) \psi  -e\bar \psi \gamma^\mu\psi A_\mu.
\label{qed23}
\eea
Now one  can show that this gauge fixing term does not change the Lagrangian or the Maxwell's equations as long as the current is 
conserved:
The new equation of motion becomes 
\bea
\Box A^\nu -\left (1-\frac{1}{\xi} \right )\partial^\nu \left(\partial_\mu A^\mu\right ) &=& -j^\nu . \label{qed24}
\eea
Operating 4-divergence and using the current conservation $\partial_\mu j^\mu=0$ one gets
\bea
\Box \left (\partial_\mu A^\mu\right)&=& -\partial_\mu j^\mu=0 \nn\\
\rightarrow \, \, \, \partial_\mu A^\mu  &=& 0 , \label{qed25}
\eea
with appropriate boundary condition on $A^\mu$ so that (\ref{qed25}) is satisfied. This implies that one can always transform $A^\mu$
according to (\ref{qed3b}) so that it satisfies (\ref{qed25}). This is called gauge fixing condition in covariant gauge.
Now one can compute canonical momenta and do the Hamiltonian formulation.



\subsection{Free Photon Partition Function}
\label{photon}

The free photon Lagrangian density can be written from (\ref{qed23}) as
\bea
{\cal L}_{\gamma}
&=& -\frac{1}{4} F_{\mu\nu}F^{\mu\nu} .\label{qed26}
\eea
 
In vacuum, photon partition function can be written as,
 \bea
 \mathcal{Z} &=& \int \mathcal{D}A_\mu \exp{[i S]}\nn\\
 &=& \int \mathcal{D}A_\mu \exp{\left[{i\int d^4X\left(-\frac{1}{4} F_{\mu\nu}F^{\mu\nu}\right)}\right]} \, , \label{qed27}
 \eea
where ${\cal D}A_\mu={\cal D}A_0{\cal D}A_1{\cal D}A_2{\cal D}A_3$. 
Gauge transformations should not change anything physically.

We now choose a covariant gauge condition as
\be
G(A) = \partial_\mu A^\mu = w(X) , \label{qed27a}
\ee
which can be imposed to (\ref{qed27}) by inserting a identity ~\cite{Yang} given as
\be
\int \mathcal{D}\alpha(x)\delta\left(G(A,\alpha)-w(X)\right)\Big\vert{\frac{\delta\left(G(A,\alpha)-w(X)\right)}{\delta\alpha}}\Big\vert = 1 \, . \label{qed28}
\ee
Lorentz gauge condition can be  recovered by taking $w=0$. There is still a residual gauge freedom as one shifts the gauge field
\be
A_\mu \rightarrow A_\mu + \partial_\mu \alpha(X) \, , \label{qed28a}
\ee
where the phase factor, $\alpha(X)$ is differentiable function of space-time.
Now we can write the gauge condition in \eqref{qed27a} as
\be
G(A,\alpha) = \partial_\mu A^\mu + \partial_\mu \partial^{\mu} \alpha \, . \label{qed28b}
\ee
The determinant term can be obtained as
\be
\frac{\delta\left(G(A,\alpha(X))-w(X)\right)}{\delta\alpha(Y)} = \partial_\mu \partial^\mu \delta^{(4)}(X-Y) \, . \label{qed28c}
\ee
So (\ref{qed27}) becomes,
\be
 \mathcal{Z} = \int \mathcal{D}A_\mu \mathcal{D}\alpha(X)\delta\left(G(A,\alpha)-w(X)\right)
 \Big\vert{\frac{\delta\left(G(A,\alpha)-w(X)\right)}{\delta\alpha}}\Big\vert  \exp{\left[{i\int d^4X\left(-\frac{1}{4} F_{\mu\nu}F^{\mu\nu}\right)}\right]} \, . \label{qed28d}
\ee
Since there is a residual gauge freedom, we shift the gauge field as \eqref{qed28a}
and write (\ref{qed28d})  as
\be
 \mathcal{Z} = \int \mathcal{D}A_\mu \,\,\mathcal{D}\alpha(X)\,\,\delta\left(G(A)-w(X)\right)\,
 \det{\partial^2} \exp{\left[{i\int d^4X\left(-\frac{1}{4} F_{\mu\nu}F^{\mu\nu}\right)}\right]} \, . \label{qed28e}
\ee
Now the integrand does not contain $\alpha$ and  the $\alpha$ integration gives diverging result. 
This is due to the redundancy of the residual gauge transformation. Now,
$w(X)$ is an arbitrary function of $X$ and the behaviour of $w(X)$ is not known, so
the integral involved in the partition function can not be solved.
One can avoid this problem by averaging over $w(X)$ around zero with a Gaussian width $\xi$
\be
\int \mathcal{D}w\frac{1}{\sqrt{2\pi\xi}}\exp{\frac{w^2(X)}{2\xi}} \, , \label{qed28f}
\ee
where $\xi$ is a gauge fixing parameter that one  chooses for the convenience of calculations and  $\frac{1}{\sqrt{2\pi\xi}}$ is normalisation factor.
Now the integration over $w$ is performed in \eqref{qed28e} and one gets
\be
 \mathcal{Z} = \int \mathcal{D}A_\mu\,\,\det{\partial^2} \exp{\left[i\int d^4X
 \left(-\frac{1}{4} F_{\mu\nu}F^{\mu\nu}-\frac{(\partial_\mu A^\mu)^2}{2\xi}\right) \right]} \, . \label{qed29}
 \ee
 Now we can write $\det{\partial^2}$ term using Grassmann property given in \eqref{GV13}  as
 \be
 \det{\partial^2} = \int \mathcal{D}\bar{C}\,\,\mathcal{D}C\, \exp({-\bar{C}\partial^2 C}) \, ,
 \ee
 where $C$ is Grassmann field which is also known as ghost field.
 This plays a crucial role to cancel the redundant gauge degrees of freedom.
 Now,
 \bea
 \mathcal{Z} &=& \int \mathcal{D}A\,\,\mathcal{D}\bar{C}\,\,\mathcal{D}C\, \exp{\left [ i\int d^4X (\mathcal{L}_\gamma + \mathcal{L}_{\textrm{gf}} + \mathcal{L}_{\textrm{gh})}\right]} \nn\\
 &=&  \mathcal{Z}_{\gamma+\textrm{gf}}\mathcal{Z}_{\textrm{gh}}\, , \label{qed29a}
 \eea
 where gauge fixing and ghost terms, respectively, are 
 \bea
 \mathcal{L}_{\textrm{gf}} &=& -\frac{(\partial_\mu A^\mu)^2}{2\xi} \, ,\nn\\
 \mathcal{L}_{\textrm{gh}} &=& -\bar{C}\,\partial^2 C \, . \nn
 \eea
 Let us calculate the contributions of gauge and gauge fixing  part of (\ref{qed29a}) to the partition function as
 \bea
\mathcal{Z}_{\gamma+\textrm{gf}}&{=\atop {t\rightarrow-i\tau}}& \int \mathcal{D}A \exp{\left(\frac{i}{2}\int d^4X A_\nu \left[\eta^{\mn}\Box-(1-\frac{1}{\xi})
 \partial^\mu \partial^\nu\right]A_\mu\right)} \, \nn \\
 &=& \int \mathcal{D}A \exp{\left(-\frac{1}{2}\int_0^\beta d\tau \, d^3\bm{\vec x} \, A_i \left[\delta^{ij}\Box_\tau+(1-\frac{1}{\xi})
 \partial^i \partial^j\right]A_j\right)} \,  , \label{qed29b}
 \eea
 where we have used  $\eta^{\mn} \leftrightarrow -\delta^{ij}$ .
 
 The Fourier transform of the gauge field is given as 
 \begin{eqnarray}
A_i(X)= A_i(\bm{\vec x},\tau) &{=\atop {t\rightarrow -i\tau}} \atop {p_0\rightarrow i\omega_n}& \frac{1}{\sqrt{V\beta}} \sum_P \, 
 e^{-i P\cdot  X }\, A_i (P)\nn\\
&=&\frac{1}{\sqrt{V\beta}} \sum_{n,\bm{\vec p}} \, 
e^{i \bm{ \vec p \cdot \vec x } }\,
\, e^{- i \omega_n \tau  } \, A_i (\omega_n,\bm{\vec p}) \, .
\label{qed29c}
\end{eqnarray}

 One can write  the partition function in \eqref{qed29b} in frequency momentum space  as
 \bea
\mathcal{Z}_{\gamma+\textrm{gf}} &=& \int \mathcal{D}A(\omega_n,\bm{\vec p})  \exp {\left(- \sum_{n,\bm{\vec p}}  \frac{1}{2} A^*_{i}(\omega_n,\bm{\vec p}) 
\left [ \delta^{ij} (\om_n^2+p^2) -(1-\frac{1}{\xi}) p^ip^j \right ]A_j (\om_n,\bm{\vec p}) \right ) }\nn \\
 &=& \int \mathcal{D}A(\omega_n,\bm{\vec p})  \exp{\left(-\sum_{n,\bm{\vec p}} \frac{1}{2} A^*_{i}(\om_n,\bm{\vec p})
 D^{-1}_{ij}A_j(\om_n,\bm{\vec p})\right)} \nn\\
 &=& \prod_{n,\bm{\vec p}}\sqrt{\frac{\pi^D}{\det{D^{-1}_{ij}(\xi)}}}
 \eea
 where $D^{-1}_{ij}$ is a $4\times4$ matrix and inverse of the gauge boson propagator. Let us set $p=(0,0,p)$ as three space part of $p$ are in equal footing now~\cite{Yang}.
 In Feynman gauge we have $\xi=1$.
 \[
   D^{-1}_{ij}=
  \left( {\begin{array}{cccc}
   p^2+\om_n^2 & 0 & 0 & 0 \\
   0 & p^2+\om_n^2 & 0 & 0 \\
   0 & 0 & p^2+\om_n^2 & 0\\
   0 & 0 & 0 & p^2+\om_n^2\\
  \end{array} } \right)
\]
\be
\therefore \det{D^{-1}}_{ij} = ( p^2+\om_n^2)^4 . \label{phot_det}
\ee
\bea
\mbox{So,}\,\,\,\ln{\mathcal{Z}_{\gamma+\textrm{gf}}} &=&  -\frac{1}{2} \sumintb_{n,\bm{\vec p}} \ln \textrm{det} [D_{ij}^{-1}] \nn \\
&=& -4\times \frac{1}{2}\sumintb_{P_E}\ln{[ P_E^2]} \, , \label{qed29cc}
\eea
where $P_E$ is Euclidean momentum and $\sum\!\!\!\!\!\!\!\!\int\limits_{P_E}$ is a bosonic sum-integral.

Now we have to calculate ghost contribution to the partition function from \eqref{qed29a} as
\be
{\mathcal Z}_{\textrm{ gh}}=\int \mathcal{D}{\bar{C}}\,\,\mathcal{D}C\, \exp{\left(-i\int d^4X \,\bar{C}\,\partial^2\, C\right)} 
= \int \mathcal{D}{\bar{C}}\,\,\mathcal{D}C\,\, \exp{\left(\int_0^{\beta} d\tau \,d^3x\, \bar{C}\,\Box_\tau\, C\right)} \, . \label{qed29d}
\ee
The Fourier transform of the ghost field is given as
\bea
C(\tau,\bm{\vec x})= \frac{1}{\sqrt{V\beta}} \sum_{n,\bm{\vec p}} \, e^{i \bm{ \vec p \cdot \vec x } }\, e^{- i \omega_n \tau  } \, C(\omega_n,\bm{\vec p})\, . \label{qed29e}
\eea
Combining \eqref{qed29d} and \eqref{qed29e}, one can get the ghost partition function in
frequency momentum space as
\bea
{\mathcal Z}_{\textrm{ gh}} &=& \int \mathcal{D}{\bar{C}(\omega_n,\bm{\vec p})}\mathcal{D}C(\omega_n,\bm{\vec p})\exp{\left(-\sum_{n,\bm{\vec p}} 
 \bar C(\omega_n,\bm{\vec p}) (\om_n^2+p^2) C(\omega_n,\bm{\vec p})\right)} \nn \\
&=& \prod_{n,\bm{\vec p}}(\om_n^2+p^2).  \label{qed29f}
\eea
\be
\mbox{So,}\,\,\,\ln{\mathcal{Z}_{\textrm{gh}}} = 2\times \frac{1}{2}\sumintb_{P_E}\ln{[ P_E^2]} \, . \label{qed29g}
\ee
The logarithm of the photon partition function can be written as
\bea
\ln{\mathcal{Z}} &=& \ln{\mathcal Z}_{\gamma+ \textrm{gf}} +  \ln{\mathcal Z}_{\textrm{gh}}  \nn \\
&=& -4 \times \frac{1}{2}\sumintb_{P_E}\ln{[ P_E^2]} +2 \times \frac{1}{2}\sumintb_{P_E}\ln{[ P_E^2]}  \nn\\
&=& -\sumintb_{P_E}\ln{[ P_E^2]} \, . 
\eea
So, ghost contribution cancels two unphysical degrees of freedom of photon. Now there are two physical transverse degrees of freedom of photon.

By calculating the frequency sum one gets,
\be
\ln{\mathcal{Z}} = -2V\int \frac{d^3p}{(2\pi)^3}\left[\frac{\beta \om_p}{2} + \ln{(1-\exp{(-\beta \om_p)})}\right] \, . \label{qed29i}
\ee
which agrees with one bosonic degree of freedom in \eqref{eq10}.

So free energy density of non-interacting photon at finite temperature is given by (ignoring vacuum part)
\bea
F_0 &=& -\frac{T}{V} \ln{\mathcal{Z}}
= 2 \int \frac{d^3p}{(2\pi)^3} \ln{(1-\exp{(-\beta \om_p)})} \nn \\
&=& -\frac{\pi^2 T^4}{45} \, . \label{qed29j}
\eea
The pressure for non-interacting photon
\be
{\cal P}_0=-F_0 =\frac{\pi^2 T^4}{45} \, . \label{qed29k}
\ee

\subsection{One-loop Fermion Self-energy $\Sigma$ and Structure Functions ${\cal A}$ and ${\cal B}$ in HTL approximation}
\label{ssf}
We have electron-photon interaction Lagrangian from \eqref{qed23} as
\be
\mathcal{L}_{int} = -e\bar{\psi}\gamma_\mu \psi A^\mu . \label{se0_a}
\ee

\begin{figure}[htb]
\begin{center}
\includegraphics[width=11cm,height=4cm]{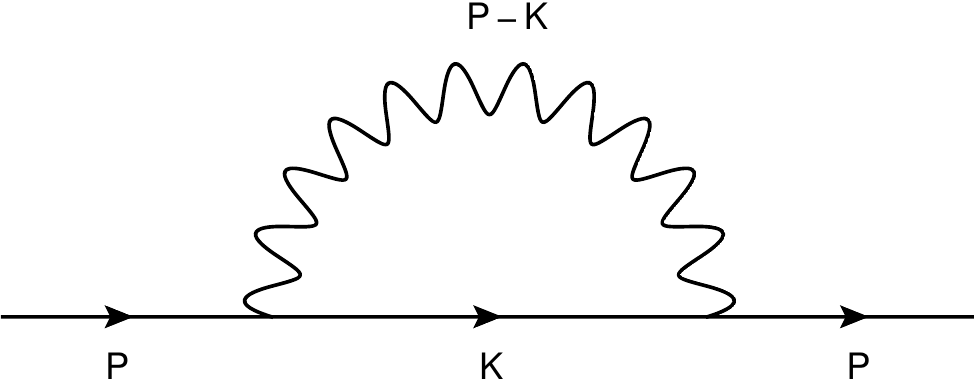}
 \caption{One loop fermion self-energy diagram.}
 \label{ele_se}
 \end{center}
 \end{figure}
 
The one loop fermion self-energy in Fig.~\ref{ele_se} can be written in Feynman gauge as 
\bea
\Sigma(P) &=& x^2 C_F T \sumintb_{\{K\}} \gamma_\mu \frac{K\!\!\!\! \slash}{K^2}\gamma^\mu \frac{1}{(P-K)^2 } \nn \\
&=& -2x^2 C_F T \sumintb_{\{K\}}  \frac{K\!\!\!\! \slash}{K^2 \  Q^2 }\, ,  \label{se0}
\eea
where $Q=(P-K)$ and $\sum\!\!\!\!\!\!\!\!\int\limits_{\{K\}}$ is a fermionic sum-integral. Also $x$ and $C_F$ are, respectively, coupling and Casimir invariant for a given group. For $U(1)$ group $x=e$ and $C_F=1$ and Eq.\eqref{se0} corresponds to electron self-energy.
For $SU(3)$ group $x=g$ and $C_F=4/3$ and Eq.\eqref{se0} corresponds to quark self-energy with internal photon line should be replaced by gluon line.

We now would like to compute the structure functions as given in the rest frame of the heat bath, respectively, in \eqref{gse7} and \eqref{gse10} as 
\begin{subequations}
 \begin{align}
{\cal A}(\omega,p) &= \frac{1}{4p^2} \left ({\rm{Tr}} \left [ \Sigma P\!\!\!\! \slash \, \right ] - \omega {\rm{Tr}} \left [\Sigma  u\!\!\!\slash \right ] \right ), \label{se1a}\\
{\cal B}(\omega,p) &= \frac{1}{4p^2} \left ( P^2 {\rm{Tr}} \left [\Sigma u\!\!\!\slash \right ] - (P\cdot u) {\rm{Tr}} \left [ \Sigma P\!\!\!\! \slash \, \right ]  \right ) .\label{se1b}
\end{align}
\end{subequations}
Now, we can write
 \begin{subequations}
 \begin{align}
 {\textrm{Tr}}\left [ \Sigma P\!\!\!\! \slash \, \right ] & 
\, = \, -8x^2C_FT  \sumintb_{\{K\}}  \frac{k_0\omega -\bm{{\vec k}\cdot {\vec p}}}{K^2Q^2} \, ,
 \label{se1c}\\ 
 {\textrm{Tr}} \left [\Sigma u\!\!\!\slash \right ] & \, =\ -8x^2C_F T \sumintb_{\{K\}} \frac{k_0}{K^2Q^2} \,  .
\label{se1d} 
\end{align}
\end{subequations}
Using \eqref{se1c} and \eqref{se1d} in \eqref{se1a}, we get
\be
{\cal A}(\omega,p) = \frac{2x^2C_F}{p^2} T  \sumintb_{\{K\}}  \frac{\bm{{\vec k}\cdot {\vec p}}}{K^2Q^2} \, . \label{se1e}
\ee
We use the following frequency sum in mixed representation 
\bea
T\sum_{k_0} \frac{1}{K^2Q^2} &=& \frac{1}{4kq} \left [ \left ( 1-n_F(k)+n_B(q)\right ) \left( \frac{1}{\omega +k+q}-\frac{1}{\omega-k-q}\right ) \right. \nn \\
&& \left. +\left(n_B(q)+n_F(k)\right ) \left( \frac{1}{\omega+k-q}-\frac{1}{\omega-k+q} \right )\right ] \, , \label{se1f}
\eea
where $n_{F\atop B}(y)= (e^{y/T}\pm 1)^{-1}$.
When loop momentum is hard, $K\sim T$ , compare to the external momentum $P$ is called hard thermal loop (HTL) approximation~\cite{BP}. 
Under this HTL approximation\footnote{The details of HTL approximation will be discussed in section~\ref{htl_app}.} one can make following simplifications as
\begin{subequations}
 \begin{align}
 q 
\, = \,\left | \bm{\vec p} - \bm{\vec k} \right | &\, = \, \sqrt{p^2+k^2-2pk\cos\theta} =\sqrt{p^2+k^2-2pk c}\approx k-pc =k-{\bm{\vec p \cdot \hat k}}\, ,
 \label{se1g}\\ 
 n_B(q) & \, = n_B(k-{\bm{\vec p \cdot \hat k}}) \approx n_B(k) - {\bm{\vec p \cdot \hat k}} \frac{d n_B(k)}{d k }\,  ,
\label{se1h}  \\
\om \pm k\pm q & \, = \om \pm k \pm k \mp \,  {\bm{\vec p \cdot \hat k}} \approx \pm 2k \, ,
\label{se1i}  \\
\omega \pm k\mp q & \, = \omega \pm k \mp k \pm \,  {\bm{\vec p \cdot \hat k}} \approx \omega \pm  {\bm{\vec p \cdot \hat k}} \ .
\label{se1j} 
\end{align}
\end{subequations}
Using \eqref{se1g} to \eqref{se1j} in \eqref {se1f}, one can write
\bea
T\sum_{k_0} \frac{1}{K^2Q^2} &=& \frac{1}{4k^2} \left [ \left ( 1-n_F(k)+n_B(k) -{\bm{\vec p \cdot \hat k}} \frac{d n_B(k)}{d k } \right ) 
\left( \frac{1}{k}\right ) \right. \nn \\
&& \left. +\left(n_B(k)+n_F(k)- {\bm{\vec p \cdot \hat k}} \frac{d n_B(k)}{d k }  \right ) \left( \frac{1}{\omega +{\bm{\vec p \cdot \hat k}}  }-\frac{1}{\omega-{\bm{\vec p \cdot \hat k}} } \right )\right ] \, , \label{se1k}
\eea
Combining \eqref{se1k} and \eqref{se1e},  we get
 \bea
 {\cal A}(\omega,p) &=&\frac{2x^2C_F}{p^2} \int \frac{d^3k}{(2\pi)^3} \frac{\bm{{\vec p}\cdot {\vec k}}}{4k^2} 
 \left [ \left ( 1-n_F(k)+n_B(k) -{\bm{\vec p \cdot \hat k}} \frac{d n_B(k)}{d k } \right ) 
\left( \frac{1}{k}\right ) \right. \nn \\
&& \left. +\left(n_B(k)+n_F(k) - {\bm{\vec p \cdot \hat k}} \frac{d n_B(k)}{d k }  \right ) 
\left( \frac{1}{\omega +{\bm{\vec p \cdot \hat k}}  }-\frac{1}{\omega-{\bm{\vec p \cdot \hat k}} } \right )\right ] \, . \label{se1l}
\eea
 We note that the term $1/k$ inside the square bracket contributes as proportional to $T$, which is sub-leading in $T$. This term can be neglected. We now 
 evaluate the second term only that contributes as $T^2$, which is leading order in $T$. We further note that the term  
 ${\bm{\vec p \cdot \hat k}} \frac{d n_B(k)}{d k } $ has soft momentum and is neglected.  With this the \eqref{se1l} reduces as 
  \bea
 {\cal A}(\omega,p) &=&\frac{2x^2C_F}{4p^2} \int \frac{k \, dk \, d\Omega}{(2\pi)^3}  
  \left(n_B(k)+n_F(k) \right ) 
\left[ \frac{{\bm{{\vec p}\cdot {\hat k}}}}{\omega +{\bm{\vec p \cdot \hat k}}  }-\frac{{\bm{{\vec p}\cdot {\hat k}}}}{\omega-{\bm{\vec p \cdot \hat k}} } \right ] \, . \label{se1m}
\eea
 Now putting $\cos\theta \rightarrow -\cos\theta$ in the first term inside the square bracket, the above equation  then becomes
  \bea
 {\cal A}(\omega,p) &=&-\frac{x^2C_F}{p^2} \int_0^\infty  k \, dk  \left(n_B(k)+n_F(k) \right ) \  \int \frac{ d\Omega}{(2\pi)^3}  
\frac{{\bm{{\vec p}\cdot {\hat k}}}}{\omega-{\bm{\vec p \cdot \hat k}} }  \, . \label{se1n}
\eea
 
 We can  perform the $k$-integration  as
  \be
 \int_0^\infty  k \, dk  \left(n_B(k)+n_F(k) \right ) =\frac{\pi^2T^2}{4}. \label{se1o}
\ee
Using \eqref{se1o} we can obtain   
 \bea
 {\cal A}(\omega,p) &=&-\frac{m^2_{\textrm{th}}}{p^2}   \int \frac{ d\Omega}{4\pi}  
\frac{{\bm{{\vec p}\cdot {\hat k}}}}{\omega-{\bm{\vec p \cdot \hat k}} } \nn \\  
&=& -\frac{m^2_{\textrm{th}}}{p^2}   \int \frac{ d\Omega}{4\pi}  
\frac{{\bm{{\vec p}\cdot {\hat k}}}}{P\cdot \hat K}  
\, , \label{se1}
\eea
 where ${\hat K}=(1, \bm{\hat  k})$ is a light like vector and the thermal mass of the fermion is given as
\bea
m_{\textrm{th}}^2 = \frac{x^2 C_F T^2}{8} \, .   \label{se3} 
\eea
For electron the thermal mass becomes $m_{\textrm{th}}^2 = \frac{e^2 T^2}{8} $ whereas for quark $m_{\textrm{th}}^2 = \frac{g^2 T^2}{6} $.

The angular integration can be performed using the following integrals
 \begin{subequations}
 \begin{align}
 \int_{-1}^{1} \frac{dy}{a-by} &= \frac{1}{b} \ln \frac{a+b}{a-b}\,  , \label{ang_int1} \\
\int_{-1}^{1} \frac{y\ dy}{a-by} &= -\frac{2}{b}+\frac{a}{b^2} \ln \frac{a+b}{a-b} . \label{ang_int2}
\end{align}
\end{subequations}

After performing the angular integration in \eqref{se1} we finally get~\cite{HAW,VVK}
\be
\mathcal{A}(\omega,p) 
\, = \,  \frac{m_{\textrm{th}}^2}{p^2} \left [ 1- \frac{\omega}{2p} \ln \left (\frac{\omega+p}{\omega-p}\right )\right ]  \, , 
 \label{se4}\\ 
\ee
Using \eqref{se1c} and \eqref{se1d} in \eqref{se1b} we write the structure function ${\cal B}(\omega,p)$ as
\bea
{\cal B}(\omega,p) &=& \frac{2x^2C_F}{p^2} \left [ p^2 T \sumintb_{\{K\}}  \frac{k_0}{K^2Q^2} - (P\cdot u) T  \sumintb_{\{K\}}  \frac{\bm{\vec p \cdot \vec k}}{K^2Q^2}\right ] \, . \label{se2a}
\eea
The frequency sum in the second term inside square bracket is already done in \eqref{se1f} . The frequency sum needed for the first term is given in mixed representation  as
\bea
T\sum_{k_0} \frac{k_0}{K^2Q^2} &=& - \frac{1}{4q} \left [ \left ( 1-n_F(k)+n_B(q)\right ) \left( \frac{1}{\omega +k+q}+\frac{1}{\omega-k-q}\right ) \right. \nn \\
&& \left. +\left(n_B(q)+n_F(k)\right ) \left( \frac{1}{\omega+k-q}+\frac{1}{\omega-k+q} \right )\right ] \, , \label{se2b}
\eea
Using HTL approximations in \eqref{se1g} to \eqref{se1j}, the frequency sum becomes
\bea
T\sum_{k_0} \frac{k_0}{K^2Q^2} &=& - \frac{1}{4k} \left [\left(n_B(k)+n_F(k) -{\bm{\vec p \cdot \hat k}} \frac{d n_B(k)}{d k } \right ) 
\left( \frac{1}{\omega+\bm{\vec p\cdot \hat k}}+\frac{1}{\omega-\bm{\vec p\cdot \hat k}} \right )\right ] \, , \label{se2c}
\eea

Substituting \eqref{se1k} and \eqref{se2c} in \eqref{se2a}, we get
\bea
{\cal B}(\omega,p)&=&\frac{2x^2C_F}{p^2} \left [-p^2 \int \frac{d^3k}{(2\pi)^3} 
 \frac{1}{4k} \left(n_B(k)+n_F(k) -{\bm{\vec p \cdot \hat k}} \frac{d n_B(k)}{d k } \right ) 
\left( \frac{1}{\omega+\bm{\vec p\cdot \hat k}}+\frac{1}{\omega-\bm{\vec p\cdot \hat k}} \right )  \right. \nn \\
&&\left. -(P\cdot u)  \int \frac{d^3k}{(2\pi)^3}  \frac{\bm{\vec p}\cdot \bm{\vec k}}{4k^2}  \left(1-n_F(k)+n_B(k) -{\bm{\vec p \cdot \hat k}} \frac{d n_B(k)}{d k } \right ) \frac{1}{k}  \right. \nn \\
&&\left. -(P\cdot u)  \int \frac{d^3k}{(2\pi)^3}  \frac{1}{4k^2} \left(n_B(k)+n_F(k) -{\bm{\vec p \cdot \hat k}} \frac{d n_B(k)}{d k } \right ) 
\left( \frac{\bm{\vec p \cdot \vec k}}{\omega+\bm{\vec p\cdot \hat k}}-\frac{\bm{\vec p \cdot \vec k}}{\omega-\bm{\vec p\cdot \hat k}} \right )  
\right ] \label{se2d}
\eea
As before we neglect the  second term inside the square bracket that gives a contribution proportional to $T$, which is sub-leading in $T$. Also 
${\bm{\vec p \cdot \hat k}} \frac{d n_B(k)}{d k } $ term is neglected as the soft momentum is associated with it. Then, the above equation can be written as
\bea
{\cal B}(\omega,p)&=&\frac{x^2C_F}{p^2} \left [ \int_0^\infty  k\ dk  \left(n_B(k)+n_F(k)\right ) \int \frac{d\Omega} {(2\pi)^3} 
\frac{(P\cdot u) (\bm{\vec p\cdot \hat k}) -p^2}{\omega-\bm{\vec p\cdot \hat k}} 
\right ] \, .\label{se2e}
\eea
After performing the $k$-integration using \eqref{se1o}, we get
 \bea
\mathcal{B}(\omega,p)& \, =\, & \frac{m_{\textrm{th}}^2}{p^2} \int\frac{d\Omega}{4\pi}\frac{(P\cdot u)(\bm{\vec p\cdot\hat{k}})-p^2}{P\cdot\hat{K}} \, , 
\label{se2} 
\eea

Now performing the angular integrations using \eqref{ang_int1} and \eqref{ang_int2}, we finally get~\cite{HAW,VVK}
\bea
\mathcal{B}(\omega,p)
&\, = \, &  \frac{m_{\textrm{th}}^2}{p} \left [ -\frac{\omega}{p} + \left ( \frac{\omega^2}{p^2} -1\right ) \frac{1}{2} \ln \left (\frac{\omega+p}{\omega-p}\right ) \right] \, .
\label{se5} 
\eea
The fermion self-energy can be written from \eqref{se0} as
\bea
\Sigma(P) &=&-2x^2C_F T \sumintb_{K}  \left [ \frac{k_0\gamma_0}{K^2Q^2} - \frac {\vec \gamma \cdot \bm{\vec k}}{K^2Q^2} \right ]  \, . \label{se6a}
\eea
Following the same procedures as those structure functions, the fermion self-energy contribution within HTL approximation can be written  as
\bea
\Sigma(P) &=& x^2C_F \int_0^\infty  k\ dk  \left(n_B(k)+n_F(k)\right ) \int \frac{d\Omega} {(2\pi)^3} \left [\frac{\gamma_0}{\omega-\bm{\vec p\cdot \hat k}} 
-  \frac {\vec \gamma \cdot \bm{\hat k}} {\omega-\bm{\vec p\cdot \hat k}} \right ] \, , \label{se6b} 
\eea
which after $k$-integration becomes
\bea
\Sigma(P) &=& 
m_{\textrm{th}}^2\int\frac{d\Omega}{4\pi}
\frac{\hat{\slashed{K}}}{P\cdot \hat{K}} .   \label{se6c}
\eea
After explicit calculations, we obtain
\bea
\Sigma(P)&=&  \frac{m_{\textrm{th}}^2}{2p}  \ln \left (\frac{\omega+p}{\omega-p}\right )  \gamma_0  +
 \frac{m_{\textrm{th}}^2}{p} \left [ 1- \frac{\omega}{2p} \ln \left (\frac{\omega+p}{\omega-p}\right ) \right ] \left({\vec \gamma} \cdot \bm{\hat p} \right )\, . 
\label{se6}
\eea
This expression can also be obtained directly by combining \eqref{gse11}, \eqref{se4} and \eqref{se5}.

\subsection{Dispersion of Fermionic Quasiparticles and Collective Excitations in HTL Approximation}
\label{disp_quasi}
\begin{figure}[h]
\begin{center}
 \vspace*{-0.2in}
\includegraphics[width=8cm,height=7cm]{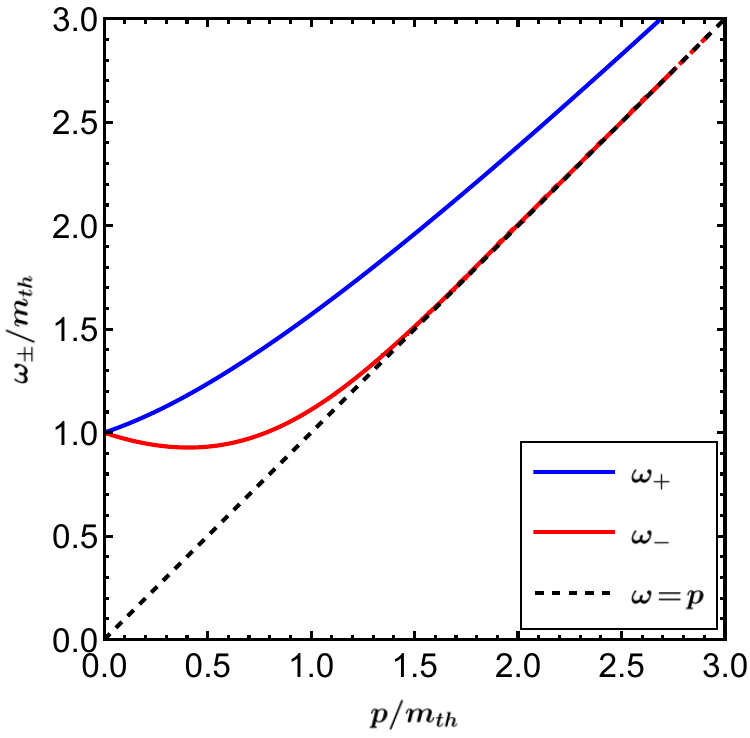}
 \caption{Plot of quasiparticles dispersion in HTL approximation.}
 \label{disp_plot}
 \end{center}
 \end{figure}
We can obtain ${\cal D}_\pm(\omega,p)$ by combining \eqref{gse19}, \eqref{se4} and \eqref{se5} as 
\bea
{\cal D}_\pm(\omega,p)&=& \left [ 1+\frac{m_{\textrm{th}}^2}{p^2} \left ( 1- \frac{\omega}{2p} \ln \left (\frac{\omega+p}{\omega-p}\right )\right ) \right ] (\omega\mp p) 
+\frac{m_{\textrm{th}}^2}{p} \left [ -\frac{\omega}{p} + \left ( \frac{\omega^2}{p^2} -1\right ) \frac{1}{2} \ln \left (\frac{\omega+p}{\omega-p}\right ) \right] \nn \\ 
&=&(\omega \mp p) -  \frac{m_{\textrm{th}}^2}{p}  \left [ \frac{1}{2}
 \left ( 1\mp \frac{\omega}{p}\right ) \ln \left (\frac{\omega+p}{\omega-p}\right )\pm 1\right ]\, . 
\label{dfq}
\eea
One can obtain the in-medium fermion propagator reads from \eqref{gse24} 
\be
S^\star(P) =  \frac{1}{2} \frac{(\gamma_0 - {\vec \gamma}\cdot \bm {\hat { p}})} {{\cal D}_+(\omega,p)}+ \frac{1}{2} \frac{(\gamma_0 
+ {\vec \gamma}\cdot \bm {\hat { p}})} {{\cal D}_- (\omega,p)} ,
 \label{dfqp}
 \ee
The charge invariance demands that ${\cal D}_\pm(-\omega, p)=-{\cal D}_\mp(\omega, p)$ which 
implies that ${\cal A}(-\omega,p) ={\cal A}(\omega,p) $ and ${\cal B}(-\omega,p) =-{\cal B}(\omega,p) $. ${\cal D}(\omega,p)$ has imaginary part for space like 
momenta $P \, (p_0^2<p^2)$, it is also useful to define the parity properties for both real and imaginary parts of ${\cal D}(\omega,p)$  as
${\mathrm {Re}} {\cal D}_+(-\omega,p) =  -{\mathrm  {Re}} {\cal D}_-(\omega,p) $ and ${\mathrm {Im}} {\cal D}_+(-\omega,p) =  {\mathrm  {Im}} {\cal D}_-(\omega,p) $.

Although the effective propagator in \eqref{dfqp} manifests the chiral symmetry, the poles of ${\cal D}_\pm(\omega,p)$ do not occur at light cone, $\omega=\pm p$. This means that the poles of the effective fermion propagator will be away from the light cone in the time like domain. This is because the extra term ${\cal B}\gamma_0$ appears in self-energy in \eqref{gse11} due to the breaking of Lorentz invariance at finite temperature~\cite{HAW}.  The zeros of ${\cal D}_\pm(\om,p)$ define dispersion property
of a quark in the thermal bath.
${\cal D}_+(\omega,p)=0$ has two ploes at $\omega=\omega_+(p)$ and $\omega=-\omega_-(p)$ whereas 
 ${\cal D}_-(\omega,p)=0$ has two ploes at $\omega=\omega_-(p)$ and $\omega=-\omega_+(p)$.  
  Only the positive energy solutions are displayed in Fig.~\ref{disp_plot}.   A mode with energy $\om_+$  represents the in-medium propagation  of a particle excitation 
  This is a Dirac spinors and eigenstate of $(\gamma_0 - {\vec \gamma}\cdot \bm{\hat { p}})$ with chirality to helicity ratio $+1$.
On the other hand  there is a new long wavelength mode known as {\em plasmino} with energy $\omega_-$  and eigenstate of 
$(\gamma_0 + {\vec \gamma}\cdot \bm{\hat { p}})$ with chirality to helicity ratio $-1$. The $\om_-$ branch has a minimum at low momentum 
and then approaches free dispersion curve at large momentum. It is important to note that the minimum leads to Van Hove singularities 
in soft dilepton rate~\cite{BPY} and meson spectral function~\cite{KMT}.

Below we present the approximate analytic solutions of $\omega_\pm(p)$ for small and large values of momentum $p$. 
For small values of momentum ($p<< m_{\textrm{th}}$), the dispersion relations are
 \begin{subequations}
 \begin{align}
\omega_+(p) &\, \approx \, m_{\textrm{th}} +\frac{1}{3}p + \frac{1}{3}\frac{p^2}{m_{\textrm{th}}}-\frac{16}{135}\frac{p^3}{m^2_{\textrm{th}}} \, ,  \label{dfq1}\\ 
\omega_-(p)
&\, \approx \, m_{\textrm{th}} -\frac{1}{3}p + \frac{1}{3}\frac{p^2}{m_{\textrm{th}}}+\frac{16}{135}\frac{p^3}{m^2_{\textrm{th}}} \, , \label{dfq2} 
\end{align}
\end{subequations}
whereas for large values of momentum ($ m_{\textrm{th}} <<p<<T$), one obtains
 \begin{subequations}
 \begin{align}
\omega_+(p) &\, \approx \, p +\frac{m^2_{\textrm{th}}}{p }-  \frac{m^4_{\textrm{th}}}{2p^3}\ln \frac{m^2_{\textrm{th}}}{2p^2}
+ \frac{m^6_{\textrm{th}}}{4p^5}\left[ \ln^2  \frac{m^2_{\textrm{th}}}{2p^2} +\ln  \frac{m^2_{\textrm{th}}}{2p^2} -1\right ] \, ,\label{dfq3}\\ 
\omega_-(p)
&\, \approx \,  p +2p\exp\left (-\frac{2p^2+m^2_{\textrm{th}}}{m^2_{\textrm{th}}}\right )  \, . \label{dfq4} 
\end{align}
\end{subequations}
We note that at large (hard) momentum the collective mode with chirality to helicity ratio $+1$ resembles the free particle in vacuum whereas the long wave length mode, 
plasmino with chirality to helicity ratio $-1$ decouples from the plasma. These are clearly evident from~\eqref{dfq3} and \eqref{dfq4}. 
At small momenta both collective modes are equally important which could be seen from~\eqref{dfq1} and \eqref{dfq2}. 
In addition to the pole contributions coming from time like domain ($\om^2>p^2$), ${\cal D}_\pm(\omega,p)$ contains a discontinuous
part corresponding to {\em Landau damping} coming from space like domain ($\om^2<p^2$) due to the presence
of logarithmic term in (\ref{dfq}). 

\subsection{Spectral Representation of Fermion Propagator}
\label{spec_prop}
From \eqref{gse24} the in-medium fermion propagator reads as
\be
S^\star(P) =  \frac{1}{2} \frac{(\gamma_0 - {\vec \gamma}\cdot \bm {\hat { p}})} {{\cal D}_+(\omega,p)}+ \frac{1}{2} \frac{(\gamma_0 + {\vec \gamma}\cdot \bm {\hat { p}})} {{\cal D}_- (\omega,p)} ,
 \label{spec0}
 \ee
where ${\cal D}_\pm(\omega,p)$ are given in \eqref{dfq} as
\bea
{\cal D}_\pm(\omega,p) &=&p(\frac{\omega}{p} \mp 1) -  \frac{m_{\textrm{th}}^2}{p}  \left [ \frac{1}{2}
 \left ( 1\mp \frac{\omega}{p}\right ) \ln \left (\frac{\frac{\omega}{p}+1}{\frac{\omega}{p}-1}\right )\pm 1\right ]\, . 
\label{dfq0a}
\eea
According to \eqref{gse19a},  ${\cal D}_\pm(\omega,p)=d_\pm(\omega,p)=\omega\pm p$, for free massless case. The corresponding free spectral function 
can be obtained following \eqref{bpy_04} as
\bea
\rho^f_\pm(\omega,k) &=& \lim_{\epsilon \rightarrow 0} \frac{1}{\pi}{\textrm{Im}}\left . \frac{1}{d_\pm(\omega,p)}\right |_{\omega\rightarrow \omega+i\epsilon} \nn \\
&=&  \lim_{\epsilon \rightarrow 0} \frac{1}{\pi}{\textrm{Im}} \frac{1}{\omega\mp p+i\epsilon} 
=\frac{\delta(\omega\mp p)}{\left | \frac{\ d(\omega\mp p)}{d\omega}\right |} = \delta(\omega\mp p) \, . \label{spec1}
\eea
As discussed in subsec\ref{disp_quasi} that ${\cal D}_\pm(\omega,p)$ has solutions at $\omega_\pm(k)$ and $-\omega_\mp(p)$ and a cut part due to space like momentum , 
$\omega^2<p^2$. In-medium spectral function corresponding to the effective fermion propagator in \eqref{spec0} will have both pole and cut contribution as
\be
\rho_\pm(\omega,p) = \rho^{\textrm{pole}}_\pm (\omega,p) + \rho^{\textrm{cut}}_\pm (\omega,p) \, . \label{spec2}
\ee
The pole part of the spectral function can be obtained following \eqref{bpy_04} as
\bea
 \rho^{\textrm{pole}}_\pm (\omega,p) &=& \lim_{\epsilon \rightarrow 0} \frac{1}{\pi}{\textrm{Im}}\left . \frac{1}{{\cal D}_\pm(\omega,p)}\right |_{\omega\rightarrow \omega+i\epsilon} 
 = \frac{\delta(\omega -\omega_\pm)}{\left | \frac{d{\cal D_\pm}}{d\omega}\right |_{\omega=\omega_\pm}} 
 +  \frac{\delta(\omega +\omega_\mp)}{\left | \frac{d{\cal D_\mp}}{d\omega}\right |_{\omega=-\omega_\mp}} \,  \nn \\
 &=& \frac{(\omega^2-p^2)}{2m^2_{\textrm{th}} }\delta(\omega -\omega_\pm) + \frac{(\omega^2-p^2)}{2m^2_{\textrm{th}} }\delta(\omega +\omega_\mp)\, . \label{spec3}
 \eea

For $\omega^2<p^2$, there is a discontinuity in $\ln\frac{\omega+p}{\omega-p}$
as $\ln{y}=\ln\left |y \right |-i\pi \ ,$ which leads to the spectral 
function, $\rho^{\textrm{cut}}_\pm(\omega,p)$, corresponding 
to the discontinuity in ${\cal D}_\pm(\omega,p)$ can be obtained from \eqref{bpy_02} as
\begin{eqnarray}
\rho^{\textrm{cut}}_\pm(\omega,p) &=& \frac{1}{2\pi i}\textrm{Disc}\frac{1}{{\cal D}_\pm(\omega,p)} 
=\frac{1}{\pi} \lim_{\epsilon \rightarrow 0} \textrm{Im}\left.\frac{1}{{\cal D}_\pm(\omega,p)}
\right |_{{\omega\rightarrow\omega+i\epsilon}\atop {\omega < p}}
\nonumber \\
&=&\!\! \frac{\frac{ m^2_{\textrm{th}}}{2p}\left(\pm\frac{\omega}{p}-1\right)
\Theta(p^2-\omega^2)}
{\left[\omega\mp 
p-\frac{m^2_{\textrm{th}}}{p}\left(\pm 1-\frac{p\mp \omega}{2p}\ln\frac{p+\omega}{p-\omega}
\right)\right]^2+\left[\pi\frac{m^2_{\textrm{th}}}{2p}\left(1\mp\frac{\omega}{ p}\right)
\right]^2} \nn \\
&=&\beta_\pm(\omega,p)\Theta(p^2-\omega^2)  . \label{spec4}
\end{eqnarray}

\subsection{Calculation of $\Pi_L$ and $\Pi_T$  from One-loop Photon Self-energy in HTL Approximation}
\label{pse1l}

\begin{figure}[htb]
\begin{center}
\includegraphics[scale=0.5]{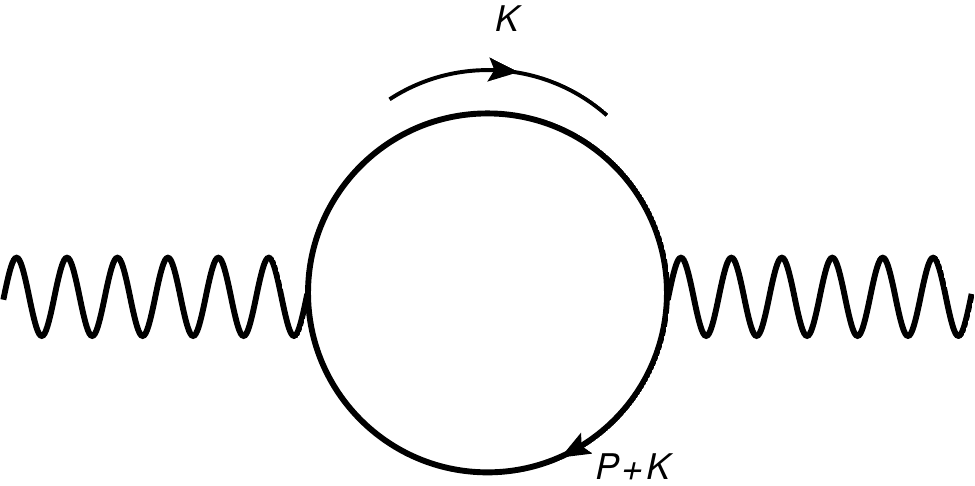}
\end{center}
\caption{One loop photon self-energy diagram.}
\label{photon_self}
\end{figure}
We have QED interaction Lagrangian from \eqref{qed23} 
\be
\mathcal{L}_{int} = -e\bar{\psi}\gamma_\mu \psi A^\mu . \label{psl1}
\ee
So using Fig.~\ref{photon_self} photon self energy can be written as
\bea
\Pi_{\mn} &=& -\int \frac{d^4K}{(2\pi)^4} \Tr \left [(-ie\gamma_\mu) \frac{i\slashed{K}}{K^2}(-ie\gamma_\nu) \frac{i(\slashed{P}+\slashed{K})}{(K+P)^2}\right] \nn \\
&=&-e^2\int \frac{d^4K}{(2\pi)^4}\Tr{\left[\frac{\gamma_\mu \slashed{K}\gamma_\nu(\slashed{P}+\slashed{K})}{K^2(P+K)^2}\right]}\nn\\
&=& -4e^2\int \frac{d^4K}{(2\pi)^4} \left[\frac{2 K_\mu K_\nu   - \eta_{\mn}K^2 -  \eta_{\mn}K\cdot P+ (K_\mu P_\nu + P_\mu K_\nu)}{K^2(P+K)^2}\right] \, . \label{pel2}
\eea
Now in HTL approximation one can neglect external soft momentum~\cite{BP,Weldon:1982aq,Fradkin,KK}, i.e.,  one or higher power of $P$. So we get,
\bea
\Pi_{\mn} &\approx& -8e^2\int \frac{d^4K}{(2\pi)^4} \left[\frac{K_\mu K_\nu}{K^2(P+K)^2}\right] + 4 e^2 \eta_{\mn}\int \frac{d^4K}{(2\pi)^4} \frac{1}{(P+K)^2}\nn\\
&\approx& -8e^2\int \frac{d^4K}{(2\pi)^4} \left[K_\mu K_\nu \Delta_F(K)\Delta_F(Q)\right] + 4e^2 \eta_{\mn}\int \frac{d^4K}{(2\pi)^4} \Delta_F(Q) \, , \label{psl3}
\eea
where $\Delta_F(K)=\frac{1}{K^2}$  and $\Delta_F(Q)=\frac{1}{(P+K)^2}=\frac{1}{Q^2}$ with $Q=P+K$. We also note that $F$ stands for fermionic part of the propagator. 

We take the time-time part of $\Pi_{\mn}$ as~\cite{Aritra,Bithika}
\bea
\Pi_{00} &=& -8e^2 T \sumintb_{\{K\}} k_0^2 \Delta_F(K) \Delta_F(Q) + 4e^2 T\sumintb_{\{ K\}} \Delta_F(Q) \nn \\
&=& -8e^2T \sumintb_{\{K\}}  (K^2+k^2) \Delta_F(K) \Delta_F(Q) + 4e^2 T\sumintb_{\{ K\}} \Delta_F(Q) \nn \\
&=& -4e^2 T\sumintb_{\{ K\}} \Delta_F(K) -8e^2T \sumintb_{\{K\}}  k^2 \Delta_F(K) \Delta_F(Q)  \, , \label{psl4}
\eea
where $\sum\!\!\!\!\!\!\!\!\int\limits_{\{K\}}$ is a fermionic sum-integral and  we have used $\Delta_F(Q)=1/(P+K)^2 \approx 1/K^2 =\Delta_F(K)$ in the first term.
 
 Also we obtain ~\cite{Aritra,Bithika}
 \bea
 \Pi_\mu^\mu &=& -8e^2T \sumintb_{\{K\}}  K^2 \Delta_F(K) \Delta_F(Q) + 4e^2\delta_\mu^\mu T\sumintb_{\{ K\}} \Delta_F(Q) \nn \\
 &=& 8e^2T \sumintb_{\{K\}}  \Delta_F(Q)  \approx 8e^2T \sumintb_{\{K\}}  \Delta_F(K) .\label{psl5} 
 \eea

We know the results of frequency sums:
 \begin{subequations}
 \begin{align}
T\sum_{k_0} \Delta_F(K) &=T \sum_{k_0} \frac{1}{K^2} = \frac{1}{2k}\left (1-2n_F(k)\right ) ,  \label{psl6} \\
T\sum_{k_0} \Delta_F(K) \Delta_F(Q) &= \frac{1}{4kq} \left [ \left(1-n_F(k)-n_F(q) \right )\left( \frac{1}{\om-k-q}-\frac{1}{\om+k+q}\right)\right. \nn \\
&\left. - \left( n_F(k)-n_F(q)\right) \left( \frac{1}{\om+k-q}-\frac{1}{\om-k+q}\right) \right ] \, , \label{psl7}
\end{align}
\end{subequations}
where $k=|\bm{\vec k}|$, $q=|\bm{\vec k +\vec p}|$ and the Fermi distribution is given as $n_F(x) =1/(\exp(\beta x)+1)$.  

Now using \eqref{psl6} in \eqref{psl5}, one gets
\be
\Pi_\mu^\mu= 8e^2 \int \frac{d^3k}{(2\pi)^3} \frac{1}{2k} \left (1-2n_F(k)\right ) \, . \label{psl8}
\ee
Neglecting vacuum part, one can have 
\bea
\Pi_\mu^\mu&=& -\frac{4e^2}{\pi^2} \int_0^\infty k \ n_F(k) \ dk  \nn \\
&=&-\frac{4e^2}{\pi^2}  \times \frac{\pi^2T^2}{12} \nn \\
&=& -\frac{e^2T^2}{3} = -m_D^2 \, , \label{psl9}
\eea
where the Debye mass in QED is given as $m_D^2 = e^2T^2/3$.

Now using \eqref{psl6} and \eqref{psl7} in \eqref{psl4}, one can write the time-time component as
\bea
\Pi_{00} &=&  -4e^2\int \frac{d^3k}{(2\pi)^3} \frac{1}{2k} \left (1-2n_F(k)\right )  -8e^2  \int \frac{d^3k}{(2\pi)^3} \frac{k^2}{4kq} \nn \\
&&\times \left [ \left(1-n_F(k)-n_F(q) \right )\left( \frac{1}{\om-k-q}-\frac{1}{\om+k+q}\right)\right. \nn \\
&& \left. - \left( n_F(k)-n_F(q)\right) \left( \frac{1}{\om+k-q}-\frac{1}{\om-k+q}\right) \right ] \, . \label{psl7a}
\eea
Under the HTL approximation~\cite{BP} one can make following simplifications as
\begin{subequations}
 \begin{align}
 q 
\, = \,\left | \bm{\vec p} + \bm{\vec k} \right | &\, = \, \sqrt{p^2+k^2+2pk\cos\theta} =\sqrt{p^2+k^2+2pk c}\approx k+pc =k+{\bm{\vec p \cdot \hat k}}\, ,
 \label{psl8a}\\ 
 n_F(q) & \, = n_F(k+{\bm{\vec p \cdot \hat k}}) \approx n_F(k) + {\bm{\vec p \cdot \hat k}} \frac{d n_F(k)}{d k }\,  ,
\label{psl9a}  \\
\om \pm k\pm q & \, = \om \pm k \pm k \pm \,  {\bm{\vec p \cdot \hat k}} \approx \pm 2k \, ,
\label{psl10}  \\
\omega \pm k\mp q & \, = \omega \pm k \mp k \mp \,  {\bm{\vec p \cdot \hat k}} \approx \omega \mp  {\bm{\vec p \cdot \hat k}} \ .
\label{psl11} 
\end{align}
\end{subequations}
Now using \eqref{psl8a} to \eqref{psl11} in \eqref{psl7}, one can write the time-time component as
\bea
\Pi_{00} &=&  -4e^2\int \frac{d^3k}{(2\pi)^3} \frac{1}{2k} \left (1-2n_F(k)\right ) -2e^2 \int \frac{d^3k}{(2\pi)^3}
\left [ \left(1-2n_F(k)-{\bm{\vec p \cdot \hat k}} \frac{d n_F(k)}{d k }\right )\left (-\frac{1}{k}\right)\right. \nn \\
&&\left. + {\bm{\vec p \cdot \hat k}} \frac{d n_F(k)}{d k } \left( \frac{1}{\omega - {\bm{\vec p \cdot \hat k}}}-
\frac{1}{\omega + {\bm{\vec p \cdot \hat k}}}\right )\right ] \nn \\
&=&  -4e^2\int \frac{d^3k}{(2\pi)^3} \frac{1}{2k} \left (1-2n_F(k)\right )  +4e^2\int \frac{d^3k}{(2\pi)^3} \frac{1}{2k} \left (1-2n_F(k)\right ) \nn\\
&& -2e^2 \int \frac{d^3k}{(2\pi)^3} \frac{{\bm{\vec p \cdot \hat k}}}{k} \frac{d n_F(k)}{d k } 
 - 2e^2 \int \frac{d^3k}{(2\pi)^3} \frac{d n_F(k)}{d k }  \left( \frac{{\bm{\vec p \cdot \hat k}} }{\omega - {\bm{\vec p \cdot \hat k}}}-
\frac{{\bm{\vec p \cdot \hat k}} }{\omega + {\bm{\vec p \cdot \hat k}}}\right ) \nn \\
&\approx&  - 2e^2 \int \frac{d^3k}{(2\pi)^3} \frac{d n_F(k)}{d k }  \left( \frac{{\bm{\vec p \cdot \hat k}} }{\omega - {\bm{\vec p \cdot \hat k}}}-
\frac{{\bm{\vec p \cdot \hat k}} }{\omega + {\bm{\vec p \cdot \hat k}}}\right ) \, , \label{psl12}
\eea
where we have neglected the third term in second line for soft momentum. Now considering $\cos\theta=-\cos\theta$ in the second term of the last line
we get
\bea
\Pi_{00}(\om,p) &=&  - 2e^2 \int_0^\infty  \frac{k^2\ dk}{2\pi^2} \frac{d n_F(k)}{d k } \int \frac{d\Omega}{4\pi} 
  \frac{{2 \bm{\vec p \cdot \hat k}} }{\omega - {\bm{\vec p \cdot \hat k}}}  \nn\\
&=&   \frac{4e^2}{T} \int_0^\infty  \frac{k^2\ n_F(k)(1-n_F(k))\ dk}{2\pi^2}    \int \frac{d\Omega}{4\pi}  \, \frac{\om-P\cdot {\hat K}}{P\cdot {\hat K}}  \nn\\
&=&  \frac{4e^2}{2\pi^2T}  \int_0^\infty  k^2\ n_F(k)(1-n_F(k))\ dk   \int_{-1}^{1} \frac{dc}{2}  \, \left( \frac{\om}{\om-pc}-1\right ) \nn \\
&=&  \frac{4e^2}{2\pi^2T}  \times  \frac{\pi^2T^3}{6} \times \left [ \frac{\om}{2p}\ln \frac{\om+p}{\om-p}-1\right ] 
=  \frac{e^2T^2}{3}   \left [ \frac{\om}{2p}\ln \frac{\om+p}{\om-p}-1\right ] \nn \\
&=&  m_D^2  \left [ \frac{\om}{2p}\ln \frac{\om+p}{\om-p}-1\right ] \, . \label{psl13}
\eea
Using \eqref{psl13} in \eqref{pi_L}  one obtains the longitudinal component of the photon self-energy~\cite{Aritra,Bithika} as
\bea
\Pi_L (\om,p)&=& - \frac{P^2}{p^2} \Pi_{00}(\om,p) \nn \\
&=&\frac{m_D^2P^2}{p^2} \left [ 1- \frac{\om}{2p}\ln \frac{\om+p}{\om-p}\right ] \, . \label{psl14}
\eea

Now using \eqref{psl9} and \eqref{psl14} in (\ref{pi_T}), one obtains the transverse component of the photon self-energy~\cite{Aritra,Bithika}  as
\bea
\Pi_T(\om,p) &=& \frac{1}{2} \left [\Pi_\mu^\mu -\Pi_L \right ] \, \nn \\
&=& \frac{1}{2} \left [-m_D^2 - \frac{m_D^2P^2}{p^2} \left ( 1- \frac{\om}{2p}\ln \frac{\om+p}{\om-p}\right ) \right ] \nn\\
&=& - \frac{m_D^2\om^2}{2p^2} \left [1+ \frac{p^2-\om^2}{2\om p} \ln \frac{\om+p}{\om-p} \right ] \, . \label{psl15}
\eea
The photon self-energies in \eqref{psl14} and \eqref{psl15}  in HTL approximation can also be derived from the Vlasov equation, 
i.e., a transport equation without collision term considering
a mean field. The only non-classical quantity, entering thereby, is Fermi-Dirac distribution for the electrons and positrons. The photon self-energy is obtained via the dielectric
constants~\cite{Elze,SM,TG} which follow from the solution of the Vlasov equation for the distribution function assuming a small derivation from equilibrium~\cite{LP}.
The coincidence of the self-energy, found in the HTL approximation with the one in Vlasov equation is caused by the equivalence of high temperature limit, 
$T\rightarrow \infty$, and the classical limit, $\hbar \rightarrow 0$. As a matter of fact Silin~\cite{Silin} had found in 1960  the results given  in \eqref{psl14} and \eqref{psl15} 
for studying the  properties of a relativistic but classical electromagnetic plasma.

We now note that in the IR limit ($\om \rightarrow 0$), one gets can also be derived from
\begin{subequations}
 \begin{align}
 \lim_{\om \rightarrow 0} \Pi_L(\om,p) & \, = \lim_{\om \rightarrow 0} -\frac{P^2}{p^2}\Pi_{00}(\om,p) =-m_D^2  \, , \label{psl16}  \\
 \lim_{\om \rightarrow 0} \Pi_T(\om,p) & \, = 0 \, . \label{psl17}
 \end{align}
\end{subequations}
 In QED the \eqref{psl16} is the Debye electric screening mass of photon acts as a IR regulator at the static electric scale ($\sim eT$). 
 On the other hand \eqref{psl17} indicates that there is no
 screening for magnetic fields in QED as the  1-loop photon transverse self-energy in leading order HTL approximation vanishes in 
 the IR limit and provides no magnetic screening mass for photon as it is a nonperturbative effect which can not be calculated 
 perturbatively~\cite{Linde1,Linde2}.
\subsection{Dispersion Relation and Collective Excitations of Photon in HTL Approximation}
\label{drtm}
Now we can find the dispersion relations of photon using the thermal propagator given in \eqref{gsp14}.
The poles of the propagator give the dispersion relations of photon in thermal medium and one needs to solve the following equations:
 \begin{subequations}
 \begin{align}
P^2+\Pi_L&=0\, \,\, \implies
\om^2-p^2+\frac{P^2m_D^2}{p^2}\left[1-\frac{\om}{2p}\ln{\frac{\om+p}{\om-p}}\right] =0\, ,
\label{Pi_L_disp}\\
P^2+\Pi_T&=0\, \, \,\implies
\om^2-p^2-\frac{m_D^2}{2}\frac{\om^2}{p^2}\left[1+\frac{p^2-\om^2}{2\om p}\ln{\frac{\om+p}{\om-p}}\right] =0 \, .
\label{Pi_T_disp}
\end{align}
\end{subequations}

\begin{figure}[htb]
\begin{center}
\includegraphics[scale=0.7]{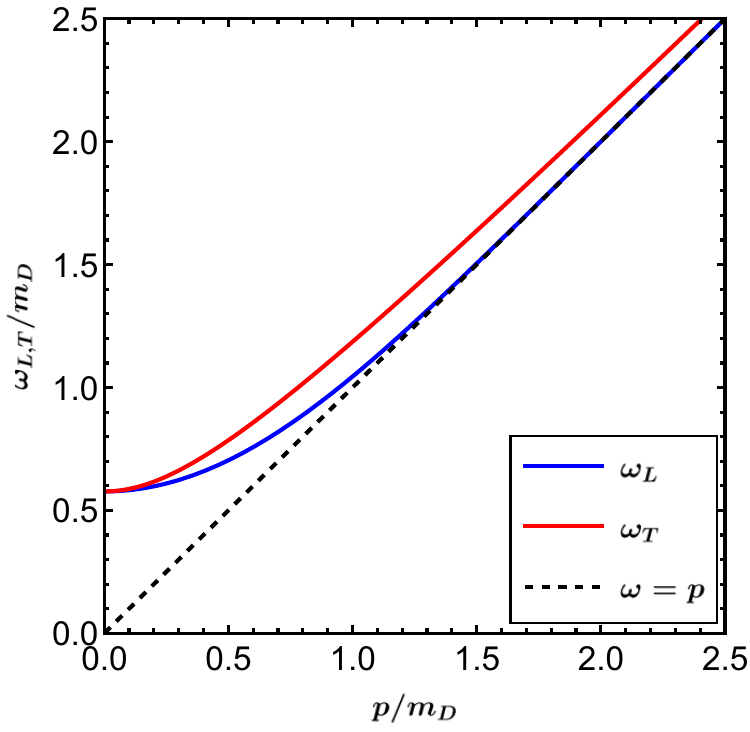}
\end{center}
\caption{Dispersion  of photon in heat bath.}
\label{disp_rel_thermal}
\end{figure}

The first condition in \eqref{Pi_L_disp}  gives the longitudinal mode of propagation with energy $\om_L$ known as plasmon and the second condition
in  \eqref{Pi_T_disp} gives the transverse mode with energy $\om_T$, which is doubly degenerate as evident from~\eqref{td14}. The plasmon mode with energy $\omega_L$ 
is a long wavelength mode that arises solely due to the 
presence of the thermal medium. The dispersion relations for photon at one loop are plotted 
in Fig~\ref{disp_rel_thermal}. Energy of photon splits in two modes due to the presence of a thermal medium. 
The longitudinal one with energy $\omega_L$ is a long wavelength mode  which 
reduces to free dispersion very fast than the transverse one with energy $\omega_T$. 
The same behaviour was also found by Silin~\cite{Silin}  for a relativistic but classical electromagnetic plasma using Vlasov equation.
We further note that at large (hard) momentum the collective mode with energy $\omega_T$ resembles the transverse photon in vacuum whereas the long wave length mode, 
the plasmon with energy $\omega_L$, decouples from the plasma. This is clearly evident from~\eqref{wl_h} and \eqref{wt_h}. 
At small momenta both collective modes are equally important which can be seen from~\eqref{wl_l} and \eqref{wt_l}.

Below we obtain approximate analytic solutions of $\om_{L,T}$ for small and large values of momentum. For small value of momentum ($p << m_D$), 
\begin{subequations}
\begin{align}
\om_L &\approx \, \frac{m_D}{\sqrt{3}} \left [1+\frac{9}{10}\frac{p^2}{m_D^2} -\frac{27}{280}\frac{p^4}{m_D^4} +\frac{9}{2000}\frac{p^6}{m_D^6}\right ], \label{wl_l} \\
\om_T&\approx \, \frac{m_D}{\sqrt{3}} \left [1+\frac{9}{5}\frac{p^2}{m_D^2} -\frac{81}{35}\frac{p^4}{m_D^4} +\frac{792}{125}\frac{p^6}{m_D^6}\right ] \, . \label{wt_l} 
\end{align}
\end{subequations}
For large value of momentum ($p>>m_D$),
\begin{subequations}
\begin{align}
\om_L &\approx \, p+ 2p\ \exp \left( -\frac{2(p^2+m_D^2)}{m_D^2}\right), \label{wl_h} \\
\om_T&\approx \, p+\frac{m_D^2}{P} + \frac{m_D^4}{32p^3} \left[3-2\ln\frac{8p^2}{m_D^2} \right] 
+\frac{m_D^6}{128p^5}\left[2\ln^2\frac{8p^2}{m_D^2}  -10\ln \frac{8p^2}{m_D^2} +7 \right]\ . \label{wt_h} 
\end{align}
\end{subequations}

In addition to the pole contributions coming from time like domain $\omega^2 > p^2$ above the light cone, there is also a discontinuous part corresponding to Landau damping 
coming from space like domain $\omega^2<p^2$ due to the presence of logarithmic terms in  \eqref{psl14} and \eqref{psl15}.

\subsection{Spectral Representation of Gauge Boson Propagator}
\label{srgbp}
 
 The effective propagator of a gauge boson in presence of thermal medium can be written from \eqref{gsp14} as
\be
{D_{\mn} = -\frac{\xi}{P^4}P_\mu P_\nu - \frac{1}{P^2+\Pi_T}A_{\mn} - \frac{1}{P^2+\Pi_L}B_{\mn}} \, , \label{srgbp1}
\ee
where
\begin{subequations}
 \begin{align}
P^2+\Pi_L&=
\om^2-p^2+\frac{(\om^2-p^2)}{p^2} \ m_D^2 \, \left[1-\frac{\om}{2p}\ln{\frac{\om+p}{\om-p}}\right] \, ,
\label{srgbp2}\\
P^2+\Pi_T&=
\om^2-p^2-\frac{m_D^2}{2}\frac{\om^2}{p^2}\left[1+\frac{p^2-\om^2}{2\om p}\ln{\frac{\om+p}{\om-p}}\right]  \, .
\label{srgbp3}
\end{align}
\end{subequations}
As discussed in subsec.~\ref{drtm} that $P^2+\Pi_{L}=0$ has solutions at $\om=\pm\om_{L}$ and $P^2+\Pi_{T}=0$ has solutions at $\om=\pm\om_{T}$.  Both also 
have a cut part due to space like momentum $\om^2<p^2$.
In-medium spectral function corresponding to the effective photon propagator in \eqref{srgbp1} will have both pole and cut contribution as
\begin{subequations}
 \begin{align}
 \rho_L(\om,p) = \rho_L^{\textrm{pole}}(\om,p) + \rho_L^{\textrm{cut}}(\om,p)\, , \label{srgbp4} \\
  \rho_T(\om,p) = \rho_T^{\textrm{pole}}(\om,p) + \rho_T^{\textrm{cut}}(\om,p)\, . \label{srgbp5}
\end{align}
\end{subequations}
The pole part of the longitudinal  spectral function can be obtained using \eqref{bpy_04} as
\bea
 \rho^{\textrm{pole}}_L (\omega,p) &=& \lim_{\epsilon \rightarrow 0} \frac{1}{\pi}{\textrm{Im}}\left . \frac{1}{P^2+\Pi_L}\right |_{\omega\rightarrow \omega+i\epsilon} \nn \\
 &=& \frac{\delta(\omega -\omega_L)}{\left | \frac{d}{d\omega}(P^2+\Pi_L)\right |_{\omega=\omega_L}} 
 +   \frac{\delta(\omega +\omega_L)}{\left | \frac{d}{d\omega}(P^2+\Pi_L)\right |_{\omega=-\omega_L}}  \,  \nn \\
 &=& \frac{\om}{p^2+m_D^2-\om^2}\left[ \delta(\om-\om_L)+\delta(\om+\om_L)\right ] . \label{srgbp6}
 \eea
Similarly, the pole part corresponding to the transverse spectral function can be obtained using \eqref{bpy_04} as
\bea
 \rho^{\textrm{pole}}_T (\omega,p) &=& \lim_{\epsilon \rightarrow 0} \frac{1}{\pi}{\textrm{Im}}\left . \frac{1}{P^2+\Pi_T}\right |_{\omega\rightarrow \omega+i\epsilon} \nn \\
 &=& \frac{\delta(\omega -\omega_T)}{\left | \frac{d}{d\omega}(P^2+\Pi_T)\right |_{\omega=\omega_T}} 
 +   \frac{\delta(\omega +\omega_T)}{\left | \frac{d}{d\omega}(P^2+\Pi_T)\right |_{\omega=-\omega_T}}  \,  \nn \\
 &=& \frac{\om(\om^2-p^2)}{m_D^2\om^2+p^2(\om^2-p^2)}\left[ \delta(\om-\om_T)+\delta(\om+\om_T)\right ] . \label{srgbp7}
 \eea
 For $\omega^2<p^2$, there is a discontinuity in $\ln\frac{\omega+p}{\omega-p}$
as $\ln{y}=\ln\left |y \right |-i\pi \ ,$ which leads to the spectral 
function, $\rho^{\textrm{cut}}_{L,T}(\omega,p)$, corresponding 
to the discontinuity in $P^2+\Pi_{L,T}$ . The cut contribution to the longitudinal spectral function can be obtained using \eqref{bpy_02}  as
\begin{eqnarray}
\rho^{\textrm{cut}}_L(\omega,p) &=& \frac{1}{2\pi i}\textrm{Disc}\frac{1}{P^2+\Pi_L}  \nonumber \\
&=&\frac{1}{\pi} \lim_{\epsilon \rightarrow 0} \textrm{Im}\left.\frac{1}{{P^2+\Pi_L}}
\right |_{{\omega\rightarrow\omega+i\epsilon}\atop {\omega < p}}
\nonumber \\
&=&\!\! \frac{(1-x^2)x\, m_D^2\, \Theta(1-x^2)/2} {\left [p^2(x^2-1)+(x^2-1)m_D^2-(x^2-1)x \, \frac{m_D^2}{2}\ln\left |\frac{x+1}{x-1} \right |\right ]^2
+\left[\pi (x^2-1)x \, \frac{m_D^2}{2} \right]^2} \nn \\
&=&\beta_L(x)\Theta(1-x^2)  \, , \label{srgbp8}
\end{eqnarray}
where $x=\om/p$.
Similarly, the cut part of the transverse spectral function can be obtained using \eqref{bpy_02} as
\begin{eqnarray}
\rho^{\textrm{cut}}_T(\omega,p) &=& \frac{1}{2\pi i}\textrm{Disc}\frac{1}{P^2+\Pi_T}  \nonumber \\
&=&\frac{1}{\pi} \lim_{\epsilon \rightarrow 0} \textrm{Im}\left.\frac{1}{{P^2+\Pi_T}}
\right |_{{\omega\rightarrow\omega+i\epsilon}\atop {\omega < p}}
\nonumber \\
&=&\!\! \frac{(x^2-1)x\, m_D^2\, \Theta(1-x^2)/4} {\left [p^2(x^2-1)-\frac{m_D^2}{2}\left(x^2-\frac{(x^2-1)x}{2} \,\ln\left |\frac{x+1}{x-1} \right |\right )\right ]^2
+\left[\pi (x^2-1)x \, \frac{m_D^2}{4} \right]^2} \nn \\
&=&\beta_T(x)\Theta(1-x^2)  \, . \label{srgbp9}
\end{eqnarray}

\section{Quantum Chromodynamics (QCD)}
\label{qcd}
 QCD is the theory of strong interaction which is a non-abelian $SU(3)$ gauge theory. It describes the quarks and gluons in the similar  
way as QED does for electrons and photons. The major difference between the two theories is that QCD  contains three colour charges
in fundamental representation. So, a  quark can be represented by a vector with  three colour states. The gauge field known as gluon mediates the interaction of
colour charges. The special unitary group $SU(3)$ has $3^2-1=8$ generators, the number of charge mediating particles, corresponding to eight gluons in QCD. 
\subsection{QCD Lagrangian}
\label{qcd_lag}
The QCD Lagrangian~\cite{pbpal,Peskin} is written as
\be
{\cal L}_{\textrm{QCD}}={\bar \psi}_{ab}(i\slashed{\partial}-m_q) \psi_{ab}-g({\bar \psi}_{ab} \lambda_i\psi_{ab})\slashed{A}^i
-\frac{1}{4} G_{\mn}^iG^{\mn}_i 
+{\cal L}_{\textrm{gf}}+{\cal L}_{\textrm{gh}} 
\, .\label{qcd1}
\ee
\begin{figure}
\begin{center}
\includegraphics[scale=0.9]{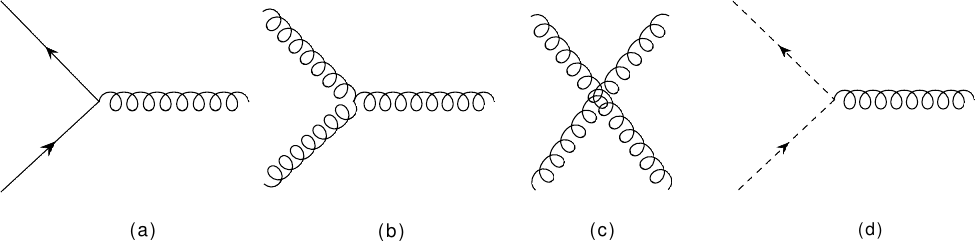}
\end{center}
\caption{The QCD vertices. (a) quark-gluon interaction, (b) three gluon interaction, (c) four gluon interaction and (d) ghost-gluon interaction.}
\label{qcd_vertices}
\end{figure}
We note that $\psi_{ab}$ is quark spinor with colour indices $a= (\textrm {r, \, g,\, b})$  and $b$ indicates the flavour index. The Lagrangian in \eqref{qcd1} is quite similar
 to QED Lagrangian in \eqref{qed23}, with the differences that   the electromagnetic field strength tensor $F^{\mn}$, is replaced by the gluonic field strength 
 tensor $G^i_{\mn}$ and an another set of indices has crept in. The first term represents the non-interacting quarks with current mass $m_q$. The second term represents quark-gluon interaction with QCD coupling $g$ and shown in  Fig.~\ref{qcd_vertices}(a). The Gell-Mann matrices $\lambda_i$ were not there in QED.  $\lambda_i$ matrices change the colour of the interacting particles. These matrices are traceless and obey the 
commutation relation $[\lambda_i,\, \lambda_j]=i\varepsilon^{ijk}\lambda_k$ with normalisation relation ${\textrm{Tr}}\lambda_i\lambda_j=2\delta_{ij}$, $\varepsilon^{ijk}$
 is the structure constant of the group which is number. It is also totally antisymmetric and vanishes  if two of the indices become same. Now, $G^i_{\mn}$ indicates the gluon 
 fields which is invariant under gauge transformation and can be defined as
\be
G_i^{\mn}=F_i^{\mn} +g\varepsilon_{ijk} A_j^\mu A_k^\nu \, , \label{qcd4}
\ee
where first term is the QED field tensor and the second term represents the self interaction of gluons. In contrast to QED, the mediator gluons  
have colour charge which enable them to interact among themselves. Now the gluonic part of the Lagrangian can be written as
\be
{\cal L}_G =-\frac{1}{4}G_{i\mn}{G^{\mn}_i}=-\frac{1}{4}F_{i\mn}{F^{\mn}_i}+g\varepsilon_{ijk}A_{i\mu}A_{j\nu}\partial^\mu A_k^\nu 
-\frac{1}{4}g^2\varepsilon_{ijk}\varepsilon_{ilm}A^\mu_j A^\nu_k A_{l\mu} A_{m\nu} \, . \label{qcd5}
\ee
The first term in \eqref{qcd5} describes  the Lagrangian for eight non-interacting, massless spin 1 gluon fields. The second and third term, respectively, in \eqref{qcd5}  
describe the self  interactions of gluon fields. This produces three- and four-point vertices in perturbation theory, displayed   in Fig.~\ref{qcd_vertices}(b) and 
 Fig.~\ref{qcd_vertices}(c), respectively. 
The last two features have no analogue in QED, as photons do not self interact.

Finally, the gauge fixing and ghost terms~\cite{Peskin} are, respectively, given as
\begin{subequations}
 \begin{align}
{\cal L}_{\textrm{gf}}&=-\frac{1}{2\xi} (\partial_\mu A^\mu_a)^2 \, , \label{qcd2} \\
{\cal L}_{\textrm{gh}}&=-{\bar C}_i\partial^2C_i - g \varepsilon_{ijk}{\bar C}_i \partial_\mu(A^\mu_j C_k) \, , \label{qcd3}
\end{align}
\end{subequations}
where $C$ is the ghost field which are Grassmann variable. The gauge fixing  Lagrangian in \eqref{qcd2} is required to eliminate the unphysical degrees of freedom present
 in the system. The ghost Lagrangian in covariant gauge is given in \eqref{qcd3} which depends on the choice of the gauge fixing term. 
 The first term in \eqref{qcd3} represents free ghost fields whereas the second term indicates interaction between ghost and gluon as shown in Fig.~\ref{qcd_vertices}(d)

We note that the previous discussion on QED in sec.~\ref{pse1l} in HTL approximation would provide a good starting point for QCD, since these two theories follow
the similar formalism. We know that QCD is a non-Abelian $SU(3)$ gauge theory whereas QED is a $U(1)$ gauge
 theory. Thus, the generalization of  QED results to QCD would  mostly involve group-theoretical factors as we will see below.

\subsection{One-loop Gluon Self-energy in HTL Approximation}
\label{gse_qcd}

\begin{figure}[htb]
\begin{center}
\includegraphics[scale=0.6]{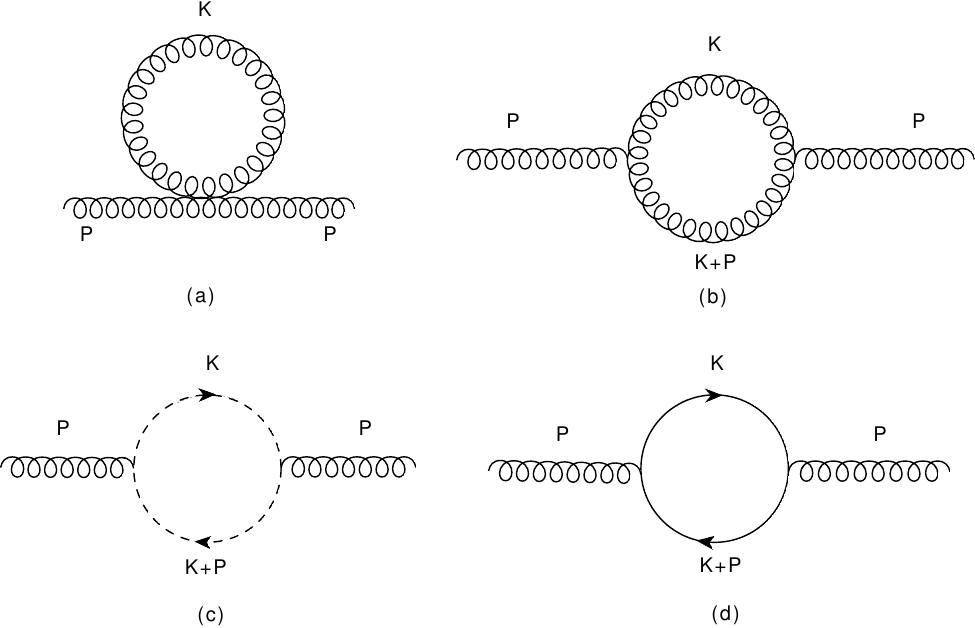}
\end{center}
\caption{One-loop gluon self-energy diagrams.}
\label{gluon_self}
\end{figure}

We calculate the one-loop gluon self-energy~\cite{Le_Bellac,Weldon:1982aq,VVK,KK} 
and the relevant diagrams are given in Fig.~\ref{gluon_self}. The contribution of the first diagram (tadpole) in HTL approximation
can be written as
\be
\Pi_{\mn}^{(a)}(P)=-3C_Ag^2\delta_{\mn} \int \frac{d^4K}{(2\pi)^4} \Delta_B(K) \, , \label{qcd6}
\ee
where $g$ is the strong or QCD coupling constant, $C_A\, (=3)$ is the group factor and $\Delta_B$ is the bosonic part of the propagator.

The contribution of the second diagram (gluon loop) in HTL approximation is obtained as
\be
\Pi_{\mn}^{(b)}(P)= g^2 C_A \int \frac{d^4K}{(2\pi)^4} 5K_\mu K_\nu \Delta_B(K) \Delta_B(K+P) +g^2 C_A \delta_{\mn}\int  \frac{d^4K}{(2\pi)^4}  \Delta_B(K)  \, . \label{qcd7}
\ee
The contribution of the third diagram (ghost loop) can be obtained as
\be
\Pi_{\mn}^{(c)}(P)= -g^2 C_A \int  \frac{d^4K}{(2\pi)^4} K_\mu K_\nu \Delta_B(K) \Delta_B(K+P)  \, . \label{qcd8}
\ee 
Combining these three diagrams, one can write
\bea
\Pi_{\mn}^{(a+b+c)}(P)&=& 4g^2C_A  \int  \frac{d^4K}{(2\pi)^4} K_\mu K_\nu \Delta_B(K) \Delta_B(K+P) 
-2g^2C_A \delta_{\mn}\int  \frac{d^4K}{(2\pi)^4}  \Delta_B(K) \nn \\
&\approx & g^2C_A  \int  \frac{d^4K}{(2\pi)^4} \left [4 K_\mu K_\nu -2K^2\delta_{\mn} \right]\Delta_B(K) \Delta_B(K+P) \, , \label{qcd9}
\eea
where we have used in the last step as $\Delta_B(K+P)=1/(K+P)^2 \approx 1/K^2$.

Now the contribution of the fourth diagram (quark loop) can be obtained similar to QED in \eqref{psl3} as
\be
\Pi_{\mn}^{(d)}(P)= -g^2\frac{N_f}{2}  \int  \frac{d^4K}{(2\pi)^4} \left [8 K_\mu K_\nu -4K^2\delta_{\mn} \right]\Delta_F(K) \Delta_F(K+P) \, , \label{qcd10}
\ee
where $N_f$ is the number of quark flavour. 
We note that one can go from fermionic loop to bosonic loop by the following transformations:
\begin{subequations}
 \begin{align}
\int  \frac{d^4K}{(2\pi)^4} \Delta_F(K) &= -\frac{1}{2} \int \frac{d^4K}{(2\pi)^4} \Delta_B(K) \, , \label{qcd11} \\
 \int  \frac{d^4K}{(2\pi)^4} \Delta_F(K) \Delta_F(K+P) &= -\frac{1}{2} \int \frac{d^4K}{(2\pi)^4} \Delta_B(K) \Delta_B(K+P)\, . \label{qcd12}
\end{align}
\end{subequations}
Using \eqref{qcd12}, one can write \eqref{qcd10} as
\be
\Pi_{\mn}^{(d)}(P)= g^2\frac{N_f}{2}  \int  \frac{d^4K}{(2\pi)^4} \left [4 K_\mu K_\nu -2K^2\delta_{\mn} \right]\Delta_B(K) \Delta_B(K+P) \, , \label{qcd13}
\ee
Now the contributions of all four diagrams in gluon self energy can be written by combining \eqref{qcd9} and \eqref{qcd13} as
\be
\Pi_{\mn}^{(a+b+c+d)}(P)= g^2\left( C_A+\frac{N_f}{2}\right)  \int  \frac{d^4K}{(2\pi)^4} \left [4 K_\mu K_\nu -2K^2\delta_{\mn} \right]\Delta_B(K) \Delta_B(K+P) \, . \label{qcd13a}
\ee
The photon self-energy can be written from \eqref{psl3} as
\be
\Pi_{\mn}(P)= -e^2  \int  \frac{d^4K}{(2\pi)^4} \left [8 K_\mu K_\nu -4K^2\delta_{\mn} \right]\Delta_F(K) \Delta_F(K+P) \, , \label{qcd14}
\ee
Using \eqref{qcd12}, one gets
\be
\Pi_{\mn}(P)= e^2  \int  \frac{d^4K}{(2\pi)^4} \left [4 K_\mu K_\nu -2K^2\delta_{\mn} \right]\Delta_B(K) \Delta_B(K+P) \, , \label{qcd15}
\ee

Now the only difference between \eqref{qcd13a} and \eqref{qcd15} is the overall factor. From QED to QCD or photon to gluon, one changes
\be
e^2 \Rightarrow g^2\left (C_A+\frac{N_f}{2}\right ) \, . \label{qcd16}
\ee
With this one can transform QED Debye mass in \eqref{psl9} to QCD Debye mass as
\be
\frac{e^2T^2}{3} \Rightarrow \frac{ g^2 T^2}{3} \left (C_A+\frac{N_f}{2}\right )=m_D^2 \, : \, \,\, {\mbox{ \, \, QCD Debye mass}} . \label{qcd17}
\ee
Therefore, we note that the HTL self-energies of photon given in  \eqref{psl13},  \eqref{psl14} and  \eqref{psl15} will be same for gluons with the replacement of
Debye mass $m_D^2$ for QCD as given in \eqref{qcd17}. We further note that in the IR limit the gluon transverse self-energy in leading order 
in HTL vanishes implying  there is no magnetic screening in QCD and provides no magnetic screening mass for gluons. This is a nonperturbative 
effect which can not be calculated perturbatively~\cite{Linde1,Linde2}. However,  the longitudinal component of gluon 
self-energy in the IR limit becomes $m_D^2$ which acts as a IR regulator in the static electric scale.

We also note that the collective excitations of gluons is same as photons which are displayed in Fig.~\ref{disp_rel_thermal}.

\subsection{One-loop Quark Self-energy in HTL Approximation}
\label{qse_qcd}
The evaluation of quark self-energy at one-loop order is even simpler. This is because there is only one diagram which is similar to Fig.~\ref{ele_se} where the internal photon 
line is to be replaced by gluon line. The one loop quark self-energy can be written as 
\bea
\Sigma(P)\delta_{ij} &=&  T \sumintb_{\{K\}}\,\,\, (-ig \gamma_\mu (\lambda^a)_{ik}) \frac{i K\!\!\!\! \slash}{K^2}(-ig\gamma^\mu (\lambda^a)_{kj}) \frac{i}{(P-K)^2 } \nn \\
&=& g^2 C_F \delta_{ij}  T \sumintb_{\{K\}}  \gamma_\mu  \frac{K\!\!\!\! \slash}{K^2} \gamma^\mu  \frac{1}{(P-K)^2 } \, ,  \label{qcd18}
\eea
where we have used the identity $(\lambda^a \lambda^a)_{ij} =\frac{Nc^2-1}{2N_c}\delta_{ij} = C_F\delta_{ij}$ with $N_c=3$.
After performing the frequency sum and $k$-integration as done  in subsec.~\ref{ssf},  the quark self-energy becomes
\bea
\Sigma(P) &=& 
\frac{g^2T^2}{8}C_F \int\frac{d\Omega}{4\pi}
\frac{\hat{\slashed{K}}}{P\cdot \hat{K}} .   \label{qcd19}
\eea
The electron self-energy is obtained in \eqref{se6c}
\bea
\Sigma(P) &=& 
\frac{e^2T^2}{8} \int\frac{d\Omega}{4\pi}
\frac{\hat{\slashed{K}}}{P\cdot \hat{K}} .   \label{qcd20}
\eea
The only difference between \eqref{qcd20} and \eqref{qcd19} is overall group factor as $C_F=1$ for $U(1)$ gauge theory whereas $C_F=4/3$ for $SU(3)$ gauge theory. 
So one obtains QCD results for quark by generalizing the QED results of electron by replacing
\be
e^2 \Rightarrow g^2  C_F\, . \label{qcd21}
\ee
With this one can transform electron thermal  mass in \eqref{se3}  to quark thermal  mass as
\be
\frac{e^2T^2}{8} \Rightarrow \frac{ g^2 T^2}{8}  C_F=  \frac{ g^2 T^2}{8}\times \frac{4}{3} = \frac{ g^2 T^2}{6} =m_{\textrm{th}}^2 : \,\, {\mbox{ \, \, quark thermal mass in QCD}}. \label{qcd22}
\ee
Now the expression for electron self-energy given in \eqref{se6c} and \eqref{se6} will be same for quark self-energy with the replacement of electron thermal mass ($e^2T^2/8$) 
by quark thermal mass ($g^2T^2/6$) as given in \eqref{qcd22}. Also the structure constants appearing in the general structure of fermion self-energy and propagator need to be
replaced by quark thermal mass to study the collective excitations of quark in thermal medium, which are same as electrons displayed in Fig.~\ref{disp_plot}.

Therefore, we  learn about the collective excitations in a QCD plasma from the acquired knowledge of QED plasma excitations by  replacing the QED Debye mass and electron thermal mass by  QCD Debye mass in \eqref{qcd17} and quark thermal mass in \eqref{qcd22}, respectively, in photon self-energy, effective photon propagator, electron self-energy and effective electron propagator.

\section{Subtleties in Finite Temperature Field Theory}
\label{SFTFT}
Let us start by introducing the parametric scales appearing in finite temperature field theory. The periodicity or anti-periodicity over Euclidean time introduces a scale present
in the non-interacting theory where momentum,  $p\sim 2\pi T$, known as hard scale and the bosonic zero modes  do not acquire any scale in the non-interacting theory.  
As we have seen in previous sections that interaction introduces softer scales corresponding to collective excitations in the thermal medium. In particular, scalar fields and the electric component of gauge fields are screened in Debye scale where momentum, $p\sim gT$.  On the other hand, the magnetic component of gauge field is screened only 
non-perturbatively~\cite{Linde1,Linde2} at the scale where momentum, $p\sim g^2T$, known as the ultra-soft scale or non-perturbative magnetic scale. 

There is an expansion parameter related to bosonic fluctuations with momentum (or mass) scale $p$ of the form
\be
\epsilon_{\textrm b} \sim \frac{1}{\pi}g^2(2\pi T) n_B(p) = \frac{g^2(2\pi T)}{\pi (e^{p/T}-1)}\,\, \stackrel{p<T}{\sim} \,\  \frac{g^2(2\pi T)T}{\pi p} \, . \label{sc1}
\ee
Thus, for the hard scale: $p\sim 2\pi T$, the expansion parameter becomes a series in $\sim g^2/\pi^2$, the even power of $g$ like $T=0$ case.
For soft (electric) scale, $p\sim gT$, the series becomes $\sim g/ \pi$. For ultra-soft (magnetic) scale $\sim g^2T$, there is no perturbative series at all~\cite{Linde1,Linde2} and
it has to be determined non-perturbatively.
Therefore, at very high temperature compared to any intrinsic mass scale  of a given theory and the coupling $g$ is less than unity, 
there appears a hierarchy of momentum (mass) scales in the system and there are three distinct scales: 
hard, soft (electric) and ultra-soft (magnetic). 

In naive perturbation theory, both static and dynamic quantities can be computed by expanding in coupling constant around the free theory. This works in hard scale
regime that uses free propagators and vertices,  and the contribution appears in even power of coupling ($g^{2n}$) as discussed above.
However, the naive application of perturbation theory would in most cases result in infrared\footnote{singularities from both electric and magnetic sectors.} and/or collinear singularities, and some cases gauge dependent results. 
It is to be noted that the infrared problems are associated with bosonic excitations but not with fermionic excitations as the fermionic expansion parameter remains finite 
for $p<T$. These in turn signal sensitivity to soft  region ($p\sim gT$)  of the phase space, where naive perturbation theory breaks down. This breakdown corresponds to 
the emergence of collective effects, arising from the dynamics of the thermal medium as discussed in previous sections. In the soft (electric) scale $\sim gT$, for which perturbation theory in principle works, does not exist in non-interacting theory but needs to be generated. This means that the perturbation theory needs to be resummed or re-organised.
This is done through the effective field theory like the hard thermal loop (HTL) resummation~ \cite{BP,BP2,BP_90,TW,FT,Barton} techniques 
and thereby HTL perturbation theory (HTLpt)~\cite{andersen1,andersen2,3loopglue1,ASS,ALSS1,HBAMSS,HAMSS}.

\section{Hard Thermal Loop (HTL) Resummation and HTL Perturbation Theory}
\label{htl_app}

\subsection{HTL Resummation}
\label{htl_resum}

As discussed in sec.~\ref{SFTFT} that the naive perturbation theory suffers  from infrared and/or collinear singularities and gauge dependence results of some quantities.
This is because certain classes of diagrams were not considered in naive perturbation theory which are higher order in the loop expansion that contribute to the 
same order in the coupling constant as the one loop diagram~\cite{BP}. These diagrams can be identified through the scale separation  as discussed below.

Considering the loop momenta to be  hard ($\sim 2\pi T$), the amplitude for higher-order loop diagrams~\cite{BP} can be written through power counting as
\be
{\bm \sim} \frac{\bm {g^2T^2}}{\bm P^2} \times { \mbox{\bf Tree level amplitude}} \, , \label{eff}
\ee
where $P$ is the external momentum. If the external momentum is hard, $P\sim 2\pi T$, then the amplitude is suppressed by $g^2$ of its tree level amplitude. If the
external momentum is soft, $P\sim gT$ then the amplitude becomes equivalent to tree level amplitude. This indicates that diagrams of higher order in loop 
expansion contribute to same order in coupling as the one-loop by distinguishing the hard ($\sim 2\pi T$ ) arising from loop momenta and soft scale ($\sim gT$) from 
external momenta. Therefore, one needs to take into account those diagrams if the external momentum is sensitive to the soft(electric) scale. The effective theory built around hard thermal loops resums those diagrams. This is as illustrated below:

\begin{enumerate}
\item
One can resum those HTL diagrams in geometrical series through  the effective propagators and vertices in one loop as done in subsections~\ref{fgp} and ~\ref{gbp}. 
They are related by the Ward-Takahashi identity in QED and by the Slanov-Taylor identity in QCD. 

\item
 At the same time medium effects: {\it viz.}, the electric screening mass, the thermal mass, the collective behaviour of quasiparticles and the 
 Landau damping, are taken into account due to resummations.
The effective $N$-point functions can be used in perturbation theory leading to an effective perturbation theory know as HTL perturbation theory (HTLpt) which will
be discussed in subsec~\ref{htlpt}. This effective perturbation theory leads to gauge independent results and also complete in certain order of the coupling.
 \item
In scalar field theory the infrared problems are cured due to appropriate resummations which take into account the presence of the electric screening (Debye) mass of order $gT$. 
In gauge theories like QED and QCD, the IR singularities are improved due to electric scale. But there exists also another sort of infrared problems associated with static magnetic fields. Static magnetic fields are not screened at leading order in HTL because the 1-loop transverse photon/gluon self-energy vanishes in all gauges in the infrared limit. 
Up to order $g^5$, the quantities can be calculated using HTLpt that takes into account the screening of the chromoelectric scale but breaks down at $g^6$ 
order due to the absence of magnetic screening~\cite{Linde1,Linde2}.
\end{enumerate}

Now we write down the HTL improved Lagrangian in non-Abelian gauge theory (QCD)~\cite{BP2,BP_90,TW,FT}  as
 \be
 {\cal L}_{\tiny{\textrm{HTL}}}=i m_\textrm{th}^2\bar\psi\gamma^\mu\left\langle\frac{y_\mu}{y\cdot\! D}\right\rangle_y\psi-\frac{1}{2}
 m_D^2 {\rm Tr}\left ( G_{\mu\alpha}\left\langle\frac{y^\alpha y_\beta}{(y\cdot\! D)^2}\right\rangle_y G^{\mu\beta}\right )\, ,
\label{htl1}
 \ee
where $G$ is the gluon field strength,  $D$ is the covariant derivative, $y^\mu = (1, {\bm{\hat y}})$ is a light-like four-vector with ${\bm{\hat y}}$= three-dimensional unit vector, and
angular braces  represent the average over the directions specified by ${\bm{\hat y}}$.   The overall trace in the second term in \eqref{htl1} is for group indices.
The two parameters $m_D$ and $m_\textrm{th}$ are,
respectively, the Debye screening mass and the thermal quark mass which take into account  the screening effects.  
The Lagrangian is non-local and gauge symmetric, which forces the presence of the covariant derivative in the denominator of \eqref{htl1}, which makes it also non-linear.
When expanded in powers of gauge field, \eqref{htl1} 
generates an infinite series of non-local self-energy and vertex corrections (viz., HTL $N$-point functions). These $N$-point functions are related by Slanov-Taylor identity.
Note that since diagrams with external ghost legs do not produce hard thermal loops, the ordinary ghost-gluon vertex remains the same. For details we refer to the review article in
Ref.~\cite{Mike}.

\subsection{HTL perturbation Theory (HTLpt)}
\label{htlpt}

QCD Lagrangian density in Minkowski space can be written from \eqref{qcd1} as
\be
{\cal L}_{\rm QCD}={\bar \psi}_{ab} i\slashed{\partial} \psi_{ab}-g({\bar \psi}_{ab} \lambda_i\psi_{ab})\slashed{A}^i
-\frac{1}{4} G_{\mn}^iG^{\mn}_i 
+{\cal L}_{\textrm{gf}}+{\cal L}_{\textrm{gh}} 
+\Delta{\cal L}_{\rm QCD} \, ,
\label{qcd_lag}
\ee
where the counterterm $\Delta{\cal L}_{\rm QCD}$ is necessary to cancel the ultraviolet (UV) divergences in 
perturbative calculations.
HTLpt is a reorganization of thermal QCD perturbation theory. 
The HTLpt Lagrangian density~\cite{andersen1,andersen2,3loopglue1,ASS,ALSS1,HBAMSS,HAMSS} can be written as
 \be
 {\cal L}_{\tiny \rm {HTLpt}}=\left.({\cal L}_{\rm {QCD}}+(1-\delta) {\cal L}_{\rm HTL})\right|_{g\rightarrow\sqrt{\delta}g}+\Delta{\cal L}_{\rm HTL}
\label{htl2} 
\ee
 where $\Delta{\cal L}_{\rm HTL}$ is additional counterterm needed to cancel the UV divergences generated in HTLpt. The HTL improved Lagrangian,
 ${\cal L}_{\rm HTL}$ is given in \eqref{htl1}. HTLpt is defined by considering $\delta$ as a formal expansion parameter. By adding the HTL
improvement term in (\ref{htl2}) to the QCD Lagrangian in (\ref{qcd_lag}), HTLpt consistently shifts the
perturbative expansion from  an ideal gas of massless particles, to a gas of massive quasiparticles
which are the more apt physical degrees of freedom at high temperature and chemical potential.
It is worth here to mention that the HTLpt Lagrangian (\ref{htl2})
becomes  the QCD Lagrangian in (\ref{qcd_lag}) if one puts $\delta=1$.

Physical quantities are computed in HTLpt through expansion in powers of $\delta$, terminating at some specified order,
and then putting $\delta = 1$. As mentioned before, this signifies a rearrangement of the
perturbation series in which the screening effects through $m_D^2$ and $m_\textrm{th}^2$ terms in (\ref{htl2}) 
have been considered to all orders but then systematically subtracted out at higher orders in perturbation theory by
the $\delta m_D^2$ and $\delta m_\textrm{th}^2$ terms in (\ref{htl2}).  
One usually expands to orders $\delta^0$, $\delta^1$, $\delta^2$, respectively, for computing 
leading order (LO), next-to-leading order (NLO), and next-to-next-leading order (NNLO) results.
Note that HTLpt is gauge invariant order-by-order in the $\delta$-expansion and, consequently, 
the results obtained would be gauge independent.

If the $\delta$-expansion  could be computed to all orders the results would not depend
on $m_D$ and $m_\textrm{th}$ when one puts $\delta=1$. However, any termination of the $\delta$-expansion   generates
 $m_D$ and $m_\textrm{th}$ dependent results. Therefore, a prescription is needed to determine $m_D$ and
$m_\textrm{th}$ as a function of $T$, $\mu$ and $\alpha_s$. There are several prescriptions and some of them
had been discussed in~\cite{andersen3} at zero chemical
potential. The HTL perturbation expansion produces UV divergences. In QCD perturbation theory,
renormalizability restricts the UV divergences in such a way that they 
can be eliminated by the counterterm Lagrangian $\Delta {\cal L}_{\rm QCD}$. Usually  the
renormalization of HTLpt can be considered by adding a counterterm Lagrangian
$\Delta{\cal L}_{\rm HTL}$  in (\ref{htl2}). However, there is no such proof yet that the HTLpt
 is renormalizable, so the general structure of the UV divergences remain
unknown. The most optimistic scenario is that HTLpt would be renormalizable, 
such that the UV divergences in the physical observables can all be eliminated using proper counterterms.

 The HTLpt has been used to study the various physical quantities relevant for understanding the properties of QGP, viz., the thermodynamic 
properties
\cite{andersen1,ABM,HM,CMT,CMT1,CMT2,HMT,ABM1,andersen2,HMS,najmul2qns,3loopglue1,ASS,ALSS,ALSS1,HBAMSS,HAMSS,andersen3,andersen4}, dilepton production rate~\cite{LT,BPS,BPY,GHMT,GM3,JG,Ghiglieri1}, photon production rate~\cite{KLS,BNNR,AGKZ,AMY,AMY1,PT,GHKMT},
single quark and quark-antiquark potentials~\cite{MTC,MTC1,CMT4,CMRT,LPRT,DYS,DGMS,THKP}, 
 fermion damping rate~\cite{RDP,PPS}, photon damping rate~\cite{MHT,AD}, gluon damping rate~\cite{BP1,BP3} and  parton energy-loss~\cite{BT,BT1,TG,CMT5,GMT}.

\subsection{One-loop  Quark Free Energy  in HTLpt}
\label{qfe}

\begin{figure}[htb]
\begin{center}
\includegraphics[scale=0.5]{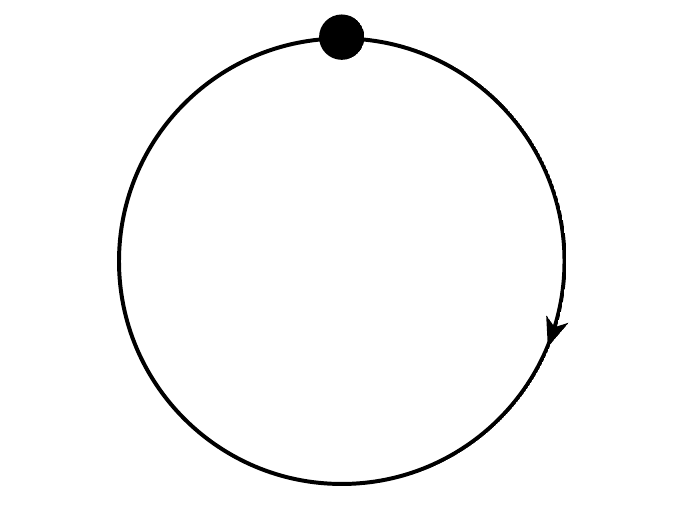}
\end{center}
\caption{one-loop quark contribution to free energy density.}
\label{golla_quark}
\end{figure}

 In thermal field theory the partition function is defined  by a 
functional determinant of the inverse propagator and by which  the quark  part of 
the free energy density in one-loop order at the leading order in the $\delta$-expansion can be written from Fig.~\ref{golla_quark} as
\bea
F^{\textrm{1-loop}}_q&=& 
-N_cN_f\int\frac{d^4P}{(2\pi)^4}~\ln\left({\mbox{det}}\left[S^{* -1}(P)\right] 
\right), 
\label{qfe1}
\eea
where  $P\equiv (p_0, \bm{\vec p})$ is the four momentum with $ p=|\bm{\vec p}|$, $N_c$ is number of colour and  $N_f$ is number of flavour. 
For ideal gas of quarks  the $\textrm{det}\left[S^{* -1}(P)\right] = P^4$ as obtained in \eqref{fpf14}  and the free energy density reads as 
\bea
F_q^\textrm{ideal} 
&=& -2N_cN_f\int\frac{d^4P}{(2\pi)^4}~\ln\left(P^2\right) \nn\\
&=& -2N_cN_f \sumintb_{\{P\}} \ln(P^2) \nn \\
&=& -\frac{7\pi^2T^4}{180}N_cN_f\left(1+\frac{120}{7}\hat{\mu}^2+\frac{240}{7}
\hat{\mu}^4\right)\, ,
\label{qfe1a}
\eea
where $\hat \mu=\mu/2\pi T$. 

Now, the effective quark 
propagator is given in \eqref{gse23} as
\bea
S^{\star-1}(P)&=&{P\!\!\!\! \slash -\Sigma(P)}\nn\\
&=& \frac{1}{2} (\gamma_0 + {\vec \gamma}\cdot \bm{\hat  p}) {\cal D}_+ + \frac{1}{2} (\gamma_0 - {\vec \gamma}\cdot \bm{\hat { p}}) {\cal D}_- \nn \\
&=& \gamma_0 p_0 \, {\cal C} - {\vec \gamma}\cdot \bm{\vec { p}}\, {\cal D} \, , \label{qfe2}
\eea
where
\begin{subequations}
 \begin{align}
{\cal C}&= \frac{1}{2p_0} \left ({\cal D}_+ + {\cal D}_-\right )  \, , \label{qfe3} \\
{\cal D}&= \frac{1}{2p} \left ({\cal D}_- - {\cal D}_+\right )  \, .\label{qfe4}
\end{align}
\end{subequations}
Using \eqref{gse19} one can have
\begin{subequations}
 \begin{align}
{\cal C}&= 1+{\cal A}+\frac{\cal B}{p_0} =1-{\cal A}' \, , \label{qfe5} \\
{\cal D}&= 1+{\cal A}  \, . \label{qfe6}
\end{align}
\end{subequations}
where
\be
{\cal A}' =-{\cal A}-\frac{\cal B}{p_0 } \, . \label{qfe7} 
\ee
From \eqref{se1} we get structure constant ${\cal A}$ as
 \bea
 {\cal A}(p_0,p) &=& -\frac{m^2_{\textrm{th}}}{p^2}   \int \frac{ d\Omega}{4\pi}  
\frac{{\bm{{\vec p}\cdot {\hat k}}}}{P\cdot \hat K}  =\frac{m^2_{\textrm {th}}}{p^2}\left [ 1- {\cal T}_P\right ]\, , \label{qfe8}
\eea
where
\be
{\cal T}_P =  \int \frac{ d\Omega}{4\pi}   \frac{p_0}{p_0 -\bm {{\vec p}\cdot {\hat k}}} \, , \label{qfe9}
\ee
and in presence of chemical potential $\mu$ the quark thermal mass of \eqref{se3} can be written as
\be
m^2_{\textrm{th}} = \frac{g^2T^2C_F}{8} \left (1+4{\hat \mu}^2\right ) . \label{qfe10}
\ee
Using \eqref{qfe8} and \eqref{se2} one can write \eqref{qfe7} as
\be
{\cal A}' = \frac{m^2_{\textrm{th}}}{p^2_0}   \int \frac{ d\Omega}{4\pi}  
\frac{p_0}{p_0 -\bm {{\vec p}\cdot {\hat k}}}  =   \frac{m^2_{\textrm{th}}}{p^2_0}  {\cal T}_P \ . \label{qfe11}
\ee
Similarly, one obtains
\begin{subequations}
 \begin{align}
{\cal C}&= 1-  \frac{m^2_{\textrm{th}}}{p^2_0}  {\cal T}_P  \, , \label{qfe12} \\
{\cal D}&= 1+ \frac{m^2_{\textrm {th}}}{p^2}\left [ 1- {\cal T}_P\right ]\  \, . \label{qfe13}
\end{align}
\end{subequations}

Now we calculate the $\textrm {det}(S^{*-1})$:
\bea
\textrm {det}({S^*}^{-1}) &=& \textrm {det} \left (\gamma_0 p_0 \, {\cal C} - {\vec \gamma}\cdot \bm{\vec { p}}\, {\cal D} \right ) \nn \\
&=& \textrm{det }\left \{
 \left[ {\begin{array}{cc}
   p_0{\cal C} & 0  \\
   0 & -   p_0{\cal C} \\
  \end{array} } \right]  
  + {\cal D}  \left[ {\begin{array}{cc}
   0 & \bm{{\vec \sigma}\cdot {\vec p}}  \\
- \bm{{\vec \sigma}\cdot {\vec p}}  & 0 \\
  \end{array} } \right]  
  \right \} \nn \\
  &=&
  \textrm{det }
 \left[ {\begin{array}{cc}
   p_0{\cal C} & {\cal D} \, (\bm{{\vec \sigma}\cdot {\vec p}})   \\
  - {\cal D}\, ( \bm{{\vec \sigma}\cdot {\vec p}})& -   p_0{\cal C} \\
  \end{array} } \right]   \, . \label{qfe14}
\eea

We know
\bea
\bm{{\vec \sigma}\cdot {\vec p}}&=&\sigma_xp_x+\sigma_y p_y + \sigma_z p_z \nn \\
&=& 
\left[ {\begin{array}{cc}
   0 & p_x  \\
   p_x & 0 \\
  \end{array} } \right]  
  +
  \left[ {\begin{array}{cc}
   0 & ip_y  \\
   -ip_y & 0 \\
  \end{array} } \right]  
  +
  \left[ {\begin{array}{cc}
   p_z& 0  \\
   0 & -p_z \\
  \end{array} } \right]   \nn \\
  &=&
  \left[ {\begin{array}{cc}
   p_z & p_x+ip_y  \\
   p_x-ip_y & -p_z \\
  \end{array} } \right]  \, . \label{qfe15} 
\eea
Using \eqref{qfe15} in \eqref{qfe14} one can write
\bea
\textrm {det}(S^{*-1}) &=&
 \textrm{det }
\left[ {\begin{array}{cccc}
   p_0\, {\cal C} & 0&-{\cal D}p_z& -{\cal D}(p_x +ip_y) \\
   0 & p_0\, {\cal C}  & -{\cal D}(p_x -ip_y)&{\cal D}p_z\\
  {\cal D}p_z\ &{\cal D}(p_x +ip_y) &- p_0\, {\cal C} &0 \\
 {\cal D}(p_x -ip_y)  &-{\cal D}p_z&0& - p_0\, {\cal C}  \\
  \end{array} } \right]   \nn \\
  &=&\left  \{ p_0{\cal C}
 \left | {\begin{array}{ccc}
    p_0\, {\cal C}  & -{\cal D}(p_x -ip_y)&{\cal D}p_z\\
  {\cal D}(p_x +ip_y) &- p_0\, {\cal C} &0 \\
 -{\cal D}p_z&0& - p_0\, {\cal C}  \\
  \end{array} } \right | 
 -{\cal D}p_z
 \left | {\begin{array}{cccc}
   0 & p_0\, {\cal C}  & {\cal D}p_z\\
  {\cal D}p_z\ &{\cal D}(p_x +ip_y) &0 \\
 {\cal D}(p_x -ip_y)  &-{\cal D}p_z& - p_0\, {\cal C}  \\
  \end{array} } \right |  \right. \nn \\
  && \left. +{\cal D}(p_x+ip_y)
  \left | {\begin{array}{cccc}
   0 & p_0\, {\cal C}  & -{\cal D}(p_x -ip_y)\\
  {\cal D}p_z\ &{\cal D}(p_x +ip_y) &- p_0\, {\cal C} \\
 {\cal D}(p_x -ip_y)  &-{\cal D}p_z&0 \\
  \end{array} } \right | 
 \right \} \nn \\
 &=& p_0^2{\cal C}^4 -p_0^2p^2 {\cal C}^2{\cal D}^2 -p_0^2p_z^2 {\cal C}^2{\cal D}^2  +p_z^2p^2 {\cal D}^4 -p_0^2(p_x^2+p_y^2){\cal C}^2{\cal D}^2 
 + (p_x^2+p_y^2)p^2{\cal D}^2  \nn \\
 &=& \left (p_0^2{\cal C}^2 -p^2 {\cal D}^2 \right )^2 \, . \label{qfe16}
\eea
Using \eqref{qfe12} and \eqref{qfe13}, we get
\bea
\textrm {det}(S^{*-1}) &=&\left [\left \{p_0-\frac{m^2_{\textrm{th}}}{p_0}{\cal T}_P \right\}^2
-\left\{p+ \frac{m^2_{\textrm{th}}}{p} \left(1-{\cal T}_P \right ) \right\}^2 \right ]^2 \nn\\
&=& \left [A_0^2-A_S^2 \right ]^2 \, , \label{qfe17}
\eea
where
\begin{subequations}
 \begin{align}
A_0&= p_0-\frac{m^2_{\textrm{th}}}{p_0}{\cal T}_P \, , \label{qfe18} \\
A_S&= p+ \frac{m^2_{\textrm{th}}}{p} \left(1-{\cal T}_P \right )  \, . \label{qfe19}
\end{align}
\end{subequations}
Combining \eqref{qfe1} and \eqref{qfe17}, the one-loop quark  free energy density in 
 HTL approximation  can be written as
\bea
F^{\textrm{1-loop}}_q&=& 
-N_cN_f \int\frac{d^4P}{(2\pi)^4}~\ln\left(A_0^2-A_S^2\right)^2
= -2N_cN_f \int\frac{d^4P}{(2\pi)^4}~\ln\left(A_0^2-A_S^2\right)\nn\\
&=& -2N_cN_f\int\frac{d^4P}{(2\pi)^4}~\ln\left(P^2\right) 
-2N_cN_f\int\frac{d^4P}{(2\pi)^4}~\ln\left(\frac{A_0^2-A_S^2}{P^2}\right) \, .\label{qfe20}
\eea
Now the argument of the logarithm in the second term in
\eqref{qfe20}  can be simplified using \eqref{qfe18} and \eqref{qfe19} as
\bea
\frac{(A^2_0-A^2_S)}{P^2} &=& \frac{1}{P^2}\left [ p_0^2 -2m^2_{\textrm{th}} {\cal T}_P +\frac{m^4_{\textrm{th}}}{p_0^2}{\cal T}_P^2 -p^2
-2m^2_{\textrm{th}} (1- {\cal T}_P) - \frac{m^4_{\textrm{th}}}{p^2}(1-{\cal T}_P)^2 \right ] \nn\\
&=& \frac{1}{P^2}\left [ P^2 -2p_0^2 {\cal A}' +p_0^2{\cal A'}^{2} -2p^2{\cal A} -p^2{\cal A}^2 \right ] \nn \\
&=& 1 + \left(\frac{\mathcal{A}'(\mathcal{A}'-2)p_0^2-\mathcal{A}(\mathcal{A}+2)p^2}{P^2}\right) \, ,
\label{qfe21}
\eea
where in the second line we have used \eqref{qfe8} and \eqref{qfe11}.
In the high temperature approximation, the logarithmic term in \eqref{qfe20} can be expanded
in a series of coupling constants $g$ and then keeping terms up to $\mathcal{O}(g^4)$ one obtains
\bea
\ln\left[\frac{(A^2_0-A^2_S)}{P^2}\right] &=& 
\frac{\mathcal{A}'^2p_0^2-\mathcal{A}^2p^2-2\mathcal{A}'p_0^2-2\mathcal{A}p^2}{P^2} 
- 2\frac{\left(\mathcal{A}'p_0^2+\mathcal{A}p^2\right)^2}{P^4}+\mathcal{O}(g^6) \, \nn\\
&=& \frac{\mathcal{A}'^2p_0^2-\mathcal{A}^2p^2}{P^2} -2\frac{\mathcal{A}'p_0^2+\mathcal{A}p^2}{P^2}
-2\frac{(\mathcal{A}'p_0^2+\mathcal{A}p^2)^2}{P^4} +\mathcal{O}(g^6) \, . \label{qfe22}
\eea
Using \eqref{qfe8} and \eqref{qfe11}, we can obtain
\begin{subequations}
 \begin{align}
(\mathcal{A}'p_0^2 +\mathcal{A}p^2) &= m_{\textrm{th}}^2, \label{qfe23} \\
(\mathcal{A}'^2p_0^2 -\mathcal{A}^2p^2) &= m_{\textrm{th}}^4\left[\frac{{\cal 
T}_P^2}{p_0^2}-\frac{\left(1-{\cal T}_P\right)^2}{p^2}\right] \, . \label{qfe24}
\end{align}
\end{subequations}

Now \eqref{qfe22} becomes
\bea
\ln\left[\frac{(A^2_0-A^2_S)}{P^2}\right] &=& -\frac{2m^2_{\textrm{th}}}{P^2} + m^4_{\textrm{th}} \left [ \frac{{\cal T}_P^2}{p_0^2P^2} -\frac{2}{P^4} 
-\frac{1}{p^2P^2} +\frac{{2\cal T}_P}{p^2P^2} -\frac{{\cal T}_P^2}{p^2P^2}\right ] 
\, . \label{qfe25}
\eea
Using \eqref{qfe25} in \eqref{qfe20}, the one-loop free energy density up to  $\mathcal{O}(g^4)$  can be written as 
\bea
F^{\textrm{1-loop}}_q &=&  N_cN_f\Bigg[ 
-2\int\frac{d^4P}{(2\pi)^4}~\ln\left(P^2\right) +4m_{\textrm{th}}^2\int\frac{d^4P}{(2\pi)^4}\frac{1}{P^2}\nn\\
&-&m_{\textrm{th}}^4\int\frac{d^4P}{(2\pi)^4}\left(\frac{2{\cal 
T}_P^2}{p_0^2P^2}-\frac{4}{P^4}-\frac{2}{p^2P^2}-\frac{2{\cal 
T}_P^2}{p^2P^2}+\frac{4{\cal T}_P}{p^2P^2}\right)\Bigg]\nn\\
&=& N_cN_f\Bigg[ 
-2\sumintb_{\{P\}} \ln\left(P^2\right)  +4m_{\textrm{th}}^2\sumintb_{\{ P\} }\frac{1}{P^2}\nn\\
&&-\ m_{\textrm{th}}^4\left(\sumintb_{\{ P\} }\frac{2{\cal T}_P^2}{p_0^2P^2}-\sumintb_{\{ P\} 
}\frac{4}{P^4}-\sumintb_{\{ P\} }\frac{2}{p^2P^2}-\sumintb_{\{ P\} }\frac{2{\cal 
T}_P^2}{p^2P^2}+\sumintb_{\{ P\} }\frac{4{\cal 
T}_P}{p^2P^2}\right)\Biggr] \, . \label{qfe26}
\eea
Using the sum-integrals of \eqref{qfe27} to \eqref{qfe33} given in appendix~\ref{fer_sum_1l}, the one-loop quark free energy density becomes
\bea
F^{\textrm{1-loop}}_q &=&  N_cN_f\Bigg[-\frac{7\pi^2T^4}{180}\left(1+\frac{120}{7}\hat{\mu}^2+\frac{240}{7}\hat{\mu}^4\right) \nn \\
&& +\frac{m_{\textrm{th}}^2T^2}{6} \left(\frac{\Lambda}{4\pi T}\right)^{2\eps}\left(1+12\hat{\mu}^2 +2\eps [1+12\hat{\mu}^2+12\aleph(1,z)]\right)\nn\\
&&+4m_{\textrm{th}}^4\left(1+\frac{1- 2\Delta_3+ \Delta_4''- \Delta_3''}{2-d}\right)\sumintb_{\{P\}}\frac{1}{P^4}\Bigg] \, . \label{qfe39}
\eea
Substituting \eqref{qfe34} to \eqref{qfe36} and \eqref{qfe29}, we get leading order quark free energy density~\cite{HMS} in thermal medium as
\bea
F^{\textrm{1-loop}}_q&=& 
N_cN_f\Bigg[-\frac{7\pi^2T^4}{180}\left(1+\frac{120}{7}\hat{\mu}^2+\frac{240}{7}
\hat{\mu}^4\right) \nn\\
&& +\frac{m_{\textrm{th}}^2T^2}{6} \left(\frac{\Lambda}{4\pi T}\right)^{2\eps}\left(1+12\hat{\mu}^2 +2\eps [1+12\hat{\mu}^2+12\aleph(1,z)]\right)\nn \\
&& +4m_{\textrm{th}}^4\left[\left(\frac{\pi^2}{3}-2\right)\eps\right] \frac{1}{\left(4\pi\right)^2}\left(\frac{\Lambda}{4\pi 
T}\right)^{2\epsilon}\Bigg[\frac{1} {\epsilon}-\aleph(z)\Bigg]\nn\\
&{=\atop \eps\rightarrow 0}&  
N_cN_f\Bigg[-\frac{7\pi^2T^4}{180}\left(1+\frac{120}{7}\hat{\mu}^2+\frac{240}{7}
\hat{\mu}^4\right)+\frac{m_{\textrm{th}}^2T^2}{6}\left(1+12\hat{\mu}^2\right)+\frac{m_{\textrm{th}}^4}{
12\pi^2}\left(\pi^2-6\right)\Bigg]\nn\\
 &=& N_c N_f\Bigg[ 
-\frac{7\pi^2T^4}{180}\left(1+\frac{120\hat\mu^2}{7}+\frac{240\hat\mu^4}{7}
\right)+\frac{g^2C_FT^4}{48}\left(1+4\hat{\mu}^2\right)\left(1+12\hat{\mu}^2\right)\nn\\
&&+\, \frac{g^4C_F^2T^4}{
768\pi^2}\left(1+4\hat{\mu}^2\right)^2\left(\pi^2-6\right)\Bigg]
. \label{free_quark}
\eea

\subsection{One-loop Gluon Free Energy  in HTLpt }
\label{gf}
\begin{figure}
\begin{center}
\includegraphics[scale=0.64]{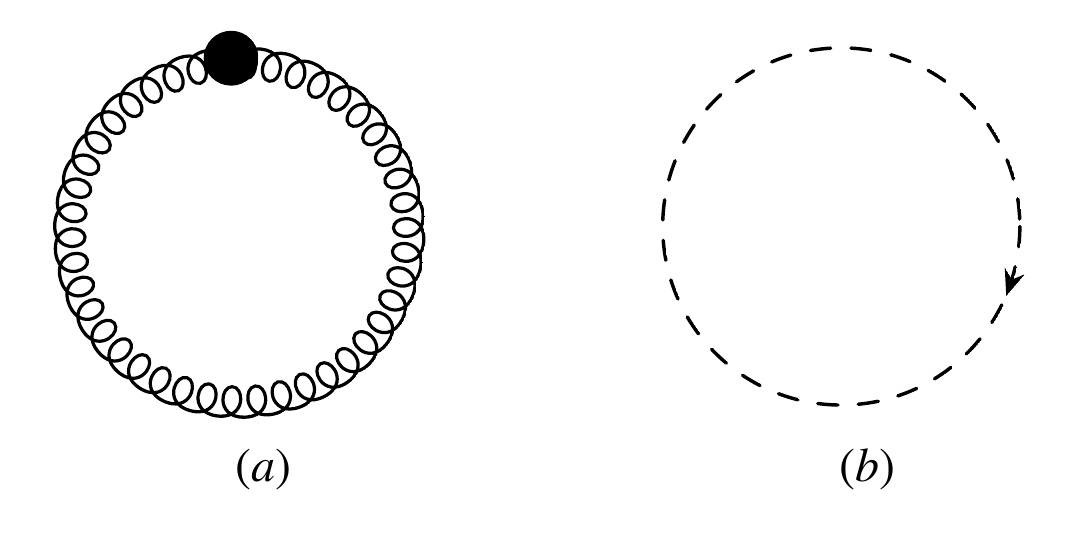}
\end{center}
\caption{One-loop gluon and ghost contribution to free energy density.}
\label{golla_gluon}
\end{figure}

Similar to photon partition function in  QED in subsec~\ref{photon},   the QCD partition function for a gluon 
can be written in Euclidean space-time as
\be 
{\cal Z}_g = {\cal Z}  {\cal Z}^{\textrm{gh}},~~
{\cal Z} = N_\xi\prod_{n,\bm{p}} \sqrt{\frac{\pi^D}{\textsf{det} D_{\mn, E}^{-1}}},~~
{\cal Z}^{\textrm{gh}} = \prod_{n,\bm{p}} P_E^2,  \label{gf1}
\ee
where the product over $n$ is for the discrete bosonic Matsubara frequencies 
($\omega_n=2\pi n\beta;\, \, n=0,1,2,\cdots $) due to
Euclidean time whereas that of $\bm{p}$ is for the spatial momentum. 
 $D_{\mn,E}^{-1}$ is the inverse gauge boson propagator in Euclidean space,
 $P_E^2=\omega_n^2+p^2$ is the square of the Euclidean four-momentum and
$D$ is the space-time dimension of the theory.  As discussed in subsec.~\ref{photon} the normalization
$N_\xi=1/(\pi^D\xi)^{1/2}$  arises due to the introduction of the  
Gaussian integral at every location of position while averaging over the gauge condition function 
with a width $\xi$, the gauge fixing parameter.  
Gluon free energy can now be written from Fig.~\ref{golla_gluon} as
\be
F_g = -(N_c^2-1)\frac{T}{V} \ln {\cal Z}_g = (N_c^2-1)\left[\frac{1}{2}\sumintb_{P_E}~\ln\Big[\textsf{det}
\left(D_{\mn, E}^{-1}(P_E)\right)\Big] -\sumintb_{P_E}~\ln P_E^2 +\frac{1}{2}\ln\xi \right ] \, .\label{gf2}
\ee
where $\sum\!\!\!\!\!\!\!\!\int\limits_{P_E}$ is a bosonic sum-integral and the gauge dependence will explicitly cancel as can be seen later.
For an ideal case $\textsf{det}\left(D_{\mn,E}^{-1}(P)\right)=(P_E^2)^4/\xi$ as obtained in \eqref{phot_det} 
and hence the free energy for $(N_c^2-1)$ massless spin one gluons yields as
\bea
F_g^{\textrm{ideal}} = (N_c^2-1)\sumintb_{P_E}~\ln P_E^2 =-(N_c^2-1) \frac{\pi^2T^4}{45}, \label{gf3}
\eea
where $P_E$ is the four-momentum in Euclidean space and can be written as $P^2=p_0^2+p^2.$

 In presence of thermal background medium one can have
\bea
\textsf{det}\left(D_{\mn,E}^{-1}(P_E)\right) = \frac{P_E^2}{\xi}\left(-P_E^2 + \Pi_T\right)^2\left(-P_E^2 + \Pi_L\right),
\label{gf4}
\eea
with four eigenvalues; respectively $P_E^2$, $(-P_E^2 + \Pi_L)$ and two fold degenerate 
$(-P_E^2 + \Pi_T)$. Here $\Pi_T$ and $\Pi_L$ are the transverse and longitudinal part of the gluon self-energy in thermal medium.  Using \eqref{gf4} in
\eqref{gf2}, one gets 1-loop gluon free energy at the leading order in $\delta$-expansion as
\bea
F_g&=& (N_c^2-1) \left [ \frac{1}{2} \sumintb_{P_E}\,\, \ln\left (1-\frac{\Pi_L}{P_E^2} \right ) + \sumintb_{P_E}\,\,\ln\left (P_E^2-\Pi_T  \right )  \right ], \label{gf5}
\eea
where a term $i\pi$ has been neglected as that would make the free energy complex. Now transforming into Minkowski space $(P_E^2\rightarrow -P^2)$ we have
\bea
F_g &=& (N_c^2-1)  \left [ \frac{1}{2} \sumintb_{P}\,\,\ln\left (1+\frac{\Pi_L}{P^2} \right ) + \sumintb_{P}\,\,\ln\left (P^2+\Pi_T  \right )  \right ] 
=(N_c^2-1) \left[ F_g^L+F_g^T\right]\, , \label{gf6}
\eea
where  $F_g^L$ and $F_g^T$ are, respectively, the  longitudinal and transverse part of the gluon free energy and are given as
\begin{subequations}
 \begin{align}
F_g^L &= \frac{1}{2} \sumintb_{P}\,\,\ln\left (1+\frac{\Pi_L}{P^2} \right )\, , \label{gf7} \\
F_g^T&= \sumintb_{P}\,\,\ln\left (P^2+\Pi_T  \right ) \, . \label{gf8}
\end{align}
\end{subequations}
The longitudinal and transverse part of the gluon self-energy can, respectively, be written from \eqref{psl13} and \eqref{psl15} as
\begin{subequations}
 \begin{align}
\Pi_L &= \frac{m_D^2 P^2}{p^2} \left (1-{\cal T}_P \right)\, , \label{gf9} \\
\Pi_T&= - \frac{m_D^2 }{2p^2} \left (p_0^2- P^2{\cal T}_P \right)\ \, ,\label{gf10}
\end{align}
\end{subequations}
where ${\cal T}_P$ is given in \eqref{qfe9} and the QCD Debye mass in presence of quark chemical potential is given as
\be
m_D^2= \frac{g^2T^2}{3} \left [ C_A +\frac{N_f}{2} (1+12{\hat \mu}^2) \right ]\, . \label{gf11} 
\ee
Now expanding the logarithm in high temperature approximation as
 \bea
F_g^L &=& \frac{1}{2} \sumintb_{P}\,\,\ln\left (1+\frac{\Pi_L}{P^2} \right ) = \frac{1}{2}\sumintb_{P} \left(\frac{\Pi_L}{P^2}-\frac{\Pi^2_L}{2P^4} \right)\,  \nn \\
&=& \frac{m_D^2}{2}\sumintb_{P} \frac{1}{p^2} - \frac{m_D^2}{2}\sumintb_{P} \frac{{\cal T}_P}{p^2} - m_D^4\sumintb_{P} \left[\frac{1}{4p^4} -\frac{{\cal T}_P}{2p^4}
+\frac{{\cal T}_P^2}{4p^4}\right]  \label{gf12} 
\eea
and
\bea
F_g^T&=& \sumintb_{P}\,\,\ln\left (P^2+\Pi_T  \right ) =\sumintb_{P} \, \, \ln(P^2) + \sumintb_{P} \,\, \ln\left(1+\frac{\Pi_T}{P^2} \right) 
= \sumintb_{P} \, \, \ln(P^2) + \sumintb_{P} \,\, \left(\frac{\Pi_T}{P^2}-\frac{\Pi^2_T}{2P^4} \right) \, \nn \\
&=&  \sumintb_{P} \, \, \ln(P^2) -m_D^2\sumintb_{P}\,\, \frac{p_0^2}{2p^2P^2} + m_D^2 \sumintb_{P}\,\, \frac{{\cal T}_P}{2p^2} 
-m_D^4 \sumintb_{P}\left [\frac{p_0^4}{8p^4P^4}-\frac{p_0^2{\cal T}_P}{4p^4P^2} +\frac{{\cal T}_P^2}{8p^4} \right ]\, , \label{gf13}
\eea
where we have kept terms up to ${\mathcal O}[m_D^4]$.
The hard contribution to gluon free energy can be obtained combining \eqref{gf12} and \eqref{gf13} with \eqref{gf6} as
\bea
F_g^{\textrm{hard}} &=& (N_c^2-1) \left(F_g^L+F_g^T\right) 
= (N_c^2-1)\left[ \sumintb_{P} \, \, \ln(P^2)- \frac{m_D^2}{2}\sumintb_{P} \, \frac{1}{P^2} \right. \nn\\
&&\left. -\frac{m_D^4}{8} \sumintb_{P} \left[\frac{1}{P^4} +\frac{2}{p^2P^2} 
-\frac{2{\cal T}_P}{p^2P^2} -\frac{6{\cal T}_P}{p^4} + \frac{3{\cal T}_P^2}{p^4}\right] \right]\, . \label{gf14}
\eea
The bosonic sum integrals have been  obtained in \eqref{gf15} to \eqref{gf20} in appendix~\ref{bos_sum_1l}. Using them in \eqref{gf14} one 
gets the hard contribution to gluon free energy as
\bea
F_g^{\textrm{hard}} &=&(N_c^2-1)\!\!\left[ - \frac{\pi^2T^4}{45} +\frac{m_D^2T^2}{24}\!\! \left(\frac{\Lambda}{4\pi T}\right)^{2\eps}\!\!\!\!\!\left(1+\mathcal{O}[\eps] \right) \right.\nn\\
&&\left. -\frac{m_D^4}{128\pi^2}\!\! \left(\frac{\Lambda}{4\pi T}\right)^{2\eps}\!\! \left(\frac{1}{\eps} +2\gamma_E+ \frac{2\pi^2}{3}-7\right)   \right].\label{gf21}
\eea
For the soft contribution the only important term in the integral is $p_0=0$. Putting $p_0=0$, we get from \eqref{gf9} and \eqref{gf10}, respectively, as
\begin{subequations}
 \begin{align}
\Pi_L&=-m_D^2 \, , \label{gf22} \\
\Pi_T&=0 \, , \label{gf23}
\end{align}
\end{subequations}
where the longitudinal mode provides the electric screening through the Debye mass $m_D$ but the transverse mode does not contribute in the soft scale 
which provides no magnetic screening in HTL.

The soft contribution from longitudinal part can be written from \eqref{gf6} as
\bea
F_g^{\textrm{soft}} &=& \frac{(N_c^2-1)}{2} \sumintb_{P} \, \, \ln(p^2+m^2_D)  \nn \\
&=& -(N_c^2-1)\frac{m_D^3T}{12\pi} \left(\frac{\Lambda}{2m_D}\right)^{2\eps} \left[1+\frac{8}{3}\eps\right] \, . \label{gf24}
\eea

Now total gluon free energy in 1-loop can be written as
\be
F_g^{\textrm{1-loop}}=F_g^{\textrm{hard}} + F_g^{\textrm{soft}} + \Delta_0{\cal E}_0 \, ,  \label{gf25}
\ee
where the HTL leading order vacuum counter term~\cite{ASS} is given as
\be
 \Delta_0{\cal E}_0= \frac{(N_c^2-1)m_D^4}{128\pi^2\eps} \, . \label{gf26}
\ee
Substituting \eqref{gf21}, \eqref{gf24} and \eqref{gf26} in \eqref{gf25}, one gets leading order gluon free energy~\cite{ASS} as
\bea
F_g^{\textrm{1-loop}}&=&(N_c^2-1)\left[ - \frac{\pi^2T^4}{45} +\frac{m_D^2T^2}{24} \left(\frac{\Lambda}{4\pi T}\right)^{2\eps} \left(1+\mathcal{O}[\eps] \right) 
-\frac{m_D^3T}{12\pi} \left(\frac{\Lambda}{2m_D}\right)^{2\eps} \left(1+\mathcal{O}[\eps] \right) \right. \nn \\
&& \left. -\frac{m_D^4}{128\pi^2}\left(\frac{\Lambda}{4\pi T}\right)^{2\eps} \left(\frac{1}{\eps} +2\gamma_E+ \frac{2\pi^2}{3}-7\right)   \right] 
+\frac{(N_c^2-1)m_D^4}{128\pi^2\eps}  \nn \\
&{=\atop \eps\rightarrow 0}&(N_c^2-1)\left[ - \frac{\pi^2T^4}{45} +\frac{m_D^2T^2}{24} 
-\frac{m_D^3T}{12\pi}   -\frac{m_D^4}{128\pi^2} \left(2\ln\left (\frac{\Lambda}{4\pi T}\right) +2\gamma_E+ \frac{2\pi^2}{3}-7\right)   \right.\nn\\
&&\left. -\frac{m_D^4}{128\pi^2\eps} \right]
+\frac{(N_c^2-1)m_D^4}{128\pi^2\eps} \nn \\
&=& -(N_c^2-1) \frac{\pi^2T^4}{45} \left[ 1-\frac{15}{2}{\hat m}_D^2+ 30 {\hat m}_D^3 
+\frac{45}{8} {\hat m}_D^4\left(2\ln\left (\frac{\hat \Lambda}{2}\right) +2\gamma_E+ \frac{2\pi^2}{3}-7\right)  \right]  \nn \\
&=&F_g^{\textrm{ideal}} \left[ 1-\frac{15}{2}{\hat m}_D^2+ 30 {\hat m}_D^3 
+\frac{45}{8} {\hat m}_D^4\left(2\ln\left (\frac{\hat \Lambda}{2}\right) +2\gamma_E+ \frac{2\pi^2}{3}-7\right)  \right] \, ,
\,   \label{gf27}
\eea
where
\begin{subequations}
 \begin{align}
{\hat m}_D &= \frac{m_D}{2\pi T} \, , \label{gf28}\\
{\hat \Lambda} &= \frac{\Lambda}{2\pi T} \, . \label{gf29}
\end{align}
\end{subequations}
\subsection{Leading Order (LO) Thermodynamics of QGP in HTLpt}
\label{lo_press}

The leading order free energy density of quarks and gluons above the deconfinement temperature is defined as
\be
F^{\textrm{LO}}= F_q^{\textrm{1-loop}} + F_g^{\textrm{1-loop}} \, , \label{lo1}
\ee
where one-loop quark and gluon free energy density are, respectively, given in \eqref{free_quark} and \eqref{gf27}. Using them the free energy 
at leading order in the $\delta$-expansion becomes
\bea
F^{\textrm{LO}} &=& -d_A \frac{\pi^2T^4} {45} \left[1+\frac{7}{4}\frac{d_F}{d_A}\left(1+\frac{120}{7}\hat{\mu}^2+\frac{240}{7}\hat{\mu}^4\right) 
-30\frac{d_F}{d_A} (1+12{\hat \mu}^2) {\hat m}^2_{\textrm{th}} \right.\nn\\
&&\left. -\frac{15}{2} {\hat m}_D^2 +30{\hat m}_D^3  -60\frac{d_F}{d_A}(\pi^2-6) {\hat m}^4_{\textrm{th}}  
+\frac{45}{8} {\hat m}_D^4\left(2\ln\left (\frac{\hat \Lambda}{2}\right) +2\gamma_E+ \frac{2\pi^2}{3}-7\right) \right]  , \nn\\
\label{lo2}
\eea
where $d_F=N_cN_f$, $d_A=N_c^2-1$ and ${\hat m}^2_{\textrm{th}} = { m}^2_{\textrm{th}}/2\pi T$.  The leading order pressure is given by
\be
{\cal P}^{\textrm{LO}}= - F^{\textrm{LO}} \, . \label{lo3}
\ee
\subsection{Next-to-leading Order (NLO) Thermodynamics of QGP in HTLpt}
\label{nlo_press}
The NLO free energy density from two-loop  HTLpt  has been obtained complete analytically in Ref.~\cite{HMS,najmul2qns} as
\bea
F^{\rm NLO}&=& F^{\rm 2-loop} = \nn\\
&-&      d_A {\pi^2 T^4\over45} \Bigg\{ 
	   1 + {7\over4} {d_F \over d_A}\left(1+\frac{120}{7}
           \hat\mu^2+\frac{240}{7}\hat\mu^4\right) - 15 \hat m_D^3 
	  - {45\over4}\left(\log\hat{\Lambda\over2}-{7\over2}+\gamma_E+{\pi^2\over3}\right)\hat m_D^4	
\nn \\
& + &
            60 {d_F \over d_A}\left(\pi^2-6\right)\hat m^4_\textrm{th}	
           + {\alpha_s\over\pi} \Bigg[ -{5\over4}\left(c_A + {5\over2}s_F\left(1+\frac{72}{5}
           \ \hat\mu^2+\frac{144}{5}\ \hat\mu^4\right)\right) 
           \nn \\
& + & 
	   15 \left(c_A+s_F(1+12\hat\mu^2)\right)\hat m_D
	- {55\over4}\left\{ c_A\left(\log{\hat\Lambda \over 2}- {36\over11}\log\hat m_D - 2.001\right) \right.
\nn \\	
&+&
      \left.  
	{4\over11} s_F \left[\left(\log{\hat\Lambda \over 2}-2.337\right)
     + (24-18\zeta(3))\left(\log{\hat\Lambda \over 2} -15.662\right)\hat\mu^2
	+ 120\left(\zeta(5) -\zeta(3)\right) \right. \right.
\nn \\
&\times&	\left.\left.  \left(\log{\hat\Lambda \over 2} -1.0811\right)\hat\mu^4 + 
         {\cal O}\left(\hat\mu^6\right)\right] \!\!\right\} \hat m_D^2
 -  	45 \, s_F \left\{\log{\hat\Lambda\over 2} + 2.198  -44.953\hat\mu^2 \right.
 \nn\\
 &-&\left.  \left(288 \ln{\frac{\hat\Lambda}{2}} 
       +19.836\right)\hat\mu^4 + {\cal O}\left(\hat\mu^6\right)\right\}\hat m^2_\textrm{th}
       +  {165\over2}\left\{ c_A\left(\log{\hat\Lambda \over 2}+{5\over22}+\gamma_E\right) \right.
\nn\\
&-& \left. {4\over11} s_F \left(\log{\hat\Lambda \over 2}-{1\over2}+\gamma_E+2\ln2 -7\zeta(3)\hat\mu^2+
         31\zeta(5)\hat\mu^4 + {\cal O}\left(\hat\mu^6\right) \right)\right\}\hat m_D^3
\nn \\
& + &
         15 s_F \left(2\frac{\zeta'(-1)}{\zeta(-1)}
         +2\ln \hat m_D\right)\left[(24-18\zeta(3))\hat\mu^2 + 120(\zeta(5)-\zeta(3))\hat\mu^4 + 
           {\cal O}\left(\hat\mu^6\right)\right] \hat m_D^3
 \nn\\
&+&
	 180\,s_F\hat m_D\hat m^2_\textrm{th} \Bigg]
\Bigg\} \; ,
\label{Omega-NLO}
\eea
where $c_A=N_c$ and $s_F=N_f/2$. The NLO pressure is given by
\be
{\cal P}^{\textrm{NLO}}= - F^{\textrm{NLO}} \, . \label{nlo1}
\ee
\subsection{Next-to-next-leading Order (NNLO) Thermodynamics of QGP in HTLpt}
\label{Nnlo_press}
The NNLO free energy density from three-loop  HTLpt  has been obtained complete analytically in Ref.~\cite{HBAMSS,HAMSS} as
\be
F^{\rm NNLO}= F_q^{\rm 3-loop} + F_g^{\rm 3-loop} ,
\ee
where the 3-loop quark contribution is given.~\cite{HBAMSS,HAMSS} as

\begin{eqnarray}
 F_q^{\rm 3-loop}
&=&-\frac{d_A\pi^2T^4}{45}\dbo\frac{7}{4}\frac{d_F}{d_A}\lb1+\frac{120}{7}\hmu^2+\frac{240}{7}\hmu^4\rb
    -\frac{s_F\alpha_s}{\pi}\bigg[\frac{5}{8}\left(1+12\hat\mu^2\right)\left(5+12\hat\mu^2\right)
    \nn \\
    &&-\frac{15}{2}\left(1+12\hat\mu^2\right)\hat m_D-\frac{15}{2}\bigg(2\ln{\frac{\hat\Lambda}{2}-1
   -\aleph(z)}\Big)\hat m_D^3
      +90\hat m_\textrm{th}^2 \hat m_D\bigg]
\nn \\
&+& s_{2F}\left(\frac{\alpha_s}{\pi}\right)^2\bigg[\frac{15}{64}\bigg\{35-32\lb1-12\hmu^2\rb\frac{\zeta'(-1)}
      {\zeta(-1)}+472 \hat\mu^2+1328  \hat\mu^4\nn \\
      &&+ 64\Big(-36i\hat\mu\aleph(2,z)+6(1+8\hat\mu^2)\aleph(1,z)+3i\hat\mu(1+4\hat\mu^2)\aleph(0,z)\Big)\bigg\}
\nn \\
&-& \frac{45}{2}\hat m_D\left(1+12\hat\mu^2\right)\bigg] +\left(\frac{s_F\alpha_s}{\pi}\right)^2\left[\frac{5}{4\hat m_D}\left(1+12\hat\mu^2\right)^2+30\left(1+12\hat\mu^2
        \right)\frac{\hat m_\textrm{th}^2}{\hat m_D}\right.\nn \\
        &+&\left.\frac{25}{12}\Bigg\{ \left(1 +\frac{72}{5}\hat\mu^2+\frac{144}{5}\hat\mu^4\right)\ln\frac{\hat\Lambda}{2}
       + \frac{1}{20}\lb1+168\hmu^2+2064\hmu^4\rb+\frac{3}{5}\lb1+12\hmu^2\rb^2\gamma_E \right.\nn\\
       && \left.
        - \frac{8}{5}(1+12\hat\mu^2)\frac{\zeta'(-1)}{\zeta(-1)} - \frac{34}{25}\frac{\zeta'(-3)}{\zeta(-3)}\right.
       -  \frac{72}{5}\Big[8\aleph(3,z)+3\aleph(3,2z)-12\hat\mu^2\aleph(1,2z)\nn\\
&& +12 i \hat\mu\,(\aleph(2,z)+\aleph(2,2z)) 
       -\left.i \hat\mu(1+12\hat\mu^2)\,\aleph(0,z)  
       - 2(1+8\hat\mu^2)\aleph(1,z)\Big]\Bigg\}\right.\nn\\
       &&-\left.\frac{15}{2}\Bigg\{\lb1+12\hat\mu^2\rb\lb2\L-1-\aleph(z)\rb\Bigg\}\hat m_D\right]
\nn\\
&+& \left(\frac{c_A\alpha_s}{3\pi}\right)\left(\frac{s_F\alpha_s}{\pi}\right)\Bigg[\frac{15}{2\hat m_D}\lb1+12\hmu^2\rb
     -\frac{235}{16}\Bigg\{\bigg(1+\frac{792}{47}\hat\mu^2+\frac{1584}{47}\hat\mu^4\bigg)\ln\frac{\hat\Lambda}{2}
     \nonumber\\
    &&-\frac{144}{47}\lb1+12\hmu^2\rb\ln\hat m_D+\frac{319}{940}\left(1+\frac{2040}{319}\hat\mu^2+\frac{38640}{319}\hat\mu^4\right)
   -\frac{24 \gamma_E }{47}\lb1+12\mu^2\rb
\nonumber\\
    &&
   -\frac{44}{47}\lb1+\frac{156}{11}\hmu^2\rb\frac{\zeta'(-1)}{\zeta(-1)}
    -\frac{268}{235}\frac{\zeta'(-3)}{\zeta(-3)}
   -\frac{72}{47}\Big[4i\hat\mu\aleph(0,z)
    \nonumber\\
    &&+\left(5-92\hat\mu^2\right)\aleph(1,z)+144i\hmu\aleph(2,z)
   +52\aleph(3,z)\Big]\Bigg\}+90\frac{\hat m_\textrm{th}^2}{\hat m_D}
\nonumber\\
   &&+\frac{315}{4}\Bigg\{\lb1+\frac{132}{7}\hmu^2\rb\L+\frac{11}{7}\lb1+12\hmu^2\rb\gamma_E+\frac{9}{14}\lb1+\frac{132}{9}\hmu^2\rb
\nn  \\
 &&+\frac{2}{7}\aleph(z)\Bigg\}\hat m_D 
\Bigg]
\dbc \, ,
\label{finalomega}
\end{eqnarray}
whereas the free energy density up to three-loop pure glue case  has been calculated
in~\cite{3loopglue1,ASS} and read as
\bea
F_g^{\rm 3-loop}&=&-\frac{d_A\pi^2T^4}{45}\dbo 1-\frac{15}{4}\hat m_D^3+\frac{c_A\alpha_s}{3\pi}\Bigg[-\frac{15}{4}
+\frac{45}{2}\hat m_D-\frac{135}{2}\hat m_D^2-\frac{495}{4}\lb\Lg+\frac{5}{22}+\gamma_E\rb \hat m_D^3 \Bigg]
\nn\\
&+&\lb\frac{c_A\alpha_s}{3\pi}\rb^2\Bigg[\frac{45}{4\hat m_D}-\frac{165}{8}\lb\Lg-\frac{72}{11}\ln\hat m_D-\frac{84}{55}-\frac{6}{11}
\gamma_E-\frac{74}{11}\Za+\frac{19}{11}\Zc\rb
\nn\\
&+&\frac{1485}{4}\lb\Lg-\frac{79}{44}+\gamma_E+\ln2-\frac{\pi^2}{11}\rb\hat m_D\Bigg]\dbc . \label{nnlo1}
\eea
It is important to note that chemical potential dependence also appears in pure glue diagrams from the
internal quark loop in effective gluon propagators and effective vertices. This chemical potential$(\mu)$ dependence are present
within Debye mass$(m_D)$. Besides the choice of the renormalization scales,
the analytic result does not contain any free fit parameters and the result is also gauge-invariant.
The NNLO pressure is given by
\be
{\cal P}^{\textrm{NNLO}}= - F^{\textrm{NNLO}} \, . \label{nnlo2}
\ee
The higher orders thermodynamical quantities~\cite{HBAMSS,HAMSS}  of hot and dense matter, such as,  the entropy density, the equation of state, 
the speed of sound, the interaction measure or the trace anomaly and various susceptibilities associated with conserved density fluctuations can be 
computed using NNLO free energy density and pressure. The equation of state is a generic quantity of a hot and dense many particle system 
and is required to investigate the expansion dynamics of hot and dense matter by using hydrodynamics. 
The obtained results on thermodynamic quantities~\cite{HBAMSS,HAMSS} are very good
agreement with lattice results within error down to 200 MeV temperature. These calculations certainly have huge impact on the thermodynamics of 
QCD matter at finite temperature and chemical potential that agree quite well with data from lattice QCD, a first principle calculation.
On the other hand the higher order thermodynamic 
quantities and various order quark number susceptibilities (QNS) are of huge interest to both theorists and experimentalists, for understanding 
the phase diagram of QCD.

Apart from QCD thermodynamics~\cite{andersen1,ABM,HM,CMT,CMT1,CMT2,HMT,ABM1,andersen2,HMS,najmul2qns,3loopglue1,ASS,ALSS,ALSS1,HBAMSS,HAMSS,andersen3,andersen4},
if readers are interested in application to other physical quantities related to QGP  created in heavy-ion collisions can go through the following extensive list of references:  for 
dilepton production rate~\cite{LT,BPS,BPY,GHMT,GM3,JG,Ghiglieri1}, photon production rate~\cite{KLS,BNNR,AGKZ,AMY,AMY1,PT,GHKMT},
single quark and quark-antiquark potentials~\cite{MTC,MTC1,CMT4,CMRT,LPRT,DYS,DGMS,THKP}, 
 fermion damping rate~\cite{RDP,PPS}, photon damping rate~\cite{MHT,AD}, gluon damping rate~\cite{BP1,BP3} and  parton energy-loss~\cite{BT,BT1,TG,CMT5,GMT}.
 
\section{Thermal  Medium with Non-perturbative Effects}
\label{nonpert}

Finite temperature QCD has been applied to study the properties of a QGP, which is believed to have existed
in the early Universe, just a few microsecond after the big bang and in the fireball created in high energy relativistic heavy-ion collisions at
RHIC in BNL  and at  LHC in CERN. Lattice QCD (lQCD) provides a first-principles-based method that can 
take into account the non-perturbative effects of QCD. lQCD has been applied to investigate the behaviour of QCD near the critical temperature $T_c$, 
where hadronic matter undergoes a phase transition  to the deconfined QGP phase. Beside lQCD also perturbation
theory has been used to investigate the phenomenologically relevant properties of QGP. In contrast to lQCD computations the purterbative method is able to deal
with dynamical quantities, a finite baryon density and non-equilibrium situations. To perturbatively understand the properties of  QGP one needs to have very good
understanding of the different collective excitations appear due to the presence of a thermal bath. There are three types of collective excitations which are associated with different thermal scales, They are (i) the energy (or hard) scale $T$, (ii) the electric scale $gT$, and  (iii) the magnetic scale $g^2T$. 
In the literature  the hard and electric scales are well studied, but not the magnetic scale since it is related to the difficult non-perturbative physics of confinement.

Based on the HTL resummations~\cite{BP2,BP,TW,FT,BP1}, a reorganization of finite-temperature and chemical potential perturbation theory
known as HTL perturbation theory (HTLpt) has been discussed in sec.~\ref{htlpt}. HTLpt deals with the hard scale $T$ and the electric scale $gT$ as the 
soft scale. This has been widely applied to compute various physical quantities associated with QGP by using HTL resummed propagators and vertices. 
HTLpt  works well at a temperature of approximately $2T_c$ and above, where $T_c\sim160$ MeV is the critical temperature for the QGP phase transition. 
Near $T_c$, the running coupling $g$ is moderately high and the QGP could therefore be completely non-perturbative in the vicinity of $T_c$. 

Given the uncertainty involved in the lattice computation of dynamical quantities and also HTLpt near $T_c$,  it is always desirable to formulate an alternative approach to consider non-perturbative effects which can be dealt in a similar way as done in HTLpt. There are some approaches available in the literature: one such approach is a semi-empirical way to include non-perturbative effects by considering a gluon condensate  within the Green functions in momentum space~\cite{lavelle,markus,MST}.  The gluon condensate has a substantial effect  on the equation of state of QCD matter,  in contrast to the quark condensate. In subsec.~\ref{gcond} the quark propagation in QGP with gluon condensate will be discussed. Another approach~\cite{nansu1,BHMS} would be to consider the non-perturbative physics involved in the QCD magnetic scale. This is taken 
into account through the non-perturbative magnetic screening scale within the Gribov-Zwanziger (GZ) action~\cite{Gribov1,zwanziger1}. The inclusion
of magnetic scale regulates the magnetic infrared (IR) behaviour of QCD, the physics associated with it is completely non-perturbative. 
The gluon propagator with the GZ action is IR regulated which mimics confinement. This also makes the calculations more compatible with the results of lQCD. 
In subsec.~\ref{gza} the quark propagation in QGP with GZ action will be discussed.

\subsection{Quark Propagation in QGP with Gluon Condensate}
\label{gcond}

\subsubsection{Quark self-energy}
Now in the rest frame of the heat bath, $u^\mu=(1,0,0,0)$, the most general ansatz for fermionic self-energy reads from \eqref{gse11} as 
\bea
\Sigma(L) &=& -{\cal A}(\om,l) L\!\!\!\! \slash - {\cal B}(\om,l)\gamma_0 , \label{gcond1} 
\eea
with structure functions
\begin{subequations}
 \begin{align}
{\cal A}(\om,l) &= \frac{1}{4l^2} \left ({\rm{Tr}} \left [ \Sigma L\!\!\!\! \slash \, \right ] - \om {\rm{Tr}} \left [\Sigma \gamma_0\right ] \right ), \label{gcond2}\\
{\cal B}(\om,l) &= \frac{1}{4l^2} \left ( L^2 {\rm{Tr}} \left [\Sigma\gamma_0 \right ] - \om {\rm{Tr}} \left [ \Sigma L\!\!\!\! \slash \, \right ]  \right ) , \label{gcond3}
\end{align}
\end{subequations}
where the four momentum of fermion is $L\equiv(\om,\bm{\vec l})$ with $l=|\bm{\vec l}|$.

\label{q_se_gc}
\begin{figure}[htb]
\begin{center}
\includegraphics[scale=0.55]{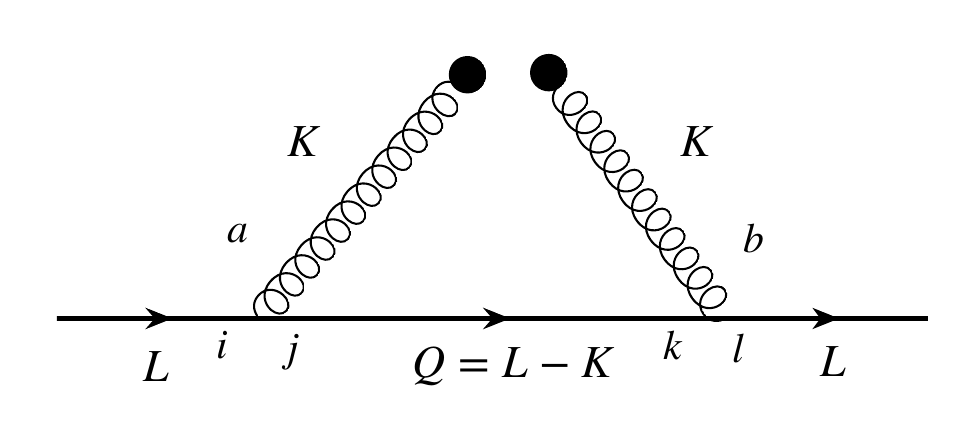}
\end{center}
\caption{Quark self-energy containing gluon condensate.}
\label{se_gc}
\end{figure}

The lowest order interaction of a quark with gluon condensate is given by self-energy diagram in Fig.~\ref{se_gc}. One can write the quark self-energy following Fig.~\ref{se_gc} as
\bea
\Sigma(L) \delta_{il} &=& T \sumintb_{K}\,\, \left(-ig\gamma^\mu \lambda^a_{ij} \right ) \left (\frac{i\slashed Q \delta_{jk}}{Q^2}\right ) \left(-ig\gamma^\nu \lambda^b_{kl} \right ) 
i{\tilde D}_{\mn}^{ab}(K) \delta_{ab} \nn \\
\Sigma(L)&=& \frac{4}{3} g^2 T \sumintb_{K}\,\, {\tilde D}_{\mn}(K) \gamma^\mu \frac{\slashed Q}{Q^2} \gamma^\nu \, \label{gcond4}
\eea
where $Q=L-K$,  $\sum\!\!\!\!\!\!\!\int\limits_{K} $ is a bosonic sum-integral, $\lambda_{ij}^a\lambda_{jl}^a =\frac{(N_c^2-1)}{2N_c}\delta_{il} =\frac{4}{3}\delta_{il}$ with $N_c=3$ and ${\tilde D}_{\mn}$ is the non-perturbative gluon propagator
containing gluon condensate. We will consider purely non-perturbative input from lQCD as parametrized by temperature dependent gluon condensates. The most general
ansatz for the non-perturbative gluon propagator at finite temperature can be written as
\be
{\tilde D}_{\mn} (K) = {\tilde D}_L(k_0,k) P^L_{\mn} + {\tilde D}_T(k_0,k)  P^T_{\mn} \, , \label{gcond5}
\ee
where the longitudinal and transverse projectors are given by
\begin{subequations}
 \begin{align}
 P^L_{\mn} &= \frac{K_\mu K_\nu}{K^2}-\eta^{\mn}-P^T_{\mn}\, , \label{gcond6} \\
 P^T_{\mu 0}&=0 \, , \label{gcond7}  \\
 P^T_{ij}&= \delta_{ij} -\frac{k_i k_j}{k^2} \, . \label{gcond7a} 
\end{align}
\end{subequations} 
In order to relate the propagator in \eqref{gcond5} to the gluon condensate one can follow the zero temperature calculation~\cite{lavelle} and expand the quark
propagator in \eqref{gcond4} for small loop momenta. Then considering  terms which  are only bilinear in $K$, one can relate the gluon condensates with 
the moments of the gluon propagator. Following this  one can obtain~\cite{markus} the structure functions in \eqref{gcond2} and \eqref{gcond3} as
\begin{subequations}
 \begin{align}
{\cal A} &= -\frac{4}{3} g^2 \frac{1}{L^6} T \sumintb_{K}\,\, \left [\frac{1}{K^2} \left \{ -\om^2k_0^4+\frac{22}{3}\om^2k_0^2k^2 -\frac{1}{3} l^2k_0^4
+\frac{34}{15} l^2k_0^2k^2 +\frac{5}{3} \om^2k^4 -\frac{1}{3} l^2k^4\right\}{\tilde D}_L(k_0,k) \right. \nn \\
& \left. + \left\{-2\om^2k_0^2-\frac{2}{3}l^2k_0^2 -2\om^2k^2 +\frac{2}{5}l^2k^2 \right \}{\tilde D}_T(k_0,k) \right ]\, , \label{gcond8} \\
{\cal B}&= -\frac{4}{3}g^2\frac{l_0}{L^6} T \sumintb_{K}\,\, \left [\frac{1}{K^2} \left \{ -\frac{8}{3}l^2k_0^4 -\left(\frac{40}{3}\om^2 +\frac{104}{15}l^2\right)k_0^2k^2
-\frac{8}{3}\om^2k^4 \right\}{\tilde D}_L(k_0,k) \right.\nn\\
&\left. -\left\{\frac{16}{3}l^2k_0^2+\frac{16}{15}l^2k^2 \right\}{\tilde D}_T(k_0,k) \right ] \, . \label{gcond9}
\end{align}
\end{subequations} 
Assuming the temperature scale to be large, $T>> l$ and one can set $k_0=2\pi i n T=0$ under the plane wave approximation. The above two equations become
\begin{subequations}
 \begin{align}
{\cal A}(\om,l) &= -\frac{4}{3} g^2 \frac{1}{L^6} T \int\frac{d^3k}{(2\pi)^3} \left [\left(\frac{1}{3}\om^2-\frac{5}{3}l^2\right)k^2{\tilde D}_L(0,k) 
+  \left( -2\om^2 +\frac{2}{5}l^2 \right )k^2{\tilde D}_T(0,k) \right ]\, , \label{gcond10} \\
{\cal B}(\om,l)&= -\frac{4}{3}g^2\frac{\om}{L^6} T \int\frac{d^3k}{(2\pi)^3}   \left [\frac{8}{3}\om^2k^2 {\tilde D}_L(0,k) 
-\frac{16}{15}l^2k^2 {\tilde D}_T(0,k) \right ] \, . \label{gcond11}
\end{align}
\end{subequations} 
The moments of the longitudinal and the transverse gluon propagators, respectively, in \eqref{gcond10} and \eqref{gcond11} can be related to the chromoelectric 
and  the chromomagnetic condensates as
\begin{subequations}
 \begin{align}
\langle E^2 \rangle_T &= 8T \int\frac{d^3k}{(2\pi)^3} k^2{\tilde D}_L(0,k)  +{\cal O}(g) \, , \label{gcond12} \\
\langle B^2 \rangle_T &= -16T \int\frac{d^3k}{(2\pi)^3} k^2{\tilde D}_T(0,k)  +{\cal O}(g) \, . \label{gcond13} 
\end{align}
\end{subequations} 
Using \eqref{gcond12} and \eqref{gcond13} in \eqref{gcond10} and \eqref{gcond11}, one can write the structure functions as
\begin{subequations}
 \begin{align}
{\cal A}(\om,l) &= -\frac{1}{6} g^2 \frac{1}{L^6}  \left [\left(\frac{1}{3}\om^2-\frac{5}{3}l^2\right) \langle E^2 \rangle_T
+  \left(  \frac{1}{5}l^2 -\om^2 \right )\langle B^2 \rangle_T\right ]\, , \label{gcond14} \\
{\cal B}(\om,l)&= -\frac{4}{9}g^2\frac{\om}{L^6}   \left [\om^2  \langle E^2 \rangle_T
+\frac{1}{5}l^2 \langle B^2 \rangle_T \right ] \, , \label{gcond15}
\end{align}
\end{subequations} 
where $g^2=4\pi \alpha_s$.
These condensates can be obtained in terms of the spacelike ($\Delta_\sigma$) 
and timelike ($\Delta_\tau$) plaquette expectation values computed on a 
lattice~\cite{boyd} in Minkowski space  as~\cite{markus}
\begin{subequations}
 \begin{align}
\frac{\alpha_s}{\pi} \langle  E^2\rangle_T &= \frac{4}{11} T^4 
\Delta_\tau
- \frac{2}{11} \langle G^2\rangle_{T=0} \ , \label{gcond16} \\ 
\frac{\alpha_s}{\pi} \langle  B^2\rangle_T &= -\frac{4}{11} T^4 
\Delta_\sigma
+ \frac{2}{11} \langle G^2\rangle_{T=0} \ . \label{gcond17}
\end{align}
\end{subequations} 
The plaquette expectation values are related to the gluon condensate above
$T_c$ as~\cite{boyd,leut}
\begin{equation}
\langle G^2 \rangle_T = \langle G^2 \rangle_{T=0} -\Delta T^4 \ , 
\label{gcond18}
\end{equation} 
where  $\Delta = \Delta_\sigma +\Delta_\tau$ and 
$\langle G^2\rangle_{T=0} = (2.5\pm 1.0)T_c^4$.

\subsubsection{Quark propagator and dispersion}
\label{q_prop_gc}
\begin{figure}
\begin{center}
\includegraphics[scale=0.65]{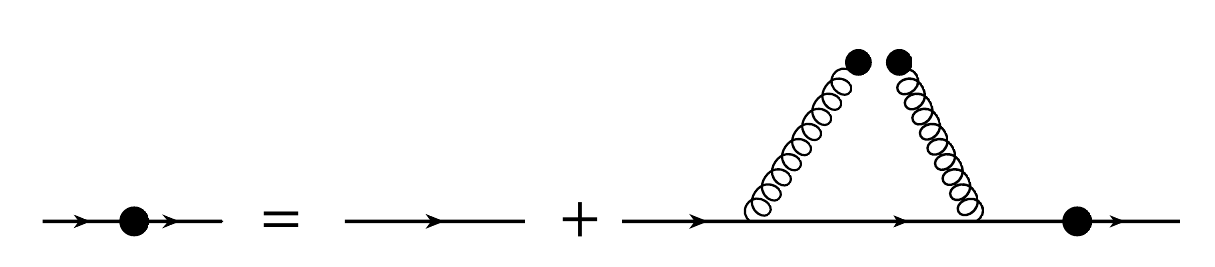}
\end{center}
\caption{Effective quark propagator containing gluon condensate.}
\label{prop_gc}
\end{figure}

The effective quark propagator containing gluon condensate follows from diagram in Fig.~\ref{prop_gc}. In helicity representation,
the effective quark propagator
is given in \eqref{gse24} as  
\begin{equation}
 S^*(L) = \frac{1}{\slashed L-\Sigma(L)}= \frac{\gamma_0-  \vec \gamma \cdot \bm{\hat l}}{2{\cal D}_{+}(\om,l)} +
\frac{\gamma_0+\vec \gamma \cdot \bm{\hat l}}{2{\cal D}_{-}(\om,l)} \ , \label{gcond19}
\end{equation}
where
\begin{equation}
{\cal D}_\pm(\om,l) = (\om \mp l)(1+{\cal A}) + {\cal B} , \label{gcond20}
\end{equation}
and  the expressions for $\cal A$ and $\cal B$ are obtained
in terms of the chromoelectric and the chromomagnetic condensates in \eqref{gcond14} and \eqref{gcond15}.

The dispersion relation of a quark interacting with the thermal gluon
condensate is obtained by the poles of ${\cal D}_\pm (\omega,l)=0$ in \eqref{gcond20}.
The functions ${\cal A}$ and ${\cal B}$  have been determined by
using \eqref{gcond16} and \eqref{gcond17}, where the plaquette expectation values  are taken from
the lattice calculations of Ref.~\cite{boyd}.
\begin{figure}
\begin{center}
\includegraphics[scale=0.5]{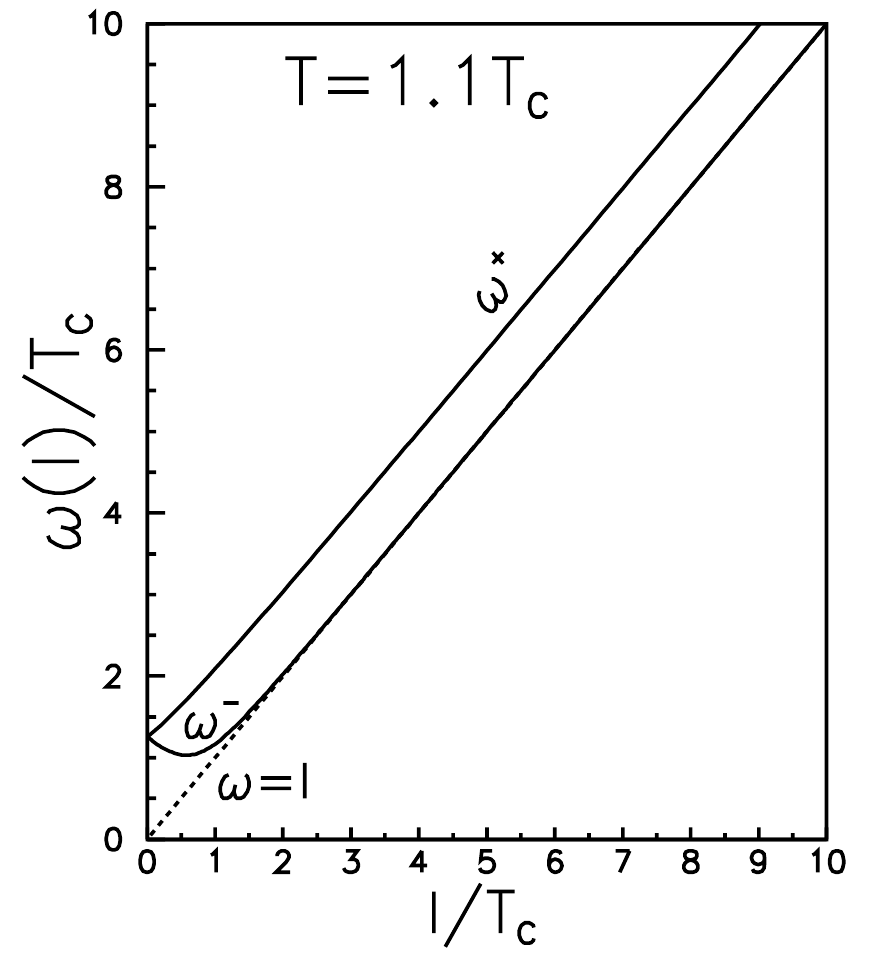}
\includegraphics[scale=0.5]{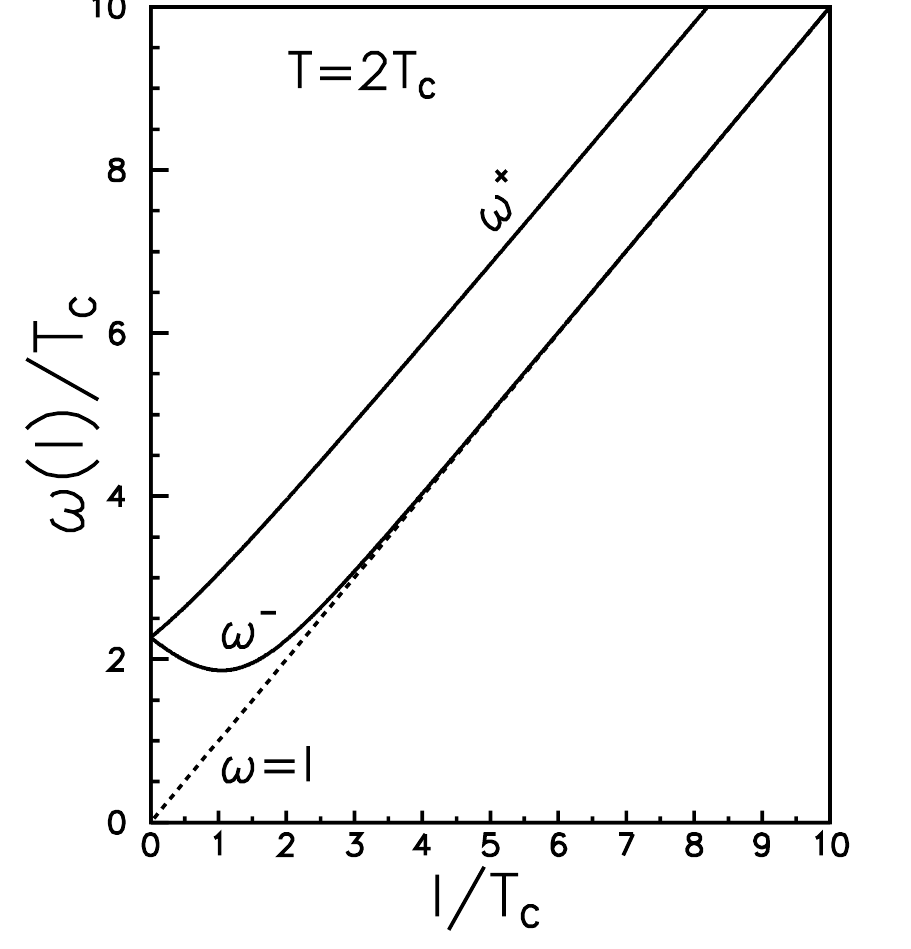}
\end{center}
\caption{Quark dispersion in presence of gluon condensate. These figures are taken from Ref.~\cite{MST}.}
\label{disp_gc}
\end{figure}
${\cal D}_+(\omega,l)=0$ has two ploes at $\omega=\omega^+(l)$ and $\omega=-\omega^-(l)$ whereas 
 ${\cal D}_-(\omega,l)=0$ has two ploes at $\omega=\omega^-(l)$ and $\omega=-\omega^+(l)$.  
 In Fig.~\ref{disp_gc} we have displayed the dispersion relation of a quark having 
momentum $l$ for $T=1.1 T_c$ (left panel) and $2T_c$ (right panel), respectively. 
Only positive solutions of ${\cal D}_\pm(L)=0$ have been displayed in Fig.~\ref{disp_gc}.
The mode with energy $\omega^+$ describes the in-medium propagation of a particle excitation.  
It is a Dirac spinors which is a eigenstate of $(\gamma_0 - {\vec \gamma}\cdot \bm{\hat { l}})$ with chirality to helicity ratio $+1$.
Also there is a new long wavelength mode known as {\em plasmino} with energy $\omega^-$. It is also a Dirac spinor and is a eigenstate of 
$(\gamma_0 + {\vec \gamma}\cdot \bm{\hat { l}})$  with chirality to 
helicity ratio $-1$. 
Both branches are situated in the time like domain (i.e., above the free
dispersion relation $\omega=l$), and they begin from a common effective
mass which is obtained in the $l\rightarrow 0$ limit as~\cite{markus}
\begin{equation}
\omega^{+}(0)=\omega^{-}(0)=m_{\rm{eff}}= \left [ \frac{2\pi\alpha_s}{3}
\left ( \langle { E}^2\rangle_T +\langle { B}^2\rangle_T 
\right )\right ]^{1/4} \ , \label{gcond21}
\end{equation}
which is given by $m_{\rm{eff}} \approx 1.15$ $T$ and found to be independent of $g$. 
For small momenta $l\rightarrow 0$, the dispersion relation behaves~\cite{markus} like
\be
\om^\pm= m_{\rm{eff}} \pm c_2l, \label{gcond22}
\ee
where
\be
c_2=\frac{3}{4} \frac{\langle { E}^2\rangle_T}{ \langle { E}^2\rangle_T +\langle { B}^2\rangle_T } \, . \label{gcond23}
\ee
It is to be noted that because of the opposite slopes of two branches, $\om^-$ branch has a minimum at low momenta then it rapidly approaches the free
dispersion for large momenta, indicating a purely long wavelength mode.  This minimum leads to Van Hove singularities in soft dilepton rate~\cite{MST} 
akin to HTL case~\cite{BPY,KMT} and will be discussed in subsec~\ref{DP_GC}. On the other hand, the $\om^+$ mode at large momenta is given by~\cite{markus} 
\be
\om^+=l+c_1, \label{gcond24}
\ee
where
\be
c_1=\left[\frac{2\pi}{9}\alpha_s \left(\langle E^2\rangle_T+\langle B^2\rangle_T \right)\right]^{1/4}. \label{gcond25}
\ee
The dispersion relation of a quark interacting with the in-medium gluon
condensate is similar to that obtained from the HTL resummed quark propagator displayed in Fig.~\ref{disp_plot}.  
It is important to note that the dispersion relations with the HTL approximation and gluon condensate, respectively, 
exhibit similar features which is the general consequence of the presence of the heat bath.

\subsubsection{Spectral representation of the quark propagator}
\label{spec_q_prop_gc}
 The spectral functions,  $\rho^\pm(\om,l)$,  corresponding to the effective propagator in \eqref{gcond19} can be obtained following  subsec.~\ref{spec_prop} or
 \eqref{bpy_04} in appendix~\ref{bpy_disc} as~\cite{MST}
\begin{equation}
\rho^\pm (\omega, l) = R^\pm (\omega , l) \delta \left ( \omega - \omega^\pm 
\right ) + R^\mp (-\omega , l) \delta \left ( \omega + \omega^\mp \right ) \ ,
\label{gcond26}  
\end{equation}
where
\begin{equation}
R^\pm= \left | \frac{\left (\omega^2 -l^2\right )^3} {C^\pm}\right |  
\label{gcond27}
\end{equation}
with
\begin{eqnarray}
C^\pm &=& - \left [ \left (1+{\cal A}\right )\left (\omega^2-l^2\right )^3 
\ +\ \frac {{\cal B} \left (\omega^2-l^2\right )^3}{\omega}
\ +\ 6\,\omega (\omega \mp l) (\omega^2-l^2)^2 \right. \nonumber \\ 
&& \left.  +\ \frac{g^2}{3} \omega (\omega \mp l) \left (\frac{5}{3} 
\langle {E}^2\rangle_T
-\langle {B}^2\rangle_T \right ) \ - \ \frac{8}{9}g^2\omega^2 
\langle { E}^2\rangle_T
 \right ] . \label{gcond28}
\end{eqnarray}
The spectral functions in (\ref{gcond26}) has only contribution from
the poles of the effective propagator.  The solutions are collective quark modes with energy $\omega^\pm$.
Since the effective quark propagator (\ref{gcond19}) does not have an 
imaginary part coming from the quark self-energy, the spectral functions
do not have a contribution from discontinuities or Landau cut.

\subsection{Quark Propagation in QGP with Gribov-Zwanziger  Action}
\label{gza}
\subsubsection{Gribov-Zwanziger action and its consequences}
\label{gribov_para}
Gribov showed in 1978~\cite{Gribov1} that in a non-Abelian gauge theory,
fixing the divergence of the potential does not commute with the gauge fixing.
Unfortunately, the solutions of the differential equations, which specify
the gauge fixing with vanishing divergence, can have several copies (Gribov copies) or none at all.
This is known as Gribov ambiguity. To resolve this ambiguity, the domain of functional integral has to be
restricted within a fundamental modular region, bounded by Gribov horizon. Following this in 1989 Zwanziger~\cite{zwanziger1} derived a 
local, renormalizable action for non-Abelian gauge theories which fulfills the idea of restriction. He also showed that
by introducing this GZ action the divergences may be absorbed by suitable field and coupling
constant renormalization. 

The GZ action is given by~\cite{NV}
\bea
S_{\textrm{GZ}}&=&S_0+S_{\gamma \textrm {G}} ; \label{gz_action} \\
S_0&=& S_{\textrm{YM}}+S_{\textrm{gf}}+\int d^Dx \left ({\bar \phi}^{ac}_\mu \partial_\nu D^{ab}_\nu \phi^{bc}_\mu 
-{\bar \om}^{ac}_\mu \partial_\nu D^{ab}_\nu \om^{bc}_\mu \right ) +\Delta S_0 ; \label{s0} \\
S_{\gamma \textrm {G}} &=&\gamma_G^2\int d^Dx \ g \ f^{abc} A_\mu^a \left (\phi^{bc}_\mu+{\bar \phi}^{bc}_\mu \right) + \Delta S_\gamma ; \label{s_gamma}
\eea
where $(\phi^{bc}_\mu+{\bar \phi}^bc_\mu ) $ and  $(\om^{bc}_\mu+{\bar \om}^{bc}_\mu ) $ are a pair of complex conjugate bosonic and Grassmann fields
respectively, introduced due to localization of the GZ action. $S_{\textrm{YM}}$ and $S_{\textrm{gf}}$ are the normal Yang-Mills and the gauge fixing term
of the action and $D$ is the dimension of the theory. $\Delta S_0 $ and $ \Delta S_\gamma$ are the corresponding counterterms of the $\gamma_G$ independent
and dependent parts of GZ action. $\gamma_G$ is called the Gribov parameter.
In reality, $\gamma_G$ is computed self-consistently  
using a one-loop\footnote{Equation (\ref{Gribov_para}) is a one-loop result.  In the 
vacuum, the two-loop result has been computed~\cite{Gracey:2005cx} and the form of
Gribov propagator in (\ref{modified_gluon_prop}) remains unaffected.  Only 
$\gamma_G$ itself is changed to take into account the two-loop correction.  It is expected that 
this would be valid also at finite temperature.}  
gap equation and at asymptotically high temperatures it becomes~\cite{nansu1,zwanziger2,nansu2}
\bea
\gamma_G = \frac{D-1}{D}\frac{N_c}{4\sqrt{2}\pi}g^2T,  \label{Gribov_para}
\eea
where $N_c$ is the number of 
colors. The 
one-loop running strong coupling, 
$g^2=4\pi\alpha_s$, is 
\bea
g^2(T)=\frac{48\pi^2}{(33-2N_f)\ln\left (\frac{Q^2_0}{\Lambda_0^2} \right )} , \label{alpha_s}
\eea
where $Q_0$ is the renormalization 
scale, which is usually chosen to be $2\pi T$ unless specified and $N_f$ is the number of quark flavors.  
The scale $\Lambda_0$ is fixed by requiring that $\alpha_s$(1.5 GeV) = 0.326, which is obtained 
from lattice calculations~\cite{alphas_lat}. For one-loop running, this 
procedure gives $\Lambda_0 = 176$ MeV. 

We know that gluons have an important role in confinement. In the GZ 
action~\cite{Gribov1,zwanziger1} the  confinement is expected to be governed
kinematically with the gluon propagator in covariant gauge~\cite{Gribov1,zwanziger1}
\bea
D^{\mu\nu}(P)=-\left[\eta^{\mu\nu}-(1-\xi)\frac{P^\mu P^\nu}{P^2}\right]\frac{P^2}{P^4+\gamma_G^4}\, ,
\label{modified_gluon_prop}
\eea
where the four-momentum $P = (p_0, \bm{\vec p})$ and $\xi$ is the gauge parameter. The term 
$\gamma_G$ in the denominator in \eqref{modified_gluon_prop} shifts the poles of the gluon propagator off the 
energy axis and there are  no asymptotic gluon modes exist.  For maintaining 
the consistency of the theory, obviously these unphysical poles should not appear
in  gauge-invariant quantities.  This indicates 
that the gluons are  unphysical excitations.  In reality, this means that the 
addition of the Gribov parameter yields the effective confinement of 
gluons.  
\subsubsection{Quark self-energy}
\label{se_gz}
 In the high-temperature limit
one can calculate the quark self-energy $\Sigma$ using the 
modified gluon propagator given in \eqref{modified_gluon_prop} as~\cite{nansu1,BHMS}
\bea
\Sigma(P) &=& C_F T \sumintb_{\{K\}} \!\! (-ig\gamma_\mu) \left(\frac{i\slashed K}{K^2} \right)(-ig\gamma_\nu) (iD^{\mu\nu}(P-K))
\approx g^2 C_F\sum_\pm\int\limits_0^\infty \frac{dk}{2\pi^2}k^2\int \frac{d\Omega}{4\pi} \nn \\
&& \times \frac{\tilde{n}_\pm(k,\gamma_G)}{4E_\pm^0}\left[\frac{\gamma_0+\bm{\hat k}\cdot\vec{\gamma}}{\om+k-E_\pm^0+
\frac{\bm{\vec p}\cdot {\vec k}}{E_\pm^0}}
 +\frac{\gamma_0-\bm{\hat k}\cdot\vec{\gamma}}{\om-k+E_\pm^0-\frac{\bm{\vec p}\cdot {\vec k}}{E_\pm^0}}\right],
\label{modified_quark_self}
\eea
where Casimir factor $C_F=4/3$,  $\sumintb_{\{K\}}$ is a fermionic sum-integral  and 
\bea
\tilde{n}_\pm(k,\gamma_G)&\equiv& n_B\!\left(\sqrt{k^2 \pm i\gamma_G^2}\right)+n_F(k) \nn ,\\
E_\pm^0 &=& \sqrt{k^2 \pm i\gamma_G^2}\ , \label{freq}
\eea
with $n_B$ and $n_F$ are Bose-Einstein and Fermi-Dirac distribution functions, respectively.
In presence  of the  Gribov term the  modified  thermal quark  mass can also be obtained  as~\cite{nansu1}
\bea
m_q^2(\gamma_G) = \frac{g^2C_F}{4\pi^2}\sum_\pm\int\limits_0^\infty dk \, \frac{k^2}{E_\pm^0} \, \tilde{n}_\pm(k,\gamma_G).
\label{tmass}
\eea

\subsubsection{Quark propagator and dispersion}
\label{quark_prop_gz}
The effective quark propagator is an important ingredient
 for  computing  various properties of a hot and dense QGP using (semi-)perturbative methods, . 
 Using the modified quark self-energy given in (\ref{modified_quark_self}), 
it would now be convenient to obtain 
the  effective quark propagator with the Gribov term.   The resummed 
quark propagator in \eqref{gse23}  can now be rearranged as
\bea
S^{\star-1}(P)&=&{P\!\!\!\! \slash -\Sigma(P)}\nn\\
&=& \frac{1}{2} (\gamma_0 + {\vec \gamma}\cdot \bm{\hat  p}) {\cal D}_+ + \frac{1}{2} (\gamma_0 - {\vec \gamma}\cdot \bm{\hat { p}}) {\cal D}_- \nn \\
&=& \gamma_0 \, {\cal A}_0 - {\vec \gamma}\cdot \bm{\hat  { p}}\, {\cal A}_s \, , \label{resum_prop}
\eea
where
\begin{subequations}
 \begin{align}
{\cal A}_0&= \frac{1}{2} \left ({\cal D}_+ + {\cal D}_-\right )  \, , \label{gz1} \\
{\cal A}_s&= \frac{1}{2} \left ({\cal D}_- - {\cal D}_+\right )  \, .\label{gz2}
\end{align}
\end{subequations}
%
%
${\cal A}_0$ and ${\cal A}_s$ are obtained
within the HTL approximation as~\cite{nansu1,BHMS}
\begin{subequations}
 \begin{align}
{\cal A}_0(\omega,p) &= \omega-\frac{2g^2C_F}{(2\pi)^2}\sum_\pm\int dk \, k \, \tilde{n}_\pm(k,\gamma_G) 
\left[Q_0(\tilde{\omega}_1^\pm,p)+Q_0(\tilde{\omega}_2^\pm,p)\right], \label{a0} \\
{\cal A}_s(\omega,p) &= p+\frac{2g^2C_F}{(2\pi)^2}\sum_\pm\int dk \, k \, \tilde{n}_\pm(k,\gamma_G) 
\left[Q_1(\tilde{\omega}_1^\pm,p)+Q_1(\tilde{\omega}_2^\pm,p)\right] . \label{as}
\end{align}
\end{subequations}
The shifted frequencies are defined here as $\tilde{\omega}_1^\pm \equiv E_\pm^0(\omega+k-E_\pm^0)/k$ and 
$\tilde{\omega}_2^\pm \equiv E_\pm^0(\omega-k+E_\pm^0)/k$. The Legendre functions of 
the second kind, $Q_0$ and $Q_1$, are given as
\begin{subequations}
 \begin{align}
Q_0(\omega,p) &=Q_0\left(\frac{\om}{p}\right)\equiv \frac{1}{2p}\ln \frac{\frac{\omega}{p}+1}{\frac{\omega}{p}-1}, \label{legd0} \\
Q_1(\omega,p) &=Q_1\left(\frac{\om}{p}\right) \equiv \frac{1}{p}\left[1-\omega Q_0\left(\frac{\omega}{p}\right)\right]. \label{legd}
\end{align}
\end{subequations}

Following \eqref{gse24}  the effective quark propagator in helicity representation can also be written  as 
\bea
 S^*(P) &=& \frac{1}{2} \frac{(\gamma_0 - {\vec \gamma}\cdot\bm{\hat p})}{{\cal D}_+}
+ \frac{1}{2} \frac{(\gamma_0 + {\vec \gamma}\cdot \bm{\hat p})}{{\cal D}_-} , \label{hprop}
\eea
where ${\cal D}_\pm$ are obtained as
\begin{subequations}
 \begin{align}
{\cal D}_+(\omega,p,\gamma_G) &= {\cal A}_0(\omega,p) - {\cal A}_s(\omega,p)\nn 
= \omega - p - \frac{2g^2C_F}{(2\pi)^2}\sum_\pm\int dk k \tilde{n}_\pm(k,\gamma_G) \nn \\
& \hspace{2cm} \times 
\left[Q_0(\tilde{\omega}_1^\pm,p)+ Q_1(\tilde{\omega}_1^\pm,p)
+Q_0(\tilde{\omega}_2^\pm,p)+ Q_1(\tilde{\omega}_2^\pm,p)\right] , \label{dpm}\\
{\cal D}_-(\omega,p,\gamma_G) &= {\cal A}_0(\omega,p) + {\cal A}_s(\omega,p)\nn 
= \omega + p - \frac{2g^2C_F}{(2\pi)^2}\sum_\pm\int dk k \tilde{n}_\pm(k,\gamma_G) \nonumber \\
& \hspace{2cm} \times 
\left[Q_0(\tilde{\omega}_1^\pm,p)- Q_1(\tilde{\omega}_1^\pm,p)
+Q_0(\tilde{\omega}_2^\pm,p)- Q_1(\tilde{\omega}_2^\pm,p)\right]. \label{dpm1}
\end{align}
\end{subequations}
\begin{figure}[t]
\begin{center}
\includegraphics[width=0.32\linewidth]{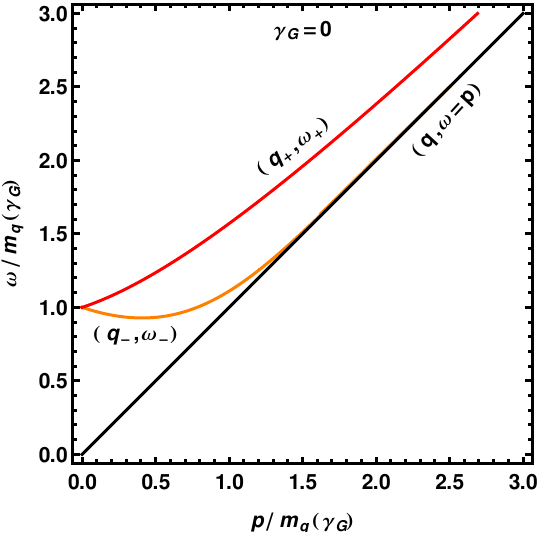}
\includegraphics[width=0.32\linewidth]{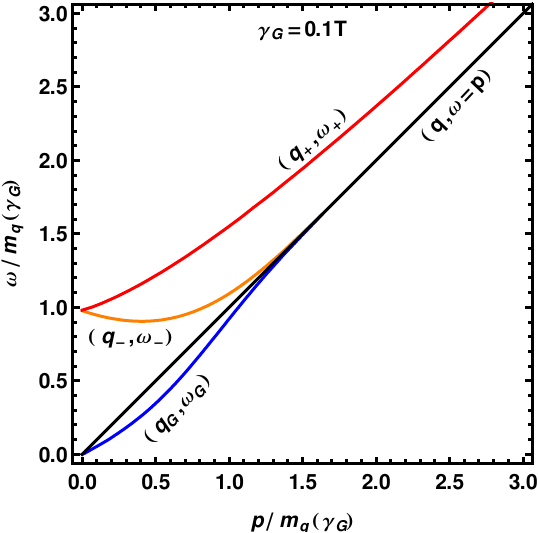}
\includegraphics[width=0.32\linewidth]{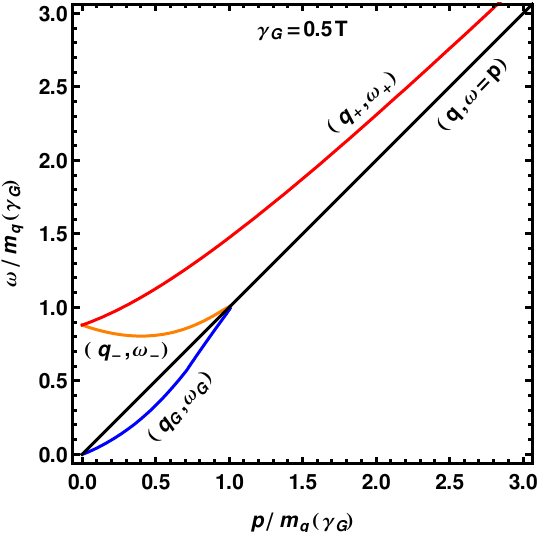} 
\end{center}
\caption{Plot of the dispersion relations for different values of $\gamma_G$. In 
the parenthesis, the first one represents a collective excitation mode whereas 
the second one is the corresponding energy of that mode. These figures are taken from Ref.~\cite{BHMS}.}
\label{disp_rel}
\end{figure}
The zeros of ${\cal D}_{\pm}(\om,p,\gamma_G)$ correspond to the dispersion 
relations for the collective excitations in the non-perturbative medium.  In Fig.~\ref{disp_rel} 
the  dispersion relations are displayed for three values of  $\gamma_G$.  For HTL case when  $\gamma_G=0$, 
one gets two massive quasiparticle modes.  One is a normal quark mode 
$q_+$ with energy $\omega_+$ and another one is a long wavelength plasmino mode $q_-$ with 
energy $\omega_-$. They are displayed  in Fig.~\ref{disp_plot} also in the left panel of Fig.~\ref{disp_rel}. 
The $q_-$ mode  has a minimum and then it quickly approaches to the non-interacting massless mode in the 
high-momentum limit.  The minimum in $q_-$ mode (plasmino mode) leads to Van Hove singularities in 
soft dilepton production rate~\cite{BHMS} which will be discussed in subsec~\ref{DP_GZ}. 
In presence of the $\gamma_G$, there appears a new
massless spacelike mode $q_G$ with energy $\omega_G$, in addition to the two massive 
modes, $q_+$ and $q_-$~\cite{nansu1} as shown in the middle and in the 
right panel of Fig.~\ref{disp_rel}. This new spacelike massless mode $q_G$ in spacelike domain
is due to the inclusion of the magnetic  scale through the GZ action. 
It becomes 
lightlike at large momentum as can also be seen from the middle and the right panel of Fig.~\ref{disp_rel}.  
The existence of this extra spacelike 
mode could affect lattice calculations of the dilepton rate because  the 
recent lQCD results~\cite{kitazawa1,kitazawa2} considered that there were only two 
poles of the in-medium propagator leading to a quark mode and a plasmino mode motivated by the HTL approximation.

\begin{figure}[t]
\begin{center}
\includegraphics[width=0.4\linewidth]{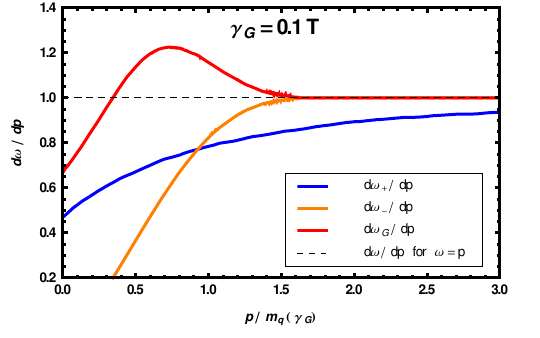}
\includegraphics[width=0.4\linewidth]{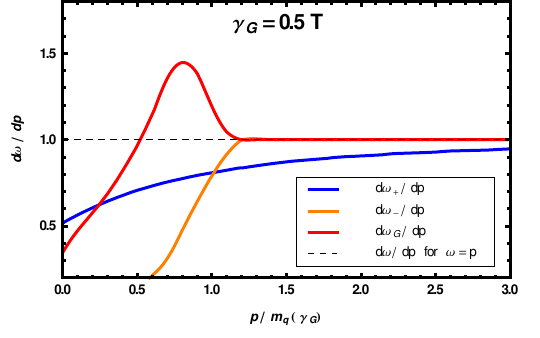}
\end{center}
\caption{Plot of the group velocity for different values of $\gamma_G$. The group velocity for the space like
Gribov mode $d\om_G/dp$ becomes superluminal, as can be seen from both plots.}
\label{group_vel}
\end{figure}

It is also to be noted that the slope of the 
dispersion curve for the new
massless  spacelike mode $q_G$ exceeds unity in some domain of momentum. 
This indicates that the group velocity, $d\omega_G/dp$, of the new mode is superluminal,
and then it approaches to the light cone ($d\omega/dp=1$)  from above as shown in Fig.~\ref{group_vel}.
Since the mode is spacelike,  there is no causality problem but could be termed 
as anomalous dispersion because  the presence $\gamma_G$  converts the Landau damping  in the spacelike domain 
into amplification of a massless spacelike dispersive mode. 

\subsubsection{Spectral representation of the quark propagator}
\label{spectral_gz}

 In absence of Gribov parameter ($\gamma_G=0$), i.e., in the  HTL approximation apart from poles
 the propagator contains a discontinuity in 
complex plane originating from the logarithmic terms in (\ref{dpm}) and \eqref{dpm1} due to 
spacelike momentum $\omega^2<p^2$.  The HTL spectral function contains  contributions from two collective excitations 
and the  Landau cut  as discussed in subsec.~\ref{spec_prop}. 
On the other hand, for $\gamma_G \neq 0$ there are three collective excitations $q_+$, $q_-$ and $q_G$, 
and no Landau cut contributions in the complex plane due to the fact that the 
poles come in complex-conjugate pairs and ultimately cancel out.
It seems that the Landau cut contribution in spacelike domain  for $\gamma_G=0$ is converted into a new 
massless spacelike dispersive mode in presence of  magnetic scale ($\gamma_G\ne 
0$). Since there is no Landau cut contribution, the spectral representation of the quark propagator ${\cal D}_{\pm}^{-1}$  
for $\gamma_G \neq 0$ has only pole contributions and obtained following subsec.~\ref{spec_prop} or
 \eqref{bpy_04} in appendix~\ref{bpy_disc} as~\cite{BHMS}

\bea
\rho_\pm^G(\omega,p)= \frac{\omega^2-p^2}{2m_q^2(\gamma_G)}\left[\delta(\omega\mp\omega_+)+\delta(\omega\pm\omega_-)
+\delta(\omega\pm\omega_G)\right], \label{gspect}
\eea
where  ${\cal D}_+$ has poles at $\omega_+$, $-\omega_-$, and $-\omega_G$ 
and ${\cal D}_-$ has poles at $\omega_-$, $-\omega_+$, and $\omega_G$ with a prefactor,  $(\omega^2-p^2)/2m_q^2(\gamma_G)$,  as the residue.  

\subsubsection{Quark-Photon vertex}
\label{qp_vert_gz}
The quark-photon three-point vertex can be obtained~\cite{BHMS} by using the Ward-Takahashi 
identity~\footnote{This procedure only constrains the longitudinal part of the vertex function.}
as
\bea
(P_1-P_2)_\mu\Gamma^\mu(P_1,P_2)= S^{-1}(P_1)-S^{-1}(P_2) \ .
\label{ward_id}
\eea
The temporal and spatial parts of the modified effective vertex can be written as
\begin{subequations}
\begin{align}
\Gamma^0 &= a_G ~\gamma^0 +b_G~\bm{\gamma \cdot \hat{p}},\label{vertex_0}\\
\Gamma^i &= c_G ~\gamma^i +b_G~\hat{p}^i\gamma_0+d_G~\hat{p}^i\left(\bm{\gamma \cdot \hat{p}}\right), \label{vertex_i}
\end{align}
\end{subequations}
where the coefficients are given by
\begin{subequations}
\begin{align}
a_G &= 1-\frac{2g^2C_F}{(2\pi)^2}\sum_\pm\int dk \, k \, \tilde{n}_\pm(k,\gamma_G)\frac{1}{\omega_1-\omega_2}
\left[\delta Q_{01}^\pm+\delta Q_{02}^\pm\right],\label{coeff_1}\\
b_G &= -\frac{2g^2C_F}{(2\pi)^2}\sum_\pm\int dk \, k \, \tilde{n}_\pm(k,\gamma_G)\frac{1}{\omega_1-\omega_2}
\left[\delta Q_{11}^\pm+\delta Q_{12}^\pm\right],\label{coeff_2}\\
c_G &= 1+\frac{2g^2C_F}{(2\pi)^2}\sum_\pm\int dk \, k \, \tilde{n}_\pm(k,\gamma_G)\frac{1}{3(\omega_1-\omega_2)}
\left[\delta Q_{01}^\pm+\delta Q_{02}^\pm-\delta Q_{21}^\pm-\delta Q_{22}^\pm\right],\label{coeff_3}\\
d_G &= \frac{2g^2C_F}{(2\pi)^2}\sum_\pm\int dk \, k \, \tilde{n}_\pm(k,\gamma_G)\frac{1}{\omega_1-\omega_2}
\left[\delta Q_{21}^\pm+\delta Q_{22}^\pm\right],\label{coeff_4}
\end{align}
\end{subequations}
with
\begin{subequations}
\begin{align}
\delta Q_{n1}^\pm &= Q_n(\tilde{\omega}_{11}^\pm,p)- Q_n(\tilde{\omega}_{21}^\pm,p){\rm{~for~}} n=0,1,2 \, \, ,\label{legd_1}\\
\omega_{m1}^\pm &= E_\pm^0(\omega_m+k-E_\pm^0)/k {\rm{~for~}} m=1,2\, \, , \label{legd_2}\\
\omega_{m2}^\pm &= E_\pm^0(\omega_m-k+E_\pm^0)/k {\rm{~for~}} m=1,2\,\, \label{legd_3}
\end{align}
\end{subequations}
Similarly, the quark-photon four-point function can be obtained from the following generalized Ward-Takahashi identity
\bea
P_\mu\Gamma^{\mu\nu}(-P_1,P_1;-P_2,P_2) = \Gamma^{\nu}(P_1-P_2,-P_1;P_2)-\Gamma^{\nu}(-P_1-P_2,P_1;P_2) \ . \label{wi_4pt}
\eea

\subsection{Dilepton Production Rate from QGP}
\label{DPQ}

Thermal dileptons ($q{\bar q}\rightarrow \gamma^*\rightarrow l^+l^-$, where $q$ and $\bar q$ are (anti)quark, $\gamma^*$ is virtual photon and $l^+l^-$ are lepton pair) 
emitted from the fireball in ultrarelativistic heavy ion collisions might serve as a promising signature~\cite{Ruuskanen:1991au} for the QGP formation 
in such collisions. In contrast to hadronic signals dileptons and photons carry direct information about the early phase of the fireball, 
since they do not interact with the surrounding medium after their production. Therefore, they can be used as a direct probe for the QGP. 
Unfortunately there is a huge background coming from hadronic decays. Hence it would be desirable to have some specific features in the dilepton spectrum 
which could signal the presence of deconfined matter. Indeed perturbative calculations~\cite{BPY,KMT} have shown distinct structures 
(van Hove peaks~\cite{VH,AM}, gaps) in the production rate of low mass dileptons caused by non-trivial in-medium quark dispersion relations. In the following 
subsec~\ref{dil_th} we briefly discuss  the dilepton production rate from a thermal medium.

\subsubsection{ Dilepton rate in presence of thermal medium}
\label{dil_th}
The dilepton multiplicity per unit space-time volume is given~\cite{Weldon:1990iw} as
\bea
\frac{dN}{d^4X}&=& 2\pi e^2 e^{-\beta 
p_0}L_{\mu\nu}\rho^{\mu\nu}\frac{d^3 \bm{\vec q}_1}{(2\pi)^3E_1}\frac{d^3\bm{\vec q}_2}{(2\pi)^3E_2}, \label{d1}
\eea
where where $e$ is the electromagnetic coupling, $\bm{\vec q}_i$ and $E_i$ with $i=1,2$ are three momentum and energy  of the lepton pairs.
The photonic tensor or the electromagnetic spectral function in thermal medium can be written as
\bea
\rho^{\mu\nu}(p_0,\bm{\vec p}) &=& -\frac{1}{\pi}\frac{e^{\beta p_0}}{e^{\beta 
p_0}-1}\textrm{Im}\left[D^{\mu\nu}(p_0,\bm{\vec p})\right]\equiv 
-\frac{1}{\pi}\frac{e^{\beta p_0}}{e^{\beta 
p_0}-1}~\frac{1}{P^4}\textrm{Im}\left[\Pi^{\mu\nu}(p_0,\bm{\vec  p})\right], \label{d2}
\eea
where $`\textrm{Im}$' stands for imaginary part,  $\Pi^{\mu\nu}$ is the two point current-current correlation function or the self-energy of photon and 
$D^{\mu\nu}$ represents the photon propagator. 
Here we have used the  relation~\cite{Weldon:1990iw} 
 \bea
D^{\mu\nu}(p_0,\bm{\vec p}) = \frac{1}{P^4}\Pi^{\mu\nu}(p_0,\bm{\vec p}) \, ,
\label{d2i}
 \eea
where $P \equiv (p_0,\bm{\vec p})$ is the four momenta of the photon. 

Also  the leptonic tensor in terms of Dirac spinors is given by
\bea
L_{\mu\nu} &=& \frac{1}{4} 
\sum\limits_{\mathrm{spins}}\mathrm{Tr}\left[\bar{u}(Q_2)\gamma_\mu 
v(Q_1)\bar{v}(Q_1)\gamma_\nu u(Q_2)\right] \nn \\
&=& Q_{1\mu}Q_{2\nu}+Q_{1\nu}Q_{2\mu}-(Q_1\cdot Q_2+m_l^2)g_{\mu\nu},
\label{d3}
\eea
where $Q_i\equiv (q_0, \bm{\vec q}_i)$ is the four momentum of the $i$-th lepton and $m_l$ is the mass of the lepton.

Now inserting $\int d^4P\, \delta^4(Q_1+Q_2-P)=1$, one can write the dilepton 
multiplicity from \eqref{d1} as
\bea
\frac{dN}{d^4X}\
&=& 2\pi e^2 e^{-\beta p_0}\int d^4P \,  \delta^4(Q_1+Q_2-P)
L_{\mu\nu}\rho^{\mu\nu}\frac{d^3 \bm{\vec q}_1}{(2\pi)^3E_1}\frac{d^3\bm{\vec q}_2}{(2\pi)^3E_2}. \label{d4} 
\eea
Using the identity
\bea
\int\frac{d^3 \bm{\vec q}_1}{E_1}\frac{d^3 \bm{\vec q}_2}{E_2}
\delta^4(Q_1+Q_2-P)\ L_{\mu\nu} &=& 
\frac{2\pi}{3} 
\left(1+\frac{2m_l^2}{P^2}\right)\sqrt{\left(1-\frac{4m_l^2}{P^2}\right)}
\left(P_\mu P_\nu-P^2g_{\mu\nu}\right) \nn\\
&=&\frac{2\pi}{3}F_1(m_l,P^2)\left(P_\mu P_\nu-P^2g_{\mu\nu}\right), \label{d5}
\eea
the dilepton production rate in \eqref{d4} comes out to be
\bea
\frac{dN}{d^4Xd^4P} = \frac{dR}{d^4P} &=&
\frac{\alpha} {12\pi^4}\frac{n_B(p_0)}{P^2} F_1(m_l,P^2) \  \textrm{Im} \left [\Pi^{\mu}_{\mu}(p_0, \bm{\vec p})\right] \nn\\
\frac{dR}{d^4P} &=& \frac{\alpha} {12\pi^4}\frac{n_B(p_0)}{P^2} F_1(m_l,P^2) \  \frac{1}{2i} {\textrm{Disc}} \left [\Pi^{\mu}_{\mu}(p_0, \bm{\vec p})\right]  , \label{d6}
\eea 
where $n_B(p_0)=(e^{p_0/T}-1)^{-1}$ and $e^2=4\pi\alpha$, $\alpha$ is the electromagnetic coupling constant. We have also used the transversality 
condition $P_\mu\Pi^{\mu\nu}=0$. The invariant mass of the lepton pair is defined as $M^2\equiv P^2(=p_0^2-|\bm{\vec p}|^2=\omega^2-|\bm{\vec p}|^2)$. 
We note that for massless lepton ($m_l=0$) $F_1(m_l,P^2)=1$.

The \eqref{d6} is the familiar result most widely used for the dilepton emission rate from a thermal medium. 
It must be emphasized that this relation is valid only to ${\cal O}(e^2)$ since it does not account for the possible reinteractions of the virtual photon 
on its way out of the thermal bath. The possibility of emission of more than one photon has also been neglected here. 
However, the expression is true to all orders in strong interaction.

\subsubsection{Dilepton production rate from QGP with gluon condensate}
\label{DP_GC}
\begin{figure}[htb]
\begin{center}
\includegraphics[height=0.3\linewidth,width=0.6\linewidth]{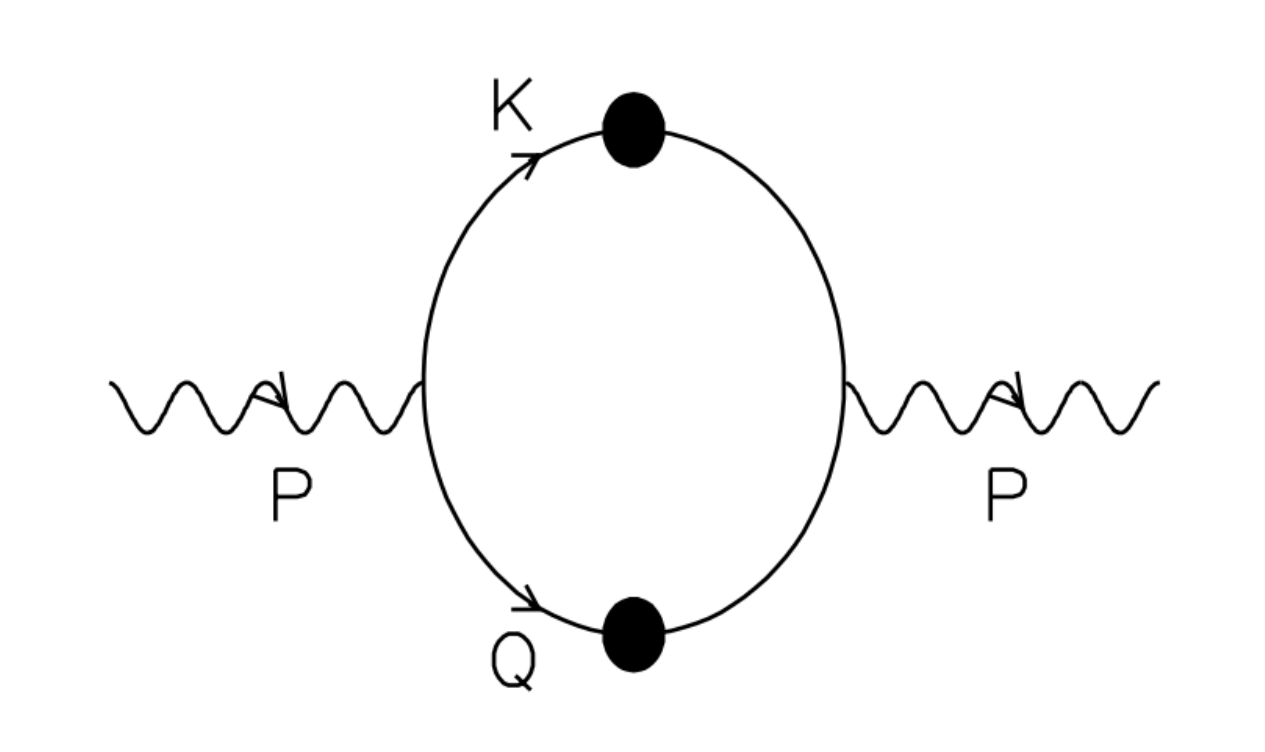}
\end{center}
\caption{One-loop photon self-energy with effective quark propagators containing gluon condensate. This figure is taken from Ref.~\cite{MST}.}
\label{phot_gc}
\end{figure}

In this subsection we calculate the effect of an in-medium gluon condensates, as discussed in subsec~\ref{gcond}, on the  production rate of lepton pairs from QGP~\cite{MST}. 
This effect can be included by using effective propagators, $S^*$,  as given in~\eqref{gcond19} containing the gluon condensate for the exchanged quarks
in photon self-energy in Fig.~\ref{phot_gc}. The photon self energy in Fig.~\ref{phot_gc} can now be 
written as
\begin{equation}
\Pi^{\mu\nu}(P) = - 3\times 2 \times \frac{5}{9} e^2 T\sum_{\{k_0\}} \int 
\frac{d^3k}{(2\pi)^3} {{\rm Tr}} \left [ S^*(K)\gamma^\mu
S^*(Q) \gamma^\nu \right ] , \label{impimunu}
\end{equation}
where $\sum_{\{k_0\}}$ is the frequency sum over fermionic Matsubara frequency,  $K$ and $Q=P-K$ are the fermionic loop four-momenta.
We have considered only massless $u$ and $d$ quarks and the total electric charge of two flavours is 
$(\frac{2}{3})^2e^2 + (\frac{1}{3})^2e^2 =\frac{5}{9}e^2$, 
the factor $2$ is for antiquarks and the color factor of quark is $3$.

Substitution of (\ref{gcond19}) in (\ref{impimunu}) and performing the
traces  one gets
\begin{eqnarray}
{\Pi}_\mu^\mu (P) &=& -\frac{10}{3} e^2 T \sum_{\{k_0\}} \int \frac{ d^3k}
{(2\pi)^3}
\left [ \frac{1}{D_{+}(K)}\left ( \frac{1-\bm{\hat k \cdot \hat q}}{D_{+}(Q)} +
\frac{1+\bm{\hat k \cdot \hat q}}{D_{-}(Q)} \right ) \right. \nonumber \\
&& + \left. \frac{1}{D_{-}(K)}\left ( \frac{1+\bm{\hat k \cdot \hat q}}{D_{+}(Q)} +
\frac{1-\bm{\hat k \cdot \hat q}}{D_{-}(Q)} \right ) \right ] \ . \label{trimpi}
\end{eqnarray}
Now according to \eqref{d6} one needs to compute the imaginary or discontinuity part of $\Pi^\mu_\mu(P)$.  The discontinuity can be obtained by using 
the Braaten-Pisarski-Yuan (BPY) prescription~\cite{BPY} obtained in \eqref{bpy12} in appendix~\ref{bpy_disc} as
\bea
\textrm{Im}  \, T\sum_{k_0}F_1(k_0)F_2(q_0)&=& \frac{1}{2i} \textmd{Disc~}T\sum_{k_0}F_1(k_0)F_2(q_0)\nn \\
&=&  \pi  (1-e^{\beta \omega})\int d\omega_1 \int d\omega_2~n_F(\omega_1)n_F(\omega_2)\nn \\
&& \times \ \delta(\omega-\omega_1-\omega_2) \ \rho_1(\omega_1)\rho_2(\omega_2), \label{bpy_pres}
\eea
where $\delta(\omega-\omega_1-\omega_2) $ is the energy conserving $\delta$-function,  $n_F$ is the Fermi-Dirac distribution function and $\rho_1$  and $\rho_2$ are the
spectral functions corresponding to the functions $F_1$ and $F_2$.

Now using \eqref{bpy_pres} one can write the imaginary part of $\Pi^\mu_\mu$ as
\begin{eqnarray}
\textrm{Im} \,\Pi_\mu^\mu(P) &=& \frac{10\pi}{3} e^2 \left (e^{E/T}-1 \right )
\int \frac {d^3 k}{(2\pi)^3} \int_{-\infty}^\infty  d\omega  
\int_{-\infty}^\infty  d\omega' \ \nonumber \\
&& \times \delta \left ( E- \omega - \omega' \right ) 
n_F(\omega) n_F(\omega') \nonumber \\
&& \times \left [ \left (1+\bm{\hat q \cdot 
\hat k}\right ) \left \{ \rho^{+} \left ( \omega, k \right ) \rho^{-}
\left ( \omega', q \right ) 
+\rho^{-} \left ( \omega, k \right ) \rho^{+}\left ( \omega', q 
\right ) \right \} \right. \nonumber \\
&& \left. + \left (1- \bm{\hat q \cdot
\hat k}\right ) \left \{ \rho^{+} \left ( \omega, k \right ) \rho^{+}
\left ( \omega', q \right )
+\rho^{-} \left ( \omega, k \right ) \rho^{-}\left ( \omega', q
\right ) \right \} \right ], \label{impi}
\end{eqnarray}
where $\rho^\pm$ are the spectral functions corresponding to $1/{\cal D}_\pm (L)$ and obtained in \eqref{gcond26}.
Inserting (\ref{gcond26}) into (\ref{impi}) and performing the 
$\omega$-integrations by exploiting the delta functions 
of the spectral functions, one finds ($x=\bm{\hat p\cdot \hat k}$)
\begin{eqnarray}
\textrm{Im} \,\Pi_\mu^\mu(P) &=&  \frac{5}{6\pi} e^2 \left (e^{E/T}-1 \right )
\int_0^\infty  d k\, k^2 \int_{-1}^{+1} d x \nonumber \\
&& \ \ \  \times \left [ \left (1 \ + \bm {\hat q \cdot \hat k}\right ) A \ + 
\ \left (  1 \ - \bm{\hat q \cdot \hat k}\right ) B \right ], \ \label{impi1}
\end{eqnarray}
where
\begin{eqnarray}
A &=& n_F\left (\omega^{+}(k)\right )  n_F\left (\omega^{-}(q)\right )
R_{+}\left (\omega^{+}(k),k\right ) R_{-}\left (\omega^{-}(q),q\right )
\delta \left (E-\omega^{+}(k) -\omega^{-}(q) \right ) \nonumber \\
&+& n_F\left (-\omega^{-}(k)\right )  n_F\left (\omega^{-}(q)\right )
R_{-}\left (\omega^{-}(k),k\right ) R_{-}\left (\omega^{-}(q),q\right )
\delta \left (E+\omega^{-}(k) -\omega^{-}(q) \right ) \nonumber \\
&+& n_F\left (\omega^{+}(k)\right )  n_F\left (-\omega^{+}(q)\right )
R_{+}\left (\omega^{+}(k),k\right ) R_{+}\left (\omega^{+}(q),q\right )
\delta \left (E-\omega^{+}(k) +\omega^{+}(q) \right ) \nonumber \\
&+& n_F\left (-\omega^{-}(k)\right )  n_F\left (-\omega^{+}(q)\right )
R_{-}\left (\omega^{-}(k),k\right ) R_{+}\left (\omega^{+}(q),q\right )
\delta \left (E+\omega^{-}(k) +\omega^{+}(q) \right ) \nonumber \\
&+& n_F\left (\omega^{-}(k)\right )  n_F\left (\omega^{+}(q)\right )
R_{-}\left (\omega^{-}(k),k\right ) R_{+}\left (\omega^{+}(q),q\right )
\delta \left (E-\omega^{-}(k) -\omega^{+}(q) \right ) \nonumber \\
&+& n_F\left (-\omega^{+}(k)\right )  n_F\left (\omega^{+}(q)\right )
R_{+}\left (\omega^{+}(k),k\right ) R_{+}\left (\omega^{+}(q),q\right )
\delta \left (E+\omega^{+}(k) -\omega^{+}(q) \right ) \nonumber \\
&+& n_F\left (\omega^{-}(k)\right )  n_F\left (-\omega^{-}(q)\right )
R_{-}\left (\omega^{-}(k),k\right ) R_{-}\left (\omega^{-}(q),q\right )
\delta \left (E-\omega^{-}(k) +\omega^{-}(q) \right ) \nonumber \\
&+& n_F\left (-\omega^{+}(k)\right )  n_F\left (-\omega^{-}(q)\right )
R_{+}\left (\omega^{+}(k),k\right ) R_{-}\left (\omega^{-}(q),q\right )
\delta \left (E+\omega^{+}(k) +\omega^{-}(q) \right ),\nonumber \\
\label{ca}
\end{eqnarray}
and
\begin{eqnarray}
B &=& n_F\left (\omega^{+}(k)\right )  n_F\left (\omega^{+}(q)\right )
R_{+}\left (\omega^{+}(k),k\right ) R_{+}\left (\omega^{+}(q),q\right )
\delta \left (E-\omega^{+}(k) -\omega^{+}(q) \right ) \nonumber \\
&+& n_F\left (-\omega^{-}(k)\right )  n_F\left (\omega^{+}(q)\right )
R_{-}\left (\omega^{-}(k),k\right ) R_{+}\left (\omega^{+}(q),q\right )
\delta \left (E+\omega^{-}(k) -\omega^{+}(q) \right ) \nonumber \\
&+& n_F\left (\omega^{+}(k)\right )  n_F\left (-\omega^{-}(q)\right )
R_{+}\left (\omega^{+}(k),k\right ) R_{-}\left (\omega^{-}(q),q\right )
\delta \left (E-\omega^{+}(k) +\omega^{-}(q) \right ) \nonumber \\
&+& n_F\left (-\omega^{-}(k)\right )  n_F\left (-\omega^{-}(q)\right )
R_{-}\left (\omega^{-}(k),k\right ) R_{-}\left (\omega^{-}(q),q\right )
\delta \left (E+\omega^{-}(k) +\omega^{-}(q) \right ) \nonumber \\
&+& n_F\left (\omega^{-}(k)\right )  n_F\left (\omega^{-}(q)\right )
R_{-}\left (\omega^{-}(k),k\right ) R_{-}\left (\omega^{-}(q),q\right )
\delta \left (E-\omega^{-}(k) -\omega^{-}(q) \right ) \nonumber \\
&+& n_F\left (-\omega^{+}(k)\right )  n_F\left (\omega^{-}(q)\right )
R_{+}\left (\omega^{+}(k),k\right ) R_{-}\left (\omega^{-}(q),q\right )
\delta \left (E+\omega^{+}(k) -\omega^{-}(q) \right ) \nonumber \\
&+& n_F\left (\omega^{-}(k)\right )  n_F\left (-\omega^{+}(q)\right )
R_{-}\left (\omega^{-}(k),k\right ) R_{+}\left (\omega^{+}(q),q\right )
\delta \left (E-\omega^{-}(k) +\omega^{+}(q) \right ) \nonumber \\
&+& n_F\left (-\omega^{+}(k)\right )  n_F\left (-\omega^{+}(q)\right )
R_{+}\left (\omega^{+}(k),k\right ) R_{+}\left (\omega^{+}(q),q\right )
\delta \left (E+\omega^{+}(k) +\omega^{+}(q) \right ). \nonumber \\
\label{cb}
\end{eqnarray}
Changing the integration variable from $x$ to 
$q=|{\vec p}-{\vec k}| = \sqrt {p^2+k^2 -2pk x}$
the dilepton production rate in (\ref{d6}) with massless leptons can be written as 
\begin{eqnarray}
\frac{dN}{d^4X d ^4P} &=& \frac{5}{18\pi^4}\frac 
{\alpha^2}{M^2} \frac{1}{p} 
\left [ \int_0^p  dk \int_{p-k}^{p+k} dq 
+ \int_p^\infty  dk \int_{k-p}^{p+k} dq \right ] \nonumber \\
&& \times \left [ \left (p^2 - (k-q)^2 \right ) A + 
\left ( (k+q)^2 -p^2 \right ) B\right ] .\label{term1} 
\end{eqnarray}
where the invariant mass of the dilepton is $M^2\equiv P^2(=p_0^2-|\bm{\vec p}|^2=E^2-|\bm{\vec p}|^2)$, where $E$ is the photon energy.

Now one can perform the $q$-integration by means of the remaining 
$\delta$-functions in $A$ and $B$
leading to
\begin{eqnarray}
\frac{ dN}{d^4X d^4P} &=& \frac{5}{18\pi^4}\frac 
{\alpha^2}{M^2} \frac{1}{p} \int_0^\infty  dk  
\left [\left (p^2 - (k-q_s)^2 \right ) 
\left ( A_1 + A_2 + A_3 + A_5 + A_6 + A_7 \right ) \right.\nonumber \\
&& \left. + \left ( (k+q_s)^2 -p^2 \right ) 
\left (B_1 + B_2 + B_3 + B_5 + B_6 +B_7 \right )\right ]_{|p-k|\le q_s \le 
p+k}\> , \label{term2} 
\end{eqnarray}
where the $q_s$ determined by the various $\delta $-functions in (\ref{ca}) 
and (\ref{cb})
can assume two different values in the case of the plasmino branch due to the
presence of the minimum and
\begin{eqnarray}
A_1 &=& n_F\left (\omega^{+}(k)\right )  n_F\left (\omega^{-}(q_s)\right )
R_{+}\left (\omega^{+}(k),k\right ) 
\frac{R_{-}\left (\omega^{-}(q_s),q_s\right )}
{| d \omega^{-}(q)/ d q|_{q_s}}, \nonumber \\
A_2 &=& n_F\left (-\omega^{-}(k)\right )  n_F\left (\omega^{-}(q_s)\right )
R_{-}\left (\omega^{-}(k),k\right ) 
\frac {R_{-}\left (\omega^{-}(q_s),q_s\right )}
{| d \omega^{-}(q)/ d q|_{q_s}}, \nonumber \\
A_3 &=& n_F\left (\omega^{+}(k)\right )  n_F\left (-\omega^{+}(q_s)\right )
R_{+}\left (\omega^{+}(k),k\right ) 
\frac {R_{+}\left (\omega^{+}(q_s),q_s\right )}
{| d \omega^{+}(q)/ d q|_{q_s}}, \nonumber \\
A_5&=& n_F\left (\omega^{-}(k)\right )  n_F\left (\omega^{+}(q_s)\right )
R_{-}\left (\omega^{-}(k),k\right ) 
\frac{R_{+}\left (\omega^{+}(q_s),q_s\right )}
{|  d \omega^{+}(q)/ d q|_{q_s}}, \nonumber \\
A_6&=& n_F\left (-\omega^{+}(k)\right )  n_F\left (\omega^{+}(q_s)\right )
R_{+}\left (\omega^{+}(k),k\right ) 
\frac{R_{+}\left (\omega^{+}(q_s),q_s\right )}
{| d \omega^{+}(q)/d q|_{q_s}}, \nonumber \\
A_7&=& n_F\left (\omega^{-}(k)\right )  n_F\left (-\omega^{-}(q_s)\right )
R_{-}\left (\omega^{-}(k),k\right ) 
\frac{R_{-}\left (\omega^{-}(q_s),q_s\right )}
{| d\omega^{-}(q)/ d q|_{q_s}}, \nonumber \\
B_1 &=& n_F\left (\omega^{+}(k)\right )  n_F\left (\omega^{+}(q_s)\right )
R_{+}\left (\omega^{+}(k),k\right ) 
\frac {R_{+}\left (\omega^{+}(q_s),q_s\right )}
{| d \omega^{+}(q)/ d q|_{q_s}}, \nonumber \\
B_2&=& n_F\left (-\omega^{-}(k)\right )  n_F\left (\omega^{+}(q_s)\right )
R_{-}\left (\omega^{-}(k),k\right ) 
\frac{R_{+}\left (\omega^{+}(q_s),q_s\right )}
{| d \omega^{+}(q)/ d q|_{q_s}}, \nonumber \\
B_3&=& n_F\left (\omega^{+}(k)\right )  n_F\left (-\omega^{-}(q_s)\right )
R_{+}\left (\omega^{+}(k),k\right ) 
\frac{R_{-}\left (\omega^{-}(q_s),q_s\right )}
{| d \omega^{-}(q)/ d q|_{q_s}}, \nonumber \\
B_5&=& n_F\left (\omega^{-}(k)\right )  n_F\left (\omega^{-}(q_s)\right )
R_{-}\left (\omega^{-}(k),k\right ) 
\frac{R_{-}\left (\omega^{-}(q_s),q_s\right )}
{| d \omega^{-}(q)/ d q|_{q_s}}, \nonumber \\
B_6&=& n_F\left (-\omega^{+}(k)\right )  n_F\left (\omega^{-}(q_s)\right )
R_{+}\left (\omega^{+}(k),k\right ) 
\frac{R_{-}\left (\omega^{-}(q_s),q_s\right )}
{| d\omega^{-}(q)/d q |_{q_s}}, \nonumber \\
B_7&=& n_F\left (\omega^{-}(k)\right )  n_F\left (-\omega^{+}(q_s)\right )
R_{-}\left (\omega^{-}(k),k\right ) 
\frac{R_{+}\left (\omega^{+}(q_s),q_s\right )}
{| d\omega^{+}(q)/ d q|_{q_s}}, \nonumber \\
\label{cabs}
\end{eqnarray}
The group velocity factors in (\ref{cabs}) follow from the dispersion
relation, ${\cal D}_\pm(L)=0$, of (\ref{gcond20}) as 
\begin{equation}
\frac{ d\omega^\pm(l)}{ d l}
= \pm \frac{F^\pm \left (\omega^{\pm}(l),{\cal A}, {\cal B}, l\right )} 
{G^\pm \left (\omega^{\pm}(l),{\cal A}, {\cal B}, l\right )} \ , \label{s}
\end{equation}
where
\begin{eqnarray}
F^\pm&=& \left (1+{\cal A}\right )^2 \left ({\omega^{\pm}}^2(l)-l^2 \right )^3 
\ \mp \ 6{\cal B} \left ({\omega^{\pm}}^2(l)-l^2 \right )^2 l 
\ \mp \ \frac{g^2}{6} {\cal B}  \left (\frac{2}{3}\langle { E}^2\rangle_T 
-\frac{2} {5}\langle { B}^2\rangle_T \right )l \nonumber \\
&& \pm \ \frac{8}{45}g^2 \left (1+{\cal A}\right ) l \omega^{\pm}(l) 
\langle {B}^2\rangle_T \ \ , \nonumber \\
G^\pm&=& -\left (1+{\cal A} \right )C_\pm \ ,
  \label{fg}
\end{eqnarray}
and ${\cal A}$, ${\cal B}$, and $C_\pm$ are given in (\ref{gcond14}),  (\ref{gcond15}) and
(\ref{gcond28}), respectively. As we will see that 
the group velocity leads to a characteristic feature of the dilepton rate.
In (\ref{term2}) we have dropped terms $A_4$,
$A_8$, $B_4$ and $B_8$ as the corresponding $\delta$-functions in (\ref{cb})
can never be satisfied by virtue of energy conservation
since $\omega^\pm$ is always positive. Now, one
can perform the $k$-integration in (\ref{term2}) numerically, and we find 
that the terms, which satisfy the energy conservation, correspond to 
various physical processes involving two quasiparticles with different
momentum $k$ and $q$. 

The 
dilepton production rate for $\bm{\vec p}=0$ is obtained by setting  $\bm{\vec q} = -\bm{\vec k}$  in (\ref{impi1}) as
\begin{eqnarray}
\frac{dN}{d^4X d^4P}(\bm{\vec p}=0) &=& \frac{20}{9\pi^4}\frac 
{\alpha^2}{M^2} \int_0^\infty  dk\, k^2 
 \left [ n_F^2\left (\omega^{+}(k)\right ) R_{+}^2\left (\omega^{+}(k)\right) 
\delta\left (E-2\omega^{+}(k)\right ) \right. \nonumber \\ 
&+& \left. 2n_F\left (\omega^{+}(k)\right )n_F\left (-\omega^{-}(k)\right )
R_{+}\left (\omega^{+}(k)\right ) R_{-}\left (\omega^{-}(k)\right )
\delta\left (E-\omega^{+}(k)+ \omega^{-}(k)\right ) \right. \nonumber \\ 
&+& \left. 2n_F\left (\omega^{-}(k)\right )n_F\left (-\omega^{+}(k)\right )
R_{+}\left (\omega^{+}(k)\right ) R_{-}\left (\omega^{-}(k)\right )
\delta\left (E+\omega^{+}(k)- \omega^{-}(k)\right ) \right. \nonumber \\   
&+& \left. n_F^2\left (\omega^{-}(k)\right ) R_{-}^2\left (\omega^{-}(k)\right) 
\delta\left (E-2\omega^{-}(k)\right ) \right ] \ .   \label{ratep0}
\end{eqnarray}    
First we would like to discuss the dilepton production from a 
QGP at momentum $\bm{\vec p=0}$ of the virtual photon. 
 The different terms in (\ref{ratep0}) correspond to
various physical processes involving two quasiparticles ${\rm q}^+$ and ${\rm q}^-$ 
with same momentum $k$. 
The first term represents the annihilation process  ${\rm q}^+\bar {\rm q}^+\rightarrow
\gamma^*$. The second term corresponds to ${\rm q}^+\rightarrow {\rm q}^-\gamma^*$, a decay
process from a ${\rm q}^+$ mode to a plasmino plus a virtual photon. 
Energy 
conservation does not allow the process given by the third term
(${\rm q}^-\rightarrow \bar{\rm q}^+\gamma^*$).
Finally, the fourth term corresponds to a process, 
${\rm q}^-\bar{\rm q}^- \rightarrow \gamma^*$, i.e. annihilation of
plasmino modes. 

The $k$-integration in \eqref{ratep0} can be performed using the standard delta function identity
\bea
\delta(f(x))&=& \sum_i \frac{\delta(x-x_i)}{\mid \! f'(x) \!\mid_{x=x_i}}, \label{delta-prop}
\eea
where $x_i$ are the solutions of $f(x_i)=0$. After performing the $k$-integration in \eqref{ratep0} 
the expression for the dilepton rate at $\bm{\vec p}=0$ becomes
\begin{eqnarray}
\frac{ dN}{ d^4X d^4P}(\bm{\vec p}=0) &=& \frac{20}{9\pi^4}\frac 
{\alpha^2}{M^2}  \sum_{k_s} k_s^2  
\left [ n_F^2\left (\omega^{+}(k_s)\right ) 
R_{+}^2\left (\omega^{+}(k_s)\right)
\frac{1}{2}\left |\frac{ d\omega^{+}(k)}{{\rm d}k}\right |_{k_s}^{-1} 
\right. \nonumber \\
&+& \left. 2n_F\left (\omega^{+}(k_s)\right )n_F\left (-\omega^{-}(k_s)\right )
R_{+}\left (\omega^{+}(k_s)\right ) R_{-}\left (\omega^{-}(k_s)\right )
\left |\frac{ d\left (\omega^{+}(k)-\omega^{-}(k)\right )}
{ d k}\right |_{k_s}^{-1}
\right. \nonumber \\ 
&+& \left. 2n_F\left (\omega^{-}(k_s)\right )n_F\left (-\omega^{+}(k_s)\right )
R_{+}\left (\omega^{+}(k_s)\right ) R_{-}\left (\omega^{-}(k_s)\right )
\left | \frac{ d\left (\omega^{-}(k)-\omega^{+}(k)\right )}
{ d k}\right |_{k_s}^{-1}
 \right. \nonumber \\   
&+& \left. n_F^2\left (\omega^{-}(k_s)\right ) 
R_{-}^2\left (\omega^{-}(k_s)\right) 
\frac{1}{2} \left |\frac{ d\omega^{-}(k)}{ d k}\right |_{k_s}^{-1} 
\right ] \ .   \label{ratep1}
\end{eqnarray}    

\begin{figure}[htb]
\begin{center}
\includegraphics[height=0.6\linewidth,width=0.85\linewidth]{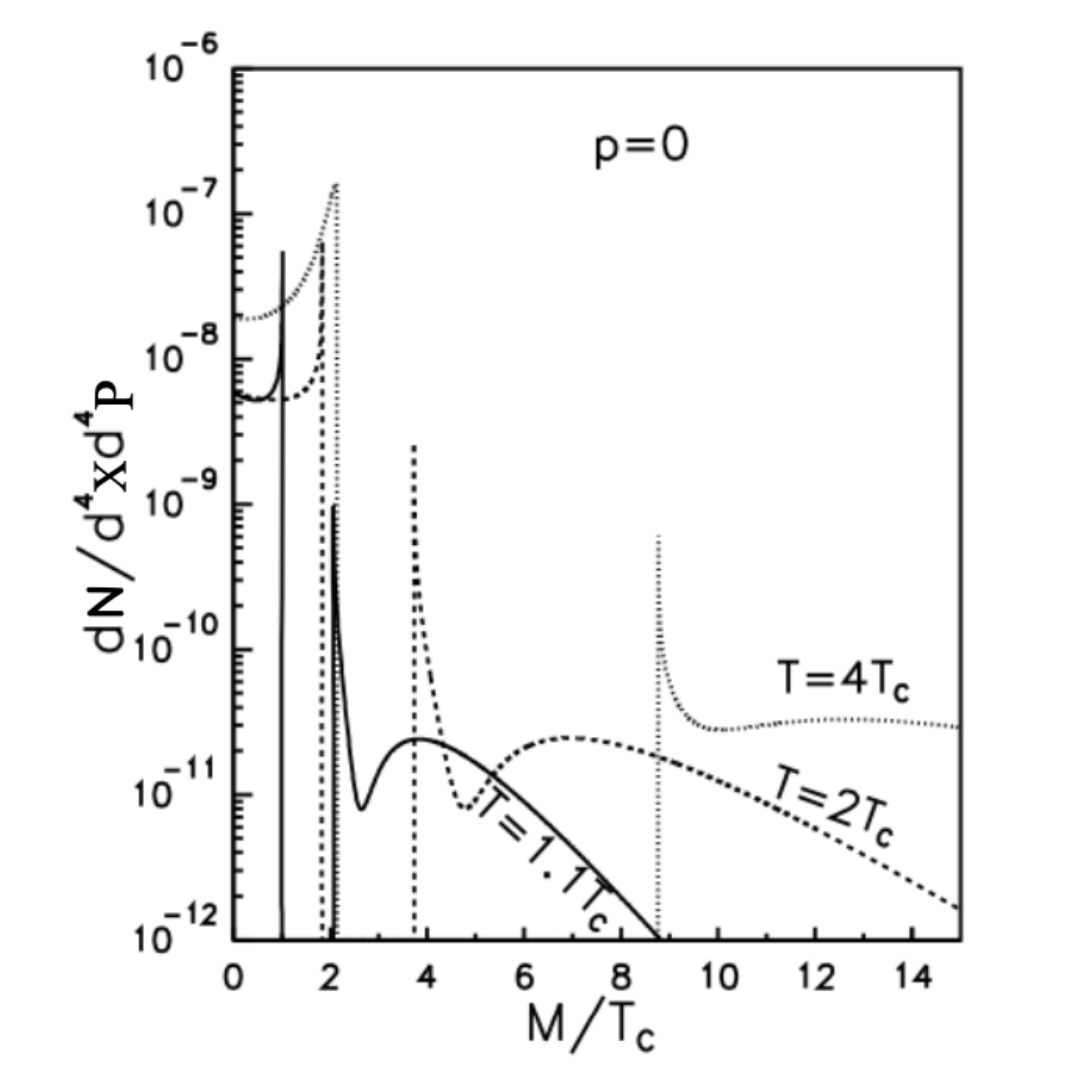}
\end{center}
\caption{The dilepton rate from QGP with gluon condensate at virtual photon momentum $\bm{\vec p}=0$. This figure is taken from Ref.~\cite{MST}.}
\label{dilep_gc_p0}
\end{figure}
The static differential rate of the aforementioned 
processes are displayed in Fig.~\ref{dilep_gc_p0} for $T=1.1T_c$ (solid line), $2 T_c$
(dashed curve) and $4T_c$ (dotted curve). Similar to the hard thermal 
loop case \cite{BPY} the differential rate in the presence of a gluon condensate also shows peaks (van Hove 
singularities~\footnote{A van Hove peak~\cite{VH,AM} appears where the density of states diverges due to the vanishing group velocity.}) 
at different invariant masses of the virtual photon. 
Now we discuss  the contributions to the rate from each process in detail.
The channel, ${\rm q}^+\rightarrow {\rm q}^-\gamma^*$, opens up at $M=0$. 
This process continues up to the first peak appears due to the vanishing
group velocity $dE/dk=0$ at the maximum $E=M=\omega^{+}(k)-\omega^-(k)$,
since the density of states is inversely proportional to the group velocity. 
The ${\rm q}^+\rightarrow {\rm q}^-\gamma^*$ channel terminates at the peak, after which 
there is a gap because neither of the other processes is possible in this
invariant mass regime. The size of the gap depends on the
temperature. For $T=1.1T_c$ it ranges from $M=1.01T_c$ to 2.07$T_c$, for 
$T=2T_c$ from $M=1.83T_c$ to 3.73$T_c$, and for $T=4T_c$  from $M=2.14T_c$ to 
8.76$T_c$.
  
The process, ${\rm q}^-\bar {\rm q}^-\rightarrow \gamma^*$, starts at an energy which is 
twice the energy of the minimum of the plasmino branch,
$E=M=2\omega_{-}(k_{min})$. The diverging density of states at that point
again causes a van Hove singularity~\cite{BPY,VH,AM}.
This process continues with increasing $M$ but
falls off very fast due to two reasons:
i) as $M$ increases the high energy plasmino modes come into the
game and the corresponding square of the residue $R_-^2(\omega^-(k),k)$,
to which the rate is proportional, becomes very
small since it is proportional to $({\omega^-}^2(k)-k^2)^6$, and
ii) with increasing $M$ the density of states decreases gradually.

At $M=E=2\omega_{+}(k)\ge 2m_{\rm{eff}}$, the process, 
${\rm q}^{+}\bar {\rm q}^+\rightarrow \gamma^*$, shows up. As $M$ increases, the
contribution from this process grows and dominates over the plasmino 
annihilation 
process, resulting in a dip in the dilepton rate. For large $M$ this 
annihilation process is solely responsible for the dilepton rate, 
which approaches the Born contribution (${\rm q}\bar {\rm q}$
annihilation of massless quarks) there
\cite{BPY,CFR}:
\begin{equation}
\frac{dN^{\rm{Born}}}{ d^4X d^4P} (\bm{\vec p}=0) 
= \frac{5}{18\pi^4} {\alpha^2} e^{-E/T}  \ . \label{born} 
\end{equation}
The reason for this is that for high energy quarks the effective propagator 
reduces to the bare one and the contribution to the dilepton rate 
comes from hard loop momenta in Fig.~\ref{dilep_gc_p0}. 

\begin{figure}[h]
\begin{center}
\includegraphics[height=0.55\linewidth,width=0.8\linewidth]{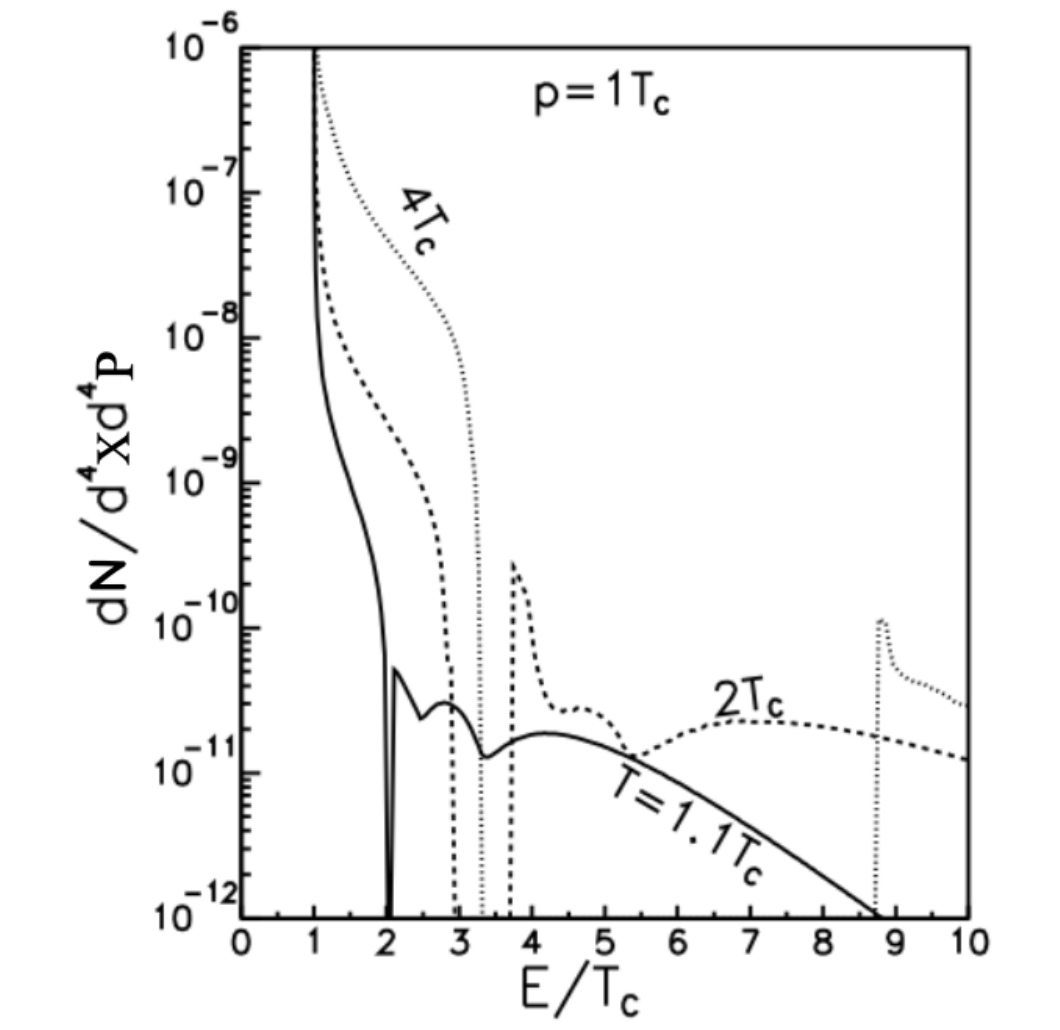}
\end{center}
\caption{The dilepton rate from QGP with gluon condensate at virtual photon momentum $\bm{\vec p}=T_c$. This figure is taken from Ref.~\cite{MST}.}
\label{dilep_gc_p1}
\end{figure}

Next,  we turn our attention to the dilepton rate at non-zero virtual photon
momentum. The corresponding rate is given in (\ref{term2}). 
The processes corresponding to terms $A_2$, $A_3$, 
$A_6$, $A_7$, $B_6$ and $B_7$, namely transitions within a branch and 
transitions from the lower to the upper branch, do not contribute to 
the rate, because they are forbidden for timelike photons decaying into
dileptons due to energy conservation \cite{SMH}.
The processes corresponding to $A_1$ and $A_5$ indicate annihilation 
between a quark (${\rm q}^+$) and a plasmino mode (${\rm q}^-$) with different momentum
to a virtual photon with energy $E$, which were absent at $\bm{\vec p}=0$. The
process given by $B_1$ is the annihilation between a quark and antiquark
(${\rm q}^+(k)\bar {\rm q}^+(q)\rightarrow \gamma^*$), whereas $B_5$ corresponds to 
the annihilation 
(${\rm q}^-(k)\bar{\rm q}^-(q)\rightarrow \gamma^*$) between two plasmino modes. 
The term $B_2$ corresponds to the
decay process, ${\rm q}^+ (q)\rightarrow {\rm q}^-(k)\gamma^*$, whereas $B_3$ corresponds to
${\rm q}_+(k)\rightarrow {\rm q}^-(q)\gamma^*$.
The differential rate involving these processes
are displayed in Fig.~\ref{dilep_gc_p1} for virtual photon momentum $p=T_c$ at different temperatures, namely $T=1.1T_c$ 
(solid line), $2T_c$ (dashed line), and $4T_c$ (dotted line).

\subsubsection{Dilepton production rate from QGP with Gribov-Zwanziger action}
\label{DP_GZ}

\begin{figure}[htb]
\begin{center}
\includegraphics[width=0.4\linewidth]{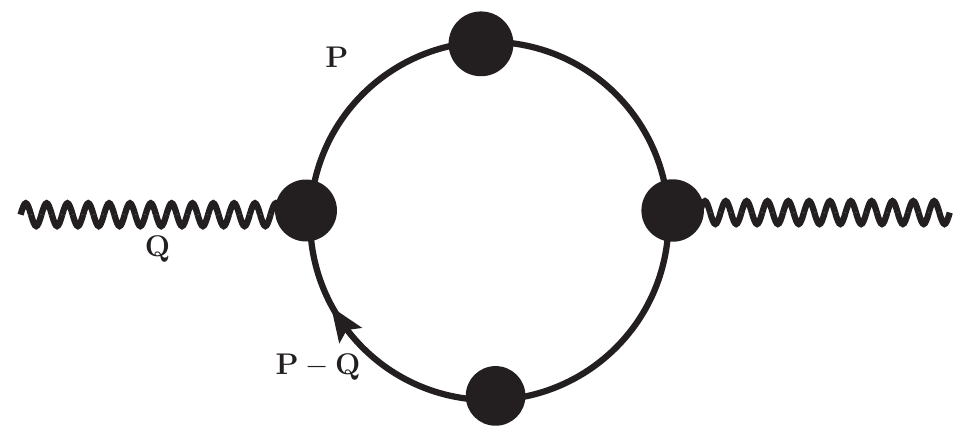} \hspace*{0.5in} 
\includegraphics[width=0.3\linewidth]{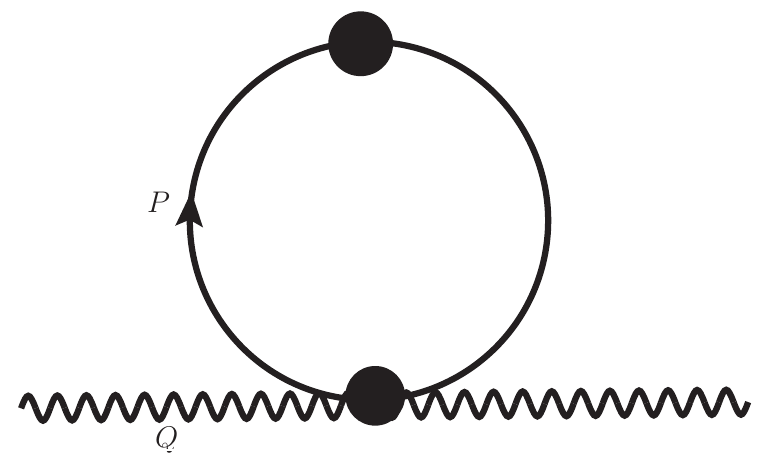}
\end{center}
\caption{One-loop photon self-energy diagram (left)  and tadpole diagram (right) with the effective propagators and vertices with GZ action. 
These diagrams are taken from Ref.~\cite{BHMS}.}
\label{photon_se_gz}
\end{figure}
The general features of non-perturbative GZ action have been discussed in subsec~\ref{gza}. Now,
in this subsec we want to compute the dilepton production rate~\cite{BHMS} with the GZ action from QGP.  At one-loop order, the dilepton production rate
is associated with photon self-energy and tadpole diagrams as shown in Fig.~\ref{photon_se_gz}.
The contributions to the one-loop photon self-energy can be written from the two diagrams in Fig.~\ref{photon_se_gz} as
\bea
\Pi_{\mu}^{\mu}(Q)&=&-\frac{10}{3} e^2 T\sum_{\{p_0\}}\int\frac{d^3p}{(2\pi)^3} \biggl\{\textmd{Tr}\biggl[S^*(P)
~\Gamma_\mu(K,Q,-P)~S^*(K)~\Gamma^\mu(-K,-Q,P)\biggr]\label{se_tr} \nn \\
&+& \textmd{Tr}\biggl[S^*(P)~\Gamma_{\mu}^{\mu}(-P,P;-Q,Q)\biggr]\biggr\}, \label{dilep_tr}
\eea
where $K=P-Q$, $S^*$ is the effective quark propagator as given in \eqref{hprop}, $\Gamma^\mu$ is the quark-photon vertex as given in \eqref{ward_id}
and $\Gamma^\mu_\mu$ are four-point quark-photon vertex as given in \eqref{wi_4pt}.
The second term in (\ref{dilep_tr}) is due to the tadpole diagram shown in Fig.~\ref{photon_se_gz} which eventually 
does not contribute  as  $\Gamma_{\mu}^{\mu}=0$. However, the tadpole diagram is essential to satisfy the Ward-Takahashi identity 
$Q_\mu\Pi^{\mu\nu}(Q)=0$ and thus the gauge  invariance and charge conservation in the system.  

Using the $N$-point functions  and performing traces, one obtains the photon self-energy with photon three momentum, $\bm{\vec q}=0$ as
\bea
\Pi_\mu^\mu(\bm{\vec q}=0)&=&-\frac{10}{3}e^2T\sum_{p_0}\int\frac{d^3p}{(2\pi)^3}  \nn \\
&& \times \Biggl[\left\{\frac{(a_G+b_G)^2}{{\cal D}_+(\omega_1,p,\gamma_G){\cal D}_-(\omega_2,p,\gamma_G)}+
\frac{(a_G-b_G)^2}{{\cal D}_-(\omega_1,p,\gamma_G){\cal D}_+(\omega_2,p,\gamma_G)}\right\}\nn\\
&& - \left\{\frac{(c_G+b_G+d_G)^2}{{\cal D}_+(\omega_1,p,\gamma_G){\cal D}_-(\omega_2,p,\gamma_G)}+
\frac{(c_G-b_G+d_G)^2}{{\cal D}_-(\omega_1,p,\gamma_G){\cal D}_+(\omega_2,p,\gamma_G)}\right\}\nn\\
&& -2c_G^2\left\{\frac{1}{{\cal D}_+(\omega_1,p,\gamma_G){\cal D}_+(\omega_2,p,\gamma_G)}+\frac{1}{{\cal D}_-(\omega_1,p,\gamma_G)
{\cal D}_-(\omega_2,p,\gamma_G)}\right\}\Biggr],
\label{dlep_trace_se}
\eea
where ${\cal D}_\pm$ are given, respectively, in \eqref{dpm} and \eqref{dpm1} whereas $a_G$, $b_G$, $c_G$ and $d_G$ are given, respectively, in
\eqref{coeff_1}, \eqref{coeff_2}, \eqref{coeff_3} and \eqref{coeff_4}.

\begin{figure}[htb]
\begin{center}
\includegraphics[width=0.5\linewidth]{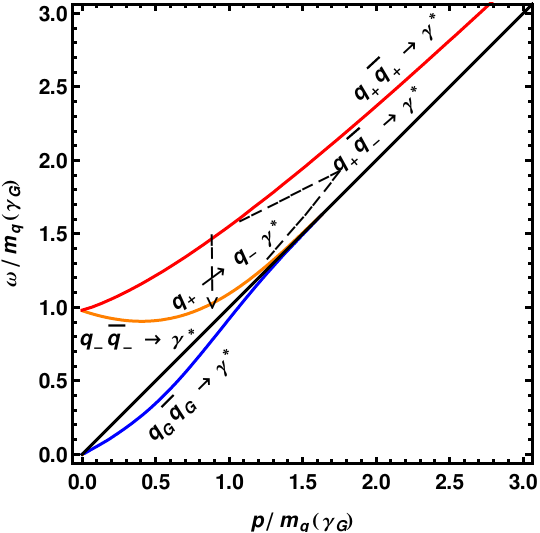} 
\end{center}
\caption{Various dilepton processes  originate from the in-medium dispersion of quasiparticles with Gribov term are displayed. 
This figure is taken from Ref.~\cite{BHMS}.}
\label{dilepton_processes}
\end{figure}

Now using the BPY prescription~\cite{BPY} given in \eqref{bpy_pres} or in \eqref{bpy12} in appendix~\ref{bpy_disc} 
we first find out  the imaginary or discontinuous part of the \eqref{dlep_trace_se} and 
then performing some more algebra, we write down the dilepton production rate with massless leptons following \eqref{d6} as
\bea
\frac{dR}{d\omega d^3q}(\bm{\vec q}=0)&=&\frac{20\alpha^2}{9\pi^4}\frac{1}{\omega^2}\int\limits_0^\infty p^2dp
\int\limits_{-\infty}^\infty d\omega_1 \int\limits_{-\infty}^\infty d\omega_2 n_F(\omega_1) n_F(\omega_2) 
\delta(\omega-\omega_1-\omega_2)\nn\\
&&\Bigg[4\left(1-\frac{\omega_1^2-\omega_2^2}{2p\,\omega}\right)^2 \rho_+^G(\omega_1,p) \rho_-^G(\omega_2,p)\nn\\
&&+\left(1+\frac{\omega_1^2+\omega_2^2-2p^2-2m_q^2(\gamma_G)}{2p\,\omega}\right)^2\rho_+^G(\omega_1,p) \rho_+^G(\omega_2,p)\nn\\
&&+\left(1-\frac{\omega_1^2+\omega_2^2-2p^2-2m_q^2(\gamma_G)}{2p\,\omega}\right)^2\rho_-^G(\omega_1,p) \rho_-^G(\omega_2,p)\Bigg]
,\label{dilep_spec}
\eea
where $\omega$ is the photon energy and $\rho^G_\pm$ are the spectral functions in presence of Gribov term given in \eqref{gspect}. 
Using \eqref{gspect} one can obtain the dilepton
production rate as
\bea
\frac{dR}{d\omega d^3q}\Big\vert^{pp}({\vec q}=0)&=&\frac{20\alpha^2}{9\pi^4}\frac{1}{\omega^2}
\int\limits_0^\infty p^2\, dp \times \nn \\
&& \Biggl [\delta(\omega-2\omega_+)\ n_F^2(\omega_+)\left(\frac{\omega_+^2-p^2}{2m_q^2(\gamma_G)}\right)^2
\left\{1+\frac{\omega_+^2-p^2-m_q^2(\gamma_G)}{p~\omega}\right\}^2\nn\\
&&+~\delta(\omega-2\omega_-)\ n_F^2(\omega_-)\left(\frac{\omega_-^2-p^2}{2m_q^2(\gamma_G)}\right)^2
\left\{1-\frac{\omega_-^2-p^2-m_q^2(\gamma_G)}{p~\omega}\right\}^2\nn\\
&&+~\delta(\omega-2\omega_G)\ n_F^2(\omega_G)\left(\frac{\omega_G^2-p^2}{2m_q^2(\gamma_G)}\right)^2
\left\{1-\frac{\omega_G^2-p^2-m_q^2(\gamma_G)}{p~\omega}\right\}^2 \nn \\
&& +4 \ \delta(\omega-\omega_+-\omega_-) \ n_F(\omega_+) \ n_F(\omega_-) 
\left(\frac{\omega_+^2-p^2}{2m_q^2(\gamma_G)}\right)
\left(\frac{\omega_-^2-p^2}{2m_q^2(\gamma_G)}\right) \nn \\
&& \times \left\{1-\frac{\omega_+^2-\omega_-^2}{2p\,\omega}\right\}^2 \nn \\
&& +\delta(\omega-\omega_++\omega_-) \ n_F(\omega_+)n_F(-\omega_-)
\left(\frac{\omega_+^2-p^2}{2m_q^2(\gamma_G)}\right)\left(\frac{\omega_-^2-p^2}{2m_q^2(\gamma_G)}\right) \nn \\
&& \times \left\{1+\frac{\omega_+^2+\omega_-^2-2p^2-2m_q^2(\gamma_G)}{2p\,\omega}\right\}^2 
~\Biggr]. \label{dilep_pp}
\eea
The momentum integration in (\ref{dilep_pp}) can be performed using the standard delta function identity given in \eqref{delta-prop}. Now,
inspecting the arguments of the various energy conserving $\delta$-functions in (\ref{dilep_pp}) one can understand the physical processes originating from the poles of the propagator.   The first three terms in (\ref{dilep_pp}) correspond to the annihilation processes of $q_+{\bar q}_+\rightarrow \gamma^*$, $q_-{\bar q}_-\rightarrow \gamma^*$, and $q_G{\bar q}_G\rightarrow \gamma^*$, respectively.  The fourth term corresponds to the annihilation of $q_+{\bar q}_-\rightarrow \gamma^*$.  On the other hand, the fifth term corresponds to a process, $q_+\rightarrow q_-\gamma*$, where a $q_+$ mode makes a transition to a $q_-$ mode along with a virtual photon.  These processes involve soft quark modes ($q_+, \, q_-$, and $q_G$ and their antiparticles) which originate by cutting the self-energy diagram in Fig.~\ref{photon_se_gz} through the internal lines without a ``blob''.  The virtual photon, $\gamma^*$, in all these five processes decays to lepton pair and can be visualized from the dispersion plot as displayed in the Fig.~\ref{dilepton_processes}.  
\begin{figure}[htb]
\begin{center}
\includegraphics[width=0.75\linewidth]{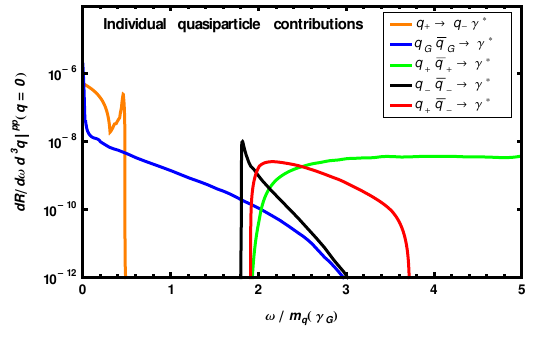} 
\end{center}
\caption{The dilepton production rates corresponding to quasiparticle processes in Fig.~\ref{dilepton_processes}. This figure is taken from Ref.~\cite{BHMS}.}
\label{dilepton_pp}
\end{figure}

\begin{figure}[htb]
\begin{center}
\includegraphics[width=0.75\linewidth]{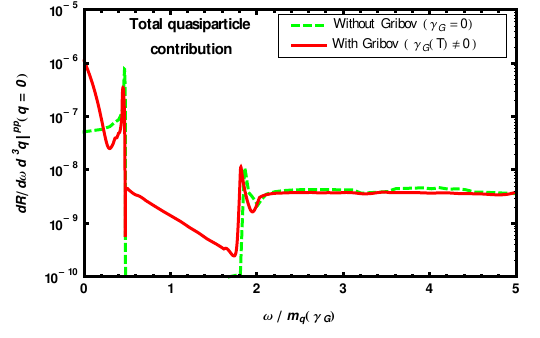} 
\end{center}
\caption{Comparison of dilepton production rates involving various quasiparticle modes with and without inclusion of $\gamma_G$. This figure is taken from Ref.~\cite{BHMS}.}
\label{dilepton_comparison}
\end{figure}

The contribution of various individual processes to  the dilepton production rate in presence of the Gribov term are displayed in the  Fig.~\ref{dilepton_pp}.  The transition process, $q_+\rightarrow q_-\gamma*$, begins at the energy $\omega=0$ and ends up with a van-Hove peak where all of the transitions from $q_+$ branch are directed towards the minimum of the $q_-$ branch. The annihilation process involving the massless spacelike Gribov modes, $q_G{\bar q}_G\rightarrow \gamma^*$, also starts at $\omega=0$ and falls-off very quickly.  The annihilation of the two plasmino modes, $q_-{\bar q}_-\rightarrow \gamma^*$, opens up with again a van-Hove peak at $\omega=2 \times $ the minimum energy of the plasmino mode. The contribution of this process decreases exponentially.  At $\omega=2m_q(\gamma_G)$, the annihilation processes involving usual quark modes, $q_+{\bar q}_+\rightarrow \gamma^*$, and that of a quark and a plasmino mode, $q_+{\bar q}_-\rightarrow \gamma^*$, begin.  However, the former one ($q_+{\bar q}_+\rightarrow \gamma^*$) grows with the energy and would converge to the usual Born rate (leading order perturbative rate)~\cite{CFR} at high mass whereas the later one ($q_+{\bar q}_-\rightarrow \gamma^*$) initially grows at a very fast rate, but then decreases slowly and finally drops very quickly. The behavior of the latter process can easily be understood from the dispersion properties of quark and plasmino mode. Summing up, the total contribution of all theses five processes is displayed in Fig.~\ref{dilepton_comparison}. This is compared with the similar dispersive contribution 
when $\gamma_G=0$~\cite{BPY}, comprising processes $q_+\rightarrow q_-\gamma*$, $q_+{\bar q}_+\rightarrow \gamma^*$, $q_-{\bar q}_-\rightarrow \gamma^*$ and 
$q_+{\bar q}_-\rightarrow \gamma^*$. 
We note that when $\gamma_G=0$, the dilepton rate contains both van-Hove peaks and an energy gap~\cite{BPY}. In presence of the Gribov term ($\gamma_G\ne 0$), 
the van-Hove peaks remain, but the energy gap disappears due to the annihilation of new massless Gribov modes, $q_G{\bar q}_G\rightarrow \gamma^*$.  This new contribution could be important for low mass dilepton spectra.

\begin{figure}[t]
\begin{center}
\includegraphics[width=0.75\linewidth]{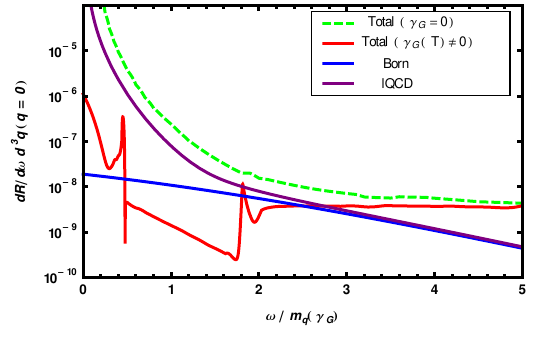} 
\end{center}
\caption{Comparison of various dilepton production  rates from  the deconfined matter. This figure is taken from Ref.~\cite{BHMS}.}
\label{dilepton_all}
\end{figure}

In Fig.~\ref{dilepton_all} we compare the rates obtained using various approximations:  leading-order perturbative (Born) rate~\cite{CFR}, quenched lattice QCD (lQCD) rate~\cite{ding,laermann}, and with and without the Gribov term.  The non-perturbative rate with the Gribov term shows important structures compared to the Born rate at low energies.  But when compared to the total HTLpt rate~\footnote{ The HTL spectral function (i.e, $\gamma_G=0$) has both pole and Landau cut contribution as obtained in~\ref{spec2}.  Therefore, the HTLpt dilepton rate~\cite{BPY} contains an additional {\textit {higher order}} contribution due to the Landau cut stemming from spacelike momenta.} it is suppressed  in the low mass region due to the absence of Landau cut contribution for $\gamma_G \neq 0$.  It seems as if the higher order Landau cut contribution due to spacelike momenta for $\gamma_G=0$ is replaced by the soft process involving spacelike Gribov modes in the collective excitations for $\gamma_G\neq 0$. We also note that the dilepton rate~\cite{kim} using the spectral function constructed with two pole ansatz by analyzing lQCD propagator in quenched approximation~\cite{kitazawa1,kitazawa2} shows similar structure  as found here for $\gamma_G\ne0$. On the other hand, such structure at low mass is also expected in the direct computation of dilepton rate from lQCD in quenched approximation~\cite{ding,laermann}.

\section{Conclusion}
\label{conc}
In this review article  some basics of the thermal field theory within the imaginary time formalism have been discussed in details. The imaginary time formalism has been 
introduced through two methods: the operatorial and the path integral methods.  The prescriptions to calculate the discrete frequency have been discussed. The Green's function
has been obtained both in  real and imaginary time.  The self-energy in $\phi^3$ theory and the tadpole diagram in $\lambda\phi^4$ theory have been calculated and their 
implications have been discussed. The partition function for non-interacting scalar, fermion, photon field and interacting scalar field have been computed using the 
functional integration approach. The general characteristics of a material medium in presence of a thermal bath have been outlined in details. 
We have computed the two-point functions for fermions and gauge bosons in HTL approximation for both QED and QCD. The collective excitations in 
both QED and QCD plasma have also been discussed. We have discussed some subtleties in finite temperature field theory and shortcomings of naive perturbation theory. Then we introduced the HTL resummation and HTL perturbation theory.  The HTLpt has been applied to calculate the LO, NLO and NNLO free energy and pressure  of deconfined QCD medium produced in high energy heavy-ion collisions. For interested readers, I have also provided an extensive list of literatures for the application of HTLpt to the various properties of deconfined QCD matter. Then we have discussed the general properties of hot QCD medium  in presence of non-perturbative effects like gluon condensate and Gribov-Zwanziger term.
The collective excitations of deconfined QCD medium have also been discussed in the presence of non-perturbative effects. Finally, we computed the dilepton production rate from deconfined QCD medium with those non-perturbative effects. Also some useful literatures have been provided  for the application of thermal field theory beyond QCD, {viz.},
the phase transitions involving symmetry restoration in theories with spontaneously broken symmetry,  the evolution of the universe at early times and cosmolog,  thermal neutrino production,  neutrino oscillations, leptogenesis, ${\cal N} = 4$ superaymmetric Yang-Mills theory, string theory and Anti de-sitter space/Conformal Field Theory (Ads/CFT) correspondence, blackhole physics, thermal axion production  and condensed matter physics.

\vspace{0.3in}

\appendix
\section{Appendix}
\label{appen}
\subsection{One-loop Fermionic Sum Integrals}
\label{fer_sum_1l}
The dimensionally regularized  fermionic sum-integrals are defined as,
\bea
\sumintb_{\{ P\} } &=& \left(\frac{e^{\gamma_E}\Lambda^2}{4\pi}\right)^\epsilon 
T\sum\limits_{\substack{\{p_0=i\omega_n\}\\ \omega_n=(2n+1)\pi 
T-i\mu}}\int\frac{d^{d-2\epsilon}p}{(2\pi)^{d-2\epsilon}},
\eea
where $d-2\eps$ is the spatial dimension, $P$ is the fermion loop momentum, 
$\Lambda$ is the $\overline{\textrm{MS}}$ renormalization scale that 
introduces the factor $\left(\frac{e^{\gamma_E}}{4\pi}\right)^\epsilon$ 
along with it, where $\gamma_E$ being the Euler-Mascheroni constant. 

The result of various fermionic sum-integrals are listed below:
\begin{subequations}
 \begin{align}
2\sumintb_{\{P\}} \ln\left(P^2\right)  &= \frac{7\pi^2 T^4}{180} + \frac{\mu^2T^2}{6} +\frac{\mu^4}{12\pi^2} 
= \frac{7\pi^2T^4}{180}\left(1+\frac{120}{7}\hat{\mu}^2+\frac{240}{7} \hat{\mu}^4\right) \, ,\label{qfe27} \\
\sumintb_{\{P\}}\frac{1}{P^2}&=\frac{T^2}{24}\left(\frac{\Lambda}{4\pi 
T}\right)^{2\epsilon}\left[1+12\hat\mu^2+2\epsilon
\left(1+12\hat\mu^2
       +12\aleph(1,z)\right)\right], \label{qfe28} \\
\sumintb_{\{ P\} 
}\frac{1}{P^4}&=\frac{1}{\left(4\pi\right)^2}\left(\frac{\Lambda}{4\pi 
T}\right)^{2\epsilon}\Bigg[\frac{1}
{\epsilon}-\aleph(z)\Bigg], \label{qfe29}\\
\sumintb_{\{P\}}\frac{1}{p^2P^2}&= -\frac{2}{d-2} \sumintb_{\{ P\} }\frac{1}{P^4}, \label{qfe30}\\
\sumintb_{\{P\}}\frac{1}{p^2P^2}{\cal T}_P 
&=-\frac{2\Delta_3}{d-2} \sumintb_{\{P\}}\frac{1}{P^4}, \label{qfe31}\\
\sumintb_{\{P\}}\frac{1}{p^2P^2}{\cal T}_P^2 
&=-\frac{2\Delta_4''}{d-2} \sumintb_{\{P\}}\frac{1}{P^4},\label{qfe32}\\
\sumintb_{\{P\}}\frac{1}{p_0^2P^2}{\cal T}_P^2 
&=-\frac{2\Delta_3''}{d-2} \sumintb_{\{P\}}\frac{1}{P^4}, \label{qfe33}
\end{align}
\end{subequations}
where the angular integrations are given as
\begin{subequations}
 \begin{align}
\Delta_3 &= \left\langle \frac{1-c^{4-d}}{1-c^2}\right\ranglec = 
\ln 2+\left(\frac{\pi^2}{6}-(2-\ln 2)\ln 2\right)\eps \nn\\
&~~~~+\left\{\frac{2}{3}(\ln 2)^2(\ln 2 -3)+\frac{\pi^2}{3}(\ln 2-1)+\zeta(3)\right\}\eps^2+\mathcal{O}[\eps]^3 \, , \label{qfe34} \\
\Delta_3'' &= \left\langle \frac{1-c_1^{4-d}}{(1-c_1^2)(c_1^2-c_2^2)}+c_1\lrarrow c_2\right\rangleci \!\!\!\!= 
-\frac{\pi^2}{12} + \left(\frac{\pi^2}{3} - \frac{\zeta(3)}{2}\right)\eps + 
\mathcal{O}(\eps^2),  \label{qfe35}\\
\Delta_4'' &= \left\langle \frac{1-c_1^{6-d}}{(1-c_1^2)(c_1^2-c_2^2)}+c_1\lrarrow c_2\right\rangleci \nn\\
&=
-\frac{\pi^2}{12} + \ln 4 + \left(\frac{\pi^2}{3} - \ln4 (2 - \ln2) -  
\frac{\zeta(3)}{2}\right) \eps + \mathcal{O}(\eps^2) \, , \label{qfe36}
\end{align}
\end{subequations}
with
\begin{subequations}
 \begin{align}
 \aleph(z)&=-2\gamma_E-4\ln 
2+14\zeta(3)\hat{\mu}^2-62\zeta(5)\hat{\mu}^4+254\zeta(7)\hat{\mu}^6+{\cal 
O}(\hat{\mu}^8),\label{qfe37}\\
 \aleph(1,z)&=-\frac{1}{12}\left(\ln2-\frac{\zeta'(-1)}{\zeta(-1)}\right) - 
\left(1-2\ln2-\gamma_E\right)\hat{\mu}^2-\frac{7}{6}\zeta(3)\hat{\mu}^4\nn\\
&\,\,\,\,\,\, + \frac{31}{15}\zeta(5)\hat{\mu}^6+{\cal O}(\hat{\mu}^8)\, . \label{qfe38}
\end{align}
\end{subequations}

\subsection{One-loop Bosonic Sum Integrals}
\label{bos_sum_1l}
The dimensionally regularized  bosonic sum-integrals are defined as,
\bea
\sumintb_{ P} &=& \left(\frac{e^{\gamma_E}\Lambda^2}{4\pi}\right)^\epsilon 
T\sum\limits_{\substack{p_0=i\omega_n\\ \omega_n=2n\pi T}}\int\frac{d^{d-2\epsilon}p}{(2\pi)^{d-2\epsilon}},
\eea
where $d-2\eps$ is the spatial dimension, $P$ is the boson loop momentum, $\Lambda$ is the $\overline{\textrm{MS}}$ renormalization scale that 
introduces the factor $\left(\frac{e^{\gamma_E}}{4\pi}\right)^\epsilon$ 
along with it, where $\gamma_E$ being the Euler-Mascheroni constant. 

Below we list various bosonic sum-integrals:
\begin{subequations}
 \begin{align}
\sumintb_P \frac{1}{P^2} &= -\frac{T^2}{12}\left(\frac{\Lambda}{4\pi T}\right)^{2\eps}\Bigg[1+2\eps\left(1+\frac{\zeta'(-1)}{\zeta(-1)}\right)
+\mathcal{O}[\eps]^2\Bigg], \label{gf15}\\
\sumintb_P \frac{1}{p^2P^2} &= -\frac{2}{(4\pi)^2}\left(\frac{\Lambda}{4\pi T}\right)^{2\eps}\Bigg[\frac{1}{\eps}+2\gamma_E+2
+\eps\left(4+4\gamma_E+\frac{\pi^2}{4}-4\gamma_1\right)+\mathcal{O}[\eps]^2 \Bigg], \label{gf16}\\
\sumintb_P \frac{1}{P^4} &=  \frac{1}{(4\pi)^2}\left(\frac{\Lambda}{4\pi T}\right)^{2\eps}\left[\frac{1}{\eps}
+2\gamma_E+\eps\left(\frac{\pi^2}{4}-4\gamma_1\right) +\mathcal{O}[\eps]^2\right]\, , \label{gf17} \\
\sumintb_{P} \frac{{\cal T}_P}{p^4} &=-\frac{1}{(4\pi)^2}\left(\frac{\Lambda}{4\pi T}\right)^{2\eps}\left[\frac{1}{\eps}
+2\gamma_E+2\ln 2+\mathcal{O}[\eps] \right]\, , \label{gf18} \\
\sumintb_{P} \frac{{\cal T}_P}{p^2P^2} &=-\frac{1}{(4\pi)^2}\left(\frac{\Lambda}{4\pi T}\right)^{2\eps}
\left[2\ln 2 \left(\frac{1}{\eps}+2\gamma_E\right)+ 2\ln^2 2 +\frac{\pi^2}{3}+\mathcal{O}[\eps]  \right]\, , \label{gf19} \\
\sumintb_{P} \frac{{\cal T}^2_P}{p^4} &=-\frac{2}{3}\frac{1}{(4\pi)^2}\left(\frac{\Lambda}{4\pi T}\right)^{2\eps}
\left[ \left(1+2\ln 2\right)\left(\frac{1}{\eps}+2\gamma_E\right)-\frac{4}{3}+\frac{22}{3}\ln 2 +2\ln^2 2 +\mathcal{O}[\eps] \right]\, .\label{gf20}
\end{align}
\end{subequations}

\subsection{Braaten-Pisarski-Yuan (BPY) Prescription}
\label{bpy_disc}

Lets consider a complex function $f(z)$ having branch cut
\be
f(z) =\frac{1}{2\pi i} \oint \frac{f(\xi)\ d\xi}{\xi-z} \, . \label{bpy_0}
\ee
considering $\xi=x+i\epsilon$, one can write
\bea
f(z) =\frac{1}{2\pi i} \int\limits_{-\infty}^{+\infty} \frac{f(x+i\epsilon) -f(x-i\epsilon)}{x-z} \,  dx 
= \frac{1}{2\pi i} \int\limits_{-\infty}^{+\infty} \frac{\textrm{Disc} f(x+i\epsilon)}{x-z} dx \, . \label{bpy_01}
\eea
where the discontinuity is related to the imaginary part of a complex function as
\be
\textrm{Disc} f(x+i\epsilon) = f(x+i\epsilon)-f(x-i\epsilon) = 2i \, \textrm{ Im} f(x+i\epsilon) \, . \label{bpy_02}
\ee
Combining \eqref{bpy_01}  and \eqref{bpy_02} one can write
\be
f(z) = \frac{1}{\pi}  \int\limits_{-\infty}^{+\infty} \frac{\textrm{\cal Im} \,  f(x+i\epsilon)}{x-z} dx 
=  \int\limits_{-\infty}^{+\infty} \frac{ \rho (x)}{x-z} dx \label{bpy_03}
\ee
where the spectral density $\rho$ is defined as
\be
\rho(x) = \frac{1}{\pi} \textrm{Im} \,  f(x+i\epsilon) \, . \label{bpy_04}
\ee
The spectral density $\rho_1 (\omega_1)$ is related to the any complex function $F_1(k_0)$ as given in \eqref{bpy_03}
\be
F_1(k_0) =\int\limits_{-\infty}^{+\infty} \frac{\rho_1(\omega_1) d\omega_1}{\omega_1-k_0-i\epsilon_1} \, . \label{bpy1}
\ee
We note that $K\equiv (k_0,\bm{\vec k})$ is the fermionic momentum with $k_0 =(2m+1) i\pi T$. 

Lets have,
\bea
\int\limits_{0}^{1/T} d\tau_1 \ e^{x\tau_1}  &=& \frac{ e^{x/T}-1} {x}  \, \, 
\Rightarrow \,\, \frac{1}{x} = \frac{1}{e^{x/T}-1} \int\limits_{0}^{1/T} d\tau_1 \  e^{x\tau_1}  \, , \label{bpy2}
\eea
where $T$ is the temperature.
Now, considering $x=(\omega_1-k_0-i\epsilon_1)$ one can write \eqref{bpy2} as
\be
\frac{1}{\omega_1-k_0-i\epsilon_1} =  \frac{1}{e^{\frac{(\omega_1-k_0)}{T}}-1} \int\limits_{0}^{1/T} d\tau_1 \  e^{(\omega_1-k_0-i\epsilon_1)\tau_1}  \, . \label{bpy3}
\ee
Combining \eqref{bpy3} with \eqref{bpy1}, one gets
\be
F_1(k_0) = \int\limits_{-\infty}^{+\infty} \frac{\rho_1(\omega_1) d\omega_1}{e^{\frac{(\omega_1-k_0)}{T}}-1} \int\limits_{0}^{1/T} 
d\tau_1 \  e^{(\omega_1-k_0-i\epsilon_1)\tau_1}  \, . \label{bpy4}
\ee 
Now, using $e^{k_0/T} = e^{(2m+1)i\pi} = -1$, one can write as
\bea
F_1(k_0) &=& - \int\limits_{-\infty}^{+\infty} \frac{\rho_1(\omega_1) d\omega_1}{e^{\frac{\omega_1}{T}}-1} \int\limits_{0}^{1/T} d\tau_1 \  e^{(\omega_1-k_0-i\epsilon_1)\tau_1} \nn \\
&=& -  \int\limits_{-\infty}^{+\infty} n_F(\omega_1) \ \rho_1(\omega_1) \, d\omega_1 \int\limits_{0}^{1/T} d\tau_1 \  e^{(\omega_1-k_0-i\epsilon_1)\tau_1}  \, . \label{bpy4}
\eea
Similarly, one can write another complex function $F_2(q_0)$ as
\bea
F_2(q_0) &=& -  \int\limits_{-\infty}^{+\infty} n_F(\omega_2) \ \rho_2(\omega_2) \, d\omega_2 \int\limits_{0}^{1/T} d\tau_2 \  e^{(\omega_2-q_0-i\epsilon_2)\tau_2}  \, , \label{bpy4}
\eea
where $q_0=(p_0-k_0)$ and $P$ is the bosonic momentum with $p_0=2mi\pi T$.

We would like to compute the imaginary part of the product of two complex functions $T\sum_{k_0} F_1(k_0)F_2(q_0)$ :
\vspace*{-0.2in}
\bea
\textrm{Im} \, \, \, T\sum_{k_0} F_1(k_0)F_2(q_0) &=& \textrm{Im} \, \, \, T\sum_{k_0} \int\limits_{-\infty}^{+\infty} d\omega_1 \int\limits_{-\infty}^{+\infty} d\omega_2\, 
n_F(\omega_1) n_F(\omega_2)  \rho_1(\omega_1)  \rho_2(\omega_2) \nn \\
&& \times \int\limits_{0}^{1/T} d\tau_1 \int\limits_{0}^{1/T} d\tau_2 \, \, e^{(\omega_1-k_0-i\epsilon_1)\tau_1}\,\, e^{(\omega_2-q_0-i\epsilon_2)\tau_2}  \nn \\
&=& \textrm{Im} \, \int\limits_{-\infty}^{+\infty} d\omega_1 \int\limits_{-\infty}^{+\infty} d\omega_2\, 
n_F(\omega_1) n_F(\omega_2)  \rho_1(\omega_1)  \rho_2(\omega_2) \nn \\
&& \times \int\limits_{0}^{1/T} d\tau_1 \int\limits_{0}^{1/T} d\tau_2 \, \, e^{(\omega_1-i\epsilon_1)\tau_1}\,\, e^{(\omega_2-p_0-i\epsilon_2)\tau_2} \, 
\underbrace{T\sum_{k_0} e^{-k_0(\tau_1-\tau_2)}}_{\delta(\tau_2-\tau_1)} . \label{bpy5}
\eea
Performing $\tau_2$-integration using $\delta$-function, one can write
\bea
\textrm{Im} \, \, \, T\sum_{k_0} F_1(k_0)F_2(q_0) 
&=& \textrm{Im} \, \int\limits_{-\infty}^{+\infty} d\omega_1 \int\limits_{-\infty}^{+\infty} d\omega_2\, 
n_F(\omega_1) n_F(\omega_2)  \rho_1(\omega_1)  \rho_2(\omega_2) \nn \\
&& \times \int\limits_{0}^{1/T} d\tau_1 \, e^{(\omega_1+\omega_2-p_0-i\epsilon)\tau_1} \, ,\label{bpy5a}
\eea
where $\epsilon=\epsilon_1+\epsilon_2 $ . Now performing the $\tau_1$-integration, one gets
\bea
\textrm{Im} \, \, \, T\sum_{k_0} F_1(k_0)F_2(q_0) 
&=& \textrm{Im} \, \int\limits_{-\infty}^{+\infty} d\omega_1 \int\limits_{-\infty}^{+\infty} d\omega_2\, 
n_F(\omega_1) n_F(\omega_2)  \rho_1(\omega_1)  \rho_2(\omega_2) \nn \\
&& \times \frac{e^{(\omega_1+\omega_2-p_0)/T} -1}{ \omega_1+\omega_2 -p_0 -i\epsilon}\,  \nn \\
&=& \, \int\limits_{-\infty}^{+\infty} d\omega_1 \int\limits_{-\infty}^{+\infty} d\omega_2\, 
n_F(\omega_1) n_F(\omega_2)  \rho_1(\omega_1)  \rho_2(\omega_2) \nn \\
&& \times \left (e^{(\omega_1+\omega_2-p_0)/T} -1\right ) \,  \textrm{Im} \left (\frac{1}{ \omega_1+\omega_2 -p_0 -i\epsilon}\right)  \, . \label{bpy6}
\eea
Now using  
\begin{subequations}
\begin{align}
e^{-{p_0}/{T}} &= e^{-2m\pi i}= 1 \,  , \label{bpy7} \\
p_0& =\omega+i\epsilon' \, , \label{bpy8} \\
\frac{1}{ \omega_1+\omega_2 -p_0 -i\epsilon} &= \frac{1}{ \omega_1+\omega_2 -\omega -i\epsilon' -i\epsilon} 
=\frac{1}{ \omega_1+\omega_2 -\omega -i\epsilon''} \, , \label{bpy9} \\
\textrm{Im} \left ( \frac{1}{\omega_1+\omega_2-\omega-i\epsilon''}\right )& = -\pi \delta \left (\omega_1+\omega_2-\omega \right ) \, , \label{bpy10}
\end{align}
\end{subequations}
one gets
\bea
\textrm{Im} \, \, \, T\sum_{k_0} F_1(k_0)F_2(q_0) 
&=& - \, \pi \,  \int\limits_{-\infty}^{+\infty} d\omega_1 \int\limits_{-\infty}^{+\infty} d\omega_2\, 
n_F(\omega_1) n_F(\omega_2)  \rho_1(\omega_1)  \rho_2(\omega_2) \nn \\
&&\times \left (e^{(\omega_1+\omega_2)/T} -1\right ) \, \delta \left(\omega_1+\omega_2-\omega \right ) \nn \\
&=& \pi  \left (1-e^{\beta \omega} \right ) \int\limits_{-\infty}^{+\infty} d\omega_1 \int\limits_{-\infty}^{+\infty} d\omega_2\, \, \,
n_F(\omega_1) n_F(\omega_2) \nn \\
&&\times  \rho_1(\omega_1)  \rho_2(\omega_2) \, \delta \left(\omega_1+\omega_2-\omega \right ) \, , \label{bpy11}
\eea
where $\beta=1/T$. 
 
We finally obtain following \eqref{bpy_02} and \eqref{bpy11},  the discontinuity or the imaginary part of a product of two complex functions~\cite{BPY} as
\bea
\textrm{Disc} \, \, T\sum_{k_0} F_1(k_0)F_2(q_0) &=&2i \,\, \textrm{Im} \, \, \, T\sum_{k_0} F_1(k_0)F_2(q_0) \nn \\
&=& 2 \pi i  \left (1-e^{\beta \omega} \right ) \int\limits_{-\infty}^{+\infty} d\omega_1 \int\limits_{-\infty}^{+\infty} d\omega_2\, \, \,
n_F(\omega_1) n_F(\omega_2) \nn \\
&&\times  \rho_1(\omega_1)  \rho_2(\omega_2) \, \delta \left(\omega_1+\omega_2-\omega \right ) \, . \label{bpy12}
\eea

\vspace{0.4in}

\noindent{\bf Acknowledgement:} 
I would like to thank Aritra Bandyopadhyay, Aritra Das, Bithika Karmakar, Chowdhury Aminul Islam, 
Najmul Haque and Ritesh Ghosh for various discussions and help received during
the preparation of this article. It is also a great pleasure to acknowledge the support received from 
Sanjay Ghosh and Rajarshi Ray who
were tutors of my lectures given at SERC Advanced School on Theoretical High 
Energy Physics, November 16-December 5, 2015 at Birla institute of Technology, Pilani, India. Finally, I would
like to thank Department of Atomic Energy, Government of India for the project TPAES in Theory division of Saha
Institute of Nuclear Physics.

\vspace*{0.3in}
\noindent{{\bf Data Availability Statement:} No Data associated in the manuscript.}


\end{document}